\documentclass[fleqn, usenatbib]{mnras}

\usepackage[T1]{fontenc}
\usepackage{graphicx}
\usepackage{amsmath}
\usepackage{subfigure}
\usepackage{newtxtext,newtxmath}

\title[Extended GRMHD using multifluids]{Beyond GRMHD: A Robust Numerical Scheme for Extended, Non-Ideal General Relativistic Multifluid Simulations}

\author[J. Gorard et al.]{Jonathan Gorard,${^{1, 2}}$\thanks{Email: gorard@princeton.edu (JG)},
James Juno,${^{2}}$
and Ammar Hakim${^{2}}$\\
${^{1}}$ Princeton University, Princeton, NJ 08544, USA\\
${^{2}}$ Princeton Plasma Physics Laboratory, Princeton, NJ 08540, USA}

\begin{document}
\maketitle

\begin{abstract}
The equations of general relativistic magnetohydrodynamics (GRMHD) have become the standard mathematical framework for modeling high-energy plasmas in curved spacetimes. However, the fragility of the primitive variable reconstruction operation in GRMHD, as well as the difficulties in maintaining strong hyperbolicity of the equations, sharply limit the applicability of the GRMHD model in scenarios involving large Lorentz factors and high magnetizations, such as around neutron stars. Non-ideal effects, such as electron inertia and Hall terms, are also neglected, and the absence of an explicitly evolved electric field precludes the self-consistent modeling of the strong poloidal fields found around spinning black holes, which are known to be crucial for jet formation. Here, we present a general relativistic \textit{multifluid} model which strictly generalizes the GRMHD equations, consisting of an arbitrary number of relativistic fluid species interacting with a shared electromagnetic field via an explicit coupling of their source terms, thus allowing for the incorporation of non-ideal effects. We sketch how our model may be derived from general relativistic kinetics (via moments of the relativistic Boltzmann-Vlasov equation), as well as how GRMHD may be recovered in the single-fluid limit as the mobility of charge carriers goes to infinity. We present a numerical scheme for solving the general relativistic multifluid equations, and validate it against the analogous scheme for the GRMHD equations. Since the primitive variable reconstruction operation for our multifluid model is purely \textit{hydrodynamic}, and therefore independent of the magnetic field, the resulting solver is highly robust, and able to simulate significantly larger Lorentz factors and higher magnetizations than GRMHD without loss of either accuracy or stability. We use 1D special relativistic Riemann problems to demonstrate that the multifluid solutions indeed converge to SRMHD solutions in the appropriate limit (using the relativistic Brio-Wu test), and that we can stably push the multifluid solver into strongly relativistic regimes involving high magnetizations in which SRMHD solvers fail to converge (using a highly magnetized variant of the relativistic Noh test). Next, we demonstrate that the multifluid solver is able to capture current sheet formation and magnetic reconnection behavior around the equatorial plane of a spinning black hole in 2D axisymmetry. Finally, we use various 3D magnetized accretion tests to demonstrate the ability of our solver to model charge separation effects and large parallel electric fields around black holes, as well as high magnetizations around neutron stars, without the limitations of the GRMHD approximation.
\end{abstract}


\section{Introduction}

Many scenarios in high-energy astrophysics depend sensitively upon the non-linear interaction between relativistic flows, strong electromagnetic fields, and the curvature of spacetime, including accretion (\cite{prieto_central_2016}), jet formation (\cite{abbott_gravitational_2017}), magnetic reconnection (\cite{ripperda_magnetic_2020}), and relativistic flaring (\cite{aharonian_fast_2006}) around black holes, neutron stars, and other compact objects. This complex interplay of processes strongly affects what we observe from these compact objects (e.g. \cite{mckinney_measurement_2004} and \cite{tchekhovskoy_efficient_2011}), and the modeling of these systems is now a critical component of understanding the influx of data from, e.g., the Event Horizon Telescope (\cite{the_event_horizon_telescope_collaboration_first_2019}, \cite{the_event_horizon_telescope_collaboration_first_2019-1}, and \cite{the_event_horizon_telescope_collaboration_first_2021}). In principle, such phenomena can be modeled self-consistently from first-principles via the formalism of general relativistic kinetics, either using general relativistic particle-in-cell (GRPIC) methods (e.g. \cite{levinson_particle--cell_2018}, 
\cite{parfrey_first-principles_2019}, and \cite{chen_physics_2020}), in which discrete particles are advected through a curved spacetime background directly, or using continuum kinetics methods (e.g. \cite{rasio_solving_1989}), in which the full covariant particle distribution function on spacetime is evolved numerically by solving the general relativistic Boltzmann-Vlasov equation. In both cases, the kinetic system may be coupled to a self-consistently generated electromagnetic field, modeled using the general relativistic Maxwell equations. In practice, however, both approaches are associated with significant drawbacks: being a particle-in-cell method, GRPIC suffers from inherent counting noise and other discretization artifacts that can substantively degrade solution quality (\cite{juno_deep_2020}), while the inherently high dimensionality of continuum kinetics (due to the 6-dimensional phase space over which the general relativistic Boltzmann-Vlasov equation is most naturally defined) severely limits the size and scale of problems that can be successfully tackled whilst still remaining computationally tractable.

For this reason, most large-scale astrophysical simulations of high-energy phenomena around compact objects employ the formalism of \textit{general relativistic magnetohydrodynamics} (GRMHD), as first formulated as a system of covariant conservation laws by \cite{anton_numerical_2006}, in which all of the particles are assumed to be in local thermodynamic equilibrium, to be highly collisional, and to exhibit infinite electrical conductivity. The former two assumptions justify the evolution of lower-order \textit{fluid moments} in place of the full continuum distribution function (with all of the velocity space dimensions having been integrated out), while the latter assumption justifies the use of an \textit{ideal magnetohydrodynamic} (MHD) model in which the electric field is fully determined by the magnetic field, and the charge-carrying particles are taken to be infinitely mobile. The GRMHD formalism has been extremely successful, and in particular has proved straightforward to augment with additional physics, for instance through the addition of electron fraction evolution equations for modeling deleptonization processes in collapsars and supernovae (e.g. \cite{miller_full_2020}), ${M_0}$ and ${M_1}$ neutrino transport equations for modeling neutrino-driven heating and pressure of nuclear matter during neutron star mergers (e.g. \cite{desai_three-dimensional_2022} and \cite{sun_jet_2022}), etc. However, the ideal MHD assumption, which necessitates in particular that ${\left\lVert \mathbf{E} \right\rVert^2 \ll \left\lVert \mathbf{B} \right\rVert^2}$, limits the proper application of the GRMHD formalism to scenarios in which there is no significant induction of (parallel) electric fields. Yet it is known from the seminal analytical work of \cite{wald_black_1974} and \cite{blandford_electromagnetic_1977} on spinning black holes immersed in magnetic fields, as well as the early numerical work of \cite{komissarov_electrodynamics_2004} on the force-free electrodynamics of black hole magnetospheres, that spinning black holes can induce very large parallel electric fields within their ergospheres due to frame-dragging effects, which is the basis of the hypothesized Blandford-Znajek mechanism of black hole jet-launching. Due to the absence of such effects, GRMHD often significantly under-predicts the strength of the poloidal magnetic field threading a spinning black hole, and so jet-launching simulations are usually not performed self-consistently (in practice, GRMHD simulations of black hole jet-launching typically \textit{impose} an initial magnetic field configuration derived from the analytical Wald equilibrium, the Blandford-Znajek perturbative solution, or from force-free electrodynamics, rather than attempting to generate the field self-consistently, see e.g. \cite{komissarov_magnetic_2007} and \cite{tchekhovskoy_black_2010} for further details). Likewise, GRMHD simulations are known to under-predict the rate of magnetic reconnection around compact objects (believed to be the mechanism responsible for relativistic flaring around black holes and neutron stars) by up to a factor of 10 (\cite{bransgrove_magnetic_2021}). Other important non-ideal effects, such as Hall terms, electron heating, and electron inertia, which are present in non-relativistic extended MHD models (such as multifluid-Maxwell systems, see e.g. \cite{hakim_high_2006} and \cite{wang_exact_2020}) are also absent from the GRMHD formalism. Perhaps most significantly for the purposes of understanding observations, the electron thermodynamics must be modeled semi-analytically with prescriptions for the evolution of the electron temperature (\cite{chael_evolving_2017}). Unfortunately, since the electrons are the source of much of the observed electromagnetic emission, significant sensitivity to the details of the electron heating model is observed in the resulting spectra and imaging across the range of observed frequencies (\cite{chael_role_2018} and \cite{chael_survey_2025}). 

However, an arguably even more serious limitation of the GRMHD formalism lies within the various mathematical and numerical pathologies of the equations themselves. The equations were shown by \cite{hilditch_hyperbolicity_2019} to violate strong hyperbolicity, and hence their Cauchy problem is not guaranteed to remain well-posed over arbitrary time-scales. Resolution studies by \cite{white_resolution_2019} have confirmed that many aspects of relativistic flows, including Lorentz factors, small-scale features, correlation lengths, and variabilities of synchrotron emission, do not appear to converge with increased grid resolution in GRMHD simulations. In terms of practical numerics, the algorithms for reconstructing the primitive variables (fluid density, three-velocity, and pressure) from the conserved variables (relativistic mass density, three-momentum, and total energy), as needed for the computation of numerical fluxes, are highly expensive and fragile due to the presence of the magnetic field and Lorentz factor terms coupling the relativistic energy and momentum variables together, making the root-finding procedure inherently high-dimensional and non-linear, especially for complex, non-analytic (e.g. tabulated) equations of state. As shown by \cite{siegel_recovery_2018}, even the most robust known primitive variable reconstruction algorithm for GRMHD, for the simplest (ideal gas) equation of state, will, for instance, not be able to converge successfully on a pressure for Lorentz factors above ${W \sim 100}$ combined with magnetizations above ${\sigma \sim 10^5}$. The magnetization near the light-cylinder of neutron stars can reach ${10^5}$ -- ${10^8}$, and only becomes larger near the surface of the star (\cite{philippov_pulsar_2022}). Furthermore, the Lorentz factor of the ejected plasma in the pulsar winds scales with the magnetization at the light cylinder, reaching at least ${W \sim 10^4}$ (\cite{philippov_ab_2014} and \cite{cerutti_modelling_2016}), so the applicability of the GRMHD formalism to neutron star simulations is clearly extremely limited. In practice, all production-scale GRMHD codes impose magnetization ceilings of around ${\sigma_{\text{max}} \sim 100}$ by injecting artificial mass density in regions where the magnetization exceeds this amount, to ensure that their primitive variable reconstruction algorithms are able to converge reliably without crashing the simulation, and the details of how this density injection is applied have been been shown by \cite{ressler_disc-jet_2017} to have a significant effect on the resulting physics. Some simulation codes employ a level set-based method, wherein regions of the domain that exceed the magnetization ceiling switch from using a GRMHD approximation to using a force-free approximation (see, e.g. \cite{chael_hybrid_2024}). In simulations of black hole jet-launching, these magnetization ceilings are invariably attained within the jet, indicating that there does not exist any known reasonable upper-bound on how large black hole jet magnetizations can get (and hence also on how unrealistic GRMHD simulations of black hole jet-launching actually are). Such thresholds not only affect the dynamics, but again affect the ability to interpret synthetic observations from simulations, as these same thresholds on magnetization will affect the amount of electron heating in electron temperature prescriptions (\cite{chael_survey_2025}). 

In the work of \cite{most_modelling_2022}, a covariant two-fluid formalism for studying electron-ion and electron-positron plasmas in general relativity was derived by computing and then evolving 14 moments of the distribution function from Boltzmann-Vlasov kinetics, which in principle circumvents many of the physical limitations of the GRMHD model. However, their equations are not presented in a form that is directly amenable to robust numerical simulation. In this paper, we derive a reduced (5-moment) \textit{general relativistic multifluid} model, consisting of a family of $n$ perfect fluid species obeying general relativistic hydrodynamics equations, interacting with a shared electromagnetic field obeying general relativistic Maxwell equations. The four-momenta of the fluid species are coupled to the four-current of the electromagnetic field via Ohm's law, allowing them to interact purely through their respective source terms. Since the equations of general relativistic hydrodynamics and electrodynamics can both be made strongly hyperbolic and well-posed, the overall multifluid system inherits these properties too. Moreover, since the primitive variable reconstruction procedure for the multifluid system is purely \textit{hydrodynamic} in nature, its convergence is independent of the magnetic field strength, and, due to the analytical work of \cite{eulderink_general_1994}, can be pushed to much lower densities/pressures and much higher Lorentz factors than the GRMHD primitive variable solve without sacrificing numerical stability. Leveraging the tetrad-first approach of \cite{gorard_tetrad-first_2025}, both the hydrodynamics equations and the electrodynamics equations can be solved in their flat spacetime form using highly robust, high-resolution shock-capturing (finite volume) schemes, with their respective source terms being integrated explicitly using a strong stability-preserving Runge-Kutta integration method. We validate the resulting numerical implementation against standard 1D Riemann problems for special relativistic magnetohydrodynamics (SRMHD), plus a standard 2D axisymmetric black hole magnetosphere problem, as well as against full 3D accretion problems for GRMHD in both black hole and neutron star spacetimes, demonstrating empirically that solutions produced using the general relativistic multifluid equations converge to the corresponding SRMHD and GRMHD solutions in the ideal MHD limit wherever such solutions exist, but also that the multifluid equations remain stable and can be used successfully to simulate magnetizations and Lorentz factors well in excess of what SRMHD and GRMHD can typically handle. These algorithms have all been implemented within the open source computational multiphysics framework \textsc{Gkeyll}\footnote{\url{https://github.com/ammarhakim/gkeyll}}, which is the software framework used to produce all simulation results presented within this paper.

In Section \ref{sec:mathematical_formulation}, we begin by describing the overall mathematical formulation of the general relativistic multifluid equations. In \ref{sec:gr_multifluids}, we show how the equations can be derived by coupling the stress-energy tensors for a family of perfect fluids to the stress-energy tensor for an electromagnetic field, and how the resulting equations can be reformulated to be solvable even in dynamic spacetimes where the a priori form of the spacetime metric tensor ${g_{\mu \nu}}$ is not known. In \ref{sec:grmhd}, we outline how the ideal GRMHD equations can be derived as a single-fluid limiting case of a two-fluid electron-ion plasma governed by the general relativistic multifluid equations, in which the ion-to-electron mass ratio ${\frac{m_i}{m_e}}$ goes to infinity, and therefore in which the mobility of the electrons also goes to infinity (corresponding to perfect electrical conductivity). In \ref{sec:gr_kinetics}, we outline, moreover, how the general relativistic multifluid equations themselves can be derived by taking moments of the covariant distribution function described by the general relativistic Boltzmann-Vlasov equation, effectively by assuming that the dissipative and anisotropic parts of the pressure tensor go to zero within the formalism of \cite{most_modelling_2022}, and that an appropriate equation of state closure relation has been provided for the heat flux. In Section \ref{sec:numerical_implementation}, we proceed to describe a numerical algorithm for solving the general relativistic multifluid equations using robust, high-resolution, shock-capturing methods. In \ref{sec:finite_volume_tetrad} we briefly summarize the local tetrad basis transformation of \cite{gorard_tetrad-first_2025} that enables us to use purely special relativistic Riemann solvers to evolve the homogeneous parts of the equations. In \ref{sec:riemann_solvers}, we describe a modification of the HLLC Riemann solver of \cite{mignone_hllc_2005} that we use to solve the homogeneous multifluid equations themselves, as well as the HLLC Riemann solver of \cite{mignone_hllc_2006} for the GRMHD equations, which we also implement in order to facilitate an \textit{apples-to-apples} comparison of the two formalisms. We also introduce the Roe-type approximate Riemann solver used for solving the general relativistic Maxwell equations, as well as the strong stability-preserving third-order Runge-Kutta method used for integrating both the electromagnetic coupling source terms and the geometric source terms arising from the spacetime curvature explicitly. In \ref{sec:primitive_variable_reconstruction}, we describe the hydrodynamic primitive variable reconstruction algorithm of \cite{eulderink_general_1994} used by the general relativistic multifluid solver, as well as the 2D and effective 1D primitive variable reconstruction algorithms of \cite{noble_primitive_2006} and \cite{newman_primitive_2014}, respectively, used by the GRMHD solver for the purposes of apples-to-apples comparisons (since these reconstruction schemes were empirically demonstrated by \cite{siegel_recovery_2018}, among all the standard reconstruction schemes for GRMHD, to remain robust and stably convergent across the widest range of Lorentz factor and magnetization parameters). In \ref{sec:divergence_correction}, we outline the schemes we use for correcting divergence errors in both the GRMHD equations (using a relativistic analog the 8-wave cleaning prescription of \cite{powell_solution-adaptive_1999}, combined with a variant of the hyperbolic divergence cleaning approach due to \cite{dedner_hyperbolic_2002}), and in the general relativistic multifluid equations (using a covariant extension of the hyperbolic divergence cleaning scheme of \cite{munz_finite-volume_2000}, \cite{munz_divergence_2000}, and \cite{munz_three-dimensional_2000} for the flat spacetime Maxwell equations).

In Section \ref{sec:1d_validation}, we present a validation of our numerical implementation of the 1D relativistic multifluid equations in the flat spacetime case, against standard SRMHD Riemann problems. In \ref{sec:convergence_brio_wu}, we present a \textit{convergence} test against a relativistic analog of the \cite{brio_upwind_1988} ideal MHD Riemann problem, as developed by \cite{vanputten_numerical_1993} and studied extensively by \cite{balsara_total_2001}. In particular, we show that, by varying the ion Larmor radius from ${r_L = \infty}$ to ${r_L = 0}$ within a two-fluid relativistic electron-ion plasma, we indeed see a continuous transition between the purely \textit{gas dynamic} solution to the problem (with no coupling between the fluids and the magnetic field) and the \textit{ideal MHD} solution to the problem (with perfect coupling between the fluids and the magnetic field), with additional electron plasma waves, such as L-modes, R-modes, and whistler waves, observed for intermediate values of ${r_L}$. In \ref{sec:robustness_noh}, we present a \textit{robustness} test against a relativistic analog of the \cite{noh_errors_1987} ideal MHD Riemann problem, as also studied by \cite{balsara_total_2001}. This is a strongly relativistic problem involving large initial Lorentz factors, and we again demonstrate appropriate convergence of the special relativistic multifluid solution to the SRMHD solution in the ${r_L \to 0}$ limit for the standard (weakly magnetized) variant of the problem. However, we also show that we are able to push the magnetization of the problem much higher, into regimes where both the \cite{noble_primitive_2006} and \cite{newman_primitive_2014} SRMHD primitive variable reconstruction algorithms fail to converge, and therefore for which no numerical SRMHD solution exists. The special relativistic multifluid solver nevertheless continues to produce stable solutions to this strongly magnetized variant of the Noh problem, for a variety of different ion Larmor radii, thereby enabling us to determine what the SRMHD solution \textit{would} look like, if it existed. In Section \ref{sec:2d_axisymmetric}, we present an intermediate validation of our numerical implementation against a black hole electrodynamics problem in 2D axisymmetry, namely the magnetospheric (plasma-filled) variant of the \cite{wald_black_1974} problem, as studied by \cite{komissarov_electrodynamics_2004} using both force-free and resistive electrodynamics. We find that a spinning black hole immersed in a uniform magnetic field forms an equatorial current sheet, with regions inside the ergosphere where ${\left\lVert \mathbf{E} \right\rVert^2 > \left \lVert \mathbf{B} \right\rVert^2}$ due to strong electric field induction resulting from the spin of the black hole (which cannot be correctly captured in GRMHD). We find that ${\mathbf{E} \cdot \mathbf{B}}$ is driven to zero within the current sheet, suggesting that magnetic reconnection has occurred, and also that the co-rotation of magnetic field lines threading the current sheet causes a much stronger poloidal field to be sustained (with a much larger corresponding Poynting flux) than in the GRMHD case, with potential implications for jet formation. We therefore find that the multifluid results are able to capture current sheet and magnetic reconnection physics that is present within the force-free and resistive electrodynamic approximations, yet absent within the GRMHD model.

In Section \ref{sec:3d_accretion}, we present a further validation of our numerical implementation of the full 3D general relativistic multifluid equations in the curved spacetime case, against 3D GRMHD accretion problems for both black hole and neutron star metrics. In \ref{sec:weakly_magnetized_black_holes}, we present a weakly magnetized variant of the 3D supersonic wind accretion problem studied by \cite{font_non-axisymmetric_1998} in the static black hole case, and by \cite{font_non-axisymmetric_1999} in the spinning black hole case, using a realistic black hole magnetization of ${\sigma \approx 100}$. We find that the general relativistic multifluid solutions are broadly in qualitative agreement with the corresponding GRMHD solutions, albeit with significant charge separations observed between the electrons and ions, indicative of the induction of substantial (parallel) electric fields and other non-ideal effects, and we quantify violations of the ${\left\lVert \mathbf{E} \right\rVert^2 \ll \left\lVert \mathbf{B} \right\rVert^2}$ assumption inherent to the GRMHD model. In \ref{sec:strongly_magnetized_neutron_stars}, we present a strongly magnetized variant of the problem studied within \ref{sec:weakly_magnetized_black_holes}, using a realistic neutron star magnetization of ${\sigma \approx 10^6}$, and a realistic exterior neutron star metric due to \cite{pappas_accurate_2017}. As in the 1D strongly magnetized Noh problem, the GRMHD primitive variable reconstruction algorithms fail to converge due to the high magnetizations and large Lorentz factors involved in this problem, yet the general relativistic multifluid model still converges stably. We find that the non-ideal effects present within the black hole accretion problems are significantly amplified in the case of neutron stars, with even stronger electric fields being generated (as quantified via the ${\left\lVert \mathbf{B} \right\rVert^2 - \left\lVert \mathbf{E} \right\rVert^2}$ and ${\mathbf{E} \cdot \mathbf{B}}$ scalar Lorentz invariants), and with further charge separations resulting from preferential clustering of the electrons near the neutron star surface as a consequence of the large differences between the electron and ion inertias.

\subsection{Notation}

In all that follows, we use the Greek indices ${\mu, \nu, \rho, \sigma}$, etc. to signify use of the \textit{spacetime} coordinate basis ${\left\lbrace t, x, y, z \right\rbrace}$ (or ${\left\lbrace t, \rho, \varphi, z \right\rbrace}$, ${\left\lbrace t, r, \theta, \phi \right\rbrace}$, etc.), while we use the Latin indices ${i, j, k, l}$, etc. to signify use of the \textit{spatial} coordinate basis ${\left\lbrace x, y, z \right\rbrace}$ (or ${\left\lbrace \rho, \varphi, z \right\rbrace}$ or ${\left\lbrace r, \theta, \phi \right\rbrace}$, etc.). Similarly, ${g_{\mu \nu}}$ is taken to refer to a spacetime metric tensor, with determinant ${g = \det \left( g_{\mu \nu} \right)}$, while ${\gamma_{i j}}$ is taken to refer to a spatial metric tensor, with determinant ${\gamma = \det \left( \gamma_{i j} \right)}$. For the avoidance of confusion, we explicitly distinguish between the spacetime Christoffel symbols, denoted as:

\begin{equation}
{}^{\left( 4 \right)} \Gamma_{\mu \nu}^{\rho} = \frac{1}{2} g^{\rho \sigma} \left( \partial_{\mu} g_{\sigma \nu} + \partial_{\nu} g_{\mu \sigma} - \partial_{\sigma} g_{\mu \nu} \right),
\end{equation}
and the spatial Christoffel symbols, denoted as:

\begin{equation}
{}^{\left( 3 \right)} \Gamma_{i j}^{k} = \frac{1}{2} \gamma^{k l} \left( \partial_i \gamma_{l j} + \partial_j \gamma_{i l} - \partial_l \gamma_{i j} \right),
\end{equation}
and likewise for the spacetime and spatial covariant derivative operators (denoted as ${{}^{\left( 4 \right)} \nabla_{\mu}}$ and ${{}^{\left( 3 \right)} \nabla_i}$, respectively). We assume Einstein summation convention, ${\left( -, +, +, + \right)}$ metric signature, and geometrized units with ${G = c = 1}$, throughout. All symbolic tensor manipulations were performed using the \textsc{Gravitas} computer algebra system (\cite{gorard_computational_2023} and \cite{gorard_computational_2024}).

\section{Mathematical Formulation}
\label{sec:mathematical_formulation}

\subsection{General Relativistic Multifluid Equations}
\label{sec:gr_multifluids}

We consider a family of $n$ relativistic fluid species (indexed by $s$) interacting with a shared electromagnetic field, embedded within an arbitrary curved spacetime with metric tensor ${g_{\mu \nu}}$. This yields an overall stress-energy tensor ${T^{\mu \nu}}$ of the general form:

\begin{equation}
T^{\mu \nu} = \left( \sum_{s = 1}^{n} T_{\text{Fluid}, s}^{\mu \nu} \right) + T_{\text{EM}}^{\mu \nu},
\end{equation}
where each ${T_{\text{Fluid}, s}^{\mu \nu}}$ is a perfect fluid stress-energy tensor of the form (neglecting heat conduction and viscous stresses):

\begin{equation}
T_{\text{Fluid}, s}^{\mu \nu} = \rho_s h_s u_{s}^{\mu} u_{s}^{\nu} + p_s g^{\mu \nu},
\end{equation}
with ${\rho_s}$ being the density of the species, ${p_s}$ being the pressure of the species, ${u_{s}^{\mu}}$ being the four-velocity of the species, ${g^{\mu \nu} = \left( g_{\mu \nu} \right)^{-1}}$ being the inverse metric tensor of the spacetime, and ${h_s}$ being the specific relativistic enthalpy of the species, which is related to the specific internal energy of the species ${\varepsilon_s \left( \rho_s, p_s \right)}$ by:

\begin{equation}
h_s = 1 + \varepsilon_s \left( \rho_s, p_s \right) + \frac{p_s}{\rho_s},
\end{equation}
which is, in turn, determined by the equation of state of the species. ${T_{\text{EM}}^{\mu \nu}}$ is an electromagnetic stress-energy tensor of the form:

\begin{equation}
T_{\text{EM}}^{\mu \nu} = \frac{1}{\mu_0} \left[ F^{\mu \alpha} F_{\alpha}^{\nu} - \frac{1}{4} g^{\mu \nu} F^{\alpha \beta} F_{\alpha \beta} \right],
\end{equation}
with ${\mu_0}$ being the vacuum permeability constant\footnote{In all that follows, we choose geometrized units where ${\mu_0 = 4 \pi}$, in order to facilitate comparison between the general relativistic multifluid and general relativistic magnetohydrodynamic formulations.} and ${F_{\mu \nu}}$ being the electromagnetic field tensor, which can be written in the manifestly covariant form:

\begin{equation}
F_{\mu \nu} = {}^{\left( 4 \right)} \nabla_{\mu} A_{\nu} - {}^{\left( 4 \right)} \nabla_{\nu} A_{\mu} = \partial_{\mu} A_{\nu} - \partial_{\nu} A_{\mu},
\end{equation}
in terms of the 1-form components ${A_{\mu} = g_{\mu \nu} A^{\nu}}$ of the electromagnetic four-potential ${\mathbf{A}}$ (with timelike component ${A^t = \phi}$ constituting the electric scalar potential, and spacelike components ${A^i}$ constituting the magnetic vector potential), such that ${F_{\mu}^{\nu} = g^{\alpha \nu} F_{\mu \alpha}}$ and ${F^{\mu \nu} = g^{\mu \alpha} g^{\beta \nu} F_{\alpha \beta}}$, and with the Christoffel symbols canceling out due to the antisymmetry of ${F_{\mu \nu}}$.

We now proceed to construct a ${3 + 1}$ decomposition of the spacetime using the ADM formalism of \cite{arnowitt_dynamical_1959}, thus foliating it into a time-ordered sequence of spacelike hypersurfaces ${\Sigma_{t_0}}$ (with ${t_0 \in \mathbb{R}}$), yielding an overall spacetime line element of the form:

\begin{equation}
d s^2 = g_{\mu \nu} d x^{\mu} d x^{\nu} = \left( - \alpha^2 + \beta_i \beta^i \right) d t^2 + 2 \beta_i dt d x^i + \gamma_{i j} d x^i d x^j,
\end{equation}
with ${\gamma_{i j}}$ being the induced (spatial) metric tensor on each hypesurface ${\Sigma_{t_0}}$, the scalar lapse function ${\alpha}$ defined in terms of the proper time distance ${d \tau}$ between corresponding points on hypersurfaces ${\Sigma_{t_0}}$ and ${\Sigma_{t_0 + dt}}$:

\begin{equation}
d \tau \left( t_0, t_0 + dt \right) = \alpha dt,
\end{equation}
and the three-dimensional shift vector field ${\beta^i}$ defined in terms of the distortion of the spatial coordinate basis ${x^i}$ between corresponding points on hypersurfaces ${\Sigma_{t_0}}$ and ${\Sigma_{t_0 + dt}}$:

\begin{equation}
x^i \left( t_0 + dt \right) = x^i \left( t_0 \right) - \beta^i dt,
\end{equation}
with ${\beta_i = \gamma_{i j} \beta^j}$. We define an \textit{Eulerian observer} to be an observer that is at rest with respect to this foliation, and whose four-velocity ${\mathbf{u}}$ is therefore always given by the four-dimensional unit normal vector ${\mathbf{n}}$ to each spacelike hypersurface ${\Sigma_{t_0}}$, which can be computed as the contravariant derivative of the time coordinate $t$:

\begin{equation}
n^{\mu} = - \alpha {}^{\left( 4 \right)} \nabla^{\mu} t = -\alpha g^{\mu \nu} {}^{\left( 4 \right)} \nabla_{\nu} t = - \alpha g^{\mu \nu} \partial_{\nu} t.
\end{equation}
In what follows, we shall express the electromagnetic field tensor in terms of ${\mathbf{D}}$ and ${\mathbf{B}}$, i.e. the electric and magnetic field vectors as perceived by an Eulerian observer, as:

\begin{equation}
F^{\mu \nu} = n^{\mu} D^{\nu} - n^{\nu} D^{\mu} - \frac{1}{\sqrt{-g}} \left( \varepsilon^{\mu \nu \alpha \beta} n_{\alpha} B_{\beta} \right),
\end{equation}
where we have used the fact that:

\begin{equation}
D^{\mu} = \alpha F^{t \mu}, \qquad \text{ and } \qquad B^{\mu} = \alpha {}^{\star} F^{\mu t},
\end{equation}
where ${{}^{\star} F^{\mu \nu}}$ is the Hodge dual of ${F_{\mu \nu}}$ (assuming vanishing electric and magnetic susceptibilities):

\begin{align}
{}^{\star} F^{\mu \nu} &= \frac{\sqrt{-g}}{2} \left( \varepsilon^{\mu \nu \alpha \beta} F_{\alpha \beta} \right),\\
F^{\mu \nu} &= - \frac{\sqrt{-g}}{2} \left( \varepsilon^{\mu \nu \alpha \beta} {}^{\star} F_{\alpha \beta} \right),
\end{align}
${B_{\mu} = g_{\mu \nu} B^{\nu}}$ and ${n_{\mu} = g_{\mu \nu} n^{\mu}}$ denote the 1-form components of the magnetic field ${\mathbf{B}}$ and the unit normal ${\mathbf{n}}$, and ${\varepsilon^{\mu \nu \alpha \beta}}$ denotes the rank-4 totally-antisymmetric Levi-Civita symbol.

This ${3 + 1}$ decomposition allows us to decompose the overall conservation law for energy-momentum within our multifluid system:

\begin{equation}
{}^{\left( 4 \right)} \nabla_{\nu} T^{\mu \nu} = \partial_{\nu} T^{\mu \nu} + {}^{\left( 4 \right)} \Gamma_{\nu \sigma}^{\mu} T^{\sigma \nu} + {}^{\left( 4 \right)} \Gamma_{\nu \sigma}^{\nu} T^{\mu \sigma} = 0,
\end{equation}
using the ${3 + 1}$ \textit{Valencia} formalism of \cite{banyuls_numerical_1997}, into a family of timelike projections\footnote{For simplicity, we treat the coupling between each relativistic fluid species and the electromagnetic field separately, in a pairwise fashion.} representing conservation of relativistic energy density ${\tau_s}$ for each fluid species:

\begin{multline}
\frac{1}{\sqrt{-g}} \left( \partial_t \left( \sqrt{\gamma} \tau_s \right) + \partial_i \left\lbrace \sqrt{-g} \left[ \tau_s \left( v_{s}^{i} - \frac{\beta^i}{\alpha} \right) + p_s v_{s}^{i} \right] \right\rbrace \right)\\
= \frac{q_s}{m_s} \left( \rho_s W_s v_{s}^{j} D_j \right) + \alpha \left( T_{\text{Fluid}, s}^{\mu t} \partial_{\mu} \alpha - T_{\text{Fluid}, s}^{\mu \nu} {}^{\left( 4 \right)} \Gamma_{\nu \mu}^{t} \right),
\end{multline}
and a family of spacelike projections representing conservation of three-momentum density ${S_{i, s}}$ for each fluid species:

\begin{multline}
\frac{1}{\sqrt{-g}} \left( \partial_t \left( \sqrt{\gamma} S_{j, s} \right) + \partial_i \left\lbrace \sqrt{-g} \left[ S_{j, s} \left( v_{s}^{i} - \frac{\beta^i}{\alpha} \right) + p_s \delta_{j}^{i} \right] \right\rbrace \right)\\
= \frac{q_s}{m_s} \left( \rho_s W_s D_j + \rho_s W_s \left[ \varepsilon_{j k l} v_{s}^{k} B^l \right] \right)\\
+ T_{\text{Fluid}, s}^{\mu \nu} \left( \partial_{\mu} g_{\nu j} - {}^{\left( 4 \right)} \Gamma_{\nu \mu}^{\sigma} g_{\sigma j} \right).
\end{multline}
In the above, the relativistic energy density ${\tau_s}$ of the species is given by:

\begin{equation}
\tau_s = \rho_s h_s W_{s}^{2} - p_s - \rho_s W_s,
\end{equation}
and the three-momentum density ${S_{i, s}}$ of the species is given by:

\begin{equation}
S_{i, s} = \rho_s h_s W_{s}^{2} v_{i, s},
\end{equation}
where the three-velocity ${v_{s}^{i}}$ of the species is obtained from its four-velocity ${u_{s}^{\mu}}$ by:

\begin{equation}
v_{s}^{i} = \frac{\perp_{\mu}^{i} u_{s}^{\mu}}{-u_{s}^{\mu} n_{\mu}} = \frac{u_{s}^{i}}{\alpha u_{s}^{t}} + \frac{\beta^i}{\alpha},
\end{equation}
with corresponding 1-form components ${v_{i, s} = \gamma_{i j} v_{s}^{j}}$, where ${\perp_{\mu}^{i}}$ is the orthogonal projector (projecting onto hypersurfaces orthogonal to ${\mathbf{n}}$):

\begin{equation}
\perp_{\mu}^{i} = \delta_{\mu}^{i} + n_{\mu} n^i,
\end{equation}
and ${W_s}$ is the Lorentz factor of the species:

\begin{equation}
W_s = \alpha u_{s}^{t} = \frac{1}{\sqrt{1 - \gamma_{i j} v_{s}^{i} v_{s}^{j}}}.
\end{equation}
${q_s}$ and ${m_s}$ denote the charge and mass of the constituent particles of the species, respectively, such that the ratio ${\frac{q_s}{m_s}}$ represents the coupling strength between the fluid species and the electromagnetic field. ${D_i = \gamma_{i j} D^j}$ represent the 1-form components of the spatial part of the electric field, and ${B^i}$ represent the vector components of the spatial part of the magnetic field, as perceived by Eulerian observers:

\begin{equation}
D^i = \alpha F^{t i}, \qquad \text{ and } \qquad B^i = \alpha {}^{\star} F^{i t}.
\end{equation}
${\varepsilon_{i j k}}$ denotes the rank-3 totally-antisymmetric Levi-Civita symbol. Finally, the conservation of the rest mass current density ${J_{s}^{\mu} = \rho_{s} u_{s}^{\mu}}$ for each species:

\begin{equation}
{}^{\left( 4 \right)} \nabla_{\mu} J_{s}^{\mu} = \partial_{\mu} J_{s}^{\mu} + {}^{\left( 4 \right)} \Gamma_{\mu \sigma}^{\mu} J_{s}^{\sigma} = 0,
\end{equation}
yields an additional family of evolution equations representing conservation of baryonic number density for each species:

\begin{equation}
\frac{1}{\sqrt{-g}} \left( \partial_t \left( \sqrt{\gamma} \rho_s W_s \right) + \partial_i \left\lbrace \sqrt{-g} \left[ \rho_s W_s \left( v_{s}^{i} - \frac{\beta^i}{\alpha} \right) \right] \right\rbrace \right) = 0.
\end{equation}

The governing equations for the magnetic field are similarly obtained by decomposing the homogeneous Maxwell equations:

\begin{equation}
{}^{\left( 4 \right)} \nabla_{\nu} {}^{\star} F^{\mu \nu} = \partial_{\nu} {}^{\star} F^{\mu \nu} + {}^{\left( 4 \right)} \Gamma_{\nu \sigma}^{\mu} {}^{\star} F^{\sigma \nu} + {}^{\left( 4 \right)} \Gamma_{\nu \sigma}^{\nu} {}^{\star} F^{\mu \sigma} = 0,
\end{equation}
into timelike and spacelike projections, yielding:

\begin{equation}
\frac{1}{\sqrt{\gamma}} \partial_i \left( \alpha \sqrt{\gamma} {}^{\star} F^{t i} \right) = 0,
\end{equation}
and:

\begin{equation}
\frac{1}{\sqrt{\gamma}} \partial_t \left( \alpha \sqrt{\gamma} {}^{\star} F^{j t} \right) + \frac{1}{\sqrt{\gamma}} \partial_i \left( \alpha \sqrt{\gamma} {}^{\star} F^{j i} \right) = 0,
\end{equation}
respectively. By introducing the spatial vector field ${\mathbf{E}}$, whose 1-form components are given by:

\begin{equation}
E_i = \frac{1}{2} \alpha \sqrt{\gamma} \varepsilon_{i j k} {}^{\star} F^{j k},
\end{equation}
with ${E^i = \gamma^{i j} E_j}$, and ${\gamma^{i j} = \left( \gamma_{i j} \right)^{-1}}$ being the inverse spatial metric tensor, these equations can be rewritten as an elliptic constraint equation and a hyperbolic evolution equation for the spatial part of the magnetic field ${\mathbf{B}}$ as perceived by Eulerian observers, namely:

\begin{equation}
{}^{\left( 3 \right)} \nabla_i B^i = \partial_i B^i = 0,
\end{equation}
and:

\begin{equation}
\partial_t B^i + \varepsilon^{i j k} {}^{\left( 3 \right)} \nabla_j E_k = \partial_t B^i + \varepsilon^{i j k} \partial_j E_k = 0,
\end{equation}
respectively. Likewise, the governing equations for the electric field can be obtained by decomposing the inhomogeneous Maxwell equations:

\begin{equation}
{}^{\left( 4 \right)} \nabla_{\nu} F^{\mu \nu} = \partial_{\nu} F^{\mu \nu} + {}^{\left( 4 \right)} \Gamma_{\nu \sigma}^{\mu} F^{\sigma \nu} + {}^{\left( 4 \right)} \Gamma_{\nu \sigma}^{\nu} F^{\mu \sigma} = I^{\mu},
\end{equation}
with four-current density ${\mathbf{I}}$:

\begin{equation}
I^{\mu} = \frac{1}{\mu_0} \partial_{\nu} \left( F^{\mu \nu} \sqrt{-g} \right),
\end{equation}
into timelike and spacelike projections, yielding:

\begin{equation}
\frac{1}{\sqrt{\gamma}} \partial_i \left( \alpha \sqrt{\gamma} F^{t i} \right) = \alpha I^t,
\end{equation}
and:

\begin{equation}
\frac{1}{\sqrt{\gamma}} \partial_t \left( \alpha \sqrt{\gamma} F^{j t} \right) + \frac{1}{\sqrt{\gamma}} \partial_i \left( \alpha \sqrt{\gamma} F^{j i} \right) = \alpha I^j,
\end{equation}
respectively. By introducing the spatial vector field ${\mathbf{H}}$, whose 1-form components are given by:

\begin{equation}
H_i = \frac{1}{2} \alpha \sqrt{\gamma} \varepsilon_{i j k} F^{j k},
\end{equation}
with ${H^i = \gamma^{i j} E_j}$, as well as the timelike and spacelike projections of the four-current density ${\mathbf{I}}$:

\begin{equation}
\rho_c = \alpha I^t, \qquad \text{ and } \qquad J_{c}^{i} = \alpha I^i,
\end{equation}
respectively, these equations can also be rewritten as an elliptic constraint equation and a hyperbolic evolution equation for the spatial part of the electric field ${\mathbf{D}}$ as perceived by Eulerian observers, namely:

\begin{equation}
{}^{\left( 3 \right)} \nabla_i D^i = \partial_i D^i = \rho_c,
\end{equation}
and:

\begin{equation}
- \partial_t D^i + \epsilon^{i j k} {}^{\left( 3 \right)} \nabla_j H_k = \partial_t D^i + \varepsilon^{i j k} \partial_j H_k = J_{c}^{i},
\end{equation}
respectively. In the above, the scalar field ${\rho_c}$ represents the charge density and the spatial vector field ${\mathbf{J}_c}$ represents the current density, and via Ohm's law we can express the four-current density ${\mathbf{I}}$ in terms of a sum over the $n$ fluid species:

\begin{equation}
I^{\mu} = \sum_{s = 1}^{n} \left( \rho_{q, s} u_{s}^{\mu} + \eta_s F^{\mu \nu} u_{\nu, s} \right),
\end{equation}
where ${\rho_{q, s}}$ is the proper charge density as measured by an observer who is comoving with the fluid species, ${\eta_s}$ is the electrical conductivity of the species (proportional to the inter-species collisional relaxation time ${\tau_{\text{coll}}}$, obtained from relativistic kinetics equations in the presence of a collision operator, as discussed briefly in \ref{sec:gr_kinetics}), and ${u_{\mu, s} = g_{\mu \nu} u_{s}^{\nu}}$ are the 1-form components of the species four-velocity. We have written the four-current density ${\mathbf{I}}$ in this way in order to match the form of equation 14 in \cite{anton_numerical_2006} more closely. However, for the purposes of the present paper, we consider the limit as the inter-species collisional relaxation time ${\tau_{\text{coll}} \to \infty}$, and therefore in which the four-current density is simply given by the flux of charge density:

\begin{equation}
I^{\mu} = \sum_{s = 1}^{n} \left( \frac{q_s}{m_s} \rho_s u_{s}^{\mu} \right),
\end{equation}
with the extension to relativistic \textit{resistive} multifluids (characterized by finite collisional relaxation time ${\tau_{\text{coll}} < \infty}$) remaining a subject for future work. This simplification allows us to rewrite the source terms for the electric field evolution equation as:

\begin{equation}
-\partial_t D^i + \varepsilon^{i j k} \partial_j H_k = \sum_{s = 1}^{n} \frac{q_s}{m_s} \left( \rho_s W_s v_{s}^{i} \right). \label{eq:current_density}
\end{equation}
In all of the above, the Christoffel symbols have canceled out due (as previously) to the antisymmetry of ${F_{\mu \nu}}$. For the purposes of solving these evolution equations numerically, the auxiliary ${\mathbf{E}}$ and ${\mathbf{H}}$ spatial vector fields are computed in terms of the spatial electric and magnetic fields ${\mathbf{D}}$ and ${\mathbf{B}}$ (as perceived by Eulerian observers) via the vacuum constitutive relations:

\begin{equation}
E^i = \alpha D^i + \varepsilon^{i j k} \beta_j B_k, \qquad \text{ and } \qquad H^i = \alpha B^i - \varepsilon^{i j k} \beta_j D_k,
\end{equation}
respectively.

Finally, since we ultimately intend for our algorithm to be capable of coupling multifluid systems to dynamic spacetimes via Einstein's equations, we cannot assume that we are able to know the partial derivatives ${\partial_{\sigma} g_{\mu \nu}}$ or the Christoffel symbols ${{}^{\left( 4 \right)} \Gamma_{\mu \nu}^{\sigma}}$ of the spacetime metric tensor ${g_{\mu \nu}}$ a priori. For this reason, we shall exploit the ADM constraint equations to rewrite the geometric source terms appearing on the right-hand side the conservation equations for relativistic energy density ${\tau_s}$ and three-momentum density ${S_{i, s}}$ for each fluid species purely in terms of the ADM gauge variables and their derivatives, as well as the overall stress-energy tensor ${T_{\text{Fluid}, s}^{\mu \nu}}$ of the fluid species. We begin by assuming that the ADM Hamiltonian:

\begin{equation}
\mathcal{H} = {}^{\left( 3 \right)} R + K^2 - K_{i j} K^{i j} - 2 \alpha^2 {}^{\left( 4 \right)} G^{t t} = 0,
\end{equation}
and momentum:

\begin{multline}
\mathcal{M}^i = {}^{\left( 3 \right)} \nabla_j \left( K^{i j} - \gamma^{i j} K \right) - \alpha {}^{\left( 4 \right)} G^{t i}\\
= \partial_j \left( K^{i j} - \gamma^{i j} K \right) + {}^{\left( 3 \right)} \Gamma_{j k}^{i} \left( K^{k j} - \gamma^{k j} K \right)\\
+ {}^{\left( 3 \right)} \Gamma_{j k}^{j} \left( K^{i k} - \gamma^{i k} K \right) - \alpha {}^{\left( 4 \right)} G^{t i} = 0,
\end{multline}
constraint equations are satisfied for our particular choice of spacetime foliation, where ${K_{i j}}$ is the extrinsic curvature tensor on spacelike hypersurfaces, defined as the Lie derivative of the spatial metric tensor ${\gamma_{i j}}$ with respect to the unit normal vector ${\mathbf{n}}$:

\begin{multline}
K_{i j} = - \frac{1}{2} \mathcal{L}_{\mathbf{n}} \gamma_{i j} = - \frac{1}{2 \alpha} \left( \partial_t \gamma_{i j} + {}^{\left( 3 \right)} \nabla_i \beta_j + {}^{\left( 3 \right)} \nabla_j \beta_i \right)\\
= - \frac{1}{2 \alpha} \left( \partial_t \gamma_{i j} + \partial_i \beta_j - {}^{\left( 3 \right)} \Gamma_{i j}^{k} \beta_k + \partial_j \beta_i - {}^{\left( 3 \right)} \Gamma_{j i}^{k} \beta_k \right),
\end{multline}
with ${K^{i j} = \gamma^{i k} \gamma^{l j} K_{k l}}$, ${K = \gamma^{i j} K_{i j}}$ is its trace, ${{}^{\left( 3 \right)} R}$ is the Ricci scalar on spacelike hypersurfaces:

\begin{multline}
{}^{\left( 3 \right)} R = \gamma^{i j} \left( \partial_k {}^{\left( 3 \right)} \Gamma_{i j}^{k} - \partial_j {}^{\left( 3 \right)} \Gamma_{i k}^{k} \right.\\
\left. + {}^{\left( 3 \right)} \Gamma_{i j}^{k} {}^{\left( 3 \right)} \Gamma_{k l}^{l} - {}^{\left( 3 \right)} \Gamma_{i k}^{l} {}^{\left( 3 \right)} \Gamma_{j l}^{k} \right),
\end{multline}
and ${{}^{\left( 4 \right)} G_{\mu \nu}}$ is the Einstein tensor on spacetime:

\begin{multline}
{}^{\left( 4 \right)} G_{\mu \nu} = \partial_{\sigma} {}^{\left( 4 \right)} \Gamma_{\mu \nu}^{\sigma} - \partial_{\nu} {}^{\left( 4 \right)} \Gamma_{\mu \sigma}^{\sigma} + {}^{\left( 4 \right)} \Gamma_{\mu \nu}^{\sigma} {}^{\left( 4 \right)} \Gamma_{\sigma \rho}^{\rho}\\
- {}^{\left( 4 \right)} \Gamma_{\mu \sigma}^{\rho} {}^{\left( 4 \right)} \Gamma_{\nu \rho}^{\sigma} - \frac{1}{2} g_{\mu \nu} g^{\alpha \beta} \left( \partial_{\sigma} {}^{\left( 4 \right)} \Gamma_{\alpha \beta}^{\sigma} - \partial_{\beta} {}^{\left( 4 \right)} \Gamma_{\alpha \sigma}^{\sigma} \right. \\
\left. + {}^{\left( 4 \right)} \Gamma_{\alpha \beta}^{\sigma} {}^{\left( 4 \right)} \Gamma_{\sigma \rho}^{\rho} - {}^{\left( 4 \right)} \Gamma_{\alpha \sigma}^{\rho} {}^{\left( 4 \right)} \Gamma_{\beta \rho}^{\sigma} \right),
\end{multline}
with ${{}^{\left( 4 \right)} G^{\mu \nu} = g^{\mu \alpha} g^{\beta \nu} {}^{\left( 4 \right)} G_{\alpha \beta}}$. In terms of the ADM gauge variables ${\alpha}$, ${\beta^i}$, and ${K_{i j}}$, their first derivatives, and the fluid species stress-energy tensor ${T_{\text{Fluid}, s}^{\mu \nu}}$, the geometric source terms for energy and momentum conservation can therefore be rewritten as (\cite{gorard_general_2024}):

\begin{multline}
\alpha \left( T_{\text{Fluid}, s}^{\mu \nu} \partial_{\mu} \alpha - T_{\text{Fluid}, s}^{\mu \nu}  \right) = T_{\text{Fluid}, s}^{t t} \left( \beta^i \beta^j K_{i j} + \beta^i \partial_i \alpha \right)\\
+ T_{\text{Fluid}, s}^{t i} \left( - \partial_i \alpha + 2 \beta^j K_{i j} \right) + T_{\text{Fluid}, s}^{i j} K_{i j},
\end{multline}
and:

\begin{multline}
T_{\text{Fluid}, s}^{\mu \nu} \left( \partial_{\mu} g_{\nu j} - {}^{\left( 4 \right)} \Gamma_{\nu \mu}^{\sigma} g_{\sigma j} \right) = T_{\text{Fluid}, s}^{t t} \left( \frac{1}{2} \beta^k \beta^l \partial_j \gamma_{k l} - \alpha \partial_j \alpha \right)\\
+ T_{\text{Fluid}, s}^{t i} \beta^k \partial_j \gamma_{i k} - \left( \frac{T_{\mu \nu, \text{Fluid}, s} n^{\mu} \perp_{k}^{\nu}}{\alpha} \right) \partial_j \beta^k,
\end{multline}
respectively, where ${T_{\mu \nu, \text{Fluid}, s} = g_{\mu \alpha} g_{\beta \nu} T_{\text{Fluid}, s}^{\alpha \beta}}$. We can, moreover, make use of the identity ${\sqrt{-g} = \alpha \sqrt{\gamma}}$ to rewrite the evolution equations for relativistic energy density ${\tau_s}$ and three-momentum density ${S_{i, s}}$ purely in terms of spatial variables, as:

\begin{multline}
\frac{1}{\alpha \sqrt{\gamma}} \left( \partial_t \left( \sqrt{\gamma} \tau_s \right) + \partial_i \left\lbrace \alpha \sqrt{\gamma} \left[ \tau_s \left( v_{s}^{i} - \frac{\beta^i}{\alpha} \right) + p_s v_{s}^{i} \right] \right\rbrace \right)\\
= \frac{q_s}{m_s} \left( \rho_s W_s v_{s}^{j} D_j \right) + T_{\text{Fluid}, s}^{t t} \left( \beta^i \beta^j K_{i j} + \beta^i \partial_i \alpha \right)\\
+ T_{\text{Fluid}, s}^{t i} \left( - \partial_i \alpha + 2 \beta^j K_{i j} \right) + T_{\text{Fluid}, s}^{i j} K_{i j},
\end{multline}
and:

\begin{multline}
\frac{1}{\alpha \sqrt{\gamma}} \left( \partial_t \left( \sqrt{\gamma} S_{j, s} \right) + \partial_i \left\lbrace \alpha \sqrt{\gamma} \left[ S_{j, s} \left( v_{s}^{i} - \frac{\beta^i}{\alpha} \right) + p_s \delta_{j}^{i} \right] \right\rbrace \right)\\
= \frac{q_s}{m_s} \left( \rho_s W_s D_j + \rho_s W_s \left[ \varepsilon_{j k l} v_{s}^{k} B^l \right] \right)\\
+ T_{\text{Fluid}, s}^{t t} \left( \frac{1}{2} \beta^k \beta^l \partial_j \gamma_{k l} - \alpha \partial_j \alpha \right) + T_{\text{Fluid}, s}^{t i} \beta^k \partial_j \gamma_{i k}\\
- \left( \frac{T_{\mu \nu, \text{Fluid}, s} n^{\mu} \perp_{k}^{\nu}}{\alpha} \right) \partial_j \beta^k,
\end{multline}
respectively. Likewise, the evolution equation for the baryonic number density ${\rho_s W_s}$ can be rewritten in purely spatial form as:

\begin{equation}
\frac{1}{\alpha \sqrt{\gamma}} \left( \partial_t \left( \sqrt{\gamma} \rho_s W_s \right) + \partial_i \left\lbrace \alpha \sqrt{\gamma} \left[ \rho_s W_s \left( v_{s}^{i} - \frac{\beta^i}{\alpha} \right) \right] \right\rbrace \right) = 0.
\end{equation}
This form of the equations is now amenable to direct numerical solution in both stationary and dynamic spacetimes.

\subsection{Relationship to GRMHD Equations}
\label{sec:grmhd}

It is possible to derive the equations of general relativistic magnetohydrodynamics (GRMHD) as a limit case of the general relativistic multifluid equations, as we shall briefly outline here. We begin by taking the limit in which the mass ${m_i}$ of a single fluid species ${s = i}$ (e.g. the mass of the ion species in a two-fluid electron-ion plasma) dominates over the masses of all other species, i.e. ${\left( \sum\limits_{s = 1}^{n} m_s \right) \to m_i}$, such that the overall stress-energy tensor converges to:

\begin{equation}
T^{\mu \nu} \to T_{\text{Fluid}}^{\mu \nu} + T_{\text{EM}}^{\mu \nu},
\end{equation}
within this limit, where we have dropped the $i$ subscript since there is now only a single fluid species. This corresponds to the limit in which the electrical conductivity of the multifluid plasma becomes infinite (i.e. ${\eta \to \infty}$, corresponding to a perfect conductor), since the mobility of the charge carriers (in this case the electrons) goes to infinity, such that ${F^{\mu \nu} u_{\nu} \to 0}$ by Ohm's law, assuming that the four-current density ${\mathbf{I}}$ remains finite. This implies that the spacetime electric field vector ${\mathbf{D}}$ as perceived by an Eulerian observer simplifies to:

\begin{equation}
D^{\mu} = \frac{1}{W \sqrt{-g}} \left( \varepsilon^{\mu \nu \alpha \beta} u_{\nu} n_{\alpha} B_{\beta} \right),
\end{equation}
or, upon taking timelike and spacelike projections:

\begin{equation}
D^t = 0, \qquad \text{ and } \qquad D^i = - \frac{\alpha}{\sqrt{-g}} \left( \varepsilon^{t i j k} v_j B_k \right),
\end{equation}
i.e. the electric field ${\mathbf{D}}$ perceived by an Eulerian observer is fully determined by the magnetic field ${\mathbf{B}}$ perceived by that observer. In particular, the electric field perceived by an observer who is comoving with the fluid (i.e. whose four-velocity is equal to the fluid four-velocity ${\mathbf{u}}$) vanishes. This corresponds to the \textit{ideal MHD} condition. Adopting the notation of \cite{anton_numerical_2006}, the spacetime magnetic field vector ${\mathbf{b}}$ perceived by such a comoving observer can be related to the spatial magnetic field vector ${\mathbf{B}}$ perceived by an Eulerian observer (upon taking timelike and spacelike projections) as:

\begin{equation}
b^t = \frac{W B^i v_i}{\alpha}, \qquad \text{ and } \qquad b^i = \frac{B^i + \alpha b^t u^i}{W},
\end{equation}
such that the electromagnetic field tensor simply becomes:

\begin{equation}
F^{\mu \nu} = - \frac{1}{\sqrt{-g}} \left( \varepsilon^{\mu \nu \alpha \beta} u_{\alpha} b_{\beta} \right),
\end{equation}
with ${b_{\mu} = g_{\mu \nu} b^{\nu}}$. In this limit, the electromagnetic stress-energy tensor is therefore:

\begin{equation}
T_{\text{EM}}^{\mu \nu} = \left( u^{\mu} u^{\nu} + \frac{1}{2} g^{\mu \nu} \right) b^2 - b^{\mu} b^{\nu},
\end{equation}
where the magnitude of ${\mathbf{b}}$ squared in the above can be computed as:

\begin{equation}
b^2 = b_{\mu} b^{\mu} = \frac{B_i B^i + \alpha^2 \left( b^t \right)^2}{W^2}.
\end{equation}

The overall stress-energy tensor in the ideal MHD limit therefore becomes (as in \cite{anton_numerical_2006}:

\begin{equation}
T^{\mu \nu} = \rho \left( h + \frac{b^2}{2} \right) u^{\mu} u^{\nu} + \left( p + \frac{b^2}{2} \right) g^{\mu \nu} - b^{\mu} b^{\nu},
\end{equation}
and so the conservation law for relativistic energy density ${\tau}$ becomes:

\begin{multline}
\frac{1}{\alpha \sqrt{\gamma}} \left( \partial_t \left( \sqrt{\gamma} \left( \tau - \left( \rho W^2 + p \right) \frac{b^2}{2} - \alpha^2 \left( b^t \right)^2 \right) \right) \right.\\
\left. + \partial_i \left\lbrace \alpha \sqrt{\gamma} \left[ \left( \tau - \left( \rho W^2 + p \right) \frac{b^2}{2} - \alpha^2 \left( b^t \right)^2 \right) \left( v^i - \frac{\beta^i}{\alpha} \right) \right. \right. \right.\\
\left. \left. \left. + \left( p + \frac{b^2}{2} \right) v^i - \frac{\alpha b^t B^i}{W} \right] \right\rbrace \right) = T^{t t} \left( \beta^i \beta^j K_{i j} + \beta^i \partial_i \alpha \right)\\
+ T^{t i} \left( - \partial_i \alpha + 2 \beta^j K_{i j} \right) + T^{i j} K_{i j},
\end{multline}
and the conservation law for three-momentum density ${S_i}$ becomes:

\begin{multline}
\frac{1}{\alpha \sqrt{\gamma}} \left( \partial_t \left( \sqrt{\gamma} \left( S_j + \frac{b^2 \rho W^2 v_j}{2} - \alpha b^t b_j \right) \right) \right.\\
\left. + \partial_i \left\lbrace \alpha \sqrt{\gamma} \left[ \left( S_j + \frac{b^2 \rho W^2 v_j}{2} - \alpha b^t b_j \right) \left( v^i - \frac{\beta^i}{\alpha} \right) \right. \right. \right.\\
\left. \left. \left. + \left( p + \frac{b^2}{2} \right) \delta_{j}^{i} - \frac{b_j B^i}{W} \right] \right\rbrace \right) = T^{t t} \left( \frac{1}{2} \beta^k \beta^l \partial_j \gamma_{k l} - \alpha \partial_j \alpha \right)\\
+ T^{t i} \beta^k \partial_j \gamma_{i k} - \left( \frac{T_{\mu \nu} n^{\mu} \perp_{k}^{\nu}}{\alpha} \right) \partial_j \beta^k.
\end{multline}
The conservation law for baryonic number density ${\rho W}$ remains unchanged from the general multifluid case:

\begin{equation}
\frac{1}{\alpha \sqrt{\gamma}} \left( \partial_t \left( \sqrt{\gamma} \rho W \right) + \partial_i \left\lbrace \alpha \sqrt{\gamma} \left[ \rho W \left( v^i - \frac{\beta^i}{\alpha} \right) \right] \right\rbrace \right) = 0.
\end{equation}
In the ideal MHD limit, the Hodge dual ${{}^{\star} F^{\mu \nu}}$ of the electromagnetic field tensor can be written purely in terms of the spacetime magnetic field ${\mathbf{B}}$ (as perceived by an Eulerian observer) as:

\begin{equation}
{}^{\star} F^{\mu \nu} = \frac{u^{\mu} B^{\nu} - u^{\nu} B^{\mu}}{W},
\end{equation}
with the timelike and spacelike projections of the inhomogeneous Maxwell equations now reducing to:

\begin{equation}
\partial_i \left( \sqrt{\gamma} B^i \right) = 0,
\end{equation}
and:

\begin{multline}
\frac{1}{\alpha \sqrt{\gamma}} \left( \partial_t \left( \sqrt{\gamma} B^k \right) \right.\\
\left. + \partial_i \left\lbrace \alpha \sqrt{\gamma} \left[ \left( v^i - \frac{\beta^i}{\alpha} \right) B^k - \left( v^k - \frac{\beta^k}{\alpha} \right) B^i \right] \right\rbrace \right) = 0,
\end{multline}
respectively.

\subsection{Relationship to General Relativistic Kinetics}
\label{sec:gr_kinetics}

Following \cite{tinti_resummed_2019} and \cite{denicol_resistive_2019}, the distribution function ${f_s \left( x^{\mu}, p^{\nu} \right)}$ for a particle species $s$ (assuming a covariant phase space parameterized by spacetime coordinates ${x^{\mu}}$ and four-momentum coordinates ${p^{\nu}}$), embedded within an arbitrary curved spacetime with metric tensor ${g_{\mu \nu}}$, evolves according to the general relativistic Vlasov equation\footnote{Note that many texts refer to this equation as the general relativistic \textit{Boltzmann} or \textit{Boltzmann-Vlasov} equation, reserving the term \textit{Vlasov} for the collisionless case ${C \left[ f_s \right] = 0}$ only. For simplicity, we refer to these instead as the \textit{Vlasov} equation and the \textit{collisionless Vlasov} equation, respectively. The overall system, when coupled to Maxwell's equations, is then referred to as the general relativistic \textit{Vlasov-Maxwell} system.}:

\begin{equation}
p^{\mu} \partial_{\mu} f_s \left[ q_s F^{\sigma \mu} p_{\mu} + {}^{\left( 4 \right)} \Gamma_{\mu \nu}^{\sigma} p^{\mu} p^{\nu} \right] \partial_{p^{\sigma}} f_s = C \left[ f_s \right],
\end{equation}
where we have included a coupling to an electromagnetic field via the ${q_s F^{\sigma \mu} p_{\mu}}$ term (with ${q_s}$ being the charge of the particle species), as well as an arbitrary collision operator ${C \left[ f_s \right]}$. The inter-species collisional relaxation time ${\tau}$ of the operator ${C \left[ f_s \right]}$, in turn, determines the bulk electrical conductivity ${\eta_s}$ of the species. The mass ${m_s}$ of the particle species is related to the four-momentum coordinates ${p^{\mu}}$ via the on-shell condition ${p_{\mu} p^{\mu} = -m_{s}^{2}}$, with ${p_{\mu} = g_{\mu \nu} p^{\nu}}$. Following \cite{most_modelling_2022}, the hydrodynamic quantities appearing within the general relativistic multifluid equations can now be derived as \textit{moments} (i.e. integrals over momentum space, with the particle four-momenta ${p^{\mu}}$ thereby replaced by a bulk fluid four-velocity ${u_{s}^{\mu}}$) of the distribution function ${f_s}$. The rest mass current density ${J_{s}^{\mu} = \rho_s u_{s}^{\mu}}$ appearing within the baryonic number density conservation equation for multifluids is derived as the first moment:

\begin{equation}
J_{s}^{\mu} = \int_{p^i} p^{\mu} f_s \frac{d^3 p}{2 p^t},
\end{equation}
where the notation ${\int_{p^i}}$ designates that the integral is evaluated over all of the \textit{spatial} momenta only. Likewise, the perfect fluid stress-energy tensor ${T_{\text{Fluid}, s}^{\mu \nu}}$, from which the relativistic energy density and three-momentum density equations for multifluids are derived, is itself derived as the second moment:

\begin{equation}
T_{\text{Fluid}, s}^{\mu \nu} = \int_{p^i} p^{\mu} p^{\nu} f_s \frac{d^3 p}{2 p^t}.
\end{equation}
These fluid quantities, as perceived by an Eulerian observer, can then be calculated explicitly from these moments as ${\rho_s W_s = - J_{\mu, s} n^{\mu}}$ (for the baryonic number density, with ${J_{\mu, s} = g_{\mu \nu} J_{s}^{\nu}}$), ${S_{s}^{i} = -T_{\text{Fluid}, s}^{\mu \nu} n_{\mu} \perp_{\nu}^{i}}$ (for the three-momentum density, with ${S_{i, s} = \gamma_{i j} S_{s}^{j}}$), and ${\tau_s = T_{\text{Fluid}, s}^{\mu \nu} n_{\mu} n_{\nu}}$ (for the relativistic energy density).

The only remaining fluid quantity appearing within the multifluid equations that has not been specified in terms of ${f_s}$ is the scalar pressure ${p_s}$. In the notation of \cite{most_modelling_2022}, the full pressure tensor ${\Pi_{s}^{\mu \nu}}$ can be obtained using a slight modification of the second moment integral used to compute the stress-energy tensor ${T_{\text{Fluid}, s}^{\mu \nu}}$ above, namely:

\begin{equation}
\Pi_{s}^{\mu \nu} = \int_{p^i} \Delta_{\alpha}^{\mu} p^{\alpha} \Delta_{\beta}^{\nu} p^{\beta} f_s \frac{d^3 p}{2 p^t},
\end{equation}
where ${\Delta_{\mu}^{\nu}}$ denotes the fluid frame projector:

\begin{equation}
\Delta_{\mu}^{\nu} = \delta_{\mu}^{\nu} + u_{s}^{\nu} u_{\mu, s}.
\end{equation}
The full pressure tensor ${\Pi_{s}^{\mu \nu}}$ can then be decomposed into a sum of an isotropic part ${\left( P_s + \Pi_s \right)}$ and an anisotropic part ${\pi_{s}^{\mu \nu}}$, as:

\begin{equation}
\Pi_{s}^{\mu \nu} = \left( P_s + \Pi_s \right) \Delta^{\mu \nu} + \pi_{s}^{\mu \nu},
\end{equation}
where ${\Delta^{\mu \nu} = \delta^{\mu \nu} + u_{s}^{\mu} u_{s}^{\nu}}$, with the isotropic part ${\left( P_s + \Pi_s \right)}$ being further decomposed into a sum of an equilibrium part ${P_s}$ and a dissipative part ${\Pi_s}$, with the sum given by:

\begin{equation}
P_s + \Pi_s = \frac{1}{3} \Delta_{\mu \nu} T_{\text{Fluid}, s}^{\mu \nu} = \frac{1}{3} \Delta_{\mu \nu} \Pi_{s}^{\mu \nu},
\end{equation}
where ${\Delta_{\mu \nu} = \delta_{\mu \nu} + u_{\mu, s} u_{\nu, s}}$, and with the anisotropic part ${\pi_{s}^{\mu \nu}}$ given by:

\begin{equation}
\pi_{s}^{\mu \nu} = \left( \frac{1}{2} \left( \Delta_{\alpha}^{\mu} \Delta_{\beta}^{\nu} + \Delta_{\alpha}^{\nu} \Delta_{\beta}^{\mu} \right) - \frac{1}{3} \Delta^{\mu \nu} \Delta_{\alpha \beta} \right) T_{\text{Fluid}, s}^{\alpha \beta}.
\end{equation}
Therefore, in order to complete the derivation of the general relativistic multifluid equations from general relativistic kinetics via the formalism of \cite{most_modelling_2022}, it is necessary to assume an equilibrium equation of state in which the equilibrium and dissipative parts of the isotropic pressure are separable (with the dissipative part vanishing, i.e. ${\Pi_s \to 0}$), as well as to assume that the anisotropic part of the pressure tensor also vanishes, i.e. ${\pi_{s}^{\mu \nu} \to 0}$. Then, the scalar pressure ${p_s}$ appearing within the multifluid equations is simply the equilibrium isotropic pressure ${P_s}$ from full relativistic kinetics. Note that full relativistic kinetics also defines a heat flux vector ${q^{\mu}}$, given by:

\begin{equation}
q^{\mu} = - \Delta_{\beta}^{\mu} u_{\alpha, s} T_{\text{Fluid}, s}^{\alpha \beta},
\end{equation}
which is not evolved directly within our relativistic multifluid equations, but is instead specified implicitly in terms of the lower (fluid) moments via an equation of state or other kind of closure relation.

\section{Numerical Implementation}
\label{sec:numerical_implementation}

\subsection{Finite Volume Algorithms in the Tetrad Basis}
\label{sec:finite_volume_tetrad}

We solve the hyperbolic evolution equations of the relativistic multifluid-Maxwell system, as well as the evolution equations of GRMHD and pure relativistic hydrodynamics (for comparison purposes), using finite volume, high-resolution, shock-capturing methods, as implemented within the \textsc{Gkeyll} computational multiphysics framework. However, to avoid the concomitant complexity and fragility of using fully general relativistic Riemann solvers, we opt to solve the evolution equations for these relativistic systems in curved spacetimes using the \textit{tetrad-first} approach of \cite{gorard_tetrad-first_2025}, based upon earlier work of \cite{pons_general_1998}, wherein a local coordinate transformation is applied at each inter-cell boundary in the computational domain, thus transforming both the conservative and primitive variables into a \textit{tetrad basis}, i.e. a locally-flat spacetime coordinate basis. Concretely, when updating cells in the $x$-direction\footnote{\textsc{Gkeyll} uses a dimensionally split algorithm for its finite volume update, so the update in each coordinate direction is performed analogously to the $x$-direction update.}, one begins by selecting a local orthonormal tetrad ${e_{a}^{\mu}}$ at the center point ${x_{0}^{\mu}}$ of the spacelike hypersurface ${\Sigma_{x_0}}$ corresponding to the inter-cell boundary; without loss of generality, we choose ${\mathbf{e}_{t} = \mathbf{n}}$, with ${\mathbf{e}_x}$, ${\mathbf{e}_y}$, and ${\mathbf{e}_z}$ being the unit normal vectors to the spacelike hypersurfaces ${\Sigma_{x_0}}$, ${\Sigma_{y_0}}$, and ${\Sigma_{z_0}}$, which bound the cells in the left/right, up/down, and front/back directions, respectively. This choice of orthonormal tetrad introduces a new spacetime coordinate basis ${\left\lbrace \widetilde{x^a} \right\rbrace}$ (indexed by $a$) which we call the \textit{tetrad basis}, and which is related to the original spacetime coordinate basis ${\left\lbrace x^{\mu} \right\rbrace}$ via the coordinate transformation:

\begin{equation}
\widetilde{x^a} = M_{\mu}^{a} \left( x^{\mu} - x_{0}^{\mu} \right),
\end{equation}
where ${M_{\mu}^{a}}$ is the change-of-basis matrix:

\begin{multline}
M_{\mu}^{a} = \frac{\partial_{\mu}}{\mathbf{e}_a}\\
= \begin{bmatrix}
\frac{1}{\sqrt{\gamma^{x x}}} & \frac{- \gamma^{x y} \gamma_{y y} - \gamma^{x z} \gamma_{y z}}{\gamma^{x x} \sqrt{\gamma_{y y}}} & \frac{- \gamma^{x z} \sqrt{\gamma_{y y} \gamma_{z z} - \left( \gamma_{y z} \right)^2}}{\gamma^{x x} \sqrt{\gamma_{y y}}}\\
0 & \sqrt{\gamma_{y y}} & 0\\
0 & \frac{\gamma_{y z}}{\sqrt{\gamma_{y y}}} & \frac{\sqrt{\gamma_{y y} \gamma_{z z} - \left( \gamma_{y z} \right)^2}}{\sqrt{\gamma_{y y}}}
\end{bmatrix}.
\end{multline}
The tetrad basis effectively diagonalizes the inverse spacetime metric tensor ${g^{\mu \nu}}$:

\begin{equation}
g^{\mu \nu} = e_{a}^{\mu} e_{b}^{\nu} \eta^{a b},
\end{equation}
with ${\eta^{\mu \nu} = \left( \eta_{\mu \nu} \right)^{-1}}$ being the inverse Minkowski metric tensor for flat spacetime, and all tensorial quantities (both primitive and conservative) can be transformed into this new coordinate basis using ${M_{\mu}^{a}}$. This transformation therefore allows us to compute the general relativistic inter-cell flux ${\mathbf{F}_{i + \frac{1}{2}}}$ through hypersurface ${\Sigma_{x_0}}$ in the curved spacetime coordinate basis ${\left\lbrace x^{\mu} \right\rbrace}$ by first computing the \textit{special relativistic} inter-cell flux in the \textit{flat spacetime} tetrad basis ${\left\lbrace \widetilde{x^a} \right\rbrace}$ (in the co-moving frame with $x$-velocity equal to ${\frac{\widetilde{\beta^x}}{\alpha}}$, where ${\widetilde{\beta^x}} = M_{i}^{x} \beta^i$), and then multiplying the result by a geometrical correction factor given by the surface integral:

\begin{equation}
I = \int_{\Sigma_{x_0}} \sqrt{\gamma^{x x}} \sqrt{-g} \, dt \, dy \, dz.
\end{equation}
This transformation implies that the \textit{homogeneous} parts of the evolution equations may be solved in their special relativistic (flat spacetime) form, although the full \textit{inhomogeneous} equations will still include geometric source terms resulting from the spacetime curvature.

As a result, after transforming all tensorial quantities into the tetrad basis, the forms of the relativistic multifluid equations that are actually solved by the finite volume algorithms within \textsc{Gkeyll} are given by:

\begin{multline}
\partial_t \left( \tau_s \right) + \partial_i \left[ \left( \tau_s + p_s \right) v_{s}^{i} \right] = \frac{q_s}{m_s} \left( \rho_s W_s v_{s}^{j} E_j \right)\\
+ T_{\text{Fluid}, s}^{t t} \left( \beta^i \beta^j K_{i j} + \beta^i \partial_i \alpha \right) + T_{\text{Fluid}, s}^{t i} \left( - \partial_i \alpha + 2 \beta^j K_{i j} \right)\\
+ T_{\text{Fluid}, s}^{i j} K_{i j},
\end{multline}
for the relativistic energy density ${\tau_s}$,

\begin{multline}
\partial_t \left( S_{j, s} \right) + \partial_i \left( S_{j, s} v_{s}^{i} + p_s \delta_{j}^{i} \right) = \frac{q_s}{m_s} \left( \rho_s W_s E_j \right.\\
\left. + \rho_s W_s \left[ \varepsilon_{j k l} v_{s}^{k} B^l \right] \right) + T_{\text{Fluid}, s}^{t t} \left( \frac{1}{2} \beta^k \beta^l \partial_j \gamma_{k l} - \alpha \partial_j \alpha \right)\\
+ T_{\text{Fluid}, s}^{t i} \beta^k \partial_j \gamma_{i k} - \left( \frac{T_{\mu \nu, \text{Fluid}, s} n^{\mu} \perp_{k}^{\nu}}{\alpha} \right) \partial_j \beta^k,
\end{multline}
for the three-momentum density ${S_{i, s}}$, and:

\begin{equation}
\partial_t \left( \rho_s W_s \right) + \partial_i \left( \rho_s W_s v_{s}^{i} \right) = 0,
\end{equation}
for the baryonic number density ${\rho_s W_s}$. Likewise, the hyperbolic parts of Maxwell's equations are solved as:

\begin{equation}
\partial_t B^i + \varepsilon^{i j k} \partial_j E_k = 0,
\end{equation}
for the magnetic field ${\mathbf{B}}$, and:

\begin{equation}
- \partial_t E^i + \varepsilon^{i j k} \partial_j B_k = - \sum_{s = 1}^{n} \frac{q_s}{m_s} \left( \rho_s W_s v_{s}^{i} \right),
\end{equation}
for the electric field ${\mathbf{E}}$. Correspondingly, the forms of the GRMHD equations that are solved by \textsc{Gkeyll}'s finite volume algorithms are given by:

\begin{multline}
\partial_t \left( \tau - \left( \rho W^2 + p \right) \frac{b^2}{2} - \left( b^t \right)^2 \right)\\
+ \partial_i \left[ \left( \tau - \left( \rho W^2 + p \right) \frac{b^2}{2} - \left( b^t \right)^2 + p + \frac{b^2}{2} \right) v^i - \frac{b^t B^i}{W} \right]\\
= T^{t t} \left( \beta^i \beta^j K_{i j} + \beta^i \partial_i \alpha \right) + T^{t i} \left( - \partial_i \alpha + 2 \beta^j K_{i j} \right) + T^{i j} K_{i j},
\end{multline}
for the relativistic energy density ${\tau}$,

\begin{multline}
\partial_t \left( S_j + \frac{b^2 \rho W^2 v_j}{2} - b^t b_j \right)\\
+ \partial_i \left[ \left( S_j + \frac{b^2 \rho W^2 v_j}{2} - b^t b_j \right) v^i + \left( p + \frac{b^2}{2} \right) \delta_{j}^{i} - \frac{b_j B^i}{W} \right]\\
= T^{t t} \left( \frac{1}{2} \beta^k \beta^l \partial_j \gamma_{k l} - \alpha \partial_j \alpha \right) + T^{t i} \beta^k \partial_j \gamma_{i k}\\
- \left( \frac{T_{\mu \nu} n^{\mu} \perp_{k}^{\nu}}{\alpha} \right) \partial_j \beta^k,
\end{multline}
for the three-momentum density ${S_i}$, and:

\begin{equation}
\partial_t \left( \rho W \right) + \partial_i \left( \rho W v^i \right) = 0,
\end{equation}
for the baryonic number density ${\rho W}$, where:

\begin{equation}
b^t = W B^i v_i, \qquad \text{ and } \qquad b^i = \frac{B^i + b^t u^i}{W},
\end{equation}
such that:

\begin{equation}
b^2 = b_{\mu} b^{\mu} = \frac{B_i B^i + \left( b^t \right)^2}{W^2}.
\end{equation}
Finally, the magnetic field ${\mathbf{B}}$ in GRMHD is simply evolved as:

\begin{equation}
\partial_t B^k + \partial_i \left( v^i B^k - v^k B^i \right) = 0.
\end{equation}

\subsection{Riemann Solvers and Source Term Coupling}
\label{sec:riemann_solvers}

We note that, following the transformation into the locally-flat tetrad basis, the hyperbolic evolution equations for relativistic hydrodynamics, electromagnetism, and magnetohydrodynamics can all be cast into the same first-order, flux-conservative form:

\begin{equation}
\partial_t \mathbf{U} + \nabla \cdot \mathbf{F} \left( \mathbf{U} \right) = \mathbf{S} \left( \mathbf{U} \right), \label{eq:flux_conservative}
\end{equation}
where ${\mathbf{U}}$ is the conserved variable vector field, ${\mathbf{F} \left( \mathbf{U} \right)}$ is the flux vector field, and ${\mathbf{S} \left( \mathbf{U} \right)}$ is the source term vector field. Such equation systems are directly amenable to solution via high-resolution shock-capturing (finite volume) methods. Upon discretizing the computational domain into cells of width ${\Delta x}$\footnote{As before, we will describe the process of solving the hyperbolic evolution equations in the $x$-direction only, since \textsc{Gkeyll}'s dimensionally split algorithm treats the updates in all other coordinate directions analogously.}, and assuming a discrete time-step of ${\Delta t}$, the \textit{homogeneous} solution to equation \ref{eq:flux_conservative} (i.e. the solution assuming vanishing source terms ${\mathbf{S} \left( \mathbf{U} \right) = \mathbf{0}}$) within cell $i$ can be evolved from time ${t^n}$ to time ${t^{n + 1} = t^n + \Delta t}$ via the explicit update formula:

\begin{equation}
\mathbf{U}_{i}^{n + 1} = \mathbf{U}_{i}^{n} + \frac{\Delta t}{\Delta x} \left[ \mathbf{F}_{i + \frac{1}{2}} - \mathbf{F}_{i - \frac{1}{2}} \right], \label{eq:conservative_update}
\end{equation}
where ${\mathbf{U}_{i}^{n}}$ is the conserved variable vector in cell $i$ at time ${t^n}$, and ${\mathbf{F}_{i + \frac{1}{2}}}$ represents the \textit{inter-cell flux} between cells $i$ and ${i + 1}$. The particular choice of exact or approximate \textit{Riemann solver}, used for the computation of the inter-cell flux ${\mathbf{F}_{i + \frac{1}{2}}}$ as a function of the left and right conserved variable vectors ${\mathbf{U}_{i}}$ and ${\mathbf{U}_{i + 1}}$, therefore determines the nature of the finite volume scheme in question. For both the (hydrodynamic part of the) general relativistic multifluid equations and the GRMHD equations, we use variants of the HLLC (Harten-Lax-van Leer with Contact, \cite{harten_upstream_1983} and \cite{toro_restoration_1994}) approximate Riemann solvers developed by \cite{mignone_hllc_2005} for relativistic hydrodynamics, and later for magnetohydrodynamics (\cite{mignone_hllc_2006}), appropriately adapted for our choices of variables and fluxes. For the general relativistic Maxwell equations (as coupled to the multifluid system), we opt to use a Roe-type approximate Riemann solver instead\footnote{Although we could use the less diffusive Roe-type approximate Riemann solver for the hydrodynamic part of the multifluid equations, as was done in \cite{gorard_tetrad-first_2025}, we choose to use HLLC in order to facilitate an \textit{apples-to-apples} comparison against GRMHD.}. In what follows, we introduce the notation ${\mathbf{U}_L = \mathbf{U}_i}$ and ${\mathbf{U}_R = \mathbf{U}_{i + 1}}$ to represent the left and right conserved variable vectors, respectively, and likewise ${\mathbf{F}_L = \mathbf{F} \left( \mathbf{U}_L \right)}$ and ${\mathbf{F}_R = \mathbf{F} \left( \mathbf{U}_R \right)}$ to represent the flux function evaluated for the left and right conserved variable vectors, respectively, and we drop the $n$ and ${n + 1}$ superscripts except in cases where there may exist ambiguity.

For the homogeneous part of the general relativistic multifluid equations in the locally-flat tetrad basis, we have:

\begin{equation}
\mathbf{U} = \begin{bmatrix}
\rho_s W_s\\
S_{x, s}\\
S_{y, s}\\
S_{z, s}\\
\tau_s
\end{bmatrix}, \qquad \text{ and } \qquad \mathbf{F} \left( \mathbf{U} \right) = \begin{bmatrix}
\rho_s W_s v_{s}^{x}\\
S_{x, s} v_{s}^{x} + p_s\\
S_{y, s} v_{s}^{x}\\
S_{z, s} v_{s}^{x}\\
\left( \tau_s + p_s \right) v_{s}^{x}
\end{bmatrix}.
\end{equation}
However, in order to be consistent with the the HLLC Riemann solver of \cite{mignone_hllc_2005}, we must transform the conserved relativistic energy density from ${\tau_s}$ to ${\tau_s + \rho_s W_s}$, thus yielding:

\begin{equation}
\mathbf{U}^{\prime} = \begin{bmatrix}
\rho_s W_s\\
S_{x, s}\\
S_{y, s}\\
S_{z, s}\\
\tau_s + \rho_s W_s
\end{bmatrix}, \qquad \text{ and } \qquad \mathbf{F}^{\prime} \left( \mathbf{U}^{\prime} \right) = \begin{bmatrix}
\rho_s W_s v_{s}^{x}\\
S_{x, s} v_{s}^{x} + p_s\\
S_{y, s} v_{s}^{x}\\
S_{z, s} v_{s}^{x}\\
S_{x, s}
\end{bmatrix}.
\end{equation}
We now estimate the left and right wave-speeds ${S_L}$ and ${S_R}$ appearing within the Riemann problem using the relativistic generalization of the prescription of \cite{davis_simplified_1988} for gas dynamics, as first employed by \cite{schneider_new_1993}, namely:

\begin{equation}
S_L = \min \left\lbrace \lambda_{-} \left( \mathbf{U}_L \right), \lambda_{-} \left( \mathbf{U}_R \right) \right\rbrace, \label{eq:left_wavespeed}
\end{equation}
and:
\begin{equation}
S_R = \max \left\lbrace \lambda_{+} \left( \mathbf{U}_L \right), \lambda_{+} \left( \mathbf{U}_R \right) \right\rbrace, \label{eq:right_wavespeed}
\end{equation}
respectively, where the ${\lambda_{\pm} \left( \mathbf{U} \right)}$ represent the maximum and minimum eigenvalues of the flux Jacobian ${\frac{\partial \mathbf{F}^{\prime} \left( \mathbf{U}^{\prime} \right)}{\partial \mathbf{U}^{\prime}}}$, given by:

\begin{equation}
\lambda_{\pm} \left( \mathbf{U} \right) = \frac{v_{s}^{x} \pm \sqrt{\sigma_s \left( 1 - v_{s}^{x} + \sigma_s \right)}}{1 + \sigma_s},
\end{equation}
with ${\sigma_s}$ being a function of the sound speed ${c_s}$ for an ideal gas:

\begin{equation}
\sigma_s = \frac{c_{s}^{2}}{W_{s}^{2} \left( 1 - c_{s}^{2} \right)}, \qquad \text{ where } \qquad c_s = \sqrt{\frac{\Gamma_s p_s}{\rho_s h_s}}.
\end{equation}
The estimate for the contact wave-speed ${S^{*}}$ is then given by the smallest solution of the following quadratic (which we solve analytically):

\begin{equation}
F_{\left( \tau_s + \rho_s W_s \right)}^{\text{HLL}} \left( S^{*} \right)^2 - \left( U_{\left( \tau_s + \rho_s W_s \right)}^{\text{HLL}} + F_{S_{x, s}}^{\text{HLL}} \right) S^{*} + U_{S_{x, s}}^{\text{HLL}} = 0,
\end{equation}
where ${U_{S_{x, s}}^{\text{HLL}}}$ and ${U_{\left( \tau_s + \rho_s W_s \right)}^{\text{HLL}}}$ are the $x$-momentum density and relativistic energy density components of the Harten-Lax-van Leer integral averaged state:

\begin{equation}
\mathbf{U}^{\text{HLL}} = \frac{S_R \mathbf{U}_{R}^{\prime} - S_L \mathbf{U}_{L}^{\prime} + \mathbf{F}_{L}^{\prime} - \mathbf{F}_{R}^{\prime}}{S_R - S_L}, \label{eq:hll_integral_average}
\end{equation}
and, likewise, ${F_{S_{x, s}}^{\text{HLL}}}$ and ${F_{\left( \tau_s + \rho_s W_s \right)}^{\text{HLL}}}$ are the $x$-momentum density and relativistic energy density components of the Harten-Lax-van Leer integral averaged flux:

\begin{equation}
\mathbf{F}^{\text{HLL}} = \frac{S_L \mathbf{F}_{L}^{\prime} - S_R \mathbf{F}_{R}^{\prime} + S_R S_L \left( \mathbf{U}_{R}^{\prime} - \mathbf{U}_{L}^{\prime} \right)}{S_R - S_L}. \label{eq:hll_flux_integral_average}
\end{equation}
Finally, the Mignone-Bodo HLLC inter-cell flux ${\mathbf{F}_{i + \frac{1}{2}}^{\text{HLLC}}}$ is computed as:

\begin{equation}
\mathbf{F}_{i + \frac{1}{2}}^{\text{HLLC}} = \begin{cases}
\mathbf{F}_{L}^{\prime}, \qquad &\text{ for } \qquad S_L \geq 0,\\
\mathbf{F}_{L}^{*}, \qquad &\text{ for } \qquad S_L < 0 \leq S^{*},\\
\mathbf{F}_{R}^{*}, \qquad &\text{ for } \qquad S^{*} < 0 \leq S_R,\\
\mathbf{F}_{R}^{\prime}, \qquad &\text{ for } \qquad S_R < 0
\end{cases}, \label{eq:hllc_intercell_flux}
\end{equation}
where the \textit{star state} flux function ${\mathbf{F}^{*} \left( \mathbf{U} \right)}$ is defined as:

\begin{equation}
\mathbf{F}^{*} \left( \mathbf{U} \right) = \begin{bmatrix}
\rho_{s}^{*} W_{s}^{*} v_{s}^{x, *}\\
S_{x, s}^{*} v_{s}^{x, *} + p_{s}^{*}\\
S_{y, s}^{*} v_{s}^{x, *}\\
S_{z, s}^{*} v_{s}^{x, *}\\
S_{x, s}^{*}
\end{bmatrix},
\end{equation}
and where the star state conserved variables ${\rho_{s}^{*} W_{s}^{*}}$, ${S_{x, s}^{*}}$, ${S_{y, s}^{*}}$, ${S_{z, s}^{*}}$, and ${\tau_{s}^{*} + \rho_{s}^{*} W_{s}^{*}}$ are given by the Rankine-Hugoniot conditions:

\begin{equation}
\rho_{s}^{*} W_{s}^{*} = \frac{\rho_s W_s \left( \lambda - v_{s}^{x} \right)}{\lambda - S^{*}},
\end{equation}
\begin{equation}
S_{x, s}^{*} = \frac{S_{x, s} \left( \lambda - v_{s}^{x} \right) + p_{s}^{*} - p_s}{\lambda - S^{*}},
\end{equation}
\begin{equation}
S_{y, s}^{*} = \frac{S_{y, s} \left( \lambda - v_{s}^{x} \right)}{\lambda - S^{*}},
\end{equation}
\begin{equation}
S_{z, s}^{*} = \frac{S_{z, s} \left( \lambda - v_{s}^{x} \right)}{\lambda - S^{*}},
\end{equation}
and:

\begin{equation}
\left( \tau_{s}^{*} + \rho_{s}^{*} W_{s}^{*} \right) = \frac{\left( \tau_s + \rho_s W_s \right) \left( \lambda - v_{s}^{x} \right) + p_{s}^{*} S^{*} - p_s v_{s}^{x}}{\lambda - S^{*}},
\end{equation}
respectively. The only remaining term appearing within the star state flux ${\mathbf{F}^{*} \left( \mathbf{U} \right)}$, namely the star state pressure ${p_{s}^{*}}$, is finally computed as:

\begin{equation}
p_{s}^{*} = \frac{- S_{x, s} \left( \lambda - v_{s}^{x} \right) + p + \left( \lambda \left( \tau_s + \rho_s W_s \right) - S_{x, s} \right) v_{s}^{x, *}}{1 + \lambda v_{s}^{x, *}}.
\end{equation}
In all of the above, ${\lambda}$ is equal to either ${S_L}$ or ${S_R}$, depending upon whether one is evaluating ${\mathbf{F}_{L}^{*}}$ or ${\mathbf{F}_{R}^{*}}$, and we assume that the normal component of the star state velocity is given by the contact wave-speed, i.e. ${v_{s}^{x, *} = S^{*}}$. Once ${\left( \mathbf{U}_{i}^{n + 1} \right)^{\prime}}$ has been computed from ${\left( \mathbf{U}_{i}^{n} \right)^{\prime}}$ using equation \ref{eq:conservative_update}, all that remains is to transform back from the Mignone-Bodo relativistic energy density variable to our original energy definition, by subtracting off the (updated) relativistic mass density term, to obtain ${\mathbf{U}_{i}^{n + 1}}$.

On the other hand, for the homogeneous part of the GRMHD equations in the locally-flat tetrad basis, we have:

\begin{equation}
\mathbf{U} = \begin{bmatrix}
\rho W\\
S_{x}^{\text{MHD}}\\
S_{y}^{\text{MHD}}\\
S_{z}^{\text{MHD}}\\
\tau^{\text{MHD}}\\
B^x\\
B^y\\
B^z
\end{bmatrix}, \qquad \mathbf{F} \left( \mathbf{U} \right) = \begin{bmatrix}
\rho W v^x\\
S_{x}^{\text{MHD}} v^x + p + \frac{b^2}{2} - \frac{b_x B^x}{W}\\
S_{y}^{\text{MHD}} v^x - \frac{b_y B^x}{W}\\
S_{z}^{\text{MHD}} v^x - \frac{b_z B^x}{W}\\
\left( \tau^{\text{MHD}} + p + \frac{b^2}{2} \right) v^x - \frac{b^t B^x}{W}\\
0\\
v^x B^y - v^y B^x\\
v^x B^z - v^z B^x
\end{bmatrix},
\end{equation}
where, to simplify notation, we have introduced the MHD form ${S_{i}^{\text{MHD}}}$ of the three-momentum density:

\begin{equation}
S_{j}^{\text{MHD}} = S_j + \frac{b^2 \rho W^2 v_j}{2} - b^t b_j,
\end{equation}
and the MHD form ${\tau^{\text{MHD}}}$ of the relativistic energy density:

\begin{equation}
\tau^{\text{MHD}} = \tau - \left( \rho W^2 + p \right) \frac{b^2}{2} - \left( b^t \right)^2.
\end{equation}
Once again, in order to match the HLLC Riemann solver of \cite{mignone_hllc_2006}, we transform our form of the MHD relativistic energy density from ${\tau^{\text{MHD}}}$ to ${\tau^{\text{MHD}} + \rho W}$, yielding:

\begin{equation}
\mathbf{U}^{\prime} = \begin{bmatrix}
\rho W\\
S_{x}^{\text{MHD}}\\
S_{y}^{\text{MHD}}\\
S_{z}^{\text{MHD}}\\
\tau^{\text{MHD}} + \rho W\\
B^x\\
B^y\\
B^z
\end{bmatrix}, \qquad \mathbf{F}^{\prime} \left( \mathbf{U}^{\prime} \right) = \begin{bmatrix}
\rho W v^x\\
S_{x}^{\text{MHD}} v^x + p + \frac{b^2}{2} - \frac{b_x B^x}{W}\\
S_{y}^{\text{MHD}} v^x - \frac{b_y B^x}{W}\\
S_{z}^{\text{MHD}} v^x - \frac{b_z B^x}{W}\\
S_{x}^{\text{MHD}}\\
0\\
v^x B^y - v^y B^x\\
v^x B^z - v^z B^x
\end{bmatrix}.
\end{equation}
The left and right wave-speeds ${S_L}$ and ${S_R}$ within the Riemann problem are computed using the same prescription of \cite{schneider_new_1993} that we used for multifluids, i.e. equations \ref{eq:left_wavespeed} and \ref{eq:right_wavespeed}, albeit now with ${\lambda_{\pm}}$ (i.e. the maximum and minimum eigenvalues of the flux Jacobian ${\frac{\partial \mathbf{F}^{\prime} \left( \mathbf{U}^{\prime} \right)}{\partial \mathbf{U}^{\prime}}}$) now corresponding to the maximum and minimum solutions of the following quartic in ${\lambda}$:

\begin{multline}
\rho h \left( 1 - c_{s}^{2} \right) \left( W \left( 1 - v^x \right) \right)^4 = \left( 1 - \lambda^2 \right)\\
\times \left[ \left( b_i b^i + \rho h c_{s}^{2} \right) \left( W \left( 1 - v^x \right) \right)^2 - c_{s}^{2} \left( b^x - \lambda b^t \right) \right],
\end{multline}
at least when the normal component of the magnetic field is non-vanishing, i.e. ${B^x \neq 0}$. Alternatively, when ${B^x = 0}$, the maximum and minimum eigenvalues ${\lambda_{\pm}}$ correspond instead to the maximum and minimum solutions of the quadratic:

\begin{multline}
\left( \rho h \left[ c_{s}^{2} + W^2 \left( 1 - c_{s}^{2} \right) \right] + \mathcal{Q} \right) \lambda^2 - \left( 2 \rho h W^2 v^x \left( 1 - c_{s}^{2} \right) \right) \lambda\\
+ \rho h \left[ - c_{s}^{2} + W^2 \left( v^x \right)^2 \left( 1 - c_{s}^{2} \right) \right] - \mathcal{Q} = 0,
\end{multline}
where we have defined:

\begin{equation}
\mathcal{Q} = b_i b^i - c_{s}^{2} \left( v^y B^y + v^z B^z \right)^2.
\end{equation}
Although the quartic in the ${B^x \neq 0}$ case could, in principle, be solved analytically, for reasons of numerical robustness we opt instead to apply a 1D Newton-Raphson method to approximate its roots iteratively; the quadratic in the ${B^x = 0}$ case, in contrast, is solved analytically. In all of the above, as before, ${c_s = \sqrt{\frac{\Gamma p}{\rho h}}}$ denotes the sound speed for an ideal gas. The estimate for the contact wave-speed ${S^{*}}$ is similarly given by the smallest solution of the following quadratic (again solved analytically):

\begin{multline}
\left( F_{\tau^{\text{MHD}} + \rho W}^{\text{HLL}} - U_{B^y}^{\text{HLL}} F_{B^y}^{\text{HLL}} - U_{B^z}^{\text{HLL}} F_{B^z}^{\text{HLL}} \right) \left( S^{*} \right)^2\\
- \left( F_{S_{x}^{\text{MHD}}}^{\text{HLL}} + U_{\tau_{\text{MHD}} + \rho W}^{\text{HLL}} - \left( U_{B^y}^{\text{HLL}} \right)^2 - \left( U_{B^z}^{\text{HLL}} \right)^2 \right.\\
\left. - \left( F_{B^y}^{\text{HLL}} \right)^2 - \left( F_{B^z}^{\text{HLL}} \right)^2 \right) S^{*}\\
+ \left( U_{S_{x}^{\text{MHD}}}^{\text{HLL}} - U_{B^y}^{\text{HLL}} F_{B^y}^{\text{HLL}} - U_{B^z}^{\text{HLL}} F_{B^z}^{\text{HLL}} \right) = 0,
\end{multline}
in the case ${B^x \neq 0}$ of non-vanishing normal component of the magnetic field, and by the smallest solution of the much simpler quadratic (also solved analytically):

\begin{equation}
F_{\tau^{\text{MHD}} + \rho W}^{\text{HLL}} \left( S^{*} \right)^2 - \left( F_{S_{x}^{\text{MHD}}}^{\text{HLL}} + U_{\tau_{\text{MHD}} + \rho W}^{\text{HLL}} \right) S^{*} + U_{S_{x}^{\text{MHD}}}^{\text{HLL}} = 0,
\end{equation}
in the case ${B^x = 0}$, where ${U_{S_{x}^{\text{MHD}}}^{\text{HLL}}}$, ${U_{\tau^{\text{MHD}} + \rho W}^{\text{HLL}}}$, ${U_{B^y}^{\text{HLL}}}$, and ${U_{B^z}^{\text{HLL}}}$ are the magnetohydrodynamic $x$-momentum density, magnetohydrodynamic relativistic energy density, $y$-magnetic field, and $z$-magnetic field components of the Harten-Lax-van Leer integral averaged state ${\mathbf{U}^{\text{HLL}}}$, given by equation \ref{eq:hll_integral_average}, respectively, and ${F_{S_{x}^{\text{MHD}}}^{\text{HLL}}}$, ${F_{\tau^{\text{MHD}} + \rho W}^{\text{HLL}}}$, ${F_{B^y}^{\text{HLL}}}$, and ${F_{B^z}^{\text{HLL}}}$ are the magnetohydrodynamic $x$-momentum density, magnetohydrodynamic relativistic energy density, $y$-magnetic field, and $z$-magnetic field components of the Harten-Lax-van Leer integral averaged flux ${\mathbf{F}^{\text{HLL}}}$, given by equation \ref{eq:hll_flux_integral_average}, respectively.

As before, we compute the Mignone-Bodo HLLC inter-cell flux ${\mathbf{F}_{i + \frac{1}{2}}^{\text{HLLC}}}$ using equation \ref{eq:hllc_intercell_flux}, where now the star state flux function ${\mathbf{F}^{*} \left( \mathbf{U} \right)}$ is given by:

\begin{equation}
\mathbf{F}^{*} \left( \mathbf{U} \right) = \begin{bmatrix}
\rho^{*} W^{*} v^{x, *}\\
S_{x}^{*} v^{x, *} - \frac{B^{x, *} B^{x, *}}{\left( W^{*} \right)^2} - B^{x, *} v^{x, *} \left( B_{i}^{*} v^{i, *} \right) + p^{*}\\
S_{y}^{*} v^{x, *} - \frac{B^{x, *} B^{y, *}}{\left( W^{*} \right)^2} - B^{x, *} v^{y, *} \left( B_{i}^{*} v^{i, *} \right)\\
S_{z}^{*} v^{x, *} - \frac{B^{x, *} B^{z, *}}{\left( W^{*} \right)^2} - B^{x, *} v^{z, *} \left( B_{i}^{*} v^{i, *} \right)\\
S_{x}^{*}\\
0\\
v^{x, *} B^{y, *} - v^{y, *} B^{x, *}\\
v^{x, *} B^{z, *} - v^{z, *} B^{x, *}
\end{bmatrix},
\end{equation}
and where the star state conserved variables ${\rho^{*} W^{*}}$, ${S_{y}^{*}}$, ${S_{z}^{*}}$, and ${\tau^{*} + \rho^{*} W^{*}}$ are again given by the Rankine-Hugoniot conditions:

\begin{equation}
\rho^{*} W^{*} = \frac{\rho W \left( \lambda - v^x \right)}{\lambda - S^{*}},
\end{equation}
\begin{equation}
S_{y}^{*} = \frac{- B^{x, *} \left[ \frac{B^{y, *}}{\left( W^{*} \right)^2} + \left( B_{i}^{*} v^{i, *} \right) v^{y, *} \right] + \lambda S_{y}^{*} - S_{y}^{\text{MHD}} v^x + \frac{b_y B^x}{W}}{\lambda - v_{x}^{*}},
\end{equation}
\begin{equation}
S_{z}^{*} = \frac{- B^{x, *} \left[ \frac{B^{z, *}}{\left( W^{*} \right)^2} + \left( B_{i}^{*} v^{i, *} \right) v^{z, *} \right] + \lambda S_{z}^{*} - S_{z}^{\text{MHD}} v^x + \frac{b_z B^x}{W}}{\lambda - v_{x}^{*}},
\end{equation}
and:

\begin{equation}
\left( \tau^{*} + \rho^{*} W^{*} \right) = \frac{\lambda \left( \tau^{\text{MHD}} + \rho W \right) - S_{x}^{\text{MHD}} + p^{*} S^{*} - \left( B_{i}^{*} v^{i, *} \right) B^{x, *}}{\lambda - S{*}},
\end{equation}
respectively, assuming a non-vanishing normal component of the magnetic field ${B^x \neq 0}$. The only remaining terms appearing within the star state flux ${\mathbf{F}^{*} \left( \mathbf{U} \right)}$ are the star state pressure ${p^{*}}$, the star state normal momentum ${S_{x}^{*}}$\footnote{Note that, in contrast to the purely hydrodynamic case, we must compute the normal component of the momentum separately since (unlike the normal component of the magnetic field) we cannot assume that it remains continuous across the interface.}, and the star state tangential velocities ${v^{y, *}}$ and ${v^{z, *}}$, which are computed as:

\begin{equation}
p^{*} = F_{S_{x}^{\text{MHD}}}^{\text{HLL}} + \left( \frac{B^{x, *}}{W^{*}} \right)^2 - \left[ F_{\tau^{\text{MHD}} + \rho W}^{\text{HLL}} - B^{x, *} \left( B_{i}^{*} v^{i, *} \right) \right] v^{x, *},
\end{equation}
\begin{equation}
S_{x}^{*} = \left( \left( \tau^{*} + \rho^{*} W^{*} \right) + p^{*} \right) v^{x, *} - \left( B_{i}^{*} v^{i, *} \right) B^{x, *},
\end{equation}
\begin{equation}
v^{y, *} = \frac{B^{y, *} v^{x, *} - F_{B^y}^{\text{HLL}}}{B^{x, *}},
\end{equation}
and:

\begin{equation}
v_{z, *} = \frac{B^{z, *} v^{x, *} - F_{B^z}^{\text{HLL}}}{B^{x, *}},
\end{equation}
respectively, where:

\begin{equation}
W^{*} = \frac{1}{\sqrt{1 - \left( v^{x, *} \right)^2 - \left( v^{y, *} \right)^2 - \left( v^{z, *} \right)^2}}.
\end{equation}
As with the multifluid equations, ${\lambda}$ in all of the above is equal to either ${S_L}$ (if one is evaluating ${\mathbf{F}_{L}^{*}}$) or ${S_R}$ (if one is evaluating ${\mathbf{F}_{R}^{*}}$), we assume that the normal component of the star state velocity is given by the contact wave-speed, i.e. ${v^{x, *} = S^{*}}$, and that the tangential components of the star state magnetic field are given by their respective Harten-Lax-van Leer integral averaged states, i.e. ${B^{y, *} = U_{B^y}^{\text{HLL}}}$ and ${B^{z, *} = U_{B^z}^{\text{HLL}}}$. As before, ${\left( \mathbf{U}_{i}^{n + 1} \right)^{\prime}}$ is computed from ${\left( \mathbf{U}_{i}^{n} \right)^{\prime}}$ using equation \ref{eq:conservative_update}, and then the Mignone-Bodo relativistic energy density variable is transformed back to our original relativistic energy definition by subtracting off the (updated) relativistic mass density, yielding ${\mathbf{U}_{i}^{n + 1}}$. In the case ${B^x = 0}$, the Rankine-Hugoniot conditions for the star state conserved variables ${\rho^{*} W^{*}}$, ${S_{y}^{*}}$, ${S_{z}^{*}}$, and ${\tau^{*} + \rho^{*} W^{*}}$ simplify to:

\begin{equation}
\rho^{*} W^{*} = \frac{\rho W \left( \lambda - v^x \right)}{\lambda - S^{*}},
\end{equation}
\begin{equation}
S_{y}^{*} = \frac{S_{y}^{\text{MHD}} \left( \lambda - v^x \right)}{\lambda - S^{*}}, \qquad S_{z}^{*} = \frac{S_{z}^{\text{MHD}} \left( \lambda - v^x \right)}{\lambda - S^{*}},
\end{equation}
and:

\begin{equation}
\left( \tau^{*} + \rho^{*} W^{*} \right) = \frac{\lambda \left( \tau^{\text{MHD}} + \rho W \right) - S_{x}^{\text{MHD}} = p^{*} S^{*}}{\lambda - S^{*}},
\end{equation}
respectively, with the star state pressure ${p^{*}}$ and star state normal momentum ${S_{x}^{*}}$ now also simplified to\footnote{We note that there is a typographical error in equation (48) of \cite{mignone_hllc_2006}, wherein $p$ is written instead of ${p^{*}}$.}:

\begin{equation}
p^{*} = F_{S_{x}^{\text{MHD}}}^{\text{HLL}} - F_{\tau^{\text{MHD}} + \rho W}^{\text{HLL}} v^{x, *},
\end{equation}
and:

\begin{equation}
S_{x}^{*} = \left( \left( \tau^{*} + \rho^{*} W^{*} \right) + p^{*} \right) v^{x, *},
\end{equation}
respectively, and with all other quantities being identical to the ${B^x \neq 0}$ case.

For the homogeneous part of the general relativistic Maxwell equations in the locally-flat tetrad basis (which, due to the manifest covariance of classical electrodynamics, are equivalent to the ordinary Maxwell equations), we have:

\begin{equation}
\mathbf{U} = \begin{bmatrix}
E^x\\
E^y\\
E^z\\
B^x\\
B^y\\
B^z\\
\end{bmatrix}, \qquad \text{ and} \qquad \mathbf{F} \left( \mathbf{U} \right) = \begin{bmatrix}
0\\
B^z\\
-B^y\\
0\\
-E^z\\
E^y
\end{bmatrix}.
\end{equation}
We solve these equations using an approximate Roe-type Riemann solver (\cite{roe_approximate_1981}), where the inter-cell flux ${\mathbf{F}_{i + \frac{1}{2}}^{\text{Roe}}}$ is given by:

\begin{equation}
\mathbf{F}_{i + \frac{1}{2}}^{\text{Roe}} = \frac{1}{2} \left[ \mathbf{F} \left( \mathbf{U}_i \right) + \mathbf{F} \left( \mathbf{U}_{i + 1} \right) \right] - \frac{1}{2} \sum_p \left\lvert \lambda_p \right\rvert \alpha_p \mathbf{r}_p,
\end{equation}
where ${\lambda_p}$ and ${\mathbf{r}_p}$ are the eigenvalues and (right) eigenvectors of the inter-cell \textit{Roe matrix} ${A \left( \mathbf{U}_i, \mathbf{U}_{i + 1} \right)}$, i.e. a linearized approximation of the flux Jacobian ${\frac{\partial \mathbf{F} \left( \mathbf{U} \right)}{\partial \mathbf{U}}}$:

\begin{equation}
\lim_{\mathbf{U}_i, \mathbf{U}_{i + 1} \to \mathbf{U}} \left[ A \left( \mathbf{U}_i, \mathbf{U}_{i + 1} \right) \right] = \frac{\partial \mathbf{F} \left( \mathbf{U} \right)}{\partial \mathbf{U}},
\end{equation}
and ${\alpha_p}$ are the coefficients of the representation of the inter-cell jump in the conserved variables, i.e. ${\mathbf{U}_{i + 1} - \mathbf{U}_i}$, in the ${\mathbf{r}_p}$ eigenbasis:

\begin{equation}
\mathbf{U}_{i + 1} - \mathbf{U}_i = \sum_p \alpha_p \mathbf{r}_p.
\end{equation}
However, since Maxwell's equations (at least in the flat spacetime coordinate basis) are themselves linear, no special averaging or linearization scheme is required, and we can use the eigendecomposition of the flux Jacobian ${\frac{\partial \mathbf{F} \left( \mathbf{U} \right)}{\partial \mathbf{U}}}$ directly, yielding eigenvalues ${\lambda{-} = -1}$, ${\lambda_0 = 0}$, and ${\lambda{+} = 1}$ (each with algebraic multiplicity 2), with the corresponding right eigenvectors:

\begin{multline}
\mathbf{r}_{-}^{1} = \begin{bmatrix}
0 & -1 & 0 & 0 & 0 & 1
\end{bmatrix}^{\intercal}, \qquad \mathbf{r}_{-}^{2} = \begin{bmatrix}
0 & 0 & 1 & 0 & 1 & 0
\end{bmatrix}^{\intercal}\\
\mathbf{r}_{0}^{1} = \begin{bmatrix}
0 & 0 & 0 & 1 & 0 & 0
\end{bmatrix}^{\intercal}, \qquad \mathbf{r}_{0}^{2} = \begin{bmatrix}
1 & 0 & 0 & 0 & 0 & 0
\end{bmatrix}^{\intercal}\\
\mathbf{r}_{+}^{1} = \begin{bmatrix}
0 & 1 & 0 & 0 & 0 & 1
\end{bmatrix}^{\intercal}, \qquad \mathbf{r}_{+}^{2} = \begin{bmatrix}
0 & 0 & -1 & 0 & 1 & 0
\end{bmatrix}^{\intercal}.
\end{multline}

Although all of the HLLC and Roe-type Riemann solvers described above are notionally first-order accurate in space, \textsc{Gkeyll} employs a wave-propagation scheme with higher-order corrections that is able to extend the spatial accuracy of the solution to second-order within all smooth regions (\cite{hakim_high_2006} and \cite{gorard_shock_2025}). Following the wave propagation formalism of \cite{leveque_finite_2011}, we rewrite the first-order hyperbolic update formula in equation \ref{eq:conservative_update} as:

\begin{equation}
\mathbf{U}_{i}^{n + 1} = \mathbf{U}_{i}^{n} - \frac{\Delta t}{\Delta x} \left[ \left( \mathcal{A}^{+} \Delta \mathbf{U} \right)_{i - \frac{1}{2}}^{n} + \left( \mathcal{A}^{-} \Delta \mathbf{U} \right)_{i + \frac{1}{2}} \right],
\end{equation}
where ${\left( \mathcal{A}^{\pm} \Delta \mathbf{U} \right)_{i \mp \frac{1}{2}}}$ represent the \textit{fluctuations}:

\begin{equation}
\left( \mathcal{A}^{+} \Delta \mathbf{U} \right)_{i - \frac{1}{2}} = \sum_{p : s_{i - \frac{1}{2}}^{p} > 0} \mathcal{Z}_{i - \frac{1}{2}}^{p} + \left( \frac{1}{2} \sum_{p : s_{i - \frac{1}{2}}^{p} = 0} \mathcal{Z}_{i - \frac{1}{2}}^{p} \right),
\end{equation}
and:

\begin{equation}
\left( \mathcal{A}^{-} \Delta \mathbf{U} \right)_{i + \frac{1}{2}} = \sum_{p : s_{i + \frac{1}{2}}^{p} < 0} \mathcal{Z}_{i + \frac{1}{2}}^{p} + \left( \frac{1}{2} \sum_{p : s_{i + \frac{1}{2}}^{p} = 0} \mathcal{Z}_{i + \frac{1}{2}}^{p} \right),
\end{equation}
with the sums in the above being taken over the positive, zero, and negative eigenvalues ${s^p}$ (indexed by $p$) of the flux Jacobian ${\frac{\partial \mathbf{F} \left( \mathbf{U} \right)}{\partial \mathbf{U}}}$, approximated at each cell boundary, and where ${\mathcal{Z}_{i \pm \frac{1}{2}}^{p}}$ are the corresponding \textit{waves}:

\begin{equation}
\mathcal{Z}_{i + \frac{1}{2}}^{p} = \left( \mathbf{l}_{i + \frac{1}{2}}^{p} \cdot \left[ \mathbf{F} \left( \mathbf{U}_{i + 1} \right) - \mathbf{F} \left( \mathbf{U}_i \right) \right] \right) \mathbf{r}_{i + \frac{1}{2}}^{p},
\end{equation}
and:

\begin{equation}
\mathcal{Z}_{i - \frac{1}{2}}^{p} = \left( \mathbf{l}_{i - \frac{1}{2}}^{p} \cdot \left[ \mathbf{F} \left( \mathbf{U}_i \right) - \mathbf{F} \left( \mathbf{U}_{i - 1} \right) \right] \right) \mathbf{r}_{i - \frac{1}{2}}^{p},
\end{equation}
with ${\mathbf{l}^p}$ and ${\mathbf{r}^p}$ being the left and right eigenvectors of the flux Jacobian ${\frac{\partial \mathbf{F} \left( \mathbf{U} \right)}{\partial \mathbf{U}}}$, respectively. Such an update formula can be extended to second-order spatial accuracy by introducing the \textit{correction fluxes} ${\widetilde{\mathcal{F}}_{i \pm \frac{1}{2}}}$ (i.e. higher-order terms in the Taylor expansion for the conserved variable vector ${\mathbf{U}}$) as follows:

\begin{multline}
\mathbf{U}_{i}^{n + 1} = \mathbf{U}_{i}^{n} - \frac{\Delta t}{\Delta x} \left[ \left( \mathcal{A}^{+} \Delta \mathbf{U} \right)_{i - \frac{1}{2}}^{n} + \left( \mathcal{A}^{-} \Delta \mathbf{U} \right)_{i + \frac{1}{2}} \right]\\
- \frac{\Delta t}{\Delta x} \left[ \widetilde{\mathcal{F}}_{i + \frac{1}{2}} - \widetilde{\mathcal{F}}_{i - \frac{1}{2}} \right],
\end{multline}
where the correction fluxes ${\widetilde{\mathcal{F}}_{i \pm \frac{1}{2}}}$ themselves are given by:

\begin{equation}
\widetilde{\mathcal{F}}_{i \pm \frac{1}{2}} = \frac{1}{2} \sum_p \left[ \mathrm{sgn} \left( s_{i \pm \frac{1}{2}}^{p} \right) \left( 1 - \frac{\Delta t}{\Delta x} \left\lvert s_{i \pm \frac{1}{2}}^{p} \right\rvert \right) \mathcal{Z}_{i \pm \frac{1}{2}}^{p} \phi \left( \theta_{i \pm \frac{1}{2}}^{p} \right) \right].
\end{equation}
In the above, ${\phi \left( \theta_{i \pm \frac{1}{2}}^{p} \right)}$ represents a \textit{flux limiter} function, intended to reduce the spatial accuracy of the scheme to first-order in the presence of sharp gradients or genuine discontinuities, to prevent spurious oscillations, where:

\begin{equation}
\theta_{i - \frac{1}{2}}^{p} = \frac{\mathcal{Z}_{i - \frac{3}{2}}^{p} \cdot \mathcal{Z}_{i - \frac{1}{2}}^{p}}{\left\lVert \mathcal{Z}_{i - \frac{1}{2}}^{p} \right\rVert^2}, \qquad \text{ and } \qquad \theta_{i + \frac{1}{2}}^{p} = \frac{\mathcal{Z}_{i - \frac{1}{2}}^{p} \cdot \mathcal{Z}_{i + \frac{1}{2}}^{p}}{\left\lVert \mathcal{Z}_{i + \frac{1}{2}}^{p} \right\rVert^2},
\end{equation}
if ${s_{i \pm \frac{1}{2}}^{p} > 0}$, and:

\begin{equation}
\theta_{i - \frac{1}{2}}^{p} = \frac{\mathcal{Z}_{i + \frac{1}{2}}^{p} \cdot \mathcal{Z}_{i - \frac{1}{2}}^{p}}{\left\lVert \mathcal{Z}_{i - \frac{1}{2}}^{p} \right\rVert^2}, \qquad \text{ and } \qquad \theta_{i + \frac{1}{2}}^{p} = \frac{\mathcal{Z}_{i + \frac{3}{2}}^{p} \cdot \mathcal{Z}_{i + \frac{1}{2}}^{p}}{\left\lVert \mathcal{Z}_{i + \frac{1}{2}}^{p} \right\rVert^2},
\end{equation}
if ${s_{i \pm \frac{1}{2}}^{p} < 0}$. For most of the 1D validation tests and 3D accretion problems presented within this paper, we opt to use the \textit{monotonized-centered} flux limiter ${\phi_{\text{MC}}}$ by default:

\begin{equation}
\phi_{\text{MC}} \left( \theta \right) = \max \left( 0, \min \left( 2 \theta, \frac{1}{2} \left( 1 + \theta \right), 2 \right) \right).
\end{equation}
On the other hand, in cases where ${\phi_{\text{MC}}}$ is found to be too oscillatory/dispersive (e.g. for the strongly magnetized variant of the relativistic Noh problem, or the strongly magnetized accretion problems onto neutron stars), we fall back to the marginally more diffusive but significantly more stable \textit{minmod} flux limiter ${\phi_{\text{MM}}}$ instead:

\begin{equation}
\phi_{\text{MM}} \left( \theta \right) = \max \left( 0, \min \left( 1, \theta \right) \right).
\end{equation}
Likewise, in the event that the higher-resolution HLLC or Roe-type Riemann solvers result in a physically invalid state (i.e. by driving ${\rho_s < 0}$, ${p_s < 0}$, or ${v_{s, i} v_{s}^{i} > 1}$ in the general relativistic multifluid case, or by driving ${\rho < 0}$, ${p < 0}$, or ${v_i v^i > 1}$ in the GRMHD case) somewhere in the domain, \textsc{Gkeyll} will automatically fall back locally to using the more diffusive Lax-Friedrichs finite-difference approximation to the inter-cell flux ${\mathbf{F}_{i + \frac{1}{2}}^{\text{LF}}}$, given by:

\begin{equation}
\mathbf{F}_{i + \frac{1}{2}}^{\text{LF}} = \frac{1}{2} \left[ \mathbf{F} \left( \mathbf{U}_i \right) + \mathbf{F} \left( \mathbf{U}_{i + 1} \right) \right] - \frac{\Delta x}{2 \Delta t} \left( \mathbf{U}_{i + 1} - \mathbf{U}_i \right),
\end{equation}
in order to facilitate recovery from the unphysical state.

Once the homogeneous part of equation \ref{eq:flux_conservative} has been solved via the high-resolution shock-capturing methods described above, all that remains is to integrate the \textit{inhomogeneous} part, i.e. the source term ordinary differential equation(s):

\begin{equation}
\frac{d \mathbf{U}}{d t} = \mathbf{S} \left( \mathbf{U} \right),
\end{equation}
which, for our purposes, simultaneously encompass the geometric source terms resulting from the coupling between fluid species and the underlying spacetime geometry, the electromagnetic source terms responsible for coupling the current density of the electric field and the momentum density of each fluid species, and any artificial source terms introduced for the purpose of correcting divergence errors in the electric and magnetic fields as described in \ref{sec:divergence_correction}. For this purpose, we opt to use the strong stability-preserving Runge-Kutta (SSP-RK) scheme of \cite{gottlieb_strong_2001}, already implemented and robustly validated within \textsc{Gkeyll}. We use, specifically, the four-stage, third-order SSP-RK3 scheme, in which the value ${f^{n + 1}}$ of the integrated function $f$ at time ${t^{n + 1} = t^n + \Delta t}$ is calculated from its value ${f^n}$ at time ${t^n}$ via the following intermediate stages:

\begin{equation}
f^{\left( 1 \right)} = \frac{1}{2} f^n + \frac{1}{2} \mathcal{F} \left[ f^n, t^n \right],
\end{equation}
\begin{equation}
f^{\left( 2 \right)} = \frac{1}{2} f^{\left( 1 \right)} + \frac{1}{2} \mathcal{F} \left[ f^{\left( 1 \right)}, t^n + \frac{\Delta t}{2} \right],
\end{equation}
\begin{equation}
f^{\left( 3 \right)} = \frac{2}{3} f^n + \frac{1}{6} f^{\left( 2 \right)} + \frac{1}{6} \mathcal{F} \left[ f^{\left( 2 \right)}, t^n + \Delta t \right],
\end{equation}
and finally:

\begin{equation}
f^{n + 1} = \frac{1}{2} f^{\left( 3 \right)} + \frac{1}{2} \mathcal{F} \left[ f^{\left( 3 \right)}, t^n + \frac{\Delta t}{2} \right],
\end{equation}
where the operator ${\mathcal{F}}$ corresponds to a first-order forward-Euler integration step of the right-hand side of the source term ordinary differential equation, i.e:

\begin{equation}
\mathcal{F} \left[ f, t \right] = f + \Delta t \text{RHS} \left[ f, t \right].
\end{equation}
We now combine the second-order hyperbolic update step with the third-order source term integration step using a second-order Strang splitting (\cite{strang_construction_1968}). Thus, in order to take a single ${\Delta t}$ time-step within a 1D simulation, we first take a ${\frac{\Delta t}{2}}$ step with the source term integrator, then a ${\Delta t}$ step with the hyperbolic update, then another ${\frac{\Delta t}{2}}$ step with the source term integrator. For a 2D simulation, the ${\Delta t}$ step with the hyperbolic update is further split into a ${\frac{\Delta t}{2}}$ step in the $x$-direction, then a ${\Delta t}$ step in the $y$-direction, then another ${\frac{\Delta t}{2}}$ step in the $x$-direction, and so on. Note that we choose to integrate the source terms explicitly in conserved variable form for simplicity and generality; however, in future work, we intend to explore the possibility of using either fully implicit or hybrid implicit-explicit (IMEX) integrators for handling the multifluid source terms, in order to circumvent the significant time-step restrictions resulting from the short plasma frequency timescales. For this purpose, it will be useful instead to integrate the source terms in their primitive variable form, as outlined in Appendix \ref{sec:primitive_variable_appendix}, although this reduces the generality of the formalism by enforcing a particular choice of equation of state.

Finally, in order to compute the maximum stable time-step ${\Delta t}$ in accordance with the CFL stability criterion ${0 \leq \frac{\left\lvert a \right\rvert \Delta t}{\Delta x} = C_{CFL} \leq 1}$, we must construct an estimate of the largest absolute eigenvalue ${\left\lvert a \right\rvert}$ of the flux Jacobian ${\frac{\partial \mathbf{F} \left( \mathbf{U} \right)}{\partial \mathbf{U}}}$ for each of our equation systems. For the hydrodynamic part of the general relativistic multifluid equations, we evaluate the eigenvalues analytically, following \cite{anile_relativistic_1990} and \cite{banyuls_numerical_1997}, yielding (in the $x$-direction) ${\lambda_0 = \alpha v_{s}^{x} - \beta^x}$, with algebraic multiplicity 3, representing the three material waves, and:

\begin{multline}
\lambda_{\pm} = \frac{\alpha}{1 - v_{i, s} v_{s}^{i} c_{s}^{2}} \left[ v_{s}^{x} \left( 1 - c_{s}^{2} \right) \right.\\
\left. \pm c_s \sqrt{\left(1 - v_{i, s} v_{s}^{i} \right) \left[ \gamma^{x x} \left( 1 - v_{i, s} v_{s}^{i} c_{s}^{2} \right) - v^x v^x \left( 1 - c_{s}^{2} \right) \right]} \right],
\end{multline}
each with algebraic multiplicity 1, representing the two acoustic waves. Likewise, for the general relativistic Maxwell equations, the eigenvalues are straightforward to evaluate analytically, yielding (in the $x$-direction) ${\lambda_{-} = -\alpha - \beta^x}$, ${\lambda_0 = 0}$, and ${\lambda_{+} = \alpha - \beta^x}$, each with algebraic multiplicity 1. However, the eigensystem for the GRMHD equations is significantly more complicated. Analytically, we again have (in the $x$-direction) ${\lambda_0 = \alpha v^x - \beta^x}$, with algebraic multiplicity 2, representing the two material waves, and:

\begin{equation}
\lambda_{\pm} = \frac{b^x \pm \sqrt{\left( \rho h + b^2 \right)} u^x}{b^t \pm \sqrt{\left( \rho h + b^2 \right)} u^t},
\end{equation}
each with algebraic multiplicity 1, representing the left- and right-going Alfv\'en waves. Yet, as shown by \cite{anile_relativistic_1990} and \cite{anton_numerical_2006}, the four possible speeds for the fast and slow magnetoacoustic waves are given by the solutions to the following quartic in ${\lambda}$:

\begin{equation}
\rho h \left( \frac{1}{c_{s}^{2}} - 1 \right) a^4 - \left( \rho h + \frac{b^2}{c_{s}^{2}} \right) a^2 G + \mathcal{B}^2 G = 0,
\end{equation}
where we have defined:

\begin{equation}
a = \frac{W}{\alpha} \left( - \lambda + \alpha v^x - \beta^x \right), \qquad \mathcal{B} = b^x - b^t \lambda,
\end{equation}
and:

\begin{equation}
G = \frac{1}{\alpha^2} \left[ - \left( \lambda + \beta^x \right)^2 + \alpha^2 \gamma^{x x} \right].
\end{equation}
Rather than attempting to solve this quartic directly, either analytically or using iterative methods, we instead use the following ``hydrodynamics-like'' approximation due to \cite{anton_numerical_2006} and \cite{porth_black_2017}:

\begin{multline}
\frac{1}{1 - v_i v^i k^2} \left[ \left( 1 - k^2 \right) v^x \right.\\
\left. \pm \sqrt{k^2 \left( 1 - v_i v^i \right) \left[ \left( 1 - v_i v^i k^2 \right) \gamma^{x x} - \left( 1 - k^2 \right) v^x v^x \right]} \right],
\end{multline}
where we have defined:

\begin{equation}
k = \sqrt{c_{s}^{2} + c_{a}^{2} - c_{s}^{2} c_{a}^{2}},
\end{equation}
and where ${c_a}$ is an approximation of the Alfv\'en wave-speed:

\begin{equation}
c_{a}^{2} = \frac{b^2}{\rho h + b^2}.
\end{equation}
To deal with the case in which the source terms for the general relativistic multifluid system are stiff, and therefore in which the stable time-step for the overall update is limited instead by the stability of the SSP-RK3 integrator rather than by the stability of the hyperbolic update (for instance because the magnetization is very large, and/or the Larmor radii of the fluid species are very small), we apply Von Mises/power iteration (\cite{mises_praktische_1929}) to approximate largest absolute eigenvalue of the source Jacobian ${\frac{\partial \mathbf{S} \left( \mathbf{U} \right)}{\partial \mathbf{U}}}$. We then use this approximation of the largest absolute eigenvalue to compute a smaller stable time-step for the SSP-RK3 integrator, and we subcycle the integration step accordingly.

\subsection{Primitive Variable Reconstruction}
\label{sec:primitive_variable_reconstruction}

Within the general relativistic multifluid solver, in order to recover the primitive fluid variables ${\rho_s}$ (rest mass density), ${v_{s}^{i}}$ (three-velocity), and ${p_s}$ (pressure) from the conserved variables ${\rho_s W_s}$ (relativistic mass density), ${S_{i, s}}$ (three-momentum density), and ${\tau_s}$ (relativistic energy density), we use the approach of \cite{eulderink_general_1994}, assuming that all fluid species obey an ideal gas equation of state, in which the specific relativistic enthalpy ${h_s}$ is given by:

\begin{equation}
h_s = 1 + \frac{p_s}{\rho_s} \left( \frac{\Gamma_s}{\Gamma_s - 1} \right),
\end{equation}
with ${\Gamma_s}$ being the adiabatic index of the species. Within their prescription, the primitive variable reconstruction operation corresponds to the solution of the following quartic in ${\xi}$:

\begin{equation}
\alpha_4 \xi^3 \left( \xi - \eta \right) + \alpha_2 \xi^2 + \alpha_1 \xi + \alpha_0 = 0,
\end{equation}
where the variable ${\xi}$ is defined by:

\begin{equation}
\xi = \frac{\sqrt{-g_{\mu \nu} T_{\text{Fluid}, s}^{t \mu} T_{\text{Fluid}, s}^{t \nu}}}{\rho_s h_s u_{s}^{t}} = \frac{\sqrt{\left( \tau_s + \rho_s W_s \right)^2 - S_{i, s} S_{s}^{i}}}{\rho_s h_s W_s},
\end{equation}
the constant ${\eta}$ by:

\begin{equation}
\eta = \frac{2 \rho_s u_{s}^{t} \left( \Gamma_s - 1 \right)}{\left( \sqrt{-g_{\mu \nu} T_{\text{Fluid}, s}^{t \mu} T_{\text{Fluid}, s}^{t \nu}} \right) \Gamma_s} = \frac{2 \rho_s W_s \left( \Gamma_s - 1 \right)}{\Gamma_s \sqrt{\left( \tau_s + \rho_s W_s \right)^2 - S_{i, s} S_{s}^{i}}},
\end{equation}
and the coefficients ${\alpha_0}$, ${\alpha_1}$, ${\alpha_2}$, and ${\alpha_4}$ by:

\begin{equation}
\alpha_0 = - \frac{1}{\Gamma_{s}^{2}},
\end{equation}
\begin{equation}
\alpha_1 = - \frac{2 \rho_s u_{s}^{t} \left( \Gamma_s - 1 \right)}{\left( \sqrt{-g_{\mu \nu} T_{\text{Fluid}, s}^{t \mu} T_{\text{Fluid}, s}^{t \nu}} \right) \Gamma_{s}^{2}} = - \frac{2 \rho_s W_s \left( \Gamma_s - 1 \right)}{\Gamma_{s}^{2} \sqrt{\left( \tau_s + \rho_s W_s \right)^2 - S_{i, s} S_{s}^{i}}},
\end{equation}
\begin{multline}
\alpha_2 = \left( \frac{\Gamma_s - 2}{\Gamma_s} \right) \left( \frac{\left( T_{\text{Fluid}, s}^{t t} \right)^2}{g^{t t} g_{\mu \nu} T_{\text{Fluid}, s}^{t \mu} T_{\text{Fluid}, s}^{t \nu}} - 1 \right) + 1\\
+ \left( \frac{\left( \rho_s u_{s}^{t} \right)^2}{g_{\mu \nu} T_{\text{Fluid}, s}^{t \mu} T_{\text{Fluid}, s}^{t \nu}} \right) \left( \frac{\Gamma_s - 1}{\Gamma_s} \right)^2\\
= \left( \frac{\Gamma_s - 2}{\Gamma_s} \right) \left( \frac{\left( \tau_s + \rho_s W_s \right)^2}{\left( \tau_s + \rho_s W_s \right)^2 - S_{i, s} S_{s}^{i}} - 1 \right) + 1\\
- \frac{\left( \rho_s W_s \right)^2 \left( \Gamma_s - 1 \right)^2}{\Gamma_{s}^{2} \left( \left( \tau_s + \rho_s W_s \right)^2 - S_{i, s} S_{s}^{i} \right)},
\end{multline}
and:

\begin{equation}
\alpha_4 = \frac{\left( T_{\text{Fluid}, s}^{t t} \right)^2}{g^{t t} g_{\mu \nu} T_{\text{Fluid}, s}^{t \mu} T_{\text{Fluid}, s}^{t \nu}} - 1 = \frac{\left( \tau_s + \rho_s W_s \right)^2}{\left( \tau_s + \rho_s W_s \right)^2 - S_{i, s} S_{s}^{i}} - 1,
\end{equation}
respectively. As before, we opt to use a 1D Newton-Raphson method to approximate the roots of this quartic iteratively, rather than attempting to solve it analytically. Once an approximate numerical value for ${\xi}$ has been calculated with the desired precision, the Lorentz factor ${W_s}$ may first be recovered as:

\begin{multline}
W_s = \frac{1}{2} \left( \frac{\tau_s + \rho_s W_s}{\sqrt{\left( \tau_s + \rho_s W_s \right)^2 - S_{i, s} S_{s}^{i}}} \right) \xi\\
\times \left( 1 + \sqrt{1 + 4 \left( \frac{\Gamma_s - 1}{\Gamma_s} \right) \left( \frac{1 - \left( \frac{\rho_s W_s}{\sqrt{\left( \tau_s + \rho_s W_s \right)^2 - S_{i, s} S_{s}^{i}}} \right) \xi}{\left( \frac{\left( \tau_s + \rho_s W_s \right)^2}{\left( \tau_s + \rho_s W_s \right)^2 - S_{i, s} S_{s}^{i}} \right) \xi^2} \right)} \right),
\end{multline}
from which the rest mass density ${\rho_s = \frac{\rho_s W_s}{W_s}}$ immediately follows, and from which the specific relativistic enthalpy ${h_s}$:

\begin{equation}
h_s = \frac{1}{\left( \frac{\rho_s W_s}{\sqrt{\left( \tau_s + \rho_s W_s \right)^2 - S_{i, s} S_{s}^{i}}} \right) \xi},
\end{equation}
can also be recovered. The pressure ${p_s}$ can be reconstructed from the equation of state, once the density ${\rho_s}$ and enthalpy ${h_s}$ are known. Finally, the 1-form components of the three-velocity ${v_{i, s}}$ are given by:

\begin{equation}
v_{i, s} = \frac{S_{i, s}}{\rho_s h_s W_{s}^{2}}.
\end{equation}
We impose a floor of ${10^{-8}}$ on the rest mass density ${\rho_s}$ and pressure ${p_s}$ (i.e. ${\rho_s \geq 10^{-8}}$, ${p_s \geq 10^{-8}}$), as well as a ceiling of ${\left( 1 - 10^{-8} \right)}$ on the square of the three-velocity ${v_{i, s} v_{s}^{i}}$ (i.e. ${v_{i, s} v_{s}^{i} \leq 1 - 10^{-8}}$), during this reconstruction step; if any of these floors or ceilings are attained, then all other primitive fluid variables are recalculated accordingly.

On the other hand, within the GRMHD solver, we will use two different reconstruction schemes in order to recover the primitive variables ${\rho}$ (rest mass density), ${v^i}$ (three-velocity), and $p$ (pressure) from the conserved variables ${\rho W}$ (relativistic mass density), ${S_{i}^{\text{MHD}}}$ (magnetohydrodynamic three-momentum density), ${\tau^{\text{MHD}}}$ (magnetohydrodynamic relativistic energy density), and ${B^i}$ (magnetic field). Here, we have reintroduced the notation ${S_{i}^{\text{MHD}}}$ and ${\tau^{\text{MHD}}}$ from the previous section, albeit now in their curved spacetime forms, namely:

\begin{equation}
S_{i}^{\text{MHD}} = \left( \rho h + b^2 \right) W^2 v_i - \alpha b^t b_i,
\end{equation}
and:

\begin{equation}
\tau^{\text{MHD}} = \left( \rho h + b^2 \right) W^2 - \left( p + \frac{b^2}{2} \right) - \alpha^2 \left( b^t \right)^2 - \rho W,
\end{equation}
respectively. The first (default) method we use is the 2D Newton-Raphson method of \cite{noble_primitive_2006}, which we note, based on the results of \cite{siegel_recovery_2018}, exhibits the ability to perform robust primitive variable reconstructions across a wide range of values of both the magnetization and the Lorentz factor, at least for ideal gas equations of state. However, since convergence of the 2D \cite{noble_primitive_2006} method is known to be less reliable in the limit of large Lorentz factors (e.g. ${10 < W < 100}$), we also implement the effective 1D method of \cite{newman_primitive_2014}, as a fallback in the event that the 2D \cite{noble_primitive_2006} method does not converge, since the effective 1D \cite{newman_primitive_2014} method has been shown to converge rapidly even for very large Lorentz factors (e.g. ${W > 100}$). For the default 2D \cite{noble_primitive_2006} method, we apply a 2D Newton-Raphson iteration to solve the following coupled system of equations in the unknowns ${z = \rho h W^2}$ and ${v^2 = v_i v^i}$:

\begin{multline}
v^2 \left( B_i B^i + z \right)^2 - \frac{\left( B^i S_{i}^{\text{MHD}} \right)^2 \left( B_i B^i + 2z \right)}{z^2}\\
- S_{i}^{\text{MHD}} S^{i, \text{MHD}} = 0,
\end{multline}
and:

\begin{multline}
\tau^{\text{MHD}} + \rho W - \frac{B_i B^i}{2} \left( 1 + v^2 \right) + \frac{\left( B^i S_{i}^{\text{MHD}} \right)^2}{2 z^2} - z\\
+ p \left( z, v^2 \right) = 0,
\end{multline}
where ${p \left( z, v^2 \right)}$ is the pressure, calculated as a function of the two unknowns, which for the case of ideal gas equation of state takes the form:

\begin{equation}
p \left( z, v^2 \right) = \left( \frac{\Gamma - 1}{\Gamma} \right) \left( h - 1 \right) \rho, \label{eq:ideal_gas_pressure}
\end{equation}
where:

\begin{equation}
W = \frac{1}{\sqrt{1 - v^2}}, \qquad \rho = \frac{\rho W}{W}, \qquad \text{ and } \qquad h = \frac{z}{\rho W^2}. \label{eq:ideal_gas_primitives}
\end{equation}
Once the value of the pressure $p$ has adequately converged (and the density ${\rho}$ computed in accordance with the prescription above), the only remaining primitive variables are the components of the three-velocity ${v^i}$, which can be recovered as:

\begin{equation}
v^i = \frac{\gamma^{i j} S_{j}^{\text{MHD}}}{z + B_i B^i} + \frac{\left( B^j S_{j}^{\text{MHD}} \right) B^i}{z \left( z + B_i B^i \right)}. \label{eq:ideal_gas_velocity}
\end{equation}

In the event that this 2D Newton-Raphson iteration does not converge on a pressure, we fall back to the effective 1D prescription of \cite{newman_primitive_2014}, in which we solve the following cubic equation in ${\varepsilon}$:

\begin{equation}
\varepsilon^3 + a \varepsilon^2 + d = 0,
\end{equation}
with the variable ${\varepsilon}$ defined by:

\begin{equation}
\varepsilon = B_i B^i + \rho h W^2,
\end{equation}
and the coefficients $a$ and $d$ by:

\begin{equation}
a = \tau^{\text{MHD}} + \rho W + \frac{B_i B^i}{2},
\end{equation}
and:

\begin{equation}
d = \frac{1}{2} \left( \frac{\left( B^i v_i \right)^2}{\sqrt{\gamma}} B_i B^i - \frac{\left( B^i S_{i}^{\text{MHD}} \right)^2}{\gamma} \right),
\end{equation}
respectively. The three solutions to this cubic can now be written as:\footnote{We note that there is a typographical error in equation (56) of \cite{siegel_recovery_2018}, wherein a factor of $a$ is omitted from the first term (i.e. they write ${\frac{1}{3}}$ rather than ${\frac{1}{3} a}$).}

\begin{equation}
\varepsilon^{\left( l \right)} = \frac{1}{3} a - \frac{2}{3} a \cos \left( \frac{2}{3} \phi + \frac{2}{3} l \pi \right),
\end{equation}
where:

\begin{equation}
\phi = \arccos \left( \frac{1}{a} \sqrt{\frac{27 d}{4 a}} \right).
\end{equation}
The physical solution then corresponds to the ${l = 1}$ case. The same variable ${z = \rho h W^2}$ from the 2D \cite{noble_primitive_2006} scheme can then be computed from this solution as ${z = \varepsilon - B_i B^i}$, with ${v^2 = v_i v^i}$ now being given by:

\begin{equation}
v^2 = \frac{\sqrt{\gamma} \left( B^i v_i \right)^2 z^2 + \left( B^i S_{i}^{\text{MHD}} \right)^2 \left( B_i B^i + 2z \right)}{z^2 \left( B_i B^i + z \right)^2 \gamma}.
\end{equation}
The reconstruction of the primitive variables $p$, ${\rho}$, and ${v^i}$ from $z$ and ${v^2}$ can now be performed in exactly the same way as for the 2D \cite{noble_primitive_2006} prescription, namely using equations \ref{eq:ideal_gas_pressure}, \ref{eq:ideal_gas_primitives}, and \ref{eq:ideal_gas_velocity}, respectively. As previously done in the case of the primitive variable reconstruction for general relativistic multifluids, we impose floors of ${10^{-8}}$ on density ${\rho}$ and pressure $p$ (${\rho > 10^{-8}}$, ${p > 10^{-8}}$), and a ceiling of ${1 - 10^{-8}}$ on the square of the three-velocity ${v_i v^i}$ (${v_i v^i < 1 - 10^{-8}}$), and if any of these floors or ceilings are attained, we recalculate all other primitive fluid variables accordingly.

\subsection{Divergence Error Correction}
\label{sec:divergence_correction}

Although, within the preceding subsections, we have described how the hyperbolic evolution equations governing both the general relativistic multifluid and GRMHD systems can be solved (using finite volume methods coupled to explicit source term integrators), the preservation of the elliptic constraint equations:

\begin{equation}
{}^{\left( 3 \right)} \nabla_i B^i = \partial_i B^i = 0, \qquad \text{ and } \qquad {}^{\left( 3 \right)} \nabla_i D^i = \partial_i D^i = \rho_c,
\end{equation}
for the multifluid system, and:

\begin{equation}
\partial_i \left( \sqrt{\gamma} B^i \right) = 0,
\end{equation}
for the GRMHD system, has not yet been addressed. Analytically, so long as the initial data satisfy these constraint equations exactly, they will remain satisfied indefinitely. However, due to the presence of finite truncation errors, this property cannot be guaranteed numerically, and in practice any hyperbolic evolution scheme will have a tendency to amplify any small initial constraint violations until they eventually grow sufficiently large in amplitude that they destabilize the scheme. To this end, we employ various hyperbolic divergence cleaning and divergence correction approaches within \textsc{Gkeyll}, wherein the underlying hyperbolic evolution equations are modified so as to control and damp any growth in these constraint violations. We briefly outline the algorithms we use here.

For correcting divergence errors within the GRMHD equations in the locally-flat tetrad basis, we first apply a relativistic extension of the \textit{8-wave} cleaning prescription of \cite{powell_solution-adaptive_1999}, as extended to special and general relativity by \cite{wu_entropy_2020}, \cite{liebling_evolutions_2010}, and others. We introduce a new ${-\alpha \left( \partial_i \left( \sqrt{\gamma} B^i \right) \right) \left( \frac{B_j}{W^2} + v_j B^k v_k \right)}$ source term into the three-momentum density evolution equation:\footnote{We note that the gauge variables ${\alpha}$ and ${\beta^i}$, as well as the spatial metric determinant term ${\sqrt{\gamma}}$, still appear in these equations, despite their transformation into the locally-flat tetrad basis, due to the need to adjust signal propagation speeds with respect to the local speed of light.}

\begin{multline}
\partial_t \left( S_j + \frac{b^2 \rho W^2 v_j}{2} - b^t b_j \right)\\
+ \partial_i \left[ \left( S_j + \frac{b^2 \rho W^2 v_j}{2} - b^t b_j \right) v^i + \left( p + \frac{b^2}{2} \right) \delta_{j}^{i} - \frac{b_j B^i}{W} \right]\\
= T^{t t} \left( \frac{1}{2} \beta^k \beta^l \partial_j \gamma_{k l} - \alpha \partial_j \alpha \right) + T^{t i} \beta^k \partial_j \gamma_{i k}\\
- \left( \frac{T_{\mu \nu} n^{\mu} \perp_{k}^{\nu}}{\alpha} \right) \partial_j \beta^k - \alpha \left( \partial_i \left( \sqrt{\gamma} B^i \right) \right) \left( \frac{B_j}{W^2} + v_j B^k v_k \right),
\end{multline}
a new ${- \alpha \left( \partial_i \left( \sqrt{\gamma} B^i \right) \right) B^k v_k}$ source term into the relativistic energy density evolution equation:

\begin{multline}
\partial_t \left( \tau - \left( \rho W^2 + p \right) \frac{b^2}{2} - \left( b^t \right)^2 \right)\\
+ \partial_i \left[ \left( \tau - \left( \rho W^2 + p \right) \frac{b^2}{2} - \left( b^t \right)^2 + p + \frac{b^2}{2} \right) v^i - \frac{b^t B^i}{W} \right]\\
= T^{t t} \left( \beta^i \beta^j K_{i j} + \beta^i \partial_i \alpha \right) + T^{t i} \left( - \partial_i \alpha + 2 \beta^j K_{i j} \right) + T^{i j} K_{i j}\\
- \alpha \left( \partial_i \left( \sqrt{\gamma} B^i \right) \right) B^k v_k,
\end{multline}
and a new ${- \alpha \left( \partial_i \left( \sqrt{\gamma} B^i \right) \right) \left( v^k - \frac{\beta^k}{\alpha} \right)}$ source term into the magnetic field evolution equation:

\begin{equation}
\partial_i B^k + \partial_i \left( v^i B^k - v^k B^i \right) = -\alpha \left( \partial_i \left( \sqrt{\gamma} B^i \right) \right) \left( v^k - \frac{\beta^k}{\alpha} \right).
\end{equation}
In all of the above, we evaluate the divergence term ${\partial_i \left( \sqrt{\gamma} B^i \right)}$ using a simple first-order finite-difference approximation. However, we note, following \cite{fedrigo_general_2025}, that although the relativistic extension of the 8-wave prescription of \cite{powell_solution-adaptive_1999} is sufficient to grant numerical stability, it is not necessarily sufficient to guarantee correctness of the results. Therefore, to ensure correctness, we supplement this prescription with the hyperbolic divergence cleaning approach of \cite{dedner_hyperbolic_2002}, extended to special and general relativity by \cite{neilsen_relativistic_2006}, \cite{mosta_grhydro:_2014}, and others. We introduce an auxiliary scalar field ${\psi}$ whose purpose is to advect any divergence errors out of the domain with the maximum possible speed, while simultaneously damping them to zero as quickly as possible. Following \cite{mosta_grhydro:_2014}, we modify the homogeneous Maxwell equations accordingly:

\begin{multline}
{}^{\left( 4 \right)} \nabla_{\nu} \left( {}^{\star} F^{\mu \nu} + g^{\mu \nu} \frac{\psi}{c_h} \right) = \partial_{\nu} \left( {}^{\star} F^{\mu \nu} + g^{\mu \nu} \frac{\psi}{c_h} \right)\\
+ {}^{\left( 4 \right)} \Gamma_{\nu \sigma}^{\mu} \left( {}^{\star} F^{\sigma \nu} + g^{\sigma \nu} \frac{\psi}{c_h} \right) + {}^{\left( 4 \right)} \Gamma_{\nu\sigma}^{\nu} \left( {}^{\star} F^{\mu \sigma} + g^{\mu \sigma} \frac{\psi}{c_h} \right)\\
= \sigma n^{\mu} \psi,
\end{multline}
where ${c_h}$ is a hyperbolic advection speed and ${\sigma}$ is a (parabolic) damping factor. Taking the timelike projection of this equation yields an evolution equation for the ${\psi}$ scalar field:

\begin{equation}
\partial_t \left( \frac{\psi}{c_h} \right) - \beta^i \partial_i \left( \frac{\psi}{c_h} \right) = - \frac{\alpha}{\sqrt{\gamma}} \partial_i \left( \sqrt{\gamma} B^i \right) - \alpha \sigma \psi,
\end{equation}
while taking spacelike projections yields a modified evolution equation for the magnetic field ${\mathbf{B}}$:

\begin{multline}
\partial_t \left( \sqrt{\gamma} B^k \right) + \partial_i \left[ \left( v^i - \frac{\beta^i}{\alpha} \right) \sqrt{\gamma} B^k - \left( v^k - \frac{\beta^k}{\alpha} \right) \sqrt{\gamma} B^i \right]\\
= - \left( \alpha \sqrt{\gamma} \right) \gamma^{i k} \partial_i \left( \frac{\psi}{c_h} \right) + \beta^k \partial_i \left( \sqrt{\gamma} B^i \right).
\end{multline}
The former equation is not (yet) written in conservation law form, but following \cite{dedner_hyperbolic_2002} we can introduce an additional advection term ${\alpha v^i \partial_i \left( \frac{\psi}{c_h} \right)}$ into the evolution equation for ${\psi}$, to obtain:

\begin{equation}
\partial_t \left( \frac{\psi}{c_h} \right) + \left( v^i - \frac{\beta^i}{\alpha} \right) \partial_i \left( \frac{\psi}{c_h} \right) = - \frac{\alpha}{\sqrt{\gamma}} \partial_i \left( \sqrt{\gamma} B^i \right) - \alpha \sigma \psi,
\end{equation}
which can then be converted into a flux-conservative form by multiplying through by the relativistic mass density ${\rho W}$, yielding:

\begin{multline}
\partial_t \left( \frac{\rho W \psi}{c_h} \right) + \partial_i \left[ \left( v^i - \frac{\beta^i}{\alpha} \right) \left( \frac{\rho W \psi}{c_h} \right) \right]\\
= - \frac{\alpha \rho W \partial_i \left( \sqrt{\gamma} B^i \right)}{\sqrt{\gamma}} - \alpha \sigma \rho W \psi.
\end{multline}
With the addition of this new advection term, the modified evolution equation for the magnetic field now takes the form:

\begin{multline}
\partial_t \left( \sqrt{\gamma} B^k \right) + \partial_i \left[ \left( v^i - \frac{\beta^i}{\alpha} \right) \sqrt{\gamma} B^k - \left( v^k - \frac{\beta^k}{\alpha} \right) \sqrt{\gamma} B^i \right]\\
= - \sqrt{\gamma} \left( \beta^k v^i + \alpha \gamma^{i k} \right) \partial_i \left( \frac{\psi}{c_h} \right) + \beta^k \partial_i \left( \sqrt{\gamma} B^i \right).
\end{multline}
We can now transform these equations back into the locally-flat tetrad basis, to obtain:

\begin{equation}
\partial_t \left( \frac{\rho W \psi}{c_h} \right) + \partial_i \left( \frac{v^i \rho W \psi}{c_h} \right) = - \frac{\alpha \rho W \partial_i \left( \sqrt{\gamma} B^i \right)}{\sqrt{\gamma}} - \alpha \sigma \rho W \psi,
\end{equation}
and:

\begin{multline}
\partial_t \left( B^k \right) + \partial_i \left( v^i B^k - v^k B^i \right)\\
= - \sqrt{\gamma} \left( \beta^k v^i + \alpha \gamma^{i k} \right) \partial_i \left( \frac{\psi}{c_h} \right) + \beta^k \partial_i \left( \sqrt{\gamma} B^i \right),
\end{multline}
respectively, which are the forms of the equations that \textsc{Gkeyll} actually solves (with the Riemann solver adjusted accordingly).

On the other hand, for correcting divergence errors within the general relativistic multifluid equations in the locally-flat tetrad basis, we use a modification of the hyperbolic divergence cleaning prescription of \cite{munz_finite-volume_2000}, \cite{munz_three-dimensional_2000}, and \cite{munz_divergence_2000} for the flat spacetime Maxwell equations. Specifically, we introduce a pair of scalar correction potentials ${\phi}$ and ${\psi}$, intended for correcting errors in the electric and magnetic fields respectively, and with corresponding propagation speeds of ${\chi}$ and ${\zeta}$, respectively. The hyperbolic parts of Maxwell's equations are now modified accordingly, yielding:

\begin{equation}
\partial_t B^i + \varepsilon^{i j k} \partial_j E_k + \zeta \gamma^{i j} \partial_j \psi = 0,
\end{equation}
for the magnetic field ${\mathbf{B}}$, and:

\begin{equation}
- \partial_t E^i + \varepsilon^{i j k} \partial_j B_k + \chi \gamma^{i j} \partial_j \phi = - \sum_{s = 1}^{n} \frac{q_s}{m_s} \left( \rho_s W_s v_{s}^{i} \right),
\end{equation}
for the electric field ${\mathbf{E}}$. The elliptic parts of Maxwell's equations now take the form of hyperbolic evolution equations for ${\phi}$ and ${\psi}$, yielding:

\begin{equation}
\frac{1}{\zeta} \left( \partial_t \psi \right) + \partial_i B^i = 0,
\end{equation}
for the magnetic field constraint equation, and:

\begin{equation}
\frac{1}{\chi} \left( \partial_t \phi \right) + \partial_i E^i = \rho_c,
\end{equation}
for the electric field constraint equation. To facilitate the computation of the charge density source term ${\rho_c}$ within the latter equation, we switch to evolving the species charge density ${\rho_{c, s} = \frac{\rho_s q_s}{m_s}}$ in place of the species mass density ${\rho_s}$ within the baryonic number density conservation equation:

\begin{equation}
\partial_t \left( \rho_{c, s} W_s \right) + \partial_i \left( \rho_{c, s} W_s v_{s}^{i} \right) = 0,
\end{equation}
such that the electric field constraint equation now becomes:

\begin{equation}
\frac{1}{\chi} \left( \partial_t \phi \right) + \partial_i E^i = \sum_{s = 1}^{n} \rho_{c, s}.
\end{equation}
We find, upon appropriate modification of the Riemann solver(s), that this switch improves numerical stability considerably.

\section{1D Validation Tests}
\label{sec:1d_validation}

\subsection{Convergence: Relativistic Brio-Wu Problem}
\label{sec:convergence_brio_wu}

\begin{figure}
\centering
\includegraphics[width=0.45\textwidth]{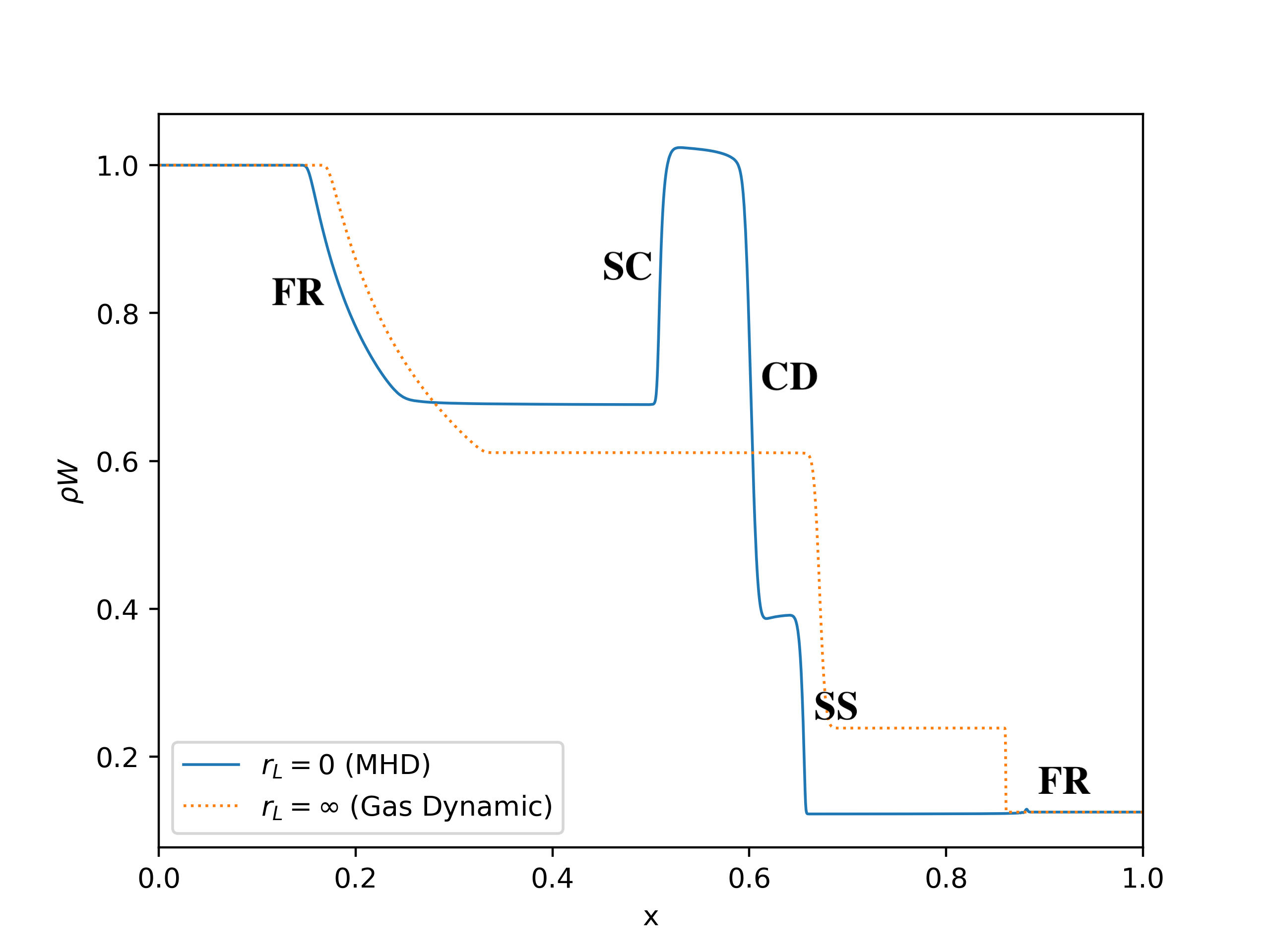}
\caption{The relativistic mass density ${\rho W}$ at time ${t = 0.4}$ for the relativistic Brio-Wu shock tube problem, showing the two asymptotic reference solutions in the ideal MHD (${r_L \to 0}$) limit and the gas dynamic (${r_L \to \infty}$) limit. The wave structure of the MHD solution has been labeled to show the left-moving fast rarefaction wave (FR), the left-moving slow compound wave (SC), the contact discontinuity (CD), the right-moving slow shock (SS), and the right-moving fast rarefaction wave (FR).}
\label{fig:balsara_mhd_reference}
\end{figure}

We begin by validating the relativistic multifluid solver against the relativistic generalization of the \cite{brio_upwind_1988} magnetohydrodynamic shock tube problem proposed by \cite{vanputten_numerical_1993}, and later studied extensively by \cite{balsara_total_2001}, in order to confirm that the relativistic MHD solution can be recovered in the appropriate limit. Note that this convergence analysis is directly analogous to the convergence analysis of the non-relativistic multifluid model to non-relativistic MHD, as performed previously by \cite{shumlak_approximate_2003} and \cite{loverich_discontinuous_2011}. This test can be formulated as a special relativistic MHD Riemann problem with the following initial conditions:

\begin{equation}
\left( \rho, p, B^y \right) = \begin{cases}
\left( 1.0, 1.0, 1.0 \right), \qquad &\text{ for } \qquad x < \frac{1}{2},\\
\left( 0.125, 0.1, -1.0 \right), \qquad &\text{ for } \qquad x \geq \frac{1}{2}
\end{cases},
\end{equation}
with ${v^x = v^y = v^z = B^z = 0.0}$, and ${B^x = 0.5}$, using an ideal gas equation of state with an adiabatic index of ${\Gamma = 2.0}$. We intend to simulate this problem as a relativistic two-fluid system consisting of an electron fluid and an ion fluid, with the rationale being that, by varying the Larmor radius ${r_L = \frac{m v_{\text{th}}}{\left\lvert q \right\rvert \left\lVert \mathbf{B} \right\rVert}}$ of the ion fluid (with ${v_{\text{th}}}$ being the ion thermal velocity), we are able to investigate how the relativistic multifluid model performs in the asymptotic limits as ${r_L \to 0}$ and ${r_L \to \infty}$. These correspond to the ideal MHD limit (with perfect coupling between the fluids and the magnetic field) and the \textit{gas dynamic} limit (with zero coupling between the fluids and the magnetic field), respectively. Both of these asymptotic reference solutions are shown at time ${t = 0.4}$ in Figure \ref{fig:balsara_mhd_reference}, which also labels the wave structure of the relativistic MHD solution, showing a left-moving fast rarefaction wave (FR), a left-moving slow compound wave (SC), a contact discontinuity (CD), a right-moving slow shock (SS), and a right-moving fast rarefaction wave (FR). These MHD and gas dynamic reference solutions have been obtained using an HLLC Riemann solver, on a computational domain ${x \in \left[ 0, 1 \right]}$ with a spatial discretization of 10,000 cells, and a CFL coefficient of 0.9.

The equivalent initial conditions for the relativistic two-fluid Riemann problem are:

\begin{multline}
\left( \rho_e, p_e, \rho_i, p_i, B^y \right) =\\
\begin{cases}
\left( 1.0 \frac{m_e}{m_i}, 1.0, 1.0, 1.0, 1.0 \right), &\text{ for } \qquad x < 1,\\
\left( 0.125 \frac{m_e}{m_i}, 0.1, 0.125, 0.1, -1.0 \right), &\text{ for} \qquad x \geq 1
\end{cases},
\end{multline}
with ${v_{e}^{x} = v_{e}^{y} = v_{e}^{z} = v_{i}^{x} = v_{i}^{y} = v_{i}^{z} = B^z = 0.0}$, and ${B^x = 0.5}$, where we use the species subscripts ${s = e}$ to denote the electron fluid and ${s = i}$ to denote the ion fluid, and with both fluids obeying an ideal gas equation of state with adiabatic indices ${\Gamma_e = \Gamma_i = 2.0}$. We have extended the computational domain to ${x \in \left[ 0, 2 \right]}$ in order to mitigate boundary effects resulting from the left-moving wave structure, and as a result we have reduced the final time of the simulation to ${t = 0.2}$ in order to keep the ratio ${\frac{x}{t}}$ constant. The ion-to-electron mass ratio is set to ${\frac{m_i}{m_e} = 1836.2}$, with the ion charge being set in accordance with the chosen ion Larmor radius ${r_L}$ (keeping ${q_e = -q_i}$), as outlined in Appendix \ref{sec:code_units_appendix}. Note that this is analogous to the limit that is taken within the derivation of \cite{kulsrud_mhd_1980} for kinetic MHD, in which the expansion is performed in powers of ${\varepsilon = \frac{m}{q}}$. In order to facilitate comparisons against the (single-fluid) MHD solutions, we compute the \textit{total} relativistic mass density ${\rho W}$ of the two-fluid system, where the total mass density ${\rho}$ and total velocity ${v^x}$ are computed as:

\begin{equation}
\rho = \rho_i + \rho_e, \qquad \text{ and } \qquad v^x = \frac{\rho_i v_{i}^{x} + \rho_e v_{e}^{x}}{\rho_i + \rho_e},
\end{equation}
respectively, yielding a total Lorentz factor ${W = \frac{1}{\sqrt{1 - \left( v^x \right)^2}}}$, which agree with the corresponding MHD variables in the limit ${\frac{m_e}{m_i} \to 0}$. The chosen values of the ion Larmor radius, normalized by the length of the domain, are ${r_L = 10, 1, 0.1, 0.01, 0.001, 0.0001}$. All solutions have been obtained using an HLLC Riemann solver, a spatial discretization of 10,000 cells, and a CFL coefficient of 0.9.

\begin{figure*}
\centering
\subfigure{\includegraphics[width=0.45\textwidth]{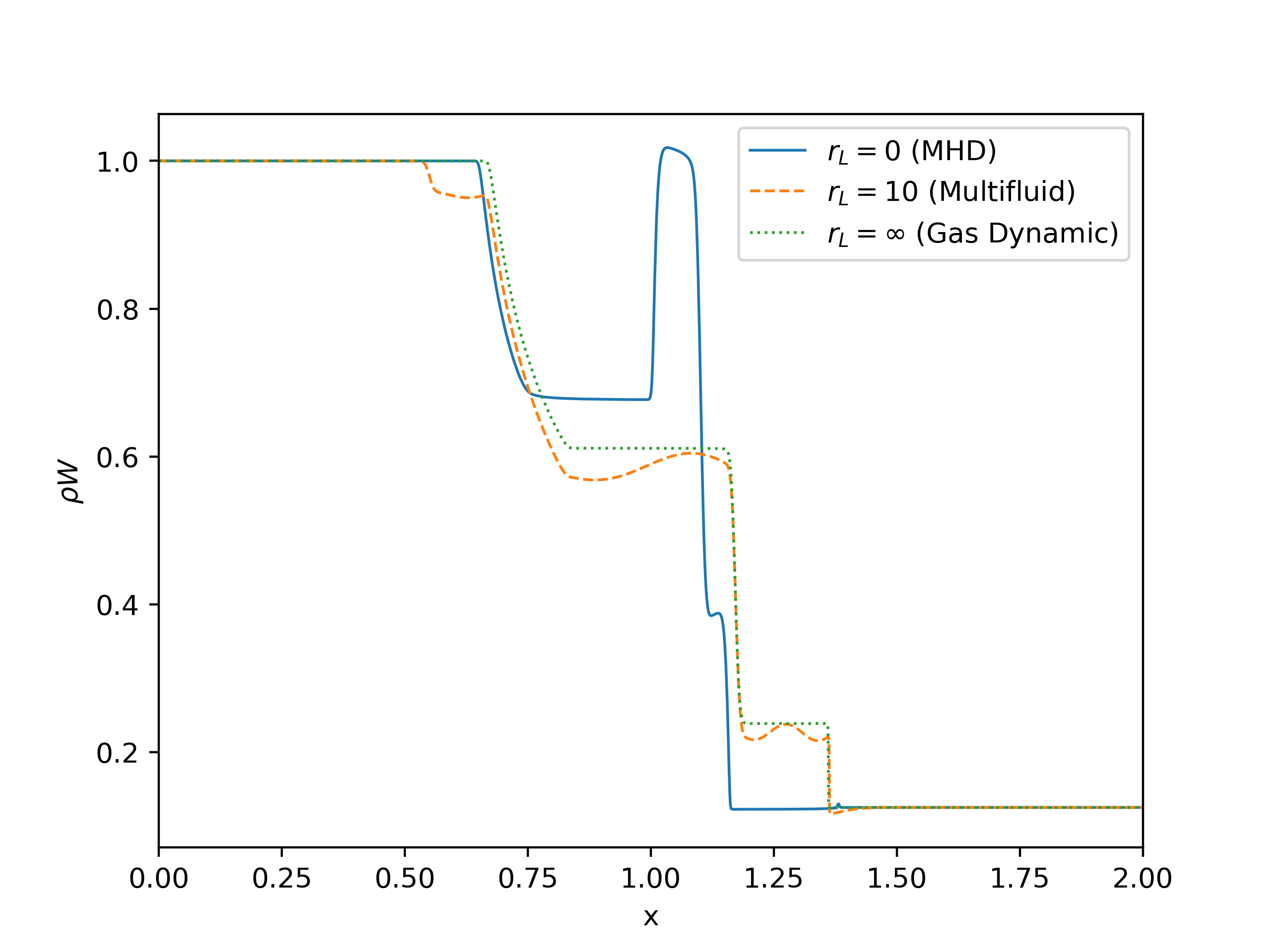}}
\subfigure{\includegraphics[width=0.45\textwidth]{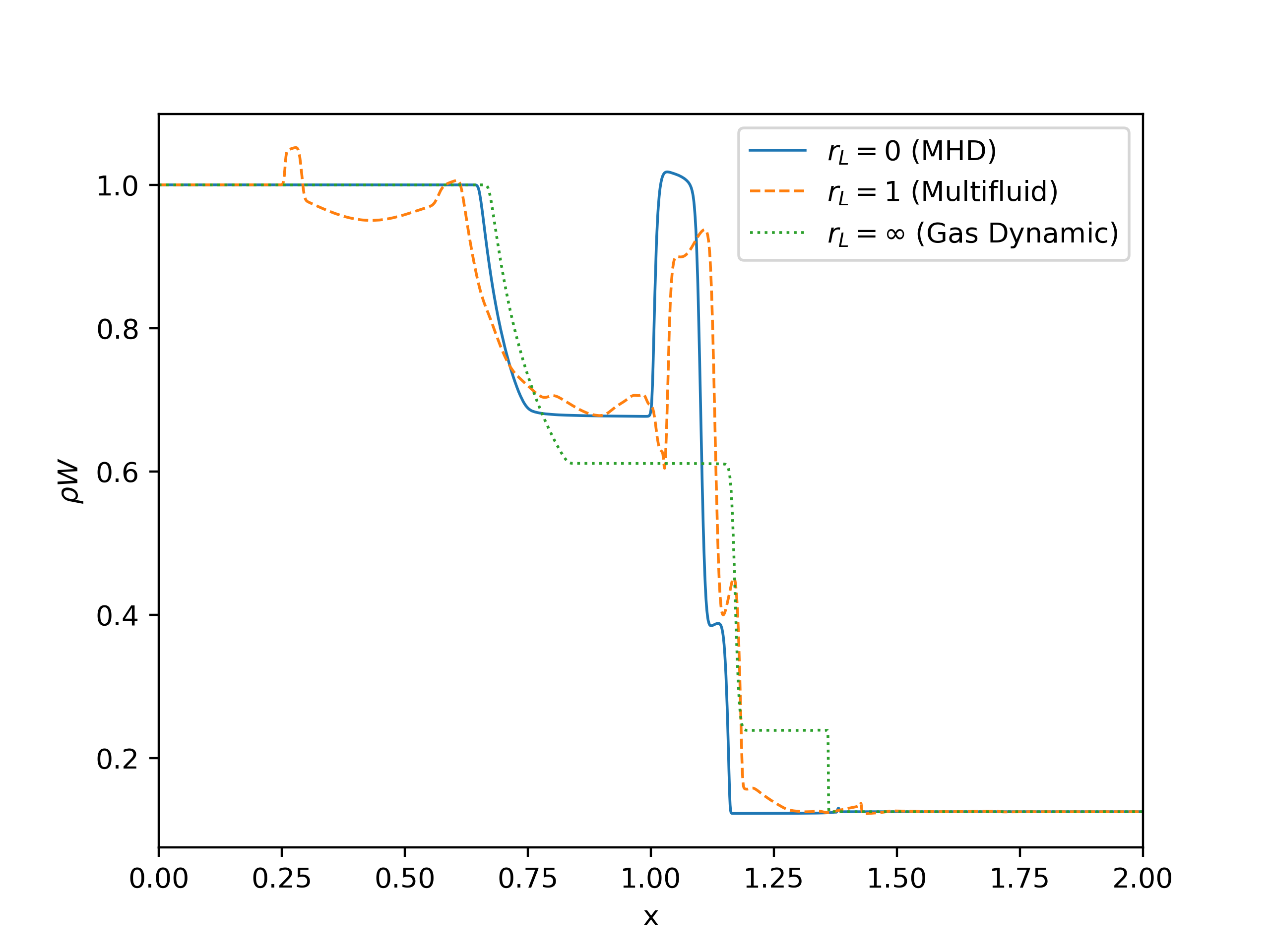}}
\subfigure{\includegraphics[width=0.45\textwidth]{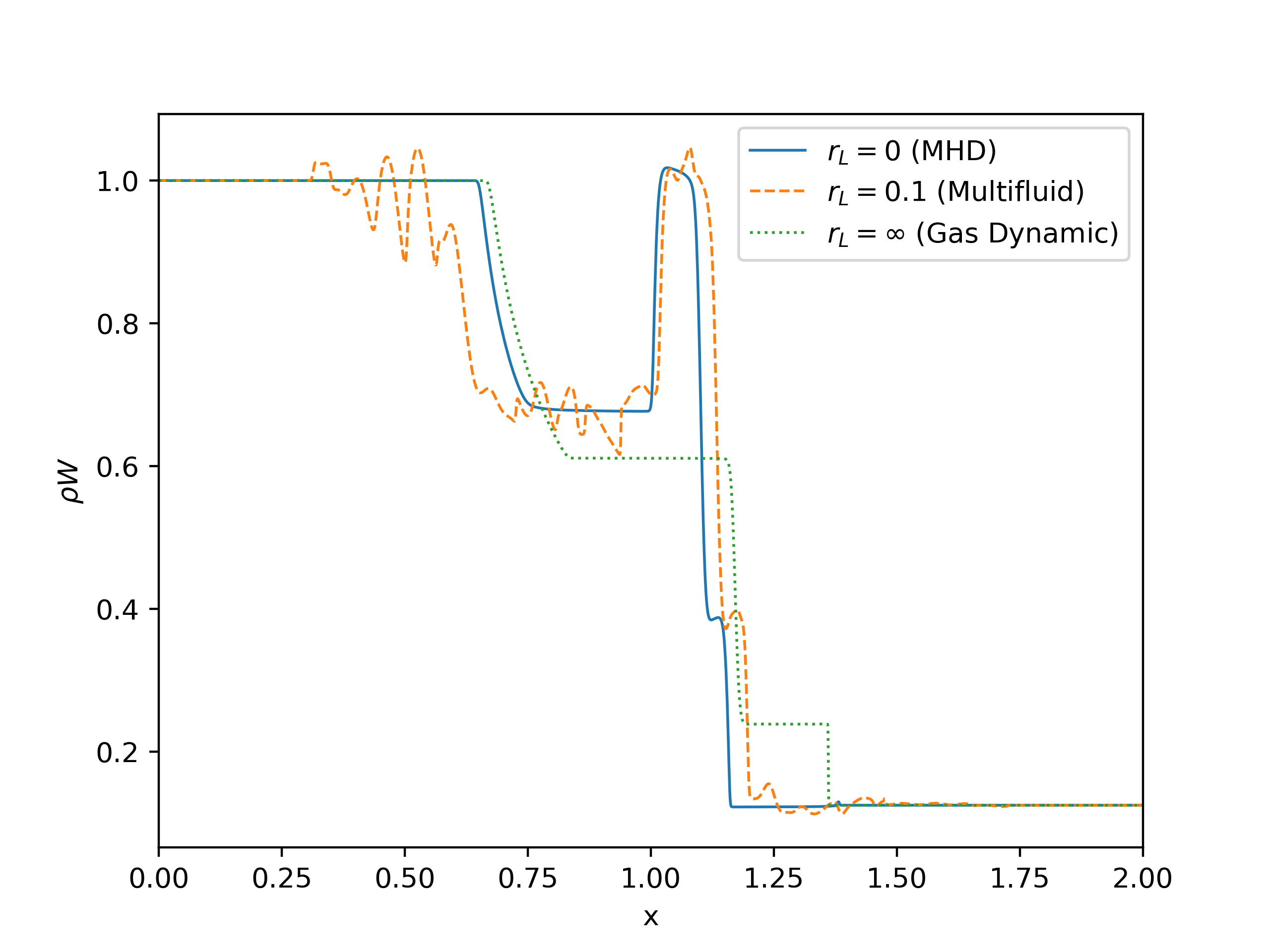}}
\subfigure{\includegraphics[width=0.45\textwidth]{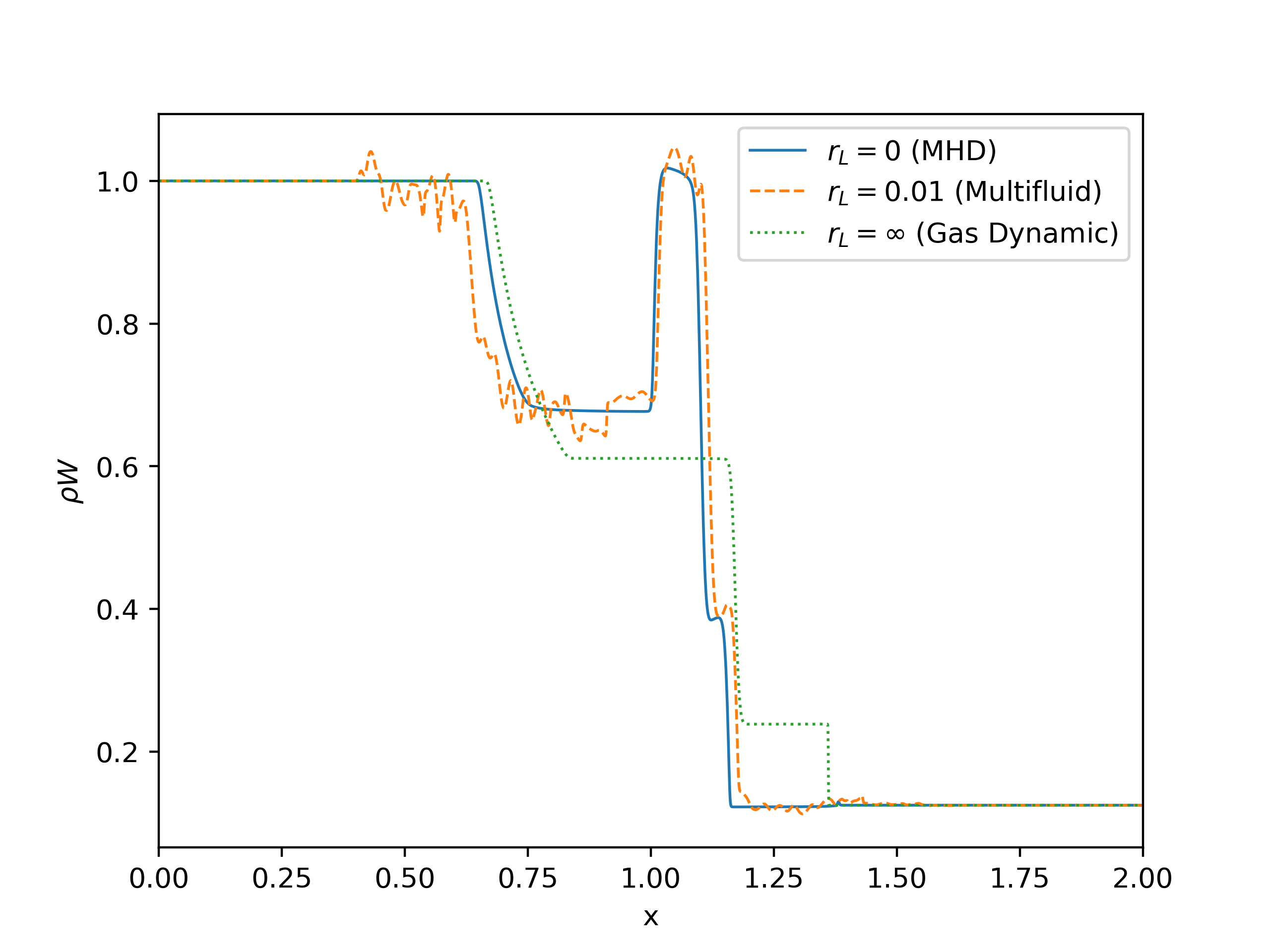}}
\subfigure{\includegraphics[width=0.45\textwidth]{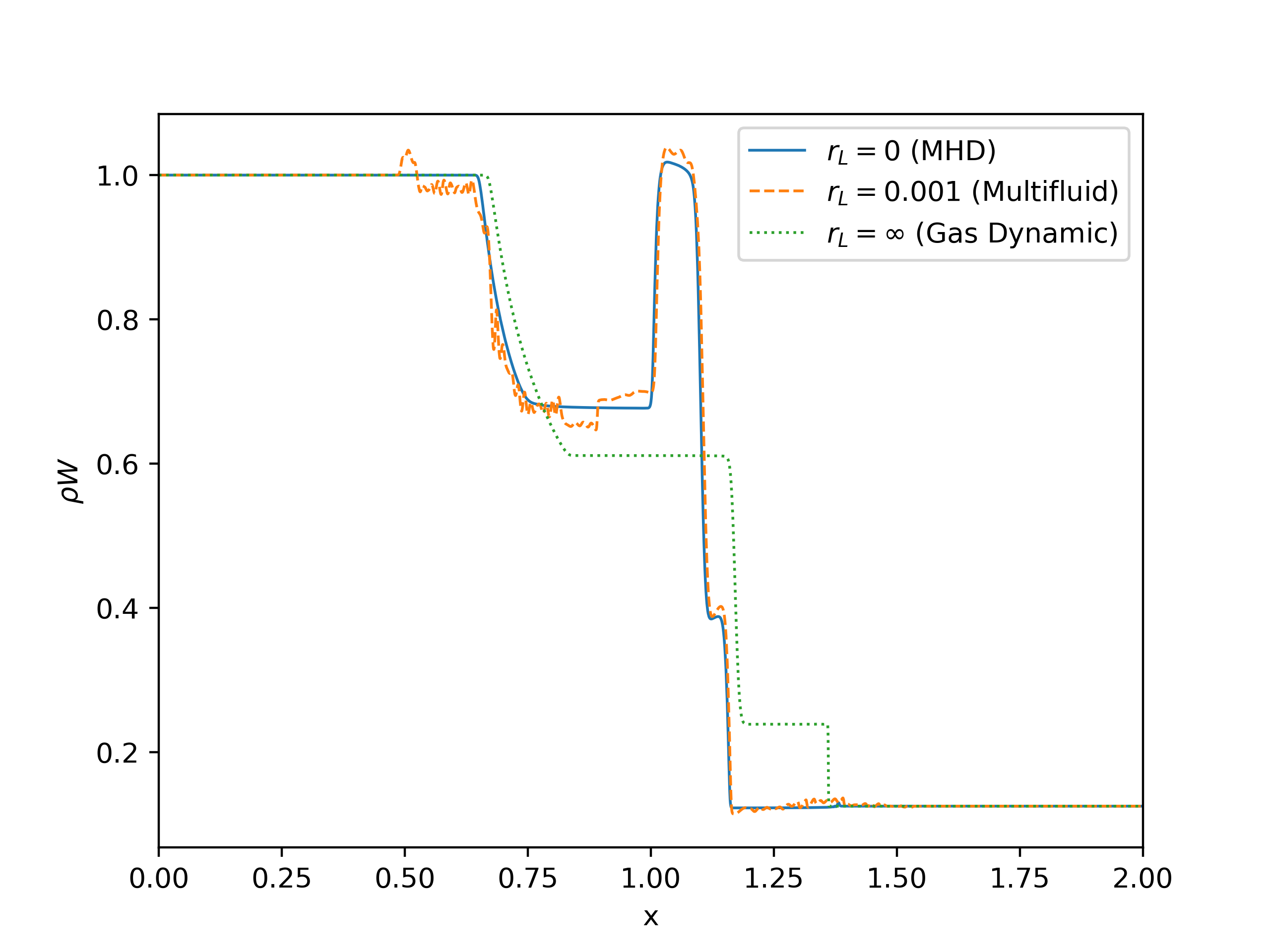}}
\subfigure{\includegraphics[width=0.45\textwidth]{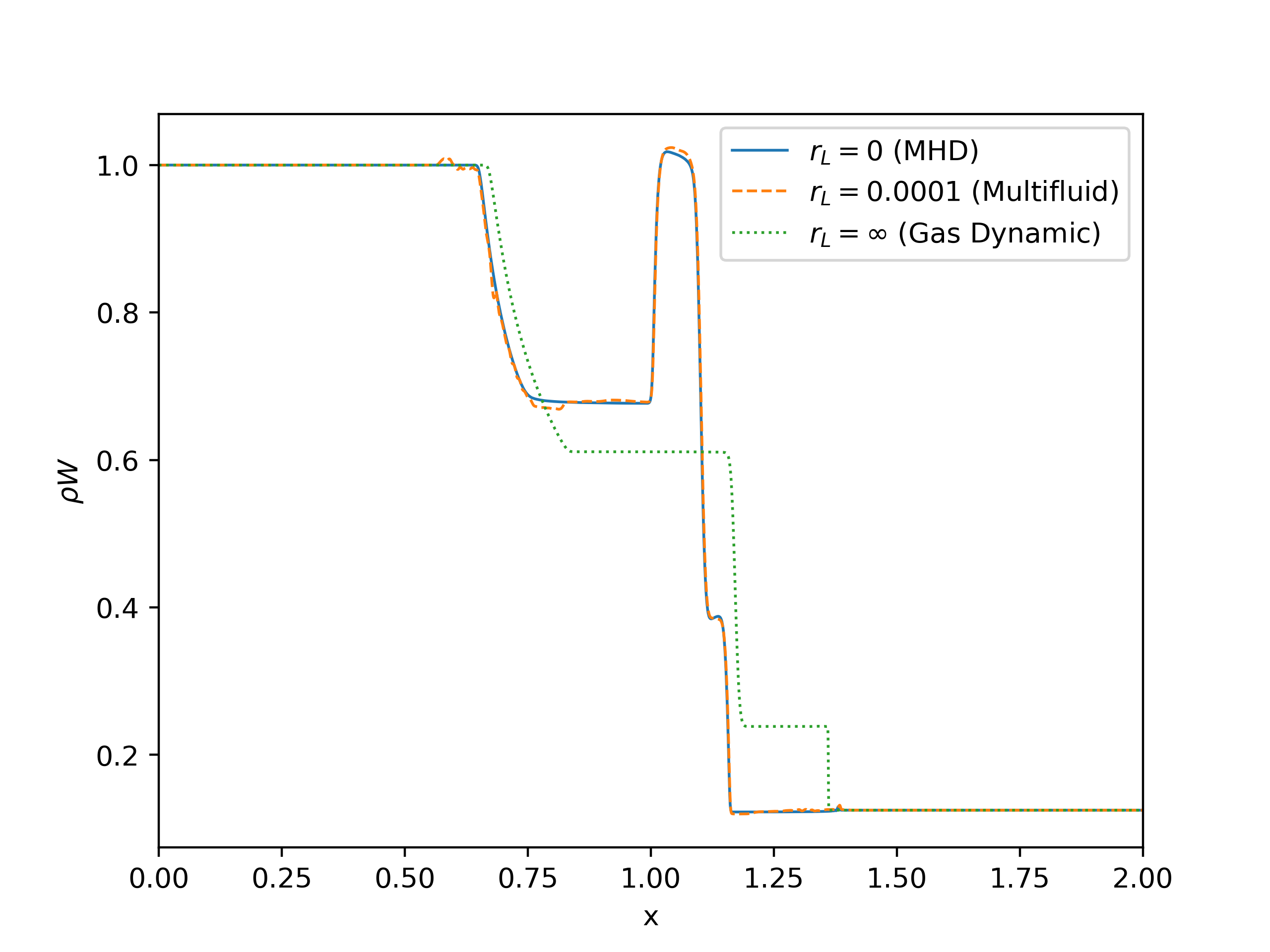}}
\caption{The total relativistic mass densities ${\rho W}$ at time ${t = 0.2}$ for the relativistic Brio-Wu shock tube problem, with ion Larmor radii ${r_L = 10, 1, 0.1, 0.01, 0.001, 0.0001}$. The ideal MHD (${r_L \to 0}$) and gas dynamic (${r_L \to \infty}$) asymptotic reference solutions are also shown in each case. At ${r_L = 10}$, the electrostatic interactions between electron and ion fluids have caused the two-fluid solution to depart slightly from the gas dynamic solution, marginally increasing the speed of the left-moving rarefaction wave and introducing an additional left-moving wave which propagates faster than the MHD waves. At ${r_L = 1}$, the right-moving shock from the gas dynamic solution has slowed considerably and is now much closer in speed to the right-moving slow MHD shock. The slow compound wave is beginning to form, and there is additional left-moving wave structure which is propagating faster than the MHD waves. At ${r_L = 0.1}$, the left-moving wave structure is propagating slower than for ${r_L = 1}$ but still faster than the MHD waves, and the speed of the left-moving fast rarefaction wave has increased even further. The slow compound wave is now approaching the MHD solution. At ${r_L = 0.01}$, both the left-moving fast rarefaction wave and the right-moving slow shock have begun to slow down, with both now approaching the MHD solution. The left-propagating wave structure has slowed further, though is still faster than the MHD waves, as compared to ${r_L = 0.1}$. By ${r_L = 0.001}$ and ${r_L = 0.0001}$, the two-fluid solution has now almost fully converged to the MHD solution.}
\label{fig:balsara_mhd_multifluid}
\end{figure*}

All results are shown in Figure \ref{fig:balsara_mhd_multifluid}. We see that, for ion Larmor radius ${r_L = 10}$, the relativistic two-fluid solution departs slightly from the gas dynamic solution due to the electrostatic interactions between the electron and ion fluids, causing the speed of the left-moving rarefaction wave in the gas dynamic solution to increase marginally (at least within the ion fluid), and also introducing a new left-moving wave which propagates faster than the MHD waves. However, the electron and ion fluids are still largely decoupled from the magnetic field, as in the gas dynamic solution. For ion Larmor radius ${r_L = 1}$, the right-moving shock from the gas dynamic solution slows considerably and becomes much closer in speed to the right-moving slow MHD shock. The electron fluid is now partially coupled to the magnetic field. The slow compound wave begins to form, and we see additional left-moving wave structure which still continues to propagate faster than the MHD waves, corresponding to the beginnings of electron plasma waves, which then couple loosely to the ion fluid. The right-moving fast rarefaction wave has formed in the ion fluid, and is propagating faster than the right-moving fast MHD rarefaction wave. Likewise, the left-moving (fast) rarefaction wave in the ion fluid is also moving slightly faster than the left-moving fast MHD rarefaction wave. In both cases, this is due to electrostatic interaction between the electron and ion fluids, resulting from increased coupling between the two. For ion Larmor radius ${r_L = 0.1}$, the speeds of both the left-moving fast rarefaction wave and the right-moving slow shock in the ion fluid have increased slightly. The electron fluid is now fully coupled to the magnetic field, and the ion fluid is electrostatically coupled to the electron fluid (and thereby loosely coupled to the magnetic field). Thus, this mild increase in wave propagation speeds within the ion fluid results from greater coupling between the ion fluid and the fast-propagating waves in both the electron fluid and the magnetic field. The left-moving wave structure has slowed slightly from the ${r_L = 1}$ case, but is still propagating faster than the MHD waves, and we can now begin to identify these electron plasma waves (coupled to the ion fluid) as corresponding to L-modes and R-modes, including whistler waves. For ion Larmor radius ${r_L = 0.01}$, both the left-moving fast rarefaction wave and the right-moving slow shock in the ion fluid have started to slow down, and to approach the corresponding waves in the MHD solution. The ion fluid is now more fully coupled to both the electron fluid and the magnetic field, and the higher mass of the ions has caused wave propagation speeds in both the electrons and the magnetic field to decrease. Finally, we see that, for ion Larmor radii ${r_L = 0.001}$ and ${r_L = 0.0001}$, the relativistic two-fluid solution has almost completely converged to the relativistic MHD solution, as the electrons, ions, and magnetic field are now almost fully coupled. This confirms empirically that the relativistic multifluid model does indeed converge to ideal relativistic MHD in the appropriate ${r_L \to 0}$ limit.

\subsection{Robustness: Relativistic Noh Problem}
\label{sec:robustness_noh}

Next, we validate the relativistic multifluid solver against the relativistic generalization of the hydrodynamic shock tube problem of \cite{noh_errors_1987}, as studied by \cite{balsara_total_2001}, in order to confirm that the increased robustness of the purely \textit{hydrodynamic} primitive variable reconstruction algorithm in the relativistic multifluid case (as compared to the more fragile \textit{magnetohydrodynamic} primitive variable reconstruction algorithm for relativistic MHD) enables our solver to obtain stable solutions in scenarios involving large Lorentz factors and high magnetizations, where conventional relativistic MHD methods are known to fail. This test can be formulated as a special relativistic MHD Riemann problem with the following initial conditions:

\begin{equation}
\left( v^x, B^y, B^z \right) = \begin{cases}
\left( 0.999, 7.0, 7.0 \right), \qquad &\text{ for } \qquad x < \frac{1}{2},\\
\left( -0.999, -7.0, -7.0 \right), \qquad &\text{ for } \qquad x \geq \frac{1}{2}
\end{cases},
\end{equation}
with ${\rho = 1.0}$, ${p = 0.1}$, and ${B^x = 10.0}$, using an ideal gas equation of state with an adiabatic index of ${\Gamma = 1.6666}$. Once again, we simulate this problem as a relativistic two-fluid system consisting of an electron fluid and an ion fluid, and vary the Larmor radius ${r_L = \frac{m v_{\text{th}}}{\left\lvert q \right\rvert \left\lVert \mathbf{B} \right\rVert}}$ of the ion fluid. The initial conditions for the corresponding relativistic two-fluid Riemann problem are:

\begin{multline}
\left( v_{e}^{x}, v_{i}^{x}, B^y, B^z \right) =\\
\begin{cases}
\left( 0.999, 0.999, 7.0, 7.0 \right), \qquad &\text{ for } x < 1,\\
\left( -0.999, -0.999, -7.0, -7.0 \right), \qquad &\text{ for } x \geq 1
\end{cases},
\end{multline}
with ${v_{e}^{y} = v_{e}^{z} = v_{i}^{y} = v_{i}^{z} = 0.0}$, ${p_e = p_i = 0.1}$, ${B^x = 10.0}$, ${\rho_e = 1.0 \frac{m_e}{m_i}}$, and ${\rho_i = 1.0}$, and with both fluids obeying an ideal gas equation of state with ${\Gamma_e = \Gamma_i = 1.6666}$. As before, we extend the computational domain to ${x \in \left[ 0, 2 \right]}$ to avoid boundary effects, reducing the final time to ${t = 0.2}$ to compensate, we set the ion-to-electron mass ratio to ${\frac{m_i}{m_e} = 1836.2}$ (and vary ${q_i}$ to adjust the Larmor radius, as in Appendix \ref{sec:code_units_appendix}), set ${q_e = - q_i}$, and compute the total relativistic mass density ${\rho W}$ of the two fluids for the purpose of comparing against the single-fluid MHD solutions. All solutions have been obtained using an HLLC Riemann solver, using a spatial discretization of 10,000 cells, and a CFL coefficient of 0.2 (matching that used by \cite{balsara_total_2001}).

\begin{figure}
\centering
\includegraphics[width=0.45\textwidth]{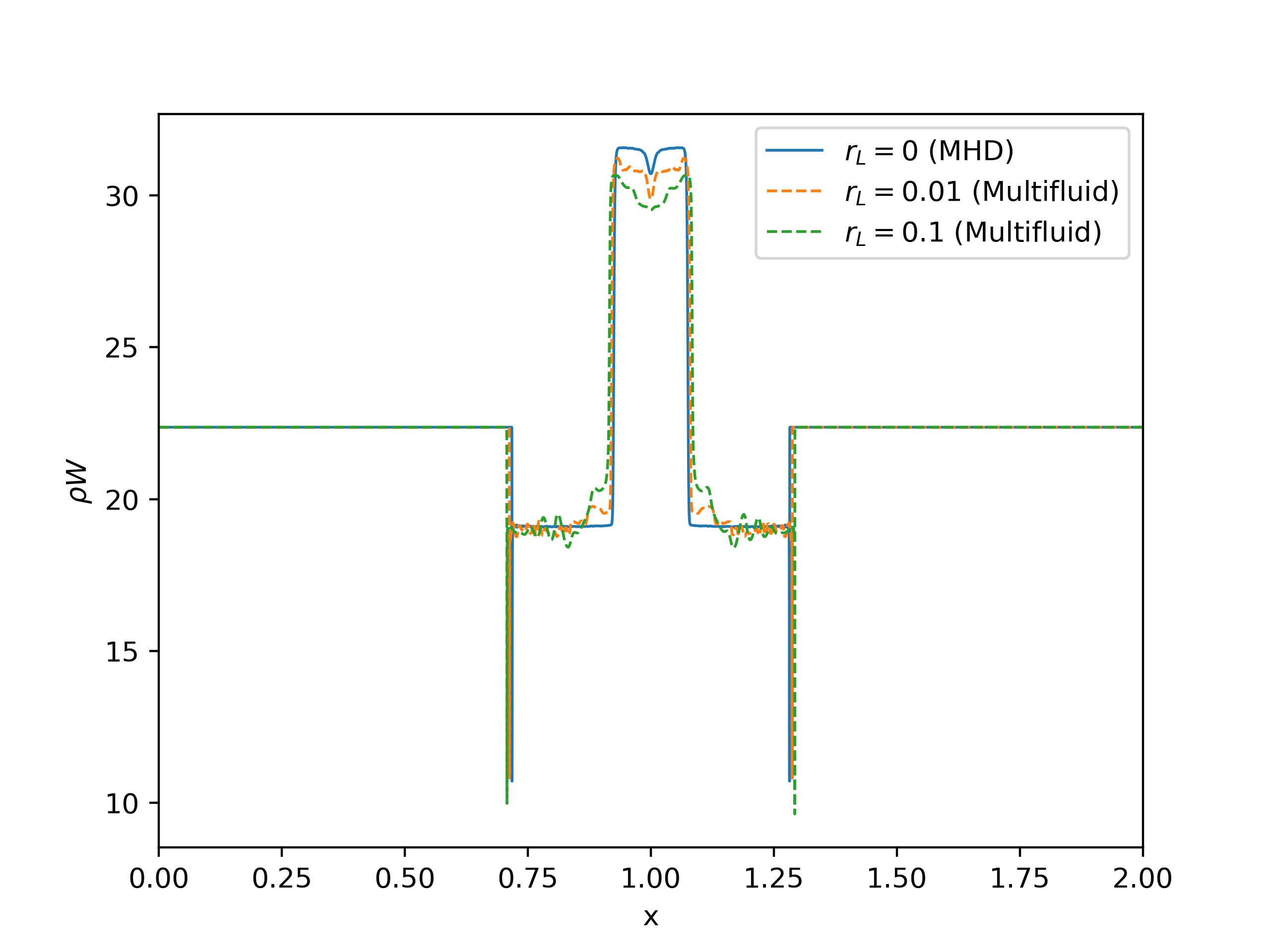}
\caption{The total relativistic mass density ${\rho W}$ at time ${t = 0.2}$ for the relativistic Noh shock tube problem, with ion Larmor radii ${r_L = 0.1}$ and ${r_L = 0.01}$. The ideal MHD (${r_L \to 0}$) asymptotic reference solution is also shown. Due to the high Lorentz factor, the 2D Noble et al. primitive variable reconstruction scheme for relativistic MHD periodically fails, and the solver falls back to the effective 1D Newman-Hamlin method.}
\label{fig:noh_mhd_multifluid}
\end{figure}

In Figure \ref{fig:noh_mhd_multifluid}, we see that, for the \cite{balsara_total_2001} form of the relativistic Noh problem, both the relativistic MHD solver and the relativistic multifluid solver (with ion Larmor radii ${r_L = 0.1}$ and ${r_L = 0.01}$) are able to simulate the problem to completion without crashing, and achieve broadly comparable results, particularly in the ${r_L = 0.01}$ case. However, we note that, due to the moderately high initial Lorentz factor of the two fluid streams (${W \approx 22.4}$), the 2D Newton-Raphson method of \cite{noble_primitive_2006} for primitive variable reconstruction encounters difficulties converging on a pressure (as previously demonstrated by \cite{siegel_recovery_2018}), and so the relativistic MHD solver must periodically fall back to using the effective 1D method of \cite{newman_primitive_2014}. Nevertheless, due to the relatively low initial magnetization of the problem (${\left\lVert \mathbf{B} \right\rVert \approx 14.1}$), the effective 1D \cite{newman_primitive_2014} method is able to recover the primitive variables without issue. The purely hydrodynamic primitive variable reconstruction for the relativistic multifluid solver occurs unproblematically.

\begin{figure}
\centering
\includegraphics[width=0.45\textwidth]{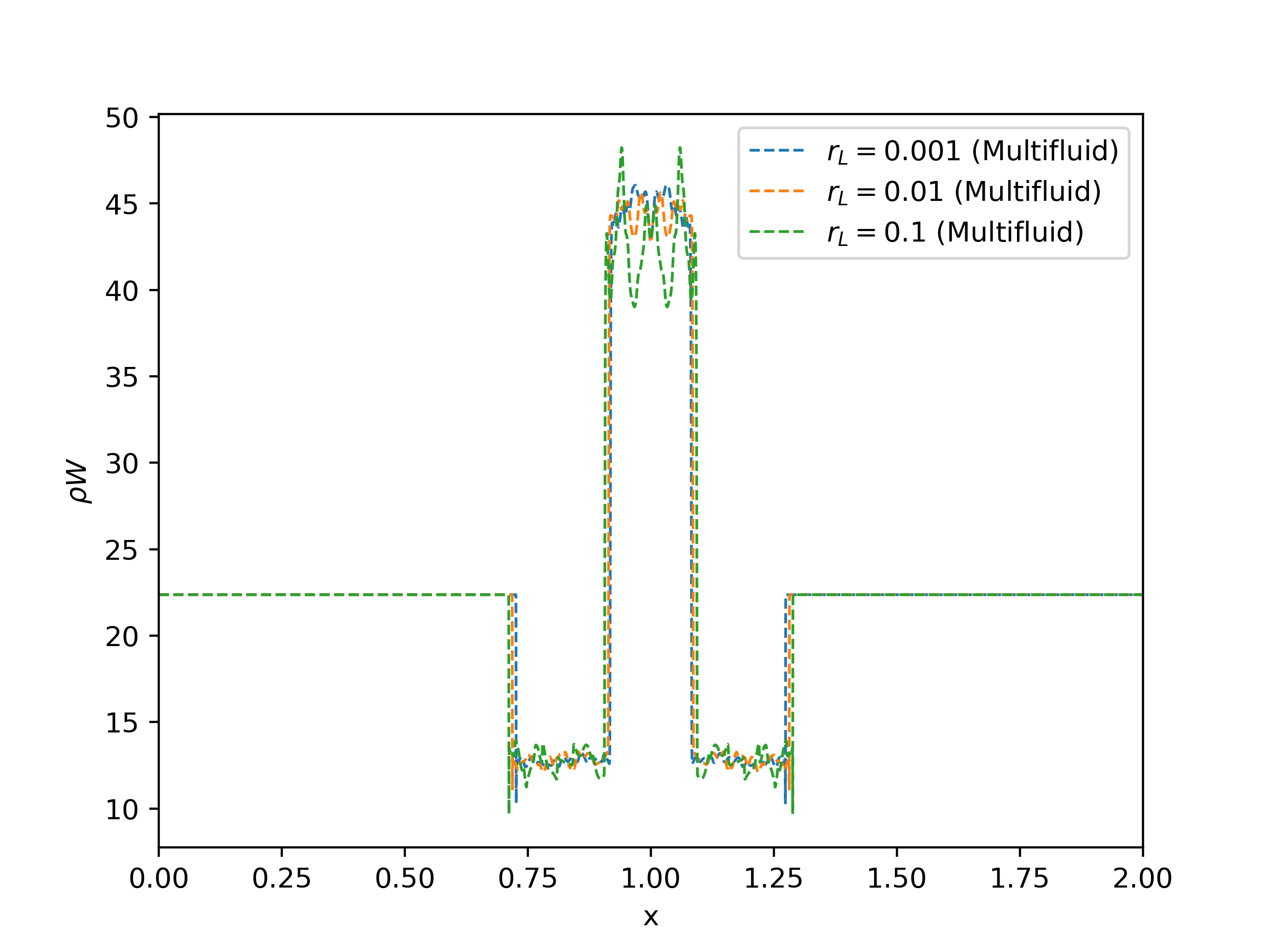}
\caption{The total relativistic mass density ${\rho W}$ at time ${t = 0.2}$ for the highly magnetized variant of the relativistic Noh shock tube problem, with ion Larmor radii ${r_L = 0.1}$, ${r_L = 0.01}$, and ${r_L = 0.001}$. Due to the combination of high magnetization and moderately high initial Lorentz factor, even the effective 1D Newman-Hamlin primitive variable reconstruction scheme for relativistic MHD fails, so no numerical MHD solution exists for this problem. However, by setting the ion Larmor radius to ${r_L = 0.001}$, we are nevertheless able to ascertain approximately what the numerical relativistic MHD solution would be, if it could be obtained.}
\label{fig:noh2_multifluid}
\end{figure}

However, if we increase the overall magnetization of the problem by a factor of 10 (i.e. such that ${\left\lVert \mathbf{B} \right\rVert = 140.7}$), then, as shown by \cite{siegel_recovery_2018}, even the effective 1D \cite{newman_primitive_2014} method may be unable to converge upon a pressure for a moderately high initial Lorentz factor of ${W \approx 22.4}$. As a result,  no numerical MHD solution exists for this highly magnetized variant of the relativistic Noh problem: the relativistic MHD solver, even with fallback, fails after just a few time-steps. However, in Figure \ref{fig:noh2_multifluid}, we see that the relativistic multifluid solver (with ion Larmor radii ${r_L = 0.1}$, ${r_L = 0.01}$, and ${r_L = 0.001}$) is still able to simulate the problem to completion, since the purely hydrodynamic primitive variable reconstruction is unaffected by the increased magnetization, and only the stiffness of the source terms is increased. The case of ion Larmor radius ${r_L = 0.001}$, in which the electron fluid, ion fluid, and magnetic field are very strongly coupled, effectively illustrates approximately what the relativistic MHD solution \textit{would} look like, if it could be obtained numerically.

\section{2D Axisymmetric Validation}
\label{sec:2d_axisymmetric}

\subsection{Magnetospheric Wald Problem}
\label{sec:magnetospheric_wald}

As an intermediate step towards the full 3D accretion problems presented in Section \ref{sec:3d_accretion}, we first attempt a 2D axisymmetric validation test in curved spacetime, namely the magnetospheric (i.e. plasma-filled) variant of the problem of \cite{wald_black_1974}, as studied by \cite{komissarov_electrodynamics_2004} using both force-free and resistive electrodynamics models. Within this problem, a Kerr (spinning) black hole is immersed within an initially uniform magnetic field, assumed to be perfectly aligned with the black hole's spin axis. As shown analytically by \cite{wald_black_1974} in the electrovacuum case, the black hole's spin forces those magnetic field lines that intersect the ergosphere to co-rotate due to frame-dragging effects, thus causing a twisting effect that induces strong electric fields within the ergosphere and generates a large Poynting flux in the poloidal direction. In the magnetospheric variant of the problem due to \cite{komissarov_electrodynamics_2004}, the black hole is also immersed within a uniform plasma atmosphere. In this regard, our setup is similar to the magnetized spherical (Bondi) accretion setup of \cite{michel_accretion_1972}, or the more recent axisymmetic magnetized accretion setup of \cite{galishnikova_collisionless_2023}. Our main objective here is to confirm that the general relativistic multifluid solver is able to reproduce the same qualitative behavior discovered by \cite{komissarov_electrodynamics_2004}, wherein a current sheet forms in the equatorial plane of the black hole and drives ${\left\lVert \mathbf{E} \right\rVert^2 > \left\lVert \mathbf{B} \right\rVert^2}$ within the ergosphere (which GRMHD simulations are, by definition, unable to capture correctly), and the co-rotation of magnetic field lines penetrating the current sheet causes a much stronger magnetic field (and a much larger corresponding Poynting flux) to be generated in the poloidal direction. For the spacetime geometry, we use the spherical Kerr-Schild coordinate system ${\left\lbrace t, r, \theta, \phi \right\rbrace}$, such the the spacetime metric tensor ${g_{\mu \nu}}$ for a Kerr black hole spacetime of mass $M$ and spin ${a = \frac{J}{M}}$ takes the form of a Kerr-Schild perturbation of the Minkowski metric ${\eta_{\mu \nu}}$ in spherical coordinates:

\begin{multline}
g_{\mu \nu} = \eta_{\mu \nu} - 2 V l_{\mu} l_{\nu},\\
\text{ with } \qquad \eta_{\mu \nu} = \mathrm{diag} \left( -1, 1, r^2, r^2 \sin^2 \left( \theta \right) \right),
\end{multline}
with scalar $V$ and null (co)vectors ${l_{\mu}}$ and ${l^{\mu}}$:

\begin{equation}
V = - \frac{M r}{r^2 + a^2 \cos^2 \left( \theta \right)},
\end{equation}
and:

\begin{equation}
l_{\mu} = \begin{bmatrix}
-1 & -1 & 0 & a \sin^2 \left( \theta \right)
\end{bmatrix}, \qquad l^{\mu} = \begin{bmatrix}
-1 & -1 & 0 & \frac{a}{r^2}
\end{bmatrix}^{\intercal},
\end{equation}
respectively. By selecting the following gauge conditions on the lapse function ${\alpha}$ and shift vector ${\beta^i}$:

\begin{equation}
\alpha = \frac{1}{\sqrt{1 - 2V}}, \qquad \text{ and } \qquad \beta^i = \frac{2 V}{1 - 2V} l^i, \label{eq:kerr_schild_gauge}
\end{equation}
we foliate the spacetime into a sequence of identical spacelike hypersurfaces whose induced (spatial) metric tensor ${\gamma_{i j}}$ takes the form of a Kerr-Schild perturbation of the flat metric ${\eta_{i j}}$ in spherical coordinates:

\begin{multline}
\gamma_{i j} = \eta_{i j} - 2 V l_i l_j,\\
\text{ with } \qquad \eta_{i j} = \mathrm{diag} \left( 1, r^2, r^2 \sin^2 \left( \theta \right) \right),
\end{multline}
and where the extrinsic curvature tensor ${K_{i j}}$ is given by:

\begin{multline}
K_{i j} = \alpha \left[ - l_i \partial_j V - l_j \partial_i V - V \partial_j l_i - V \partial_i l_j \right.\\
\left. + 2 V^2 \left( l_i l^k \partial_k l_j + l_j l^k \partial_k l_i \right) + 2 V l_i l_j l^k \partial_k V \right]. \label{eq:kerr_schild_extrinsic}
\end{multline}
To prevent prohibitive time-step restrictions and/or unphysical wave propagation from the interior of the black hole, we place an excision boundary at the Schwarzschild radius ${r = 2M}$, with a region of zero flux on the interior ${r < 2M}$.

\begin{figure*}
\centering
\subfigure{\includegraphics[trim={1cm, 0.5cm, 1cm, 0.5cm}, clip, width=0.45\textwidth]{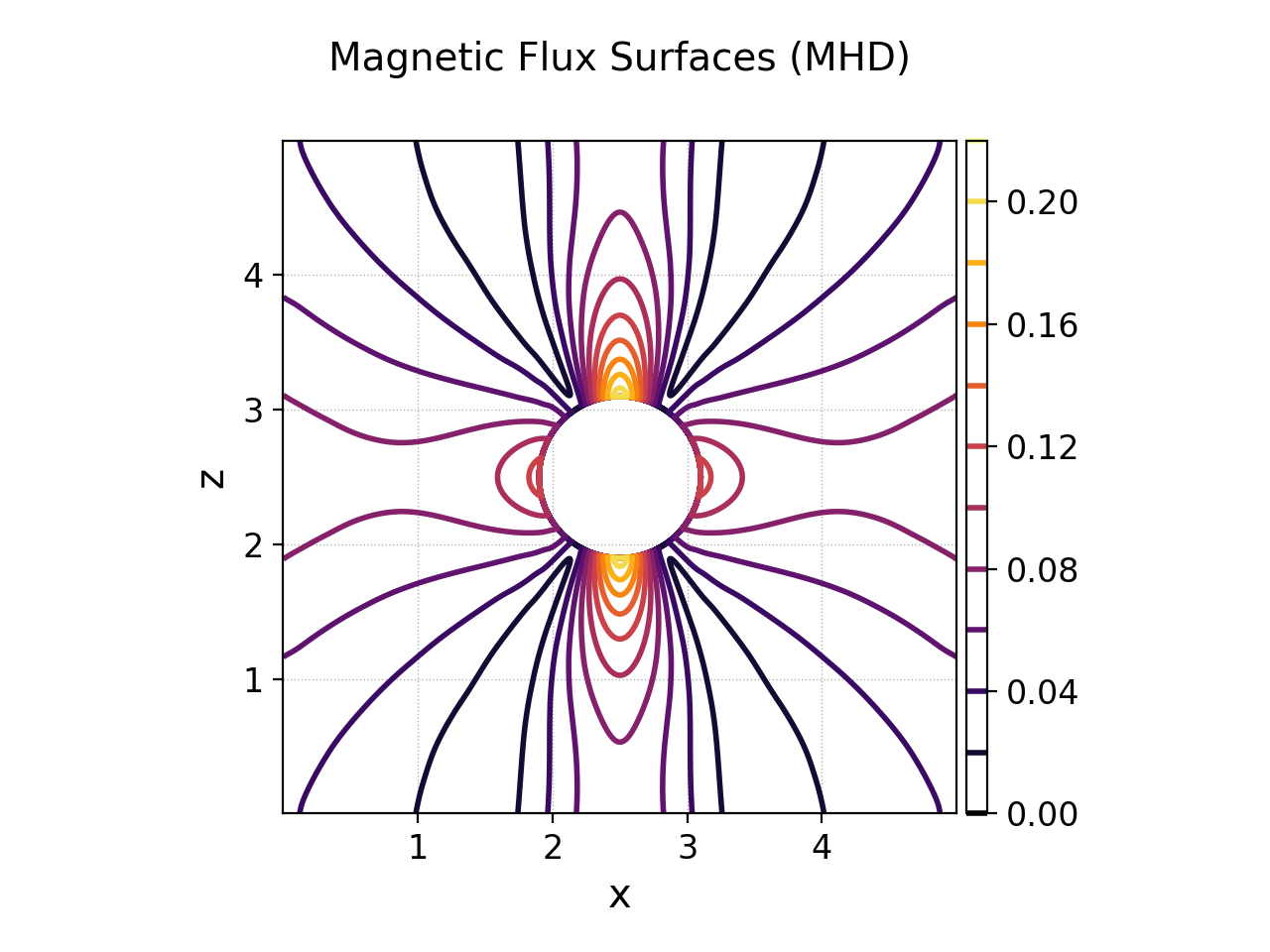}}
\subfigure{\includegraphics[trim={1cm, 0.5cm, 1cm, 0.5cm}, clip, width=0.45\textwidth]{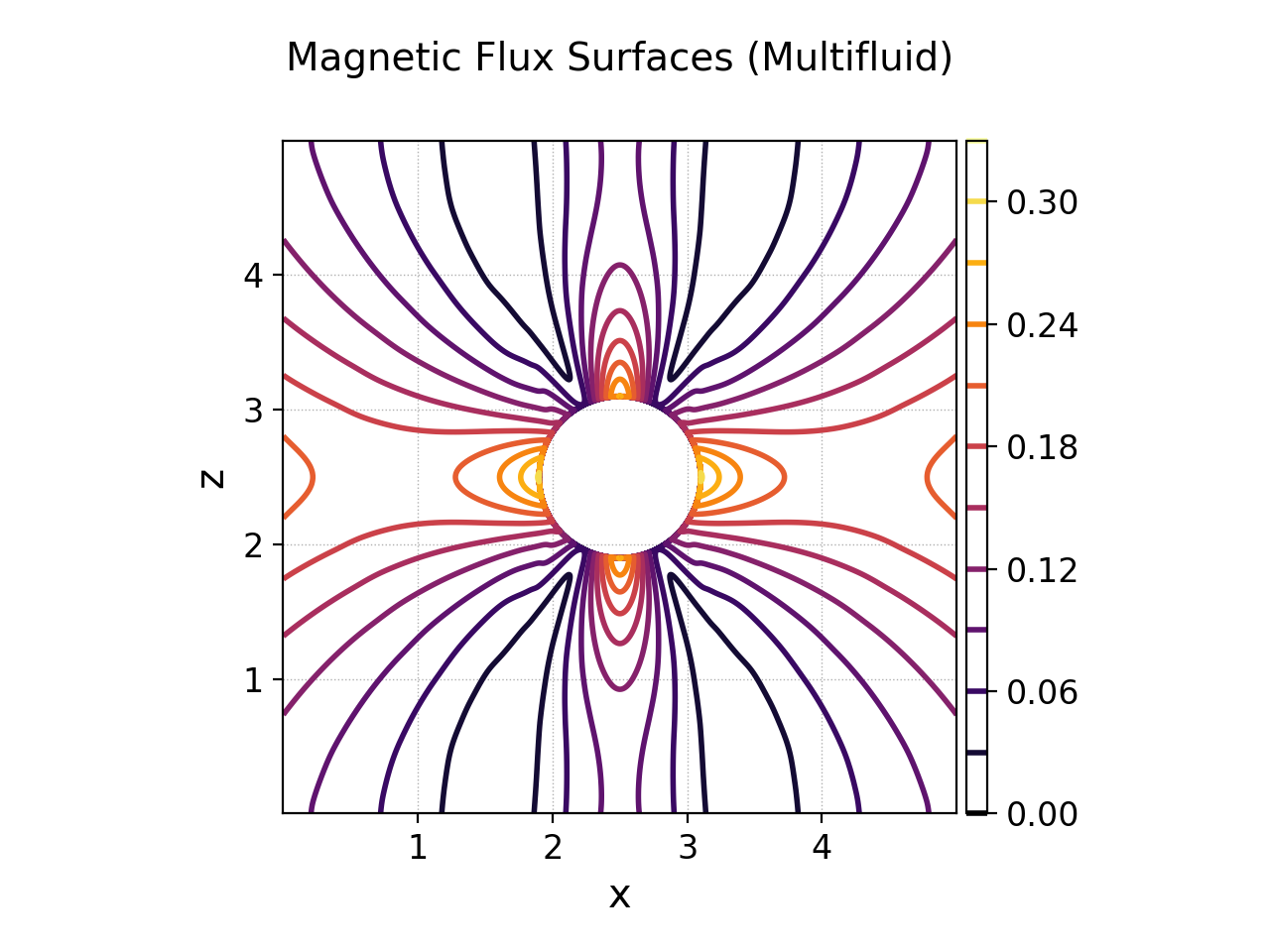}}
\caption{Magnetic flux surfaces at time ${t = 50}$ for the magnetospheric Wald problem in axisymmetry for a spinning (Kerr) black hole with ${a = 0.95}$, obtained using the GRMHD solver (i.e. ${r_L \to 0}$) on the left, and the general relativistic multifluid solver (with initial ion Larmor radius ${r_L = 0.1}$) on the right. We see a larger induced Poynting flux in the multifluid solution, particularly in the poloidal direction, with a larger proportion of the flux surfaces intersecting the horizon of the black hole as compared to the MHD solution.}
\label{fig:magnetospheric_wald_flux}
\end{figure*}

\begin{figure*}
\centering
\subfigure{\includegraphics[trim={1cm, 0.5cm, 1cm, 0.5cm}, clip, width=0.45\textwidth]{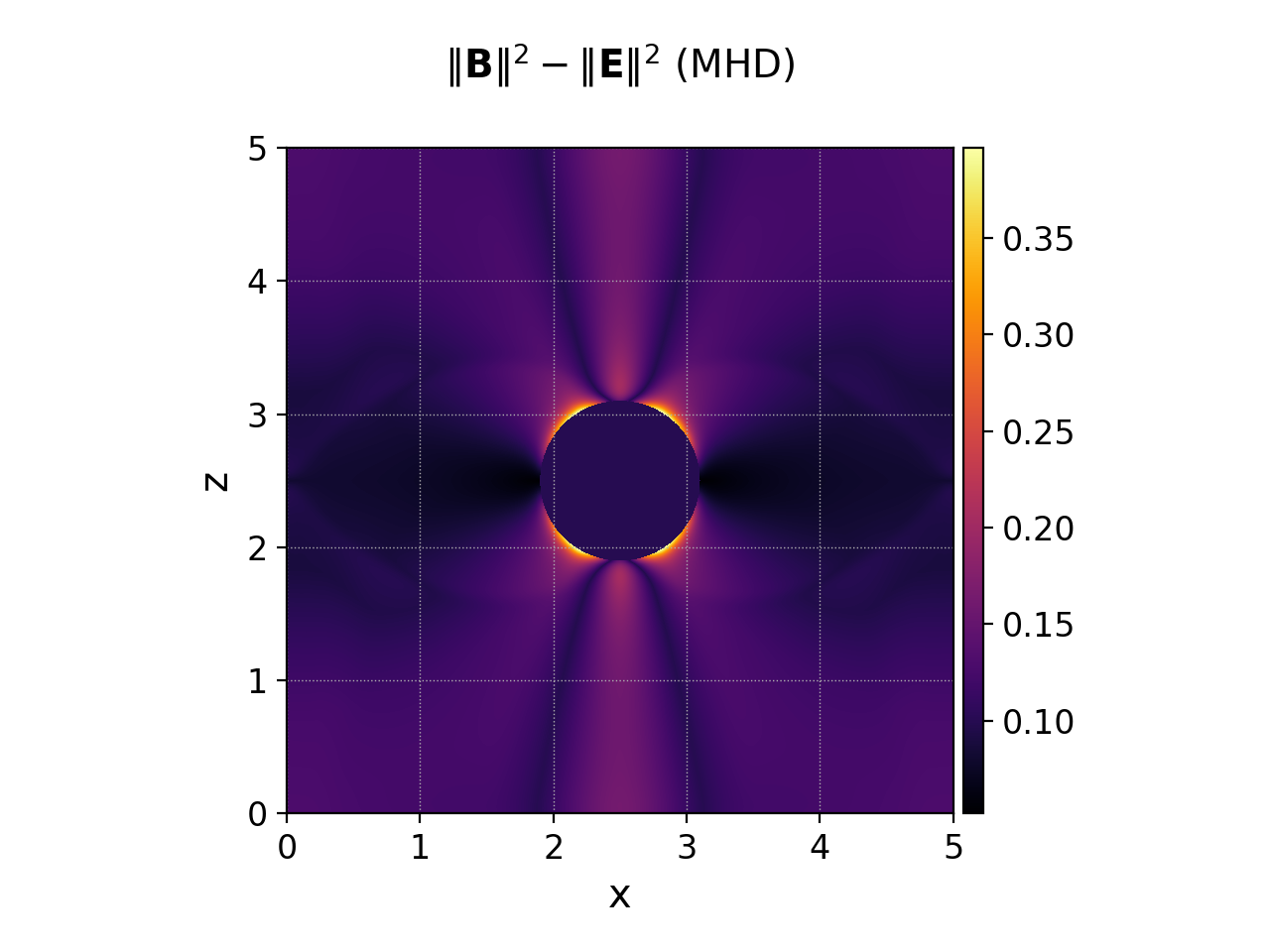}}
\subfigure{\includegraphics[trim={1cm, 0.5cm, 1cm, 0.5cm}, clip, width=0.45\textwidth]{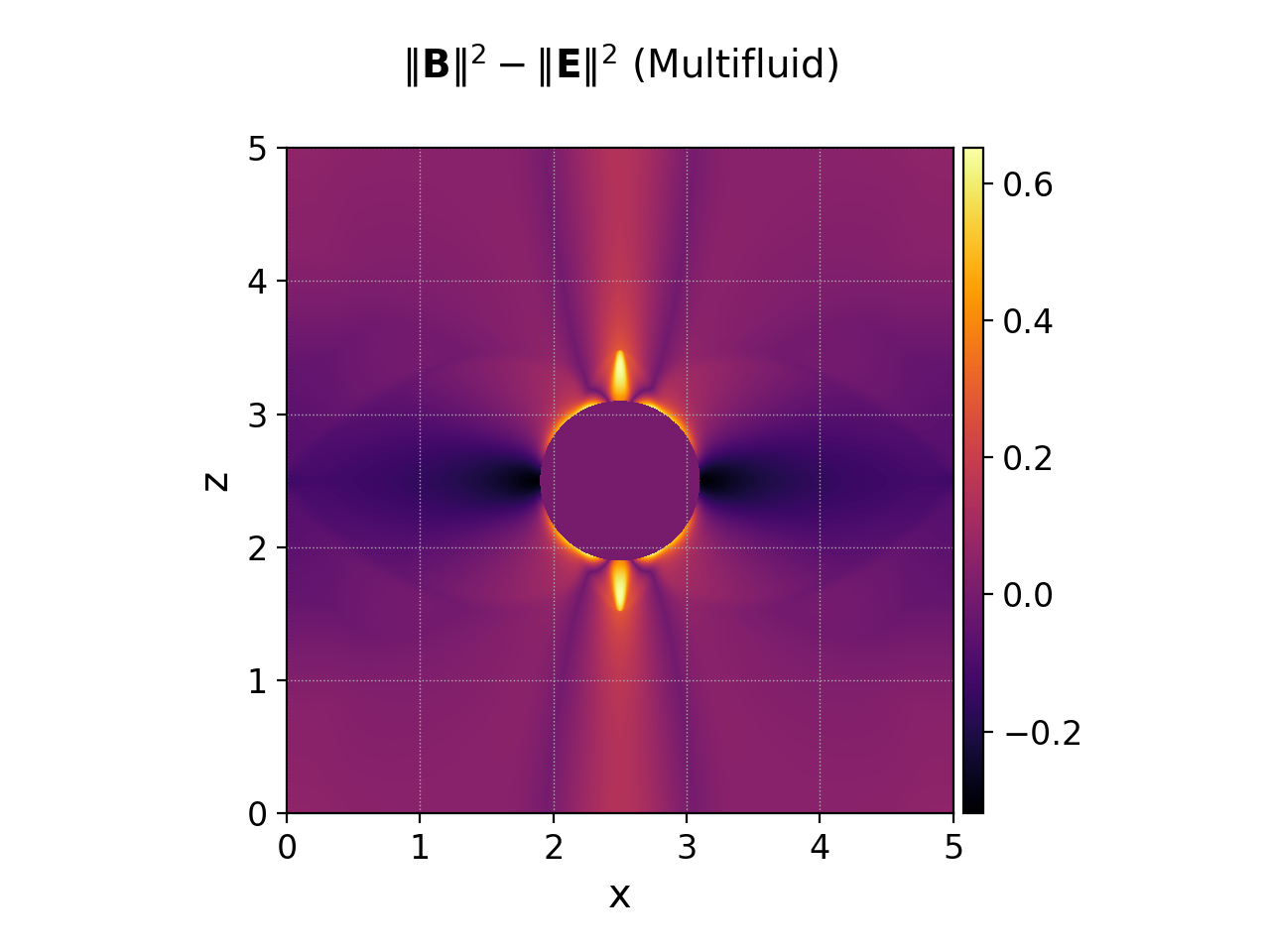}}
\caption{The Lorentz invariant scalar ${\left\lVert \mathbf{B} \right\rVert^2 - \left\lVert \mathbf{E} \right\rVert^2}$ at time ${t = 50}$ for the magnetospheric Wald problem in axisymmetry for a spinning (Kerr) black hole with ${a = 0.95}$, obtained using the GRMHD solver (i.e. ${r_L \to 0}$) on the left, and the general relativistic multifluid solver (with initial ion Larmor radius ${r_L = 0.1}$) on the right. We again see stronger magnetic fields in the poloidal direction within the multifluid solution compared to the MHD solution, as well as the formation of a clear region around the equator of the black hole (absent in the MHD solution) in which ${\left\lVert \mathbf{B} \right\rVert^2 - \left\lVert \mathbf{E} \right\rVert^2}$ is driven slightly negative as a consequence of strong electric field induction resulting from the black hole's rotation.}
\label{fig:magnetospheric_wald_field}
\end{figure*}

We set up a 2D axisymmetric domain ${\left( r, \theta \right) \in \left[ 2M, 5 \right] \times \left[ 0, 2 \pi \right]}$ containing a black hole of mass ${M = 0.3}$ and spin ${a = 0.95}$, centered at ${r = 0}$. The black hole is threaded by a uniform magnetic field in the $z$-direction (i.e. the spin axis) ${B^z = B_0}$, with strength ${B_0}$ set such that the initial magnetization is ${\sigma = \frac{\left\lVert \mathbf{B} \right\rVert^2}{4 \pi \rho} = 10}$ (see Appendix \ref{sec:code_units_appendix}). In the GRMHD case, we initialize a uniform plasma atmosphere throughout the domain, with density and pressure set to ${\rho_0 = \frac{1}{4 \pi}}$ and ${p_0 = 0.01}$ (again, see Appendix \ref{sec:code_units_appendix} for details of why this density is chosen), with vanishing velocity ${v_0 = 0}$, and assuming a perfect monatomic gas, i.e. an ideal gas with adiabatic index ${\Gamma = \frac{5}{3}}$. In the general relativistic multifluid case, this atmosphere is modeled as a two-fluid plasma (consisting of an electron fluid and an ion fluid) using the same prescription that we employed within the previous section, i.e. with ${\rho_i = \frac{1}{4 \pi}}$ and ${\rho_e = \frac{1}{4 \pi} \left( \frac{m_e}{m_i} \right)}$, etc. The two fluids are assumed to be ideal gases with adiabatic indices ${\Gamma_e = \Gamma_i = \frac{5}{3}}$, with the ion-to-electron mass ratio set to ${\frac{m_i}{m_e} = 1836.2}$, and the electron and ion charges ${q_e = -q_i}$ set in order to yield an initial ion Larmor radius of ${r_L = \frac{m v_{\text{th}}}{\left\lvert q \right\rvert \left\lVert \mathbf{B} \right\rVert} = 0.1}$, as shown in Appendix \ref{sec:code_units_appendix}. The magnetic field is initialized identically in the two cases, and in both cases we use a spatial discretization of ${4096 \times 4096}$ cells in $r$ and ${\theta}$ in the first instance, with an HLLC Riemann solver and a CFL coefficient of 0.9. Figure \ref{fig:magnetospheric_wald_flux} shows the magnetic flux surfaces obtained from both the GRMHD solver and the general relativistic multifluid solver at time ${t = 50}$. We find that the magnetic flux profiles are qualitatively different between the two solutions, particularly in the equatorial direction, and that a stronger Poynting flux is induced in the poloidal direction within the multifluid solution (as exemplified by a greater proportion of the flux surfaces intersecting the horizon of the black hole, as compared to the GRMHD solution). A stronger poloidal magnetic field is also maintained within the multifluid solution, and there is evidence of a field reversal occurring in the region beyond ${x < 1}$ and ${x > 4}$, indicating the possibility of magnetic reconnection within the equatorial current sheet, that is absent within the GRMHD solution. We can confirm that the equatorial current sheet, with ${\left\lVert \mathbf{E} \right\rVert^2 > \left\lVert \mathbf{B} \right\rVert^2}$ regions in the ergosphere, as observed by \cite{komissarov_electrodynamics_2004}, is indeed reproduced within the multifluid solution, by plotting the Lorentz invariant scalar quantity ${\left\lVert \mathbf{B} \right\rVert^2 - \left\lVert \mathbf{E} \right\rVert^2}$, as shown in Figure \ref{fig:magnetospheric_wald_field}. In order to facilitate comparisons, we compute the electric field vector ${\mathbf{E}}$ in GRMHD as:

\begin{equation}
E^i = - \varepsilon^{i j k} v_j B_k,
\end{equation}
since, within the ideal MHD approximation, the comoving electric field vanishes (and therefore ${\left\lVert \mathbf{E} \right\rVert^2 < \left\lVert \mathbf{B} \right\rVert^2}$ since ${v_i v^i < 1}$). We indeed see a region in the equatorial plane of the black hole within the multifluid solution where ${\left\lVert \mathbf{E} \right\rVert^2 - \left\lVert \mathbf{B} \right\rVert^2}$ is driven negative, indicating the induction of strong electric fields within the ergosphere due to the black hole's spin that are absent within the GRMHD solution. We also see a much stronger poloidal magnetic field being generated close to the surface of the black hole in the multifluid solution as compared to the GRMHD solution.

\begin{figure*}
\centering
\subfigure{\includegraphics[width=0.45\textwidth]{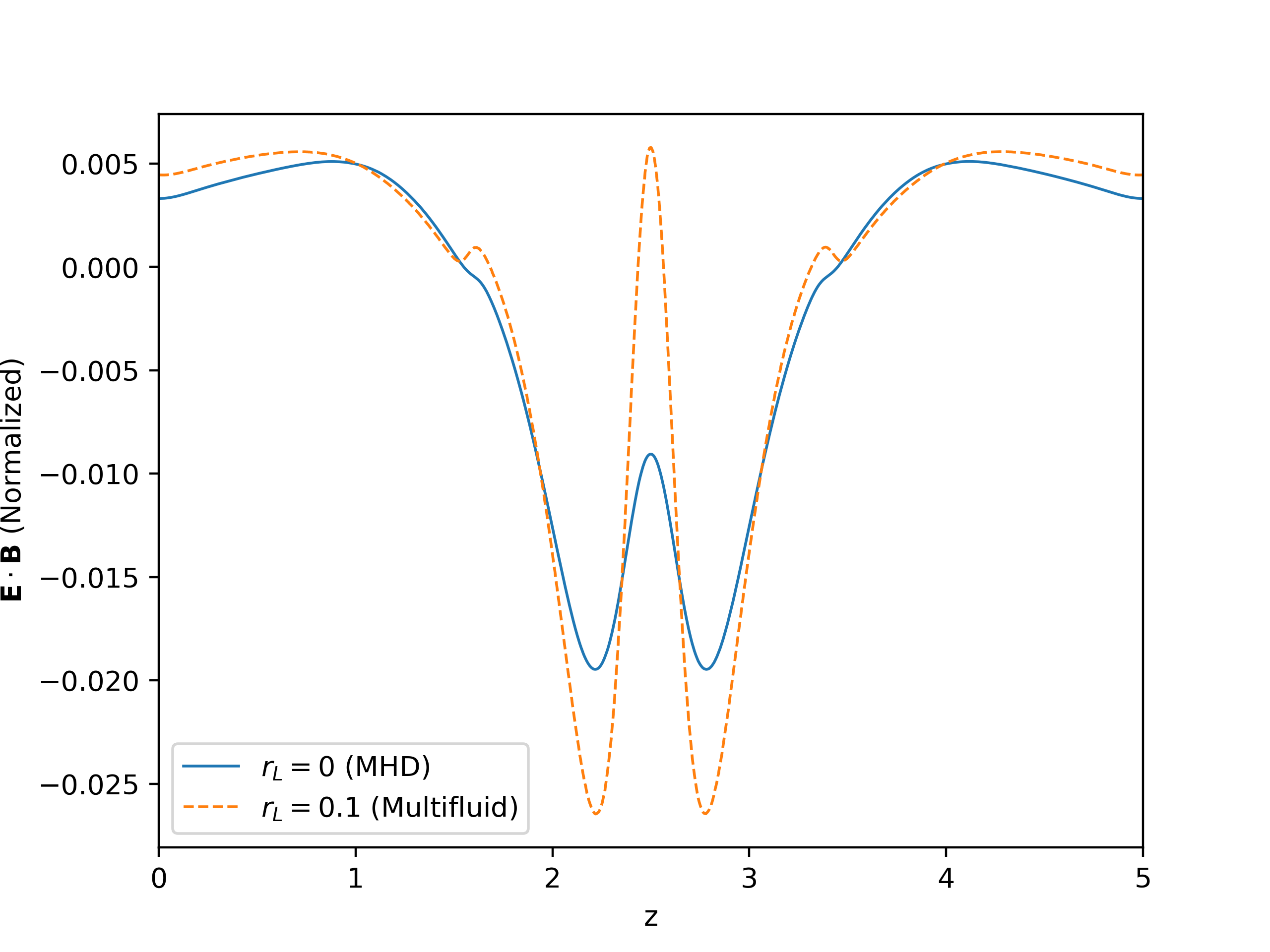}}
\subfigure{\includegraphics[width=0.45\textwidth]{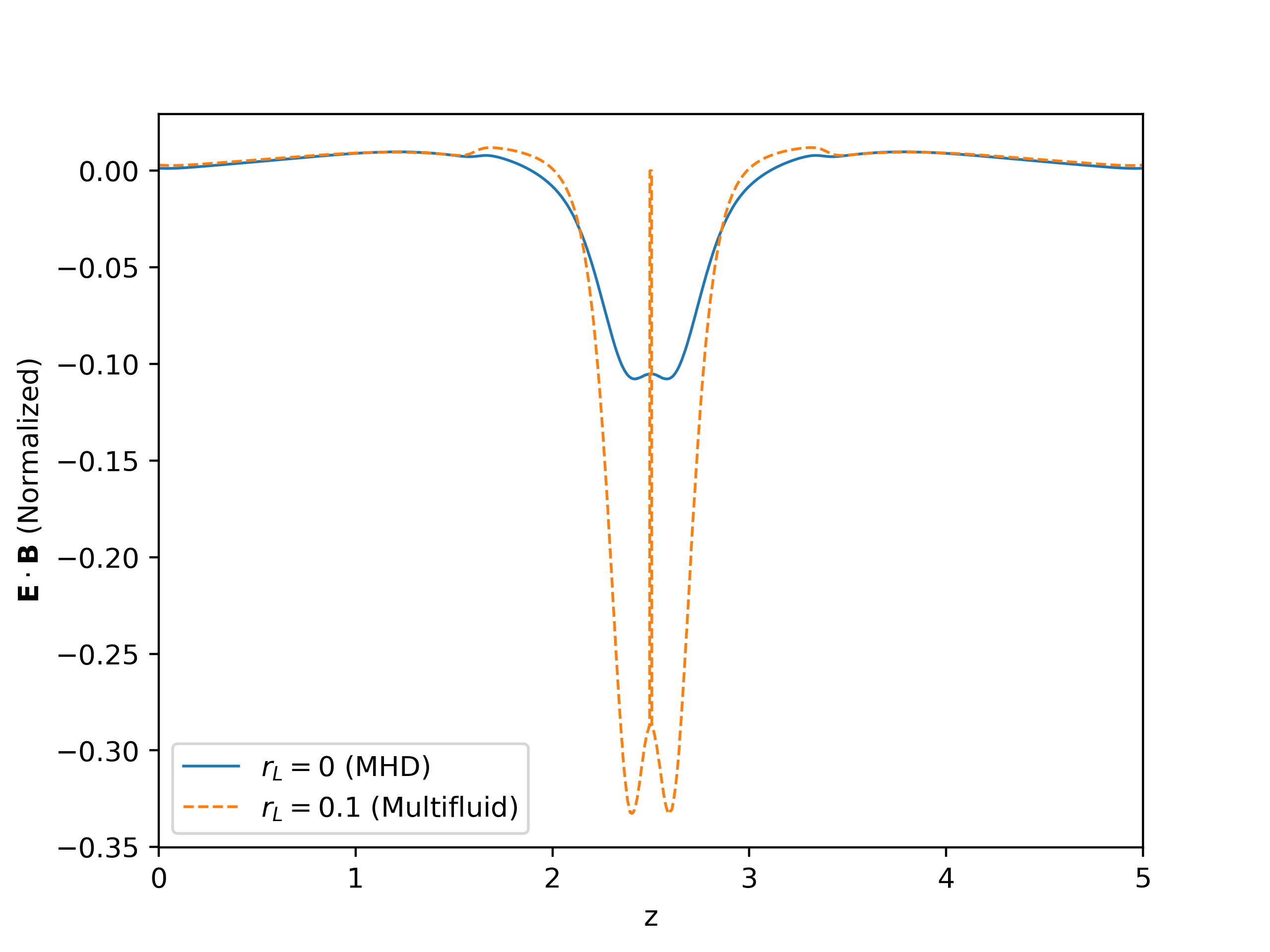}}
\caption{Cross-sectional profiles of the Lorentz invariant scalar ${\mathbf{E} \cdot \mathbf{B}}$ at time ${t = 50}$ for the magnetospheric Wald problem in axisymmetry for a spinning (Kerr) black hole with ${a = 0.95}$, taken through the poloidal plane ${x = 3.5}$ of the black hole, obtained using both the GRMHD solver (i.e. ${r_L \to 0}$) and the general relativistic multifluid solver (with initial ion Larmor radius ${r_L = 0.1}$). On the left, the simulation is run with an angular resolution of 4096 cells in ${\theta}$, and we see the beginnings of the formation of an equatorial current sheet in the multifluid solution that is largely absent in the GRMHD solution. On the right, the simulation is run with an angular resolution of 16,384 cells in ${\theta}$, and we see that the current sheet has collapsed to a width of just a few cells in the multifluid solution, with no corresponding collapse observed in the GRMHD solution.}
\label{fig:magnetospheric_wald_current_sheet}
\end{figure*}

Finally, we can confirm that the current sheet in the equatorial plane of the black hole has indeed formed correctly, by plotting the Lorentz invariant scalar quantity ${\mathbf{E} \cdot \mathbf{B}}$ through the poloidal plane ${x = 3.5}$, as shown in Figure \ref{fig:magnetospheric_wald_current_sheet} (with the corresponding GRMHD solution also shown for reference). With an angular resolution of 4096 cells in ${\theta}$, we already see the full current sheet starting to form within the multifluid solution, with only a small ``bump'' appearing within the corresponding GRMHD solution, although the current sheet still appears to have a finite width due to the presence of significant numerical diffusion. By increasing the angular resolution to 16,384 cells in ${\theta}$, we find that the current sheet fully collapses to be just a few cells in width within the multifluid solution, with no corresponding collapse within the GRMHD solution. The fact that the value of ${\mathbf{E} \cdot \mathbf{B}}$ converges to 0 in the current sheet confirms that genuine magnetic reconnection is taking place in the equatorial plane. This test demonstrates that the general relativistic multifluid solver is able to reproduce current sheet formation and magnetic reconnection behavior in the equatorial plane of the black hole, as predicted by the force-free and resistive electrodynamics simulations of \cite{komissarov_electrodynamics_2004}, which GRMHD simulations are unable to model self-consistently due to violation of the ${\left\lVert \mathbf{E} \right\rVert^2 \ll \left\lVert \mathbf{B} \right\rVert^2}$ assumption of ideal MHD within the black hole ergosphere. The ability of the general relativistic multifluid solver to sustain a strong poloidal magnetic field around the black hole, and to generate correspondingly large Poynting fluxes in the poloidal direction, is also likely to have implications for its ability to perform self-consistent modeling of black hole jets. We intend to explore this topic in detail as part of future work.

\section{3D Accretion Tests}
\label{sec:3d_accretion}

\subsection{Weakly Magnetized Accretion onto Black Holes}
\label{sec:weakly_magnetized_black_holes}

We begin our 3D accretion tests in curved spacetime by simulating the supersonic accretion of a magnetized ideal gas onto a (potentially spinning) black hole, representing a magnetized version of the gas accretion problems onto Schwarzschild (static) black holes studied by \cite{font_non-axisymmetric_1998}, and onto Kerr (spinning) black holes by \cite{font_non-axisymmetric_1999}. Our main objectives here are to confirm that the general relativistic multifluid solver is able to produce results that are in broad agreement with GRMHD simulations in the appropriate limit, while simultaneously capturing non-ideal effects, such as charge separation and large parallel electric field generation near the black hole, that the GRMHD model neglects. For the spacetime geometry, we use now the Cartesian form of the Kerr-Schild coordinate system ${\left\lbrace t, x, y, z \right\rbrace}$, such that the spacetime metric tensor ${g_{\mu \nu}}$ for a Kerr black hole spacetime of mass $M$ and spin ${a = \frac{J}{M}}$ now takes the form of a Kerr-Schild perturbation of the Minkowski metric ${\eta_{\mu \nu}}$ in Cartesian coordinates:

\begin{equation}
g_{\mu \nu} = \eta_{\mu \nu} - 2 V l_{\mu} l_{\nu}, \qquad \text{ with } \qquad \eta_{\mu \nu} = \mathrm{diag} \left( -1, 1, 1, 1 \right),
\end{equation}
with scalar $V$ and null (co)vector ${l_{\mu} = l^{\mu}}$, defined by:

\begin{equation}
V = - \frac{M R^3}{R^4 + M^2 a^2 z^2},
\end{equation}
and:

\begin{equation}
l^{\mu} = \begin{bmatrix}
-1 & - \frac{R x + a M y}{R^2 + M^2 a^2} & - \frac{R y - a M x}{R^2 + M^2 a^2} & - \frac{z}{R}
\end{bmatrix}^{\intercal},
\end{equation}
respectively, where $R$ is a generalized radial coordinate, defined implicitly via the algebraic equation:

\begin{equation}
\frac{x^2 + y^2}{R^2 + M^2 a^2} + \frac{z^2}{R^2} = 1,
\end{equation}
or explicitly through its unique positive solution:

\begin{equation}
R = \frac{1}{\sqrt{2}} \sqrt{Q},
\end{equation}
with:
\begin{multline}
Q = x^2 + y^2 + z^2 - M^2 a^2\\
+ \sqrt{\left( x^2 + y^2 + z^2 - M^2 a^2 \right)^2 + 4 M^2 a^2 z^2}.
\end{multline}
We select the same gauge conditions for the lapse function ${\alpha}$ and shift vector ${\beta^i}$ as in equation \ref{eq:kerr_schild_gauge} in the previous section, again yielding a foliation of the spacetime into a sequence of identical spacelike hypersurfaces whose induced (spatial) metric tensor ${\gamma_{i j}}$ takes the form of a Kerr-Schild perturbation of the flat metric ${\delta_{i j}}$ in Cartesian coordinates:

\begin{equation}
\gamma_{i j} = \delta_{i j} - 2 V l_i l_j, \qquad \text{ with } \qquad \delta_{i j} = \mathrm{diag} \left( 1, 1, 1 \right),
\end{equation}
with the extrinsic curvature tensor ${K_{i j}}$ taking the same form as before, i.e. given by equation \ref{eq:kerr_schild_extrinsic}. The Cartesian Kerr-Schild coordinate system ${\left( t, x, y, z \right)}$ can be obtained from the previously-used spherical Kerr-Schild coordinate system ${\left( t, r, \theta, \phi \right)}$ for a Kerr black hole of spin $a$ via the coordinate transformation:

\begin{multline}
x = \left[ r \cos \left( \phi \right) - a \sin \left( \phi \right) \right] \sin \left( \theta \right)\\
= \sqrt{r^2 + a^2} \sin \left( \theta \right) \cos \left( \phi - \arctan \left( \frac{a}{r} \right) \right),
\end{multline}
\begin{multline}
y = \left[ r \sin \left( \phi \right) + a \cos \left( \phi \right) \right] \sin \left( \theta \right)\\
= \sqrt{r^2 + a^2} \sin \left( \theta \right) \sin \left( \phi - \arctan \left( \frac{a}{r} \right) \right),
\end{multline}
with ${z = r \cos \left( \theta \right)}$. Once again, we place the excision boundary at the Schwarzschild radius ${r = 2 M}$, with a zero flux region on the interior ${r < 2 M}$.

\begin{figure*}
\centering
\subfigure{\includegraphics[trim={1cm, 0.5cm, 1cm, 0.5cm}, clip, width=0.45\textwidth]{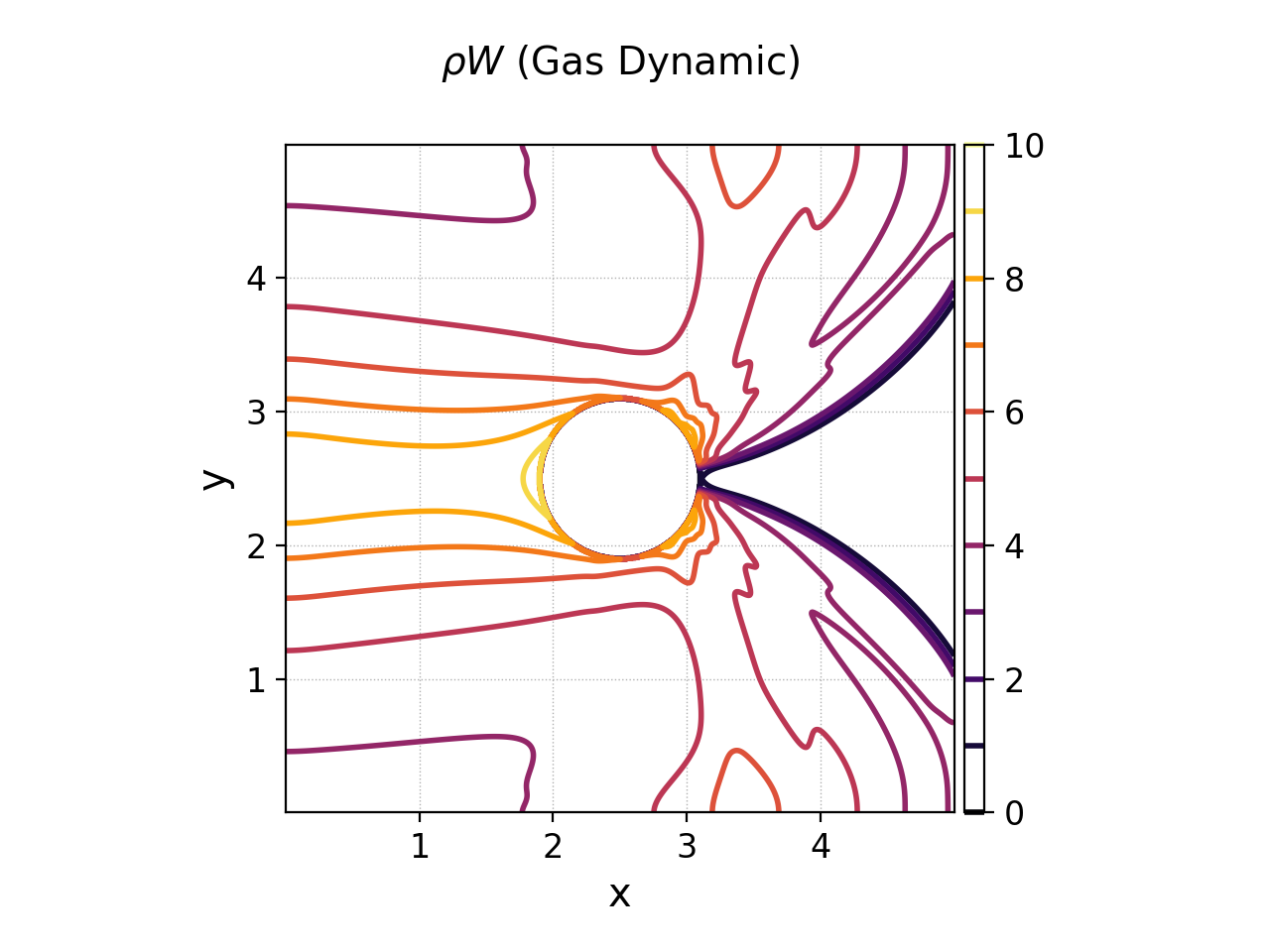}}
\subfigure{\includegraphics[trim={1cm, 0.5cm, 1cm, 0.5cm}, clip, width=0.45\textwidth]{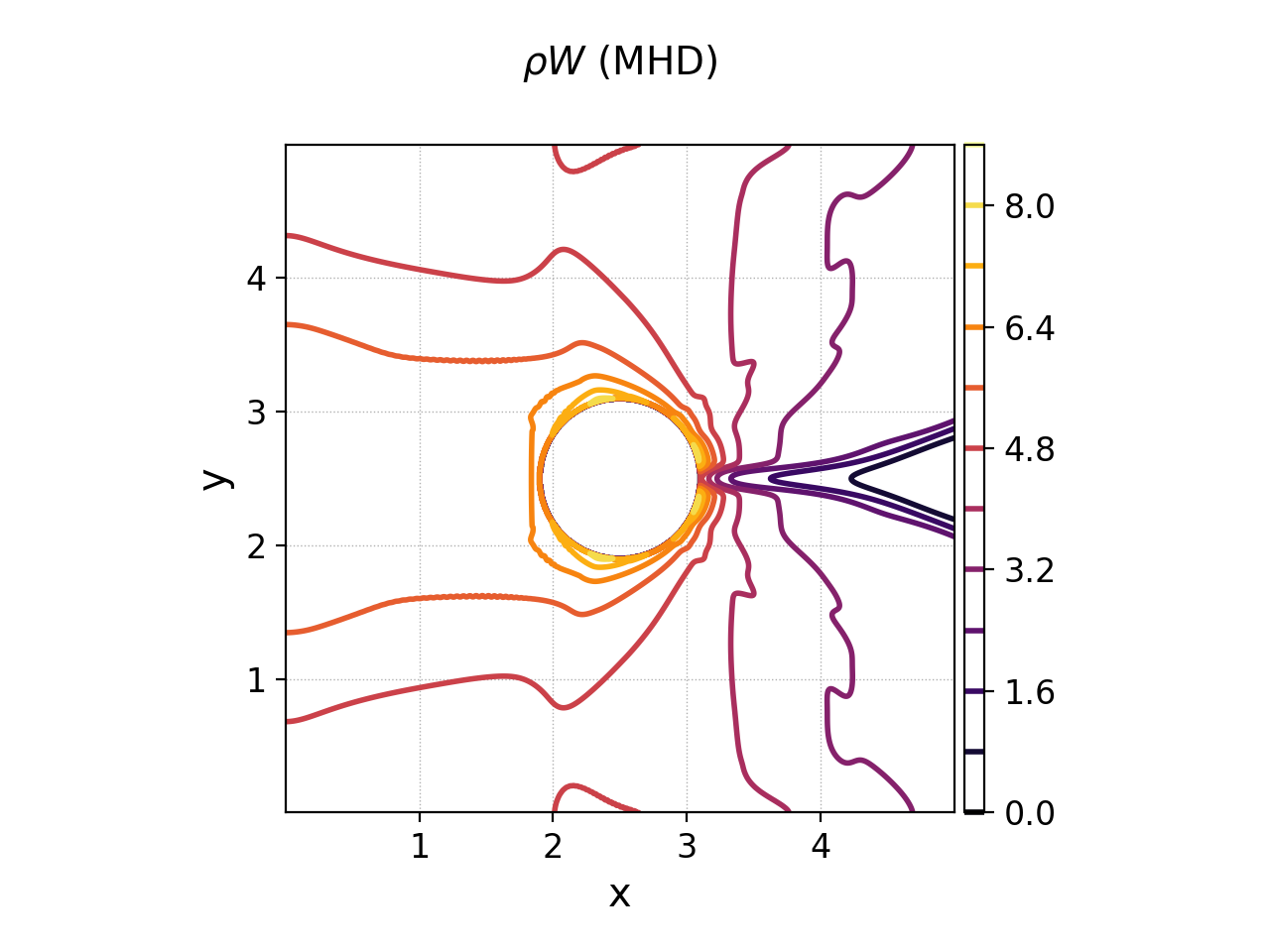}}
\caption{The relativistic mass density ${\rho W}$ contours at time ${t = 15}$ for the weakly magnetized accretion problem onto a static (Schwarzschild) black hole with ${v_{\infty} = 0.3}$, ${B_{\infty}^{z} = 1.0}$, and ${a = 0}$, showing the two asymptotic reference solutions, namely gas dynamic (${r_L \to \infty}$) on the left, and ideal MHD (${r_L \to 0}$) on the right. These accretion profiles have been taken through the equatorial plane ${z = 2.5}$ of the black hole. We observe slight numerical oscillations in the upstream fluid profile appearing within the ideal MHD solution.}
\label{fig:schwarzschild_mhd_reference}
\end{figure*}

We first consider the accretion of a weakly magnetized ideal gas onto a static (Schwarzschild) black hole, and formulate this problem as a simple 3D GRMHD wind accretion setup, with a cubical domain ${\left( x, y, z \right) \in \left[ 0, 5 \right]^3}$ containing a black hole of mass ${M = 0.3}$ and spin ${a = 0}$, centered at ${\left( x, y, z \right) = \left( 2.5, 2.5, 2.5 \right)}$. The black hole itself is initialized with a purely radial magnetic field (in spherical Kerr-Schild coordinates), based on the electrodynamic solutions of \cite{michel_rotating_1973} and \cite{blandford_electromagnetic_1977}:

\begin{equation}
B^r = \frac{1}{\sqrt{\gamma}} \left( B_0 \sin \left( \theta \right) \right),
\end{equation}
with strength ${B_0}$ set such that the initial magnetization near the black hole is ${\sigma = \frac{\left\lVert \mathbf{B} \right\rVert^2}{4 \pi \rho} = 100}$ (see Appendix \ref{sec:code_units_appendix}). We initialize a fluid velocity of ${v_{\infty} = 0.3}$ at spatial infinity, with the wind oriented directly towards the black hole, and with corresponding fluid density of ${\rho_{\infty} = 3.0}$, fluid pressure of ${p_{\infty} = 0.05}$, and a uniform $z$-magnetic field ${B_{\infty}^{z} = 1.0}$ at spatial infinity (with ${B_{\infty}^{x} = B_{\infty}^{y} = 0}$). In the remainder of the domain, the density and pressure are set to ${\rho_0 = \frac{1}{4 \pi}}$ and ${p_0 = 0.01}$ (again, see Appendix \ref{sec:code_units_appendix} for details of why this density is chosen), with the velocity vanishing. We assume a perfect monatomic gas, i.e. an ideal gas with adiabatic index of ${\Gamma = \frac{5}{3}}$, and use a spatial discretization of ${1024^3}$ cells, with an HLLC Riemann solver and a CFL coefficient of 0.9. We show the two asymptotic reference solutions at time ${t = 15}$, i.e. the gas dynamic solution (${r_L \to \infty}$, with zero coupling between the fluids and the magnetic field) and the ideal GRMHD solution (${r_L \to 0}$, with perfect coupling between the fluids and the magnetic field), in Figure \ref{fig:schwarzschild_mhd_reference}, displaying the relativistic mass density ${\rho W}$ contours through the equatorial plane ${z = 2.5}$ of the black hole. The peak magnetization near the black hole in the GRMHD solution is around ${\sigma \approx 135}$. We note that slight numerical oscillations in the upstream region of the black hole (i.e. ${x < 2.5}$) can be observed within the mass density contours ${\rho W}$ for the GRMHD solution, arising from small instabilities in the magnetic field.

\begin{figure*}
\centering
\subfigure{\includegraphics[trim={1cm, 0.5cm, 1cm, 0.5cm}, clip, width=0.45\textwidth]{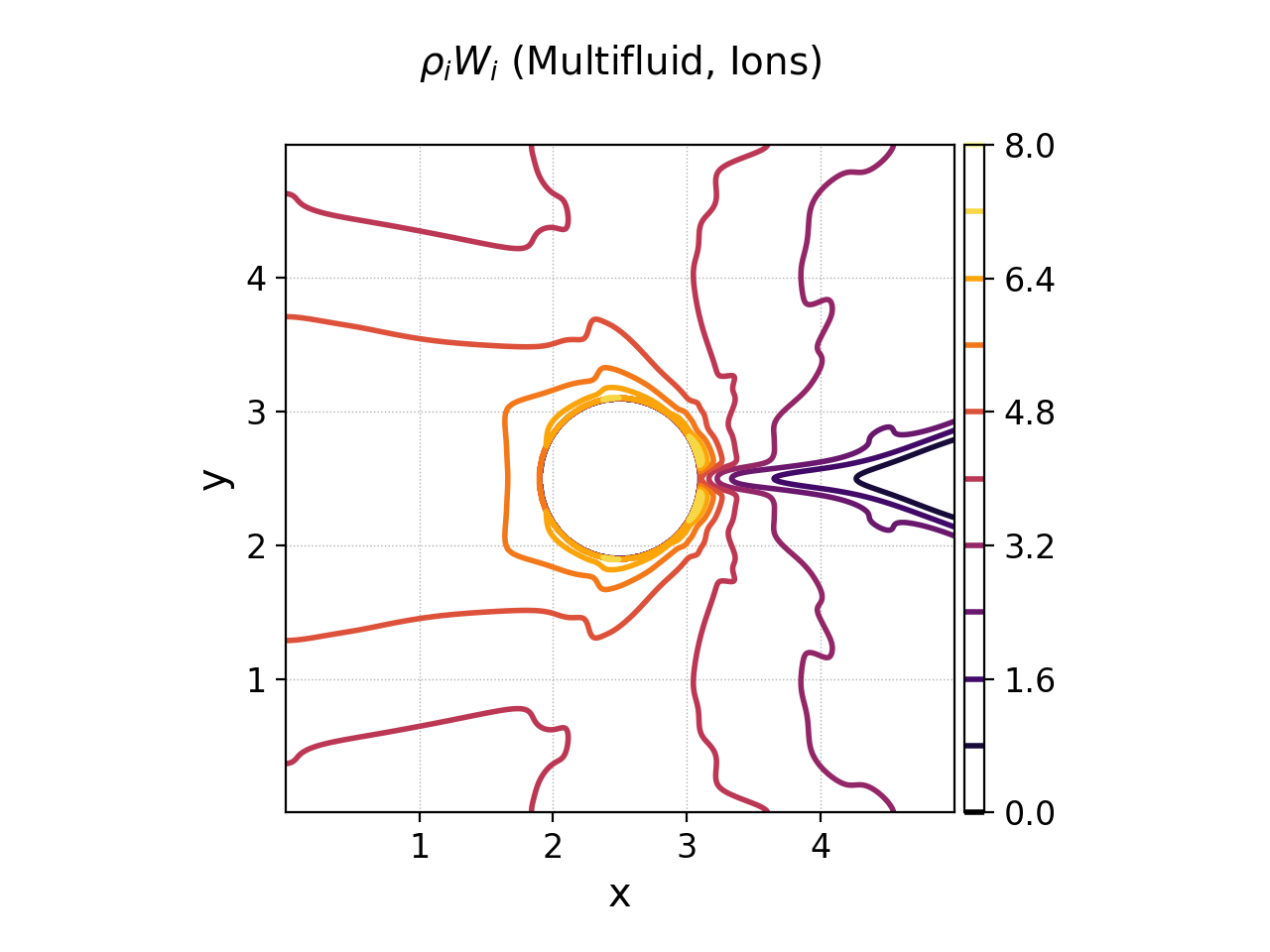}}
\subfigure{\includegraphics[trim={1cm, 0.5cm, 1cm, 0.5cm}, clip, width=0.45\textwidth]{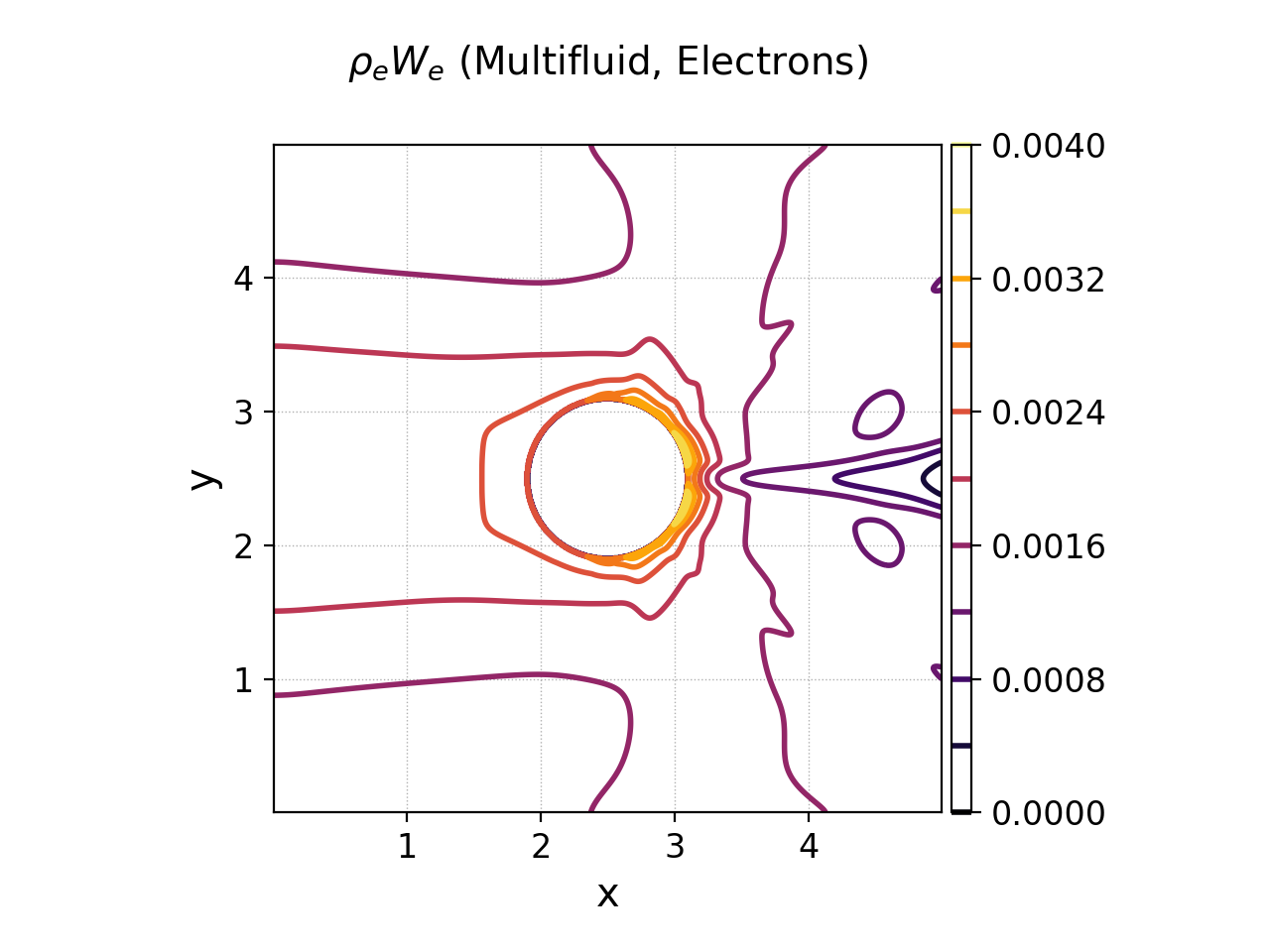}}
\caption{The relativistic mass density contours ${\rho_s W_s}$ at time ${t = 15}$ for the weakly magnetized accretion problem onto a static (Schwarzschild) black hole with ${v_{\infty} = 0.3}$, ${B_{\infty}^{z} = 1.0}$, and ${a = 0}$, showing the mass density of the ion fluid species (${\rho_i W_i}$) on the left, and the electron fluid species (${\rho_e W_e}$) on the right, obtained using the general relativistic multifluid solver with initial ion Larmor radius ${r_L = 0.1}$. These accretion profiles have been taken through the equatorial plane ${z = 2.5}$ of the black hole. The ion fluid profile broadly resembles the MHD fluid profile, although the numerical oscillations upstream of the black hole that were present within the ideal MHD solution have vanished. The differences between the ion and electron fluid profiles are indicative of the presence of non-ideal charge separation effects in the region surrounding the black hole.}
\label{fig:schwarzschild_mhd_multifluid}
\end{figure*}

\begin{figure*}
\centering
\subfigure{\includegraphics[width=0.45\textwidth]{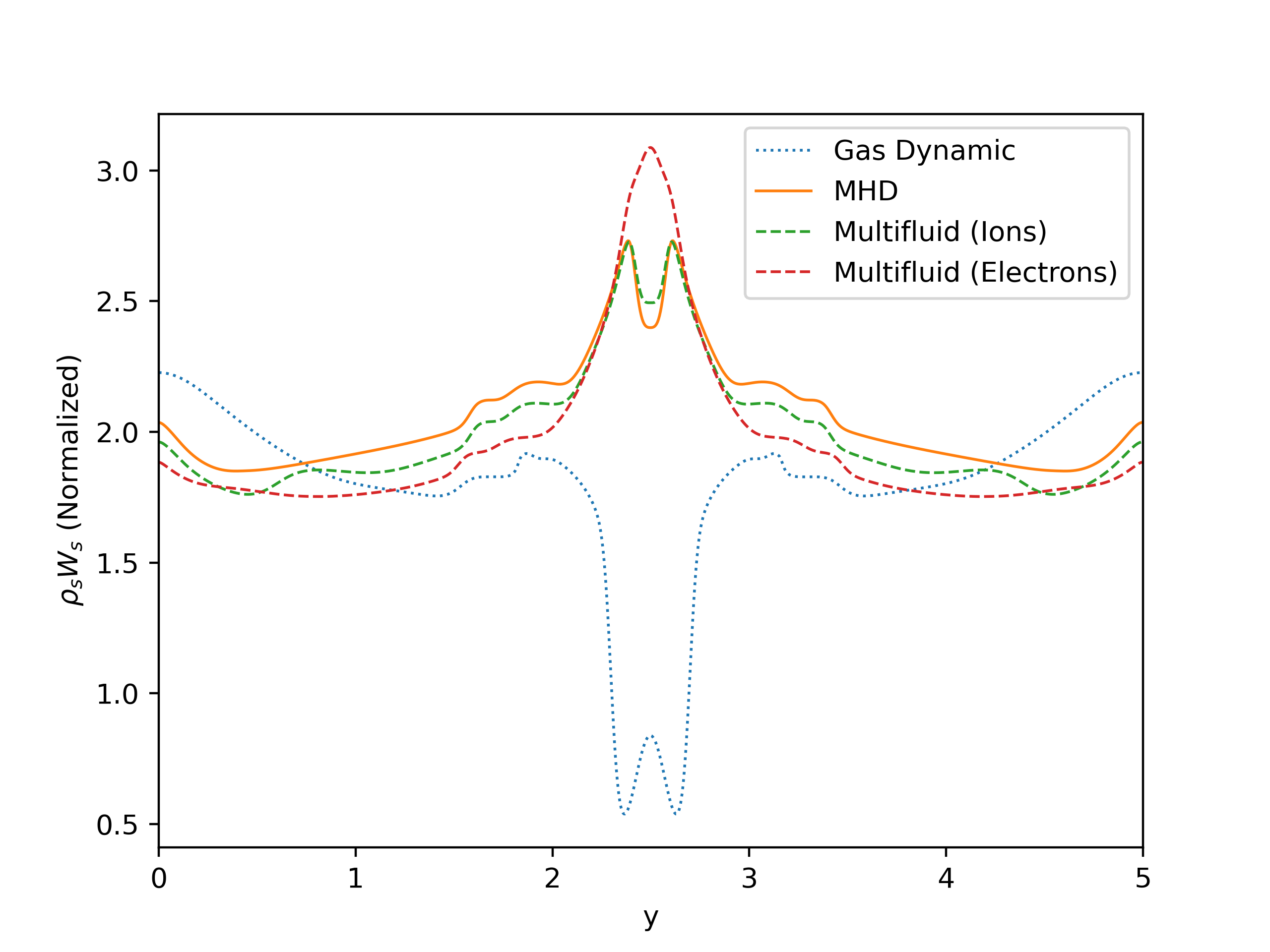}}
\subfigure{\includegraphics[width=0.45\textwidth]{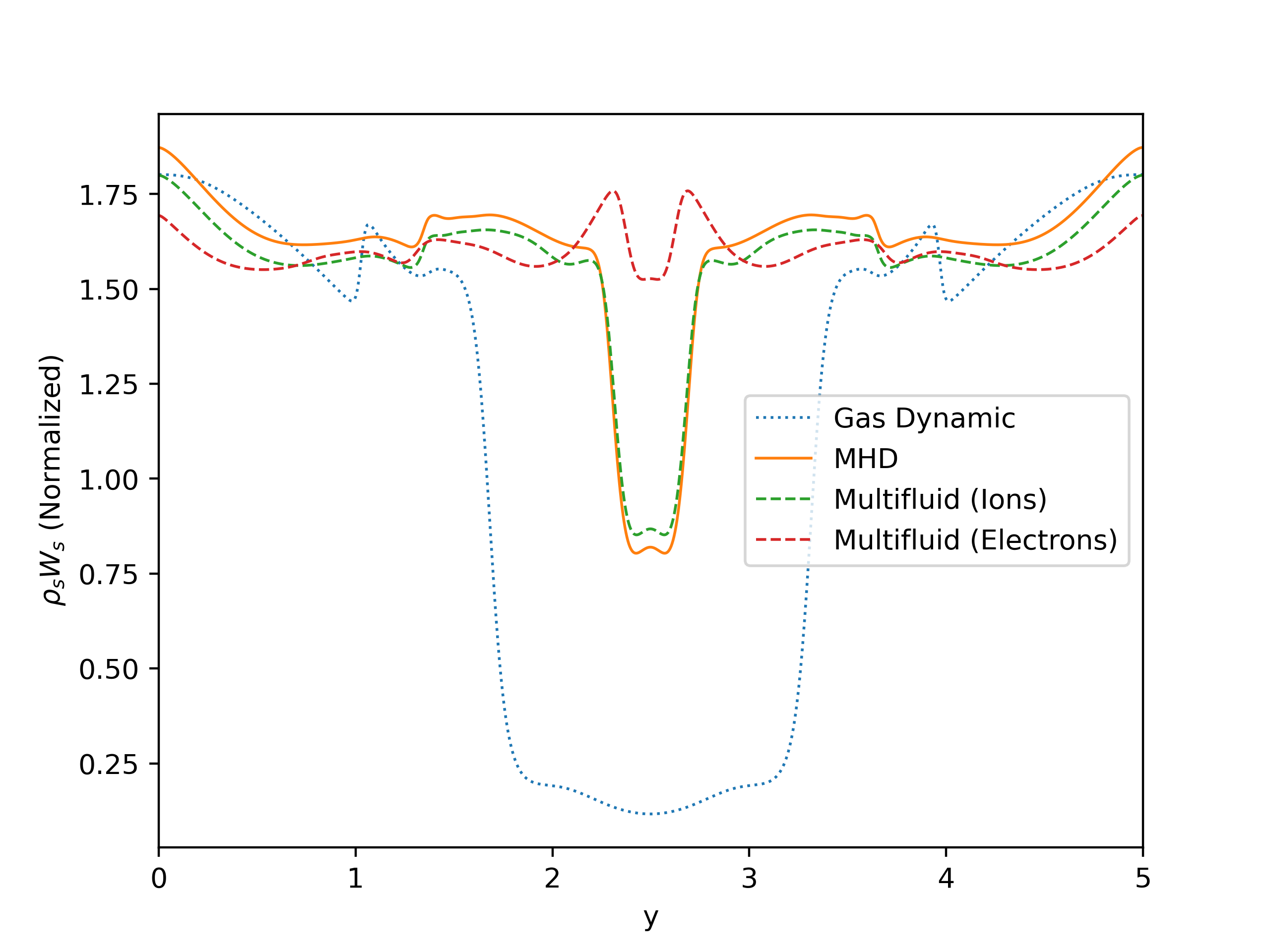}}
\caption{Cross-sectional profiles of the relativistic mass densities ${\rho_s W_s}$ for each fluid species at time ${t = 15}$ for the weakly magnetized accretion problem onto a static (Schwarzschild) black hole with ${v_{\infty} = 0.3}$, ${B_{\infty}^{z} = 1.0}$, and ${a = 0}$, taken in the near-downstream region (${x = 3.5}$) on the left, and the far-downstream region (${x = 4.5}$) on the right, of the black hole, obtained using the general relativistic multifluid solver with initial ion Larmor radius ${r_L = 0.1}$. The ideal MHD (${r_L \to 0}$) and gas dynamic (${r_L \to \infty}$) asymptotic reference solutions are also shown, and the relativistic mass density ${\rho_e W_e}$ of the electron fluid species has been multiplied by ${\frac{m_i}{m_e} = 1836.2}$ for normalization purposes. The profiles are taken in the equatorial plane ${z = 2.5}$ of the black hole. We see a small charge separation appearing around ${y = 2.5}$ in the near-downstream region of the black hole, which then becomes progressively more significant in the far-downstream region.}
\label{fig:schwarzschild_mhd_multifluid_stream}
\end{figure*}

We initialize a corresponding general relativistic two-fluid simulation (consisting of an electron fluid and an ion fluid) using the same prescription that we employed within previous sections, i.e. with ${\rho_{i, \infty} = 3.0}$ and ${\rho_{e, \infty} = 3.0 \frac{m_e}{m_i}}$, etc. The two fluids are assumed to be ideal gases with adiabatic indices ${\Gamma_e = \Gamma_i = \frac{5}{3}}$, with the ion-to-electron mass ratio set to ${\frac{m_i}{m_e} = 1836.2}$, and the electron and ion charges ${q_e = - q_i}$ set in order to yield an initial ion Larmor radius of ${r_L = \frac{m v_{\text{th}}}{\left\lvert q \right\rvert \left\lVert \mathbf{B} \right\rVert} = 0.1}$, as shown in Appendix \ref{sec:code_units_appendix}. The black hole is initialized with the same radial magnetic field as for GRMHD. Figure \ref{fig:schwarzschild_mhd_multifluid} shows the relativistic mass density contours produced by the general relativistic multifluid solver for both the ion fluid (i.e. ${\rho_i W_i}$) and the electron fluid (i.e. ${\rho_e W_e}$) at time ${t = 15}$, again through the equatorial plane ${z = 2.5}$ of the black hole. We note that the fluid profile for the ion species is broadly qualitatively similar to the fluid profile produced by GRMHD, albeit now without any oscillations in the ion mass density ${\rho_i W_i}$, or the magnetic field upstream of the black hole. However, the qualitative discrepancies between the shapes of the mass density profiles for the ion and electron fluid species are indicative of the presence of substantial (non-ideal) charge separation effects in the immediate vicinity of the black hole. This is confirmed quantitatively in Figure \ref{fig:schwarzschild_mhd_multifluid_stream}, which shows cross-sectional profiles for the relativistic mass densities of the ion fluid (i.e. ${\rho_i W_i}$) and the electron fluid (i.e. ${\rho_e W_e}$) at time ${t = 15}$, through the equatorial plane ${z = 2.5}$ of the black hole, and in both the near-downstream (${x = 3.5}$) and far-downstream (${x = 4.5}$) region. The relativistic mass density ${\rho_e W_e}$ of the electron fluid species has been normalized by multiplying it by ${\frac{m_i}{m_e} = 1836.2}$, to facilitate more direct comparison with the ion mass density. We have also included the gas dynamic (${r_L \to \infty}$) and ideal GRMHD (${r_L \to 0}$) asymptotic reference solutions for comparison, the latter of which broadly tracks the ion fluid solution. We see a small charge separation appearing around ${y = 2.5}$ in the near-downstream region of the black hole, manifesting as a peak in the electron number density, which then grows progressively more substantial towards the far-downstream region. The maintenance of such a charge separation is indicative of the induction of a significant (parallel) electric field downstream of the black hole. The peak magnetization near the black hole in the general relativistic multifluid solution is slightly higher than it is for GRMHD, at around ${\sigma \approx 155}$\footnote{Specifically, this is the peak magnetization of the \textit{ion} species. In this case, and all that follow, peak magnetization for the electrons is a factor of ${\frac{m_i}{m_e} = 1836.2}$ larger.}.

\begin{figure*}
\centering
\subfigure{\includegraphics[trim={1cm, 0.5cm, 1cm, 0.5cm}, clip, width=0.45\textwidth]{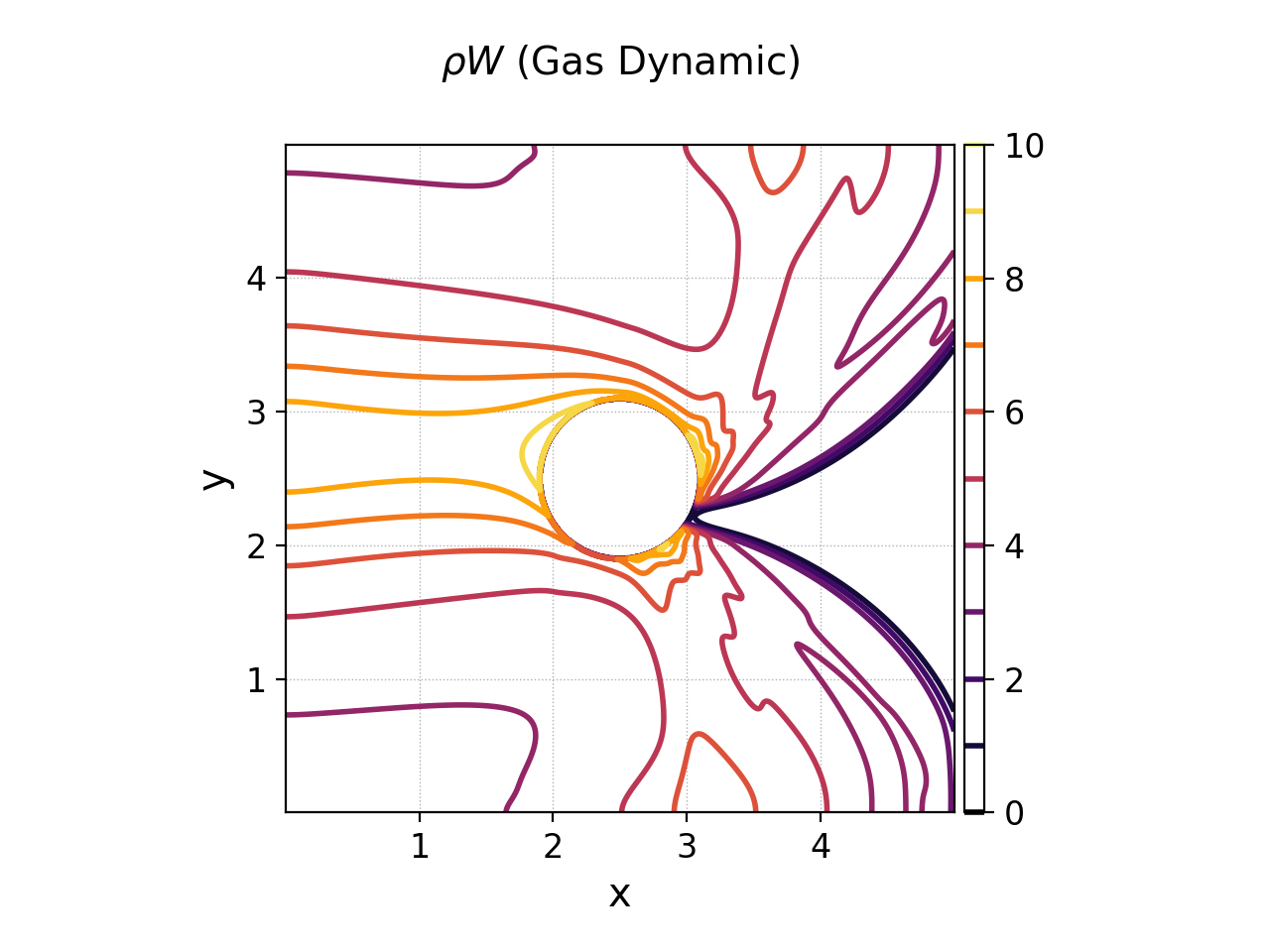}}
\subfigure{\includegraphics[trim={1cm, 0.5cm, 1cm, 0.5cm}, clip, width=0.45\textwidth]{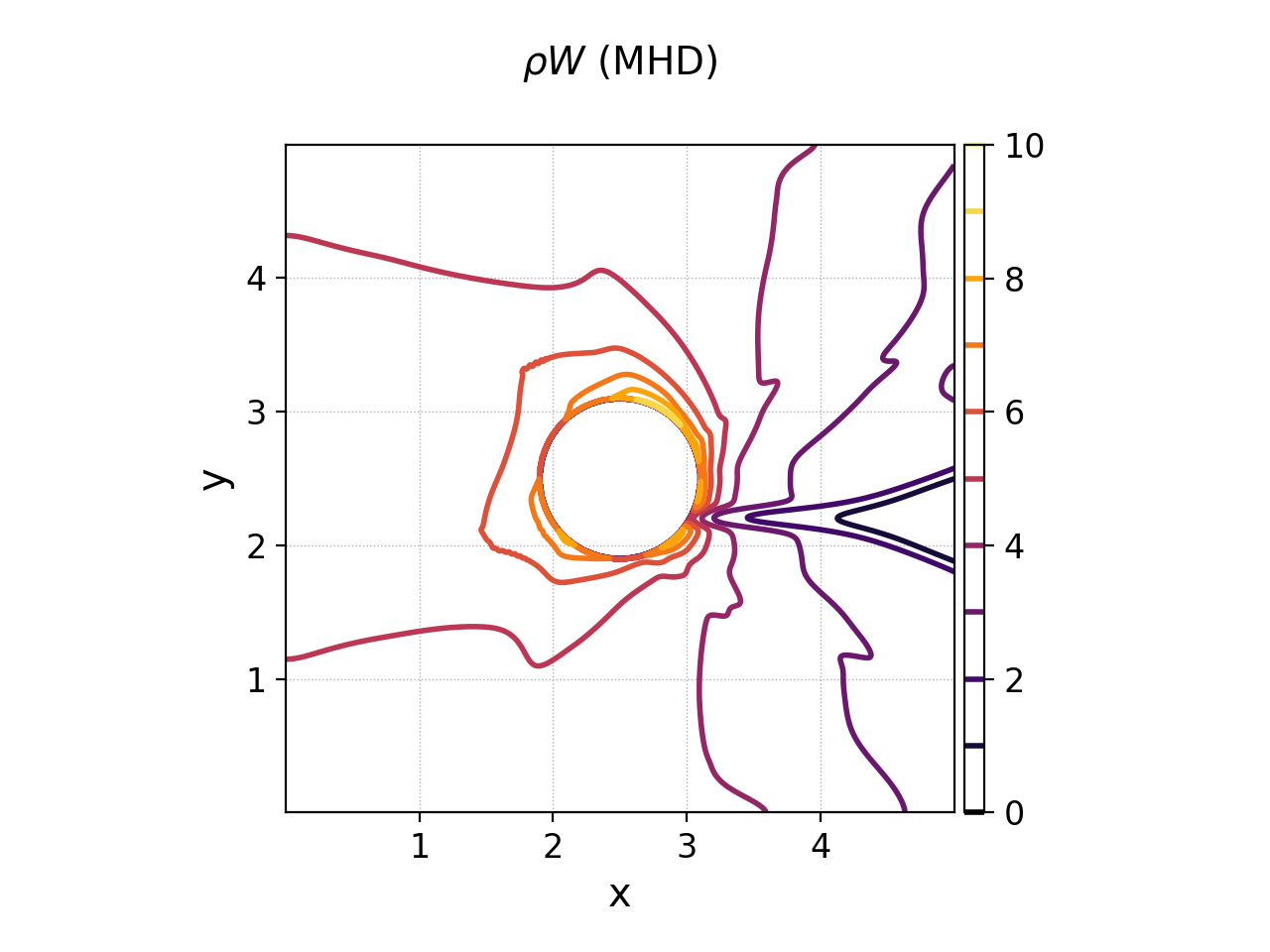}}
\caption{The relativistic mass density ${\rho W}$ contours at time ${t = 15}$ for the weakly magnetized accretion problem onto a spinning (Kerr) black hole with ${v_{\infty} = 0.3}$, ${B_{\infty}^{z} = 1.0}$, and ${a = 0.9}$, showing the two asymptotic reference solutions, namely gas dynamic ${r_L \to \infty}$ on the left, and ideal MHD ${r_L \to 0}$ on the right. These accretion profiles have been taken through the equatorial plane ${z = 2.5}$ of the black hole. As in the static (Schwarzschild) case, we observe slight numerical oscillations in the upstream fluid profile appearing within the ideal MHD solution.}
\label{fig:kerr_mhd_reference}
\end{figure*}

\begin{figure*}
\centering
\subfigure{\includegraphics[trim={1cm, 0.5cm, 1cm, 0.5cm}, clip, width=0.45\textwidth]{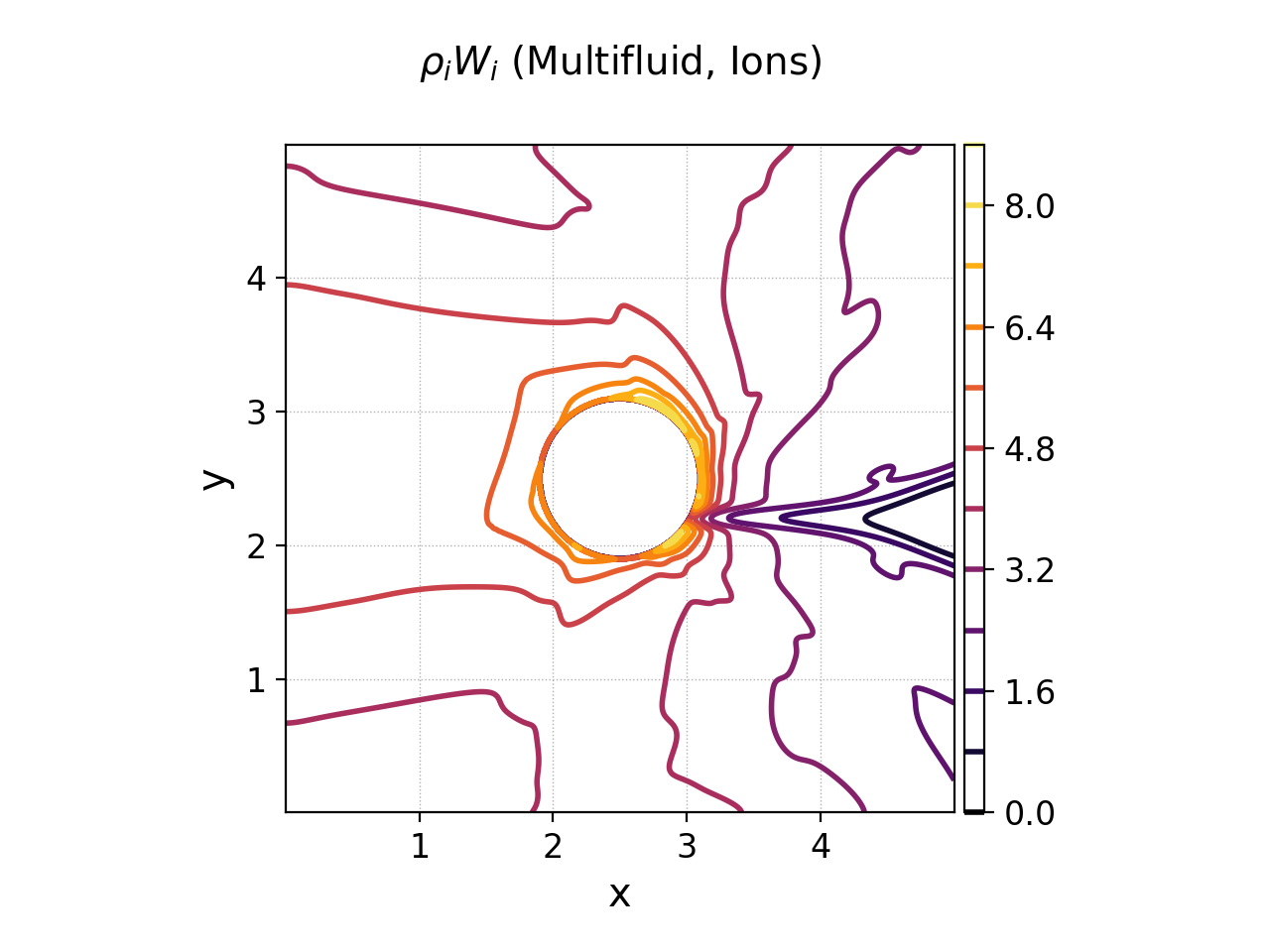}}
\subfigure{\includegraphics[trim={1cm, 0.5cm, 1cm, 0.5cm}, clip, width=0.45\textwidth]{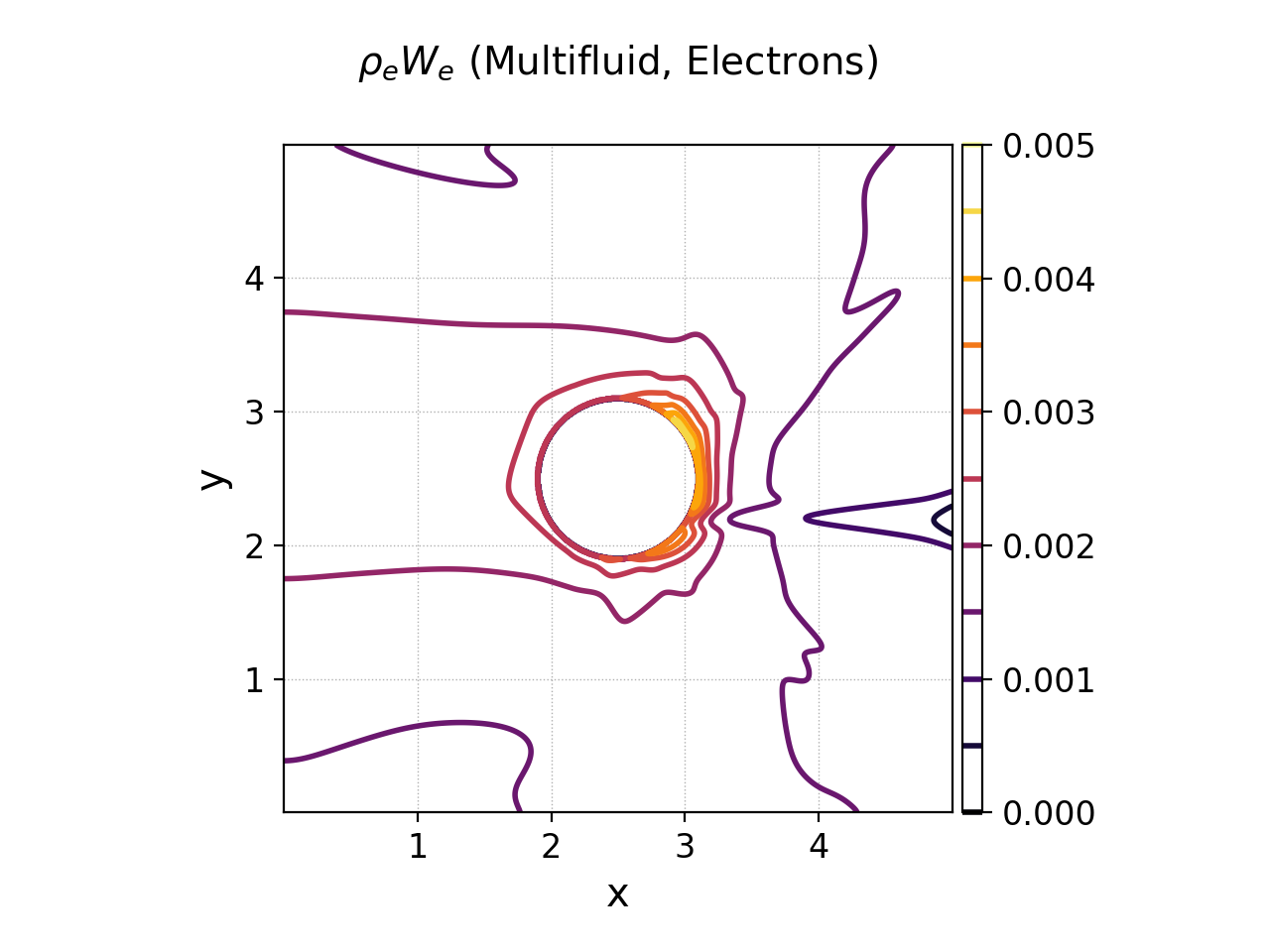}}
\caption{The relativistic mass density contours ${\rho_s W_s}$ at time ${t = 15}$ for the weakly magnetized accretion problem onto a spinning (Kerr) black hole with ${v_{\infty} = 0.3}$, ${B_{\infty}^{z} = 1.0}$, and ${a = 0.9}$, showing the mass density of the ion fluid species (${\rho_i W_i}$) on the left, and the electron fluid species (${\rho_e W_e}$) on the right, obtained using the general relativistic multifluid solver with initial ion Larmor radius ${r_L = 0.1}$. These accretion profiles have been taken through the equatorial plane ${z = 2.5}$ of the black hole. As in the static (Schwarzschild) case, the ion fluid profile broadly resembles the MHD fluid profile, with the numerical oscillations upstream of the black hole having vanished. The more substantive differences between the ion and electron fluid profiles are indicative of the presence of larger non-ideal charge separation effects near the black hole.}
\label{fig:kerr_mhd_multifluid}
\end{figure*}

We also consider wind accretion of a weakly magnetized ideal gas, obeying the same initial conditions as for the Schwarzschild case, onto a spinning (Kerr) black hole of mass ${M = 0.3}$ and spin ${a = 0.9}$. Our simulation setup is the same as before, with a cubical domain ${\left( x, y, z \right) \in \left[ 0, 5 \right]^3}$ containing the Kerr black hole at ${\left( x, y, z \right) = \left( 2.5, 2.5, 2.5 \right)}$, with the same radial magnetic field with magnetization ${\sigma = \frac{\left\lVert \mathbf{B} \right\rVert^2}{4 \pi \rho} = 100}$ as for the Schwarzschild case, using a discretization of ${1024^3}$ cells, an HLLC Riemann solver, and a CFL coefficient of 0.9. The two asymptotic reference solutions (${r_L \to \infty}$ for gas dynamic, ${r_L \to 0}$ for ideal GRMHD) at time ${t = 15}$ are shown in Figure \ref{fig:kerr_mhd_reference}, where we display the relativistic mass density ${\rho W}$ contours through the equatorial plane ${z = 2.5}$ of the black hole, as was done previously for Schwarzschild. The peak magnetization near the black hole in the GRMHD solution is now around ${\sigma \approx 150}$, with a small increase due to the twisting of magnetic field lines from frame-dragging effects within the ergosphere of the black hole. The same small numerical oscillations immediately upstream of the black hole (${x < 2.5}$) in the mass density contours ${\rho W}$, resulting from mild magnetic field instabilities, are again observed in the GRMHD solution. The corresponding general relativistic two-fluid solution is found in Figure \ref{fig:kerr_mhd_multifluid}, with the relativistic mass density contours for both the ion fluid (${\rho_i W_i}$) and the electron fluid (${\rho_e W_e}$) shown at time ${t = 15}$, again through the equatorial plane ${z = 2.5}$ of the black hole. As in the Schwarzschild case, the mass density profile of the ion species ${\rho_i W_i}$ exhibits qualitatively similar behavior to the fluid profile produced using GRMHD, but with the magnetic field-induced density oscillations in the upstream black hole region now eliminated. The qualitative discrepancies between the ion and electron fluid profiles (indicative of charge separation) appear somewhat more significant, especially downstream of the black hole, for the Kerr black hole case than they did for the Schwarzschild black hole, presumably due to the induction of larger (parallel) electric fields within the ergosphere as a consequence of the black hole's rotation.

\begin{figure*}
\centering
\subfigure{\includegraphics[width=0.45\textwidth]{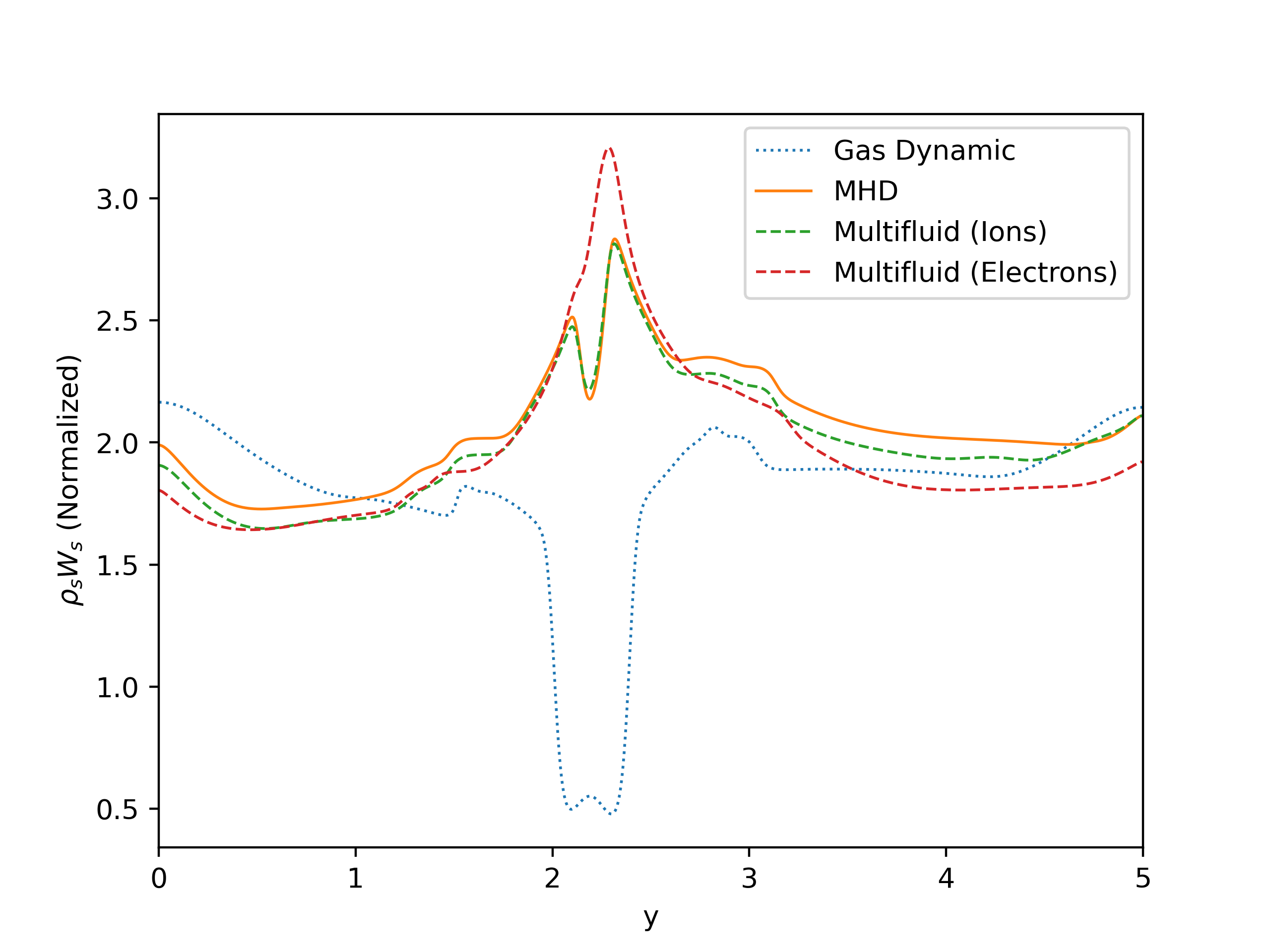}}
\subfigure{\includegraphics[width=0.45\textwidth]{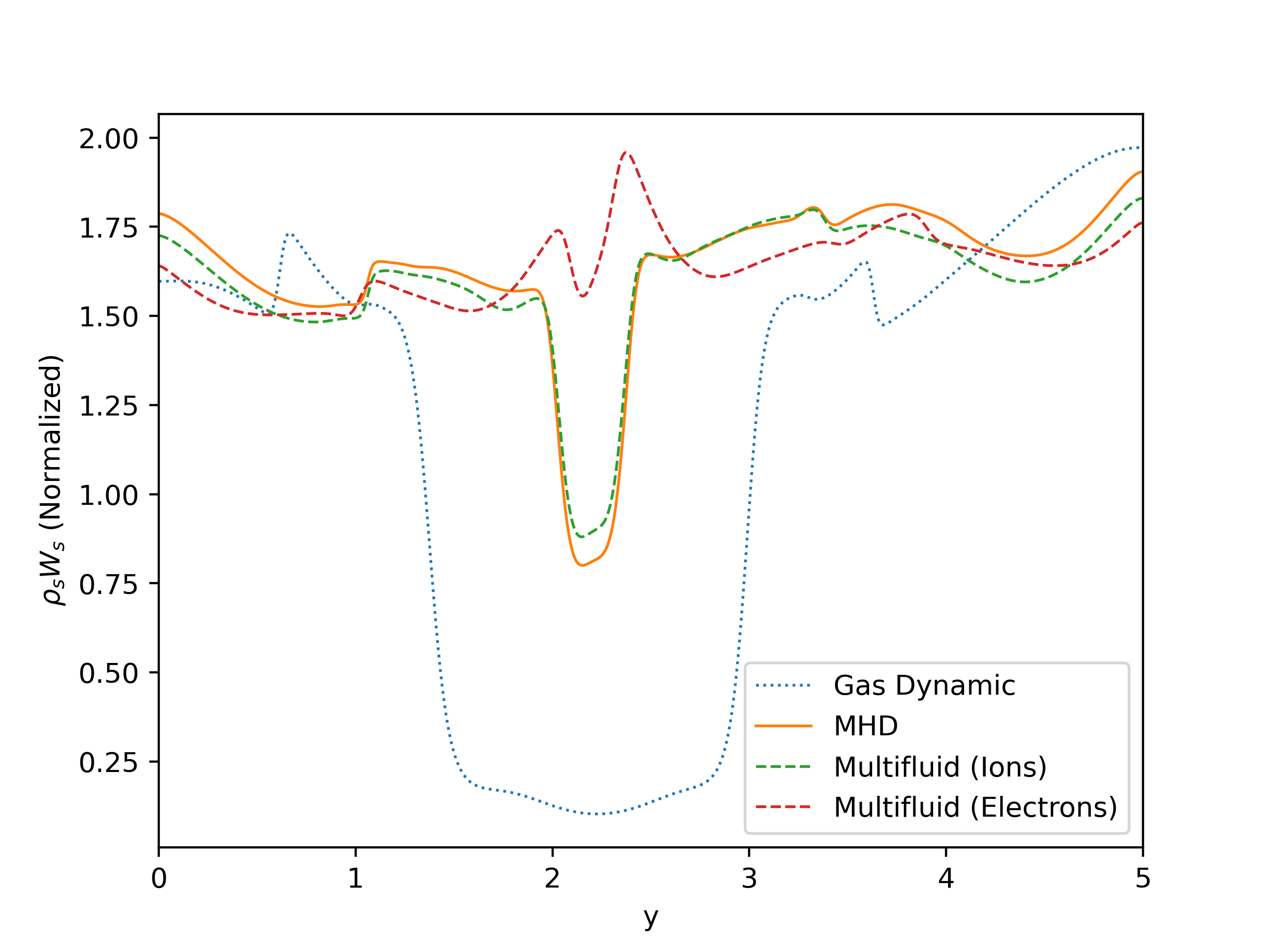}}
\caption{Cross-sectional profiles of the relativistic mass densities ${\rho_s W_s}$ for each fluid species at time ${t = 15}$ for the weakly magnetized accretion problem onto a spinning (Kerr) black hole with ${v_{\infty} = 0.3}$, ${B_{\infty}^{z} = 1.0}$, and ${a = 0.9}$, taken in the near-downstream region (${x = 3.5}$) on the left, and the far-downstream region (${x = 4.5}$) on the right, of the black hole, obtained using the general relativistic multifluid solver with initial ion Larmor radius ${r_L = 0.1}$. The ideal MHD (${r_L \to 0}$) and gas dynamic (${r_L \to \infty}$) asymptotic reference solutions are also shown, and the relativistic mass density ${\rho_e W_e}$ of the electron fluid species has been multiplied by ${\frac{m_i}{m_e} = 1836.2}$ for normalization purposes. The profiles are taken in the equatorial plane ${z = 2.5}$ of the black hole. We see a larger charge separation than in the static (Schwarzschild) case appearing around ${y = 2.2}$ in the near-downstream region of the black hole, which again becomes progressively more significant in the far-downstream region.}
\label{fig:kerr_mhd_multifluid_stream}
\end{figure*}

\begin{figure*}
\centering
\subfigure{\includegraphics[width=0.45\textwidth]{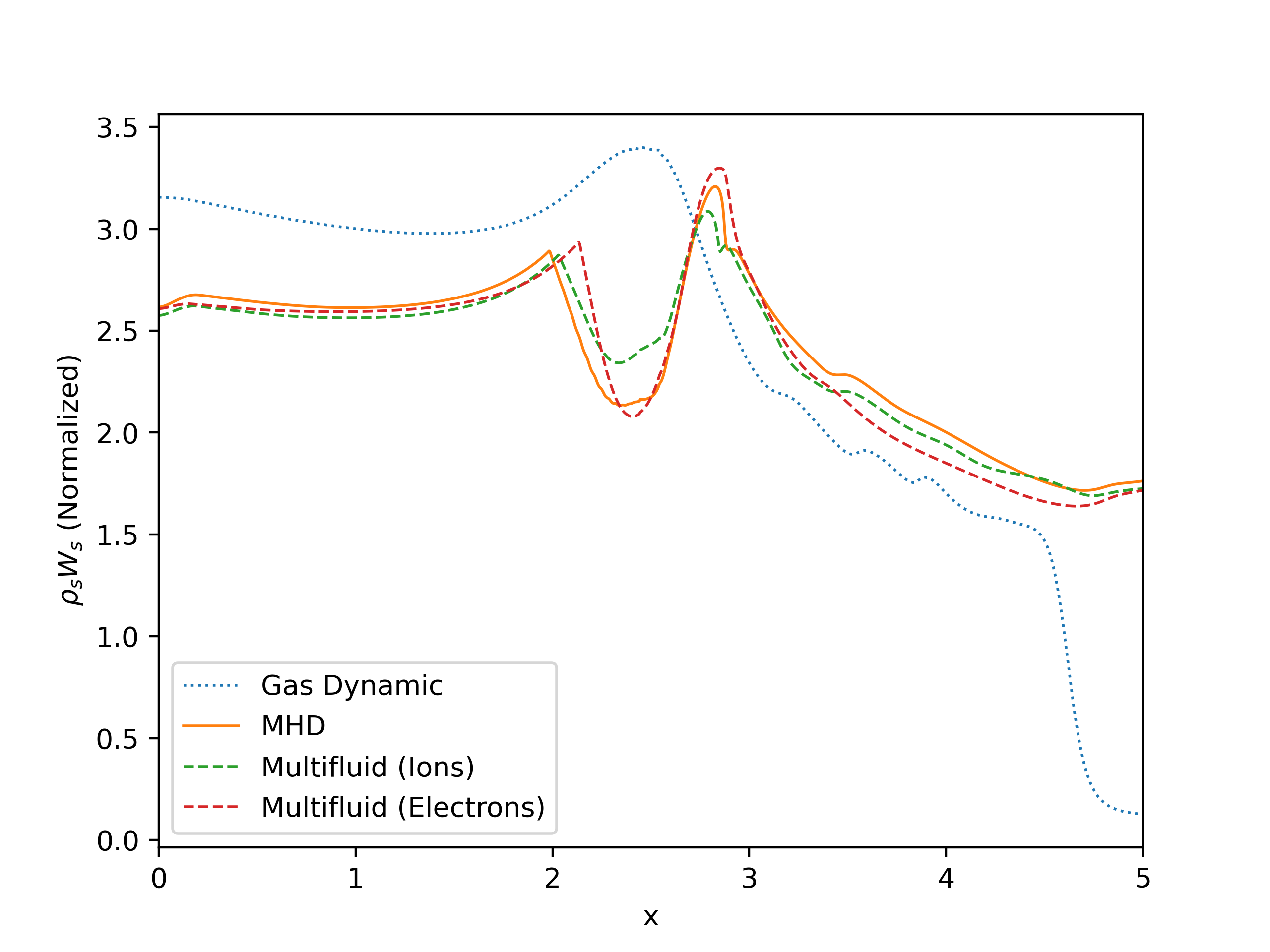}}
\subfigure{\includegraphics[width=0.45\textwidth]{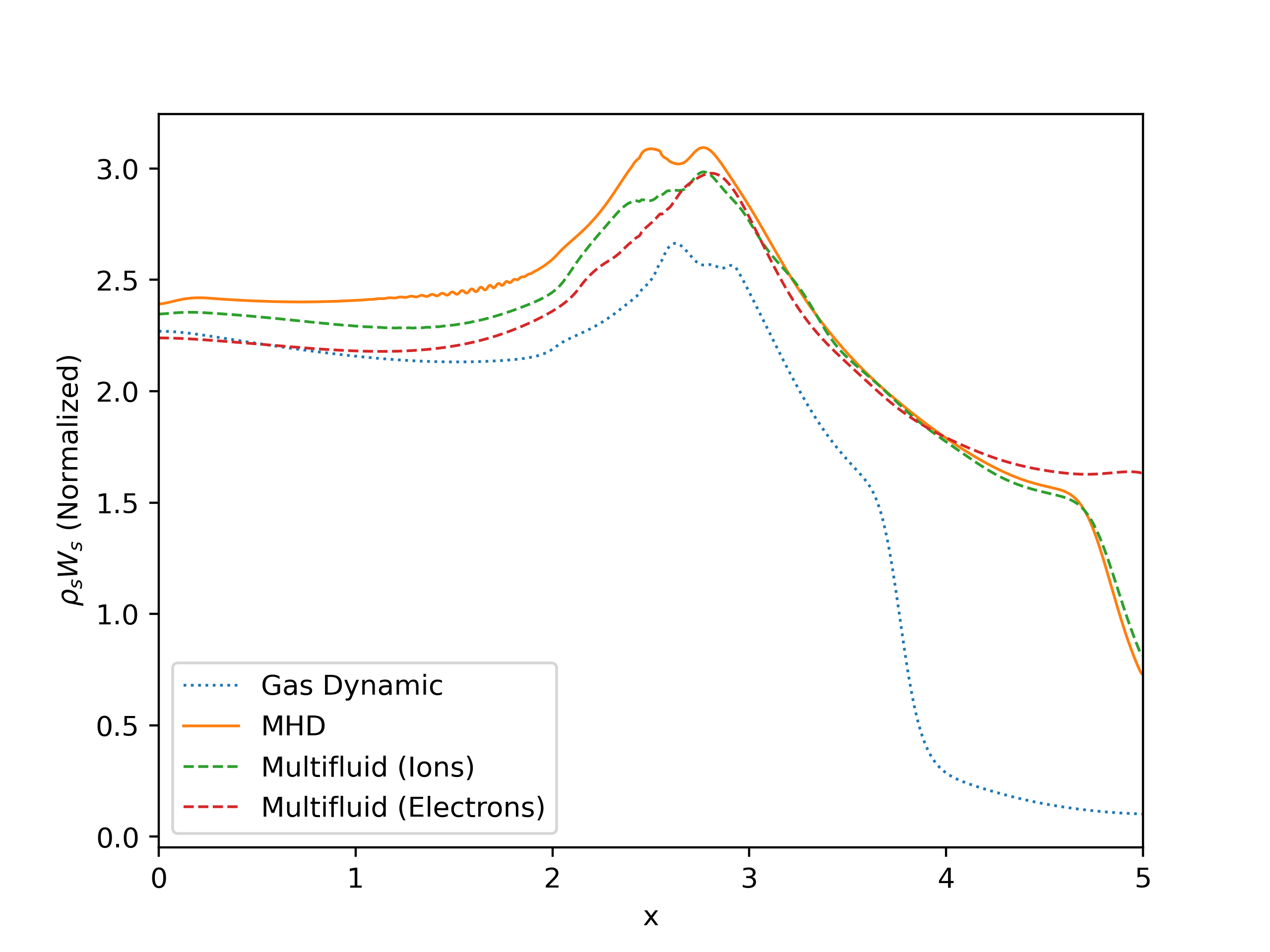}}
\caption{Cross-sectional profiles of the relativistic mass densities ${\rho_s W_s}$ for each fluid species at time ${t = 15}$ for the weakly magnetized accretion problem onto a spinning (Kerr) black hole with ${v_{\infty} = 0.3}$, ${B_{\infty}^{z} = 1.0}$, and ${a = 0.9}$, taken on the co-rotating (${y = 3.1}$) and counter-rotating (${y = 1.9}$) sides of the black hole, obtained using the general relativistic multifluid solver with initial ion Larmor radius ${r_L = 0.1}$. The ideal MHD (${r_L \to 0}$) and gas dynamic (${r_L \to \infty}$) asymptotic reference solutions are also shown, and the relativistic mass density ${\rho_e W_e}$ of the electron fluid species has been multiplied by ${\frac{m_i}{m_e} = 1836.2}$ for normalization purposes. The profiles are taken in the equatorial plane ${z = 2.5}$ of the black hole. We see very close agreement between the MHD, ion fluid, and electron fluid solutions on the co-rotating side, but on the counter-rotating side a significant charge separation arises in the far-downstream (${x > 4.5}$) region due to the effects of ion vs. electron inertia causing preferential acceleration of the electrons.}
\label{fig:kerr_mhd_multifluid_stream2}
\end{figure*}

\begin{figure*}
\centering
\subfigure{\includegraphics[trim={1cm, 0.5cm, 1cm, 0.5cm}, clip, width=0.45\textwidth]{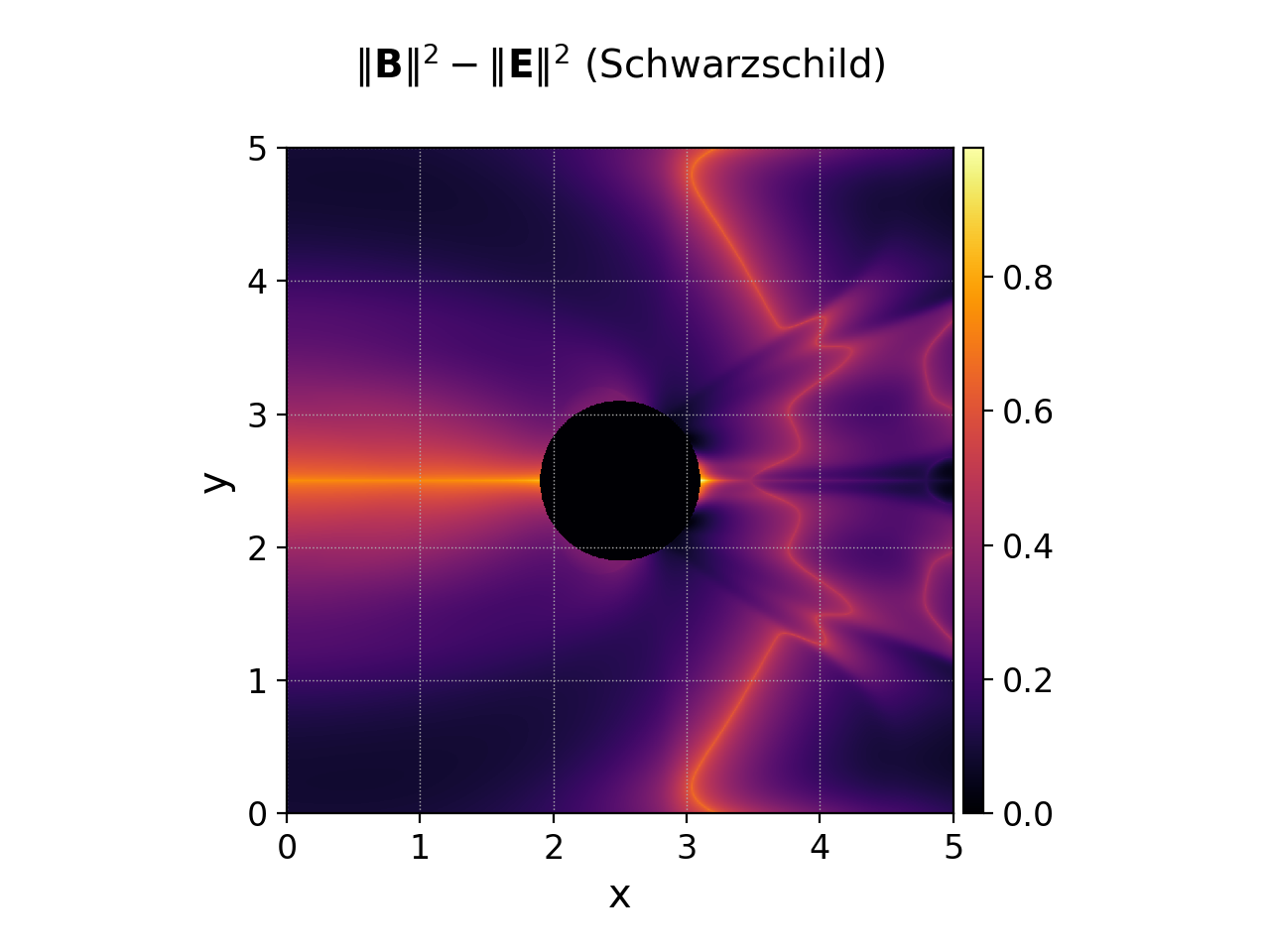}}
\subfigure{\includegraphics[trim={1cm, 0.5cm, 1cm, 0.5cm}, clip, width=0.45\textwidth]{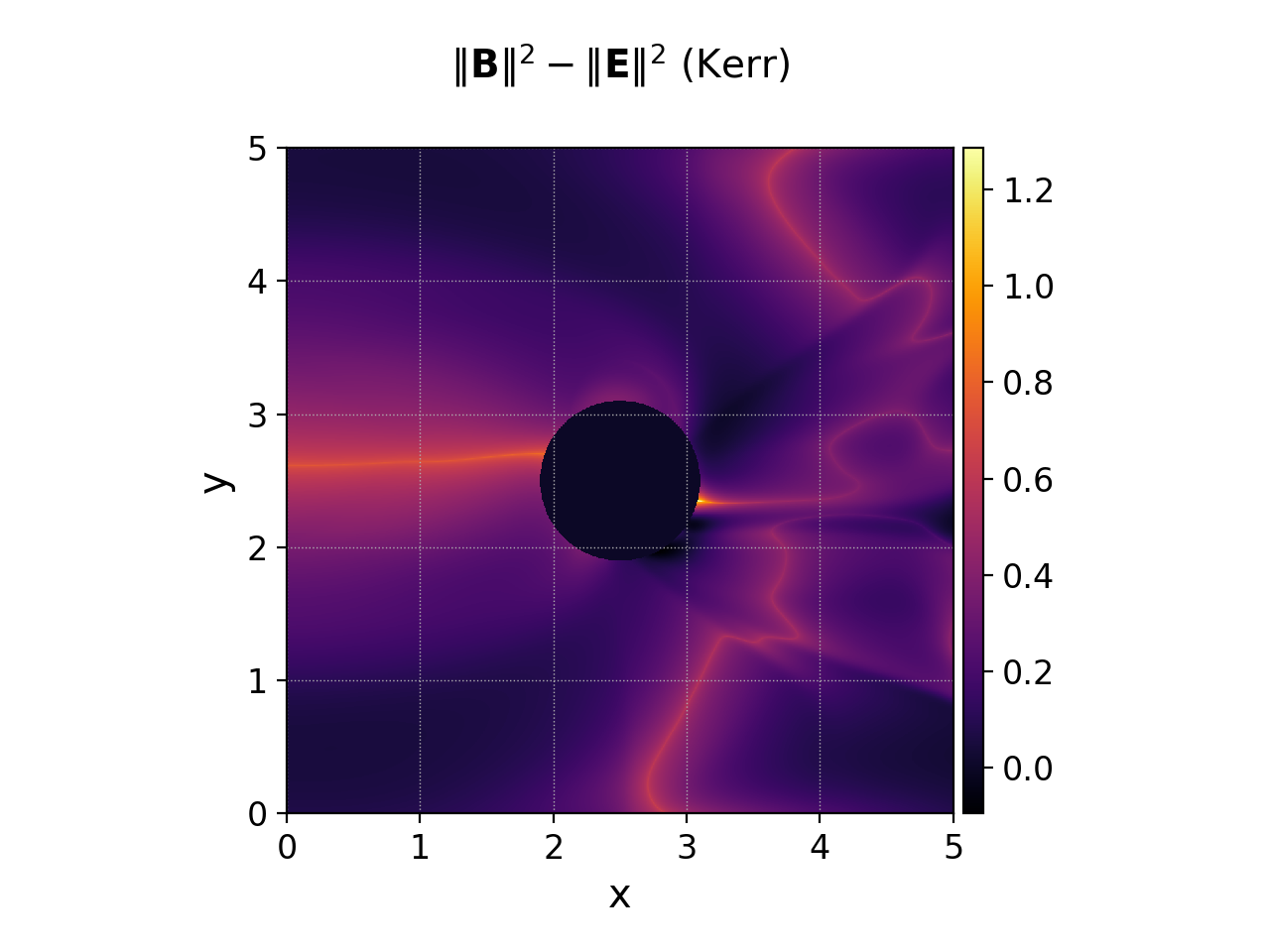}}
\caption{The Lorentz invariant scalar ${\left\lVert \mathbf{B} \right\rVert^2 - \left\lVert \mathbf{E} \right\rVert^2}$ at time ${t = 15}$ for the weakly magnetized accretion problem with ${v_{\infty} = 0.3}$ and ${B_{\infty}^{z} = 1.0}$ onto a static black hole (Schwarzschild, ${a = 0}$) on the left, and onto a spinning black hole (Kerr, ${a = 0.9}$) on the right, obtained using the general relativistic multifluid solver with initial ion Larmor radius ${r_L = 0.1}$. These field profiles have been taken through the equatorial plane ${z = 2.5}$ of the black hole. In the Schwarzschild case, we see dark regions immediately downstream of the black hole where ${\left\lVert \mathbf{B} \right\rVert^2 - \left\lVert \mathbf{E} \right\rVert^2}$ is driven close to zero, indicating the presence of large (parallel) electric fields. In the Kerr case, ${\left\lVert \mathbf{B} \right\rVert^2 - \left\lVert \mathbf{E} \right\rVert^2}$ is actually driven slightly negative in these regions as a consequence of stronger electric field induction resulting from the black hole's rotation.}
\label{fig:blackhole_mhd_multifluid_field}
\end{figure*}

Again, we can confirm this quantitatively by examining the cross-sectional profiles for the relativistic mass densities of the ion fluid (${\rho_i W_i}$) and the electron fluid (${\rho_e W_e}$) in both the near-downstream (${x = 3.5}$) and far-downstream (${x = 4.5}$) regions of the black hole at time ${t = 15}$, as shown in Figure \ref{fig:kerr_mhd_multifluid_stream}, with profiles taken through the equatorial plane ${z = 2.5}$ as usual. As before, we also include the gas dynamic (${r_L \to \infty}$) and ideal GRMHD (${r_L \to 0}$) reference solutions, and multiply the electron mass density ${\rho_e W_e}$ by ${\frac{m_i}{m_e} = 1836.2}$ to facilitate comparisons. We see that the initial charge separation now appears around ${y = 2.2}$ in the near-downstream region of the black hole, and is significantly larger on the left (counter-rotating) side than on the right (co-rotating) side, as a consequence of frame-dragging effects. The overall charge separation grows more substantive, as in the Schwarzschild case, as one moves towards the far-downstream region of the black hole, but the aforementioned asymmetry also reverses, with the charge separation now being slightly larger on the co-rotating side than on the counter-rotating one. We can explore this rotation-induced asymmetry in the accretion profile more directly by taking corresponding cross-sections in the same ${z = 2.5}$ equatorial plane, but now on the co-rotating (${y = 3.1}$) and counter-rotating (${y = 1.9}$) sides of the black hole, as shown in Figure \ref{fig:kerr_mhd_multifluid_stream2}. We find, as expected, that the fluid profiles for ions, electrons, and GRMHD agree very closely on the co-rotating side of the black hole where frame-dragging effects accelerate all fluid species uniformly, but a significant charge separation appears in the far-downstream (${x > 4.5}$) region on the counter-rotating side, as a result of the differences in electron vs. ion inertia, causing the moving electrons to be preferentially accelerated by the radial magnetic field due to their lower mass. Finally, we can confirm that these charge separations, in both the Schwarzschild and Kerr cases, are indeed sustained by a very large value of ${\mathbf{E} \times \mathbf{B}}$ downstream of the black hole (signaling the transition from magnetically-dominated to electrically-dominated flow, consistent with the presence of large parallel electric fields and other physics beyond the magnetohydrodynamic approximation), by plotting the Lorentz invariant scalar quantity ${\left\lVert \mathbf{B} \right\rVert^2 - \left\lVert \mathbf{E} \right\rVert^2}$ through the equatorial plane ${z = 2.5}$ for both black holes, as shown in Figure \ref{fig:blackhole_mhd_multifluid_field}. In the Schwarzschild case, we see the clear formation of a current sheet in the near-downstream (${3.0 < x < 4.0}$) region of the black hole, as well as dark regions where ${\left\lVert \mathbf{B} \right\rVert^2 - \left\lVert \mathbf{E} \right\rVert^2}$ is being driven close to zero, and therefore in which the (parallel) electric fields have become very strong. In the Kerr case, the current sheet has become slightly more smeared due to the presence of significant frame-dragging effects, but the rotation of the black hole has also caused ${\left\lVert \mathbf{B} \right\rVert^2 - \left\lVert \mathbf{E} \right\rVert^2}$ to be driven slightly negative in the near-downstream region, due to the increased rotational induction of parallel electric fields within the ergosphere. This indicates a strong violation of the ${\left\lVert \mathbf{E} \right\rVert^2 \ll \left\lVert \mathbf{B} \right\rVert^2}$ assumption inherent to the GRMHD approximation, and thus confirms that physics which has previously only ever been simulated using kinetic, first-principles methods, has been successfully captured using the multifluid approach.

\subsection{Strongly Magnetized Accretion onto Neutron Stars}
\label{sec:strongly_magnetized_neutron_stars}

We complete our 3D accretion tests in curved spacetime by simulating the supersonic accretion of a magnetized ideal gas onto a strongly magnetized (and potentially spinning) neutron star, representing a highly magnetized variant of the black hole accretion tests presented within the previous subsection. Our main objective here is to verify that the general relativistic multifluid solver is still able to produce qualitatively plausible results in curved spacetimes, even when magnetizations and Lorentz factors are much larger than what GRMHD primitive variable reconstruction algorithms are able to handle stably, analogous to the highly magnetized version of the special relativistic Noh test that was previously analyzed in \ref{sec:robustness_noh}. For the spacetime geometry, we use the (axisymmetric) approximate exterior neutron star metric of \cite{pappas_accurate_2017}, as implemented numerically by \cite{gorard_hydrodynamic_2025}, for a neutron star of mass $M$ and angular momentum $J$, assuming mass quadrupole ${M_2}$, spin octupole ${S_3}$, and mass hexadecapole ${M_4}$, whose spacetime line element can be written in the cylindrical Weyl-Lewis-Papapetrou coordinate system ${\left( t, \rho, \varphi, z \right)}$ as:

\begin{multline}
d s^2 = g_{\mu \nu} d x^{\mu} d x^{\nu}\\
= - f \left( dt - \omega d \varphi \right)^2 + f^{-1} \left[ e^{2 \gamma} \left( d \rho^2 + d z^2 \right) + \rho^2 d \varphi^2 \right],
\end{multline}
with the scalar metric functions ${f \left( \rho, z \right)}$, ${\omega \left( \rho, z \right)}$, and ${\gamma \left( \rho, z \right)}$ given in Appendix \ref{sec:neutron_star_appendix}. We select the following gauge conditions on the lapse function ${\alpha}$ and shift vector ${\beta^i}$:

\begin{equation}
\alpha = \frac{\sqrt{f} \rho}{\sqrt{\rho^2 - f^2 \omega^2}}, \qquad \text{ and } \qquad \beta^{\varphi} = - \frac{f^2 \omega}{f^2 \omega^2 - \rho^2},
\end{equation}
with all other components vanishing, i.e. ${\beta^{\rho} = \beta^z = 0}$, thus foliating the spacetime into identical spacelike hypersurfaces whose line elements are given by:

\begin{equation}
d l^2 = \gamma_{i j} d x^i d x^j = f^{-1} e^{2 \gamma} \left( d \rho^2 + d z^2 \right) + \rho^2 \omega^2 d \varphi^2,
\end{equation}
and where the only non-vanishing components of the extrinsic curvature tensor ${K_{i j}}$ are:

\begin{equation}
K_{\rho \varphi} = K_{\varphi \rho} = - \frac{\sqrt{f} \omega}{\sqrt{\left( \rho - f \omega \right) \left( \rho + f \omega \right)}}.
\end{equation}

Following \cite{pappas_effectively_2014}, the higher moments ${M_2}$, ${S_3}$, and ${M_4}$ appearing in the multipolar expansion can be parameterized in terms of deviations away from the Kerr metric:

\begin{equation}
M_2 = - \alpha j^2 M ^3, \qquad S_3 = -\beta j^3 M^4, \qquad M_4 = \gamma j^4 M^5,
\end{equation}
where ${j = \frac{J}{M^2}}$ is the dimensionless spin of the neutron star. We assume a physically realistic neutron star equation of state (modeled on the FPS equation of state for nuclear matter, characterized by quadrupolar deformability ${\alpha \approx 4.209}$, see \cite{cook_rapidly_1994}), whereby the coefficients ${\beta}$ and ${\gamma}$ appearing in front of the higher moments can be expressed, following \cite{yagi_effective_2014}, in terms of the quadrupolar deformability parameter ${\alpha}$ as:

\begin{equation}
\beta = \left( -0.36 + 1.48 \left( \sqrt{\alpha} \right)^{0.65} \right)^3,
\end{equation}
and:

\begin{equation}
\gamma = \left( 4.749 + 0.27613 \left( \sqrt{\alpha} \right)^{1.5146} + 5.5168 \left( \sqrt{\alpha} \right)^{0.22229} \right)^4,
\end{equation}
respectively. The spherical Kerr-Schild coordinate system ${\left( t, r, \theta, \phi \right)}$ for a neutron star of dimensionless spin $j$, within which we represent the initial magnetic field configuration, can be converted to the Weyl-Lewis-Papapetrou cylindrical coordinate system ${\left( t, \rho, \varphi, z \right)}$ via the transformation:

\begin{equation}
\rho = \left( \sqrt{r^2 + \left( j M \right)^2} \right) \sin \left( \theta \right), \qquad \varphi = \phi, \qquad z = r \cos \left( \theta \right).
\end{equation}
We place the excision boundary for the simulation, as in the black hole case, at the Schwarzschild radius ${r = 2M}$ (with ${r < 2M}$ corresponding to a region of zero flux), although we note that the physical surface of a neutron star above approximately one solar mass is likely to lie in the interval ${3M \leq r \leq 4M}$ (\cite{haensel_apparent_2001}). For this reason, we shall set the physical boundary of the neutron star at ${r = 3M}$, and simply treat all behavior in the interior neutron star region ${2M \leq r \leq 3M}$ as being unphysical, hence excluding it from our analysis of the simulation results.

We initially consider the accretion of a weakly magnetized ideal gas onto a very strongly magnetized static (non-rotating) neutron star, formulated as a 3D GRMHD wind accretion setup analogous to the black hole case previously considered, with a cubical domain ${\left( x, y, z \right) \in \left[ 0, 5 \right]^3}$ containing a neutron star of mass ${M = 0.3}$, dimensionless spin ${j = 0}$, and quadrupolar deformability ${\alpha = 5.0}$ (corresponding to a moderately stiff equation of state for the nuclear matter), centered at ${\left( x, y, z \right) = \left( 2.5, 2.5, 2.5 \right)}$. The neutron star is strongly magnetized, with the same radial magnetic field profile of \cite{michel_rotating_1973} and \cite{blandford_electromagnetic_1977} (in Kerr-Schild coordinates):

\begin{equation}
B^r = \frac{1}{\sqrt{\gamma}} \left( B_0 \sin \left( \theta \right) \right),
\end{equation}
as in the black hole case, but with strength ${B_0}$ now set such that the initial magnetization near the neutron star is ${\sigma = \frac{\left\lVert \mathbf{B} \right\rVert^2}{4 \pi \rho} = 10^6}$ (i.e. around 4 orders of magnitude stronger than for the black hole simulations, see Appendix \ref{sec:code_units_appendix}). Once again, the fluid velocity is set to ${v_{\infty} = 0.3}$ at spatial infinity, with wind oriented directly towards the neutron star, with fluid density ${\rho_{\infty} = 3.0}$, fluid pressure ${p_{\infty} = 0.05}$, and uniform $z$-magnetic field ${B_{\infty}^{z} = 1.0}$ at spatial infinity (with ${B_{\infty}^{x} = B_{\infty}^{y} = 0}$). The density and pressure are set to ${\rho_0 = \frac{1}{4 \pi}}$ and ${p_0 = 0.01}$ in the remainder of the domain (again, see Appendix \ref{sec:code_units_appendix} for details of why this density is chosen), with zero velocity. As before, we assume a perfect monatomic gas (adiabatic index ${\Gamma = \frac{5}{3}}$) for the equation of state, using a spatial discretization of ${1024^3}$ cells, with an HLLC Riemann solver, and a CFL coefficient of 0.9. Due to the high magnetizations near the neutron star, both the 2D Newton-Raphson method of \cite{noble_primitive_2006} and the effective 1D method of \cite{newman_primitive_2014}, quickly fail to converge upon a pressure as the Lorentz factors near the neutron star begin to grow large. As a consequence, no numerical GRMHD solution exists for this test, although as with the highly magnetized variant of the relativistic Noh test considered previously, we are able to ascertain approximately what the GRMHD solution \textit{would} be, if it could be obtained, by considering the zero ion Larmor radius limit of the general relativistic multifluid solution, i.e. ${r_L \to 0}$.

\begin{figure*}
\centering
\subfigure{\includegraphics[width=0.45\textwidth]{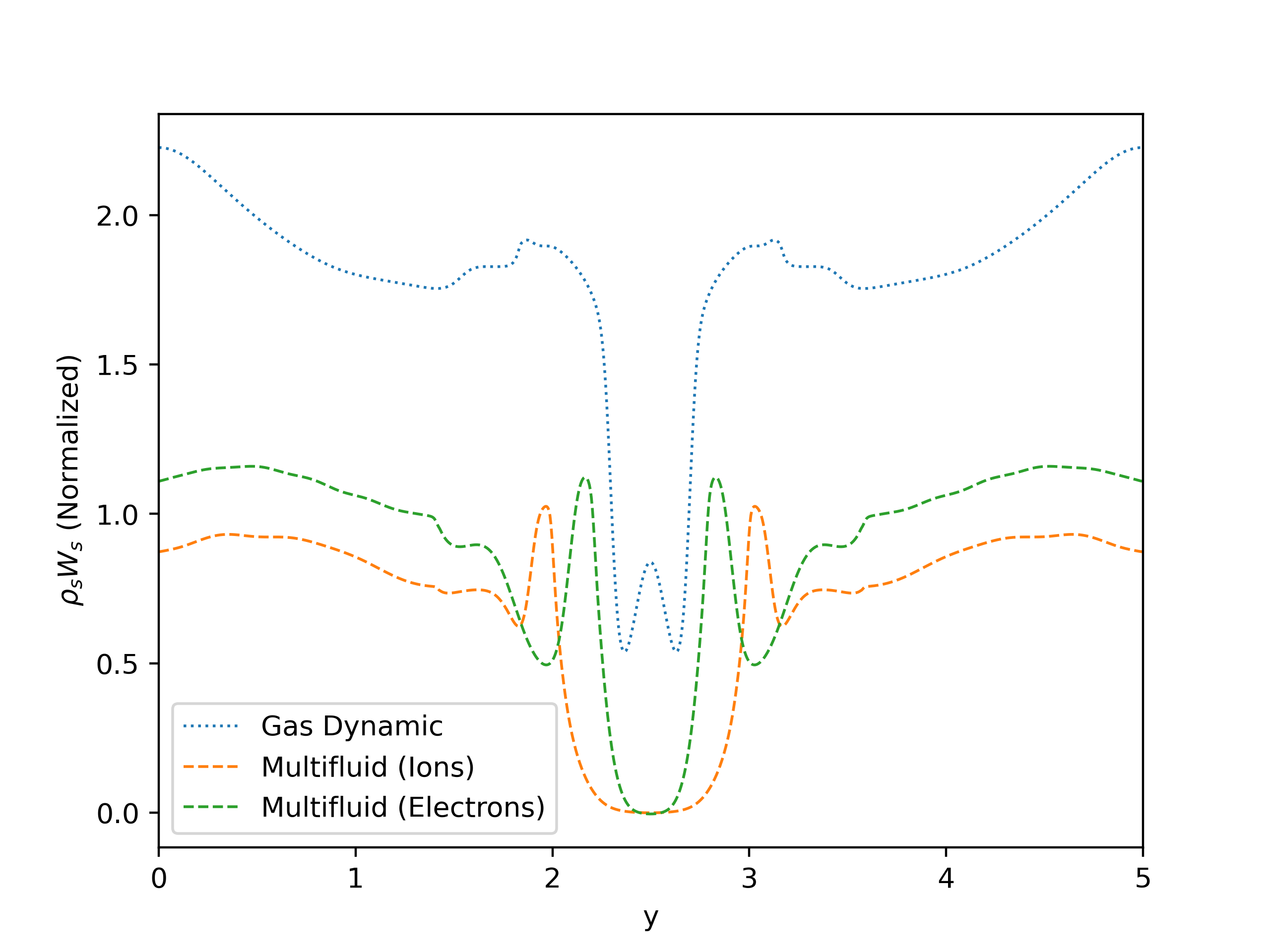}}
\subfigure{\includegraphics[width=0.45\textwidth]{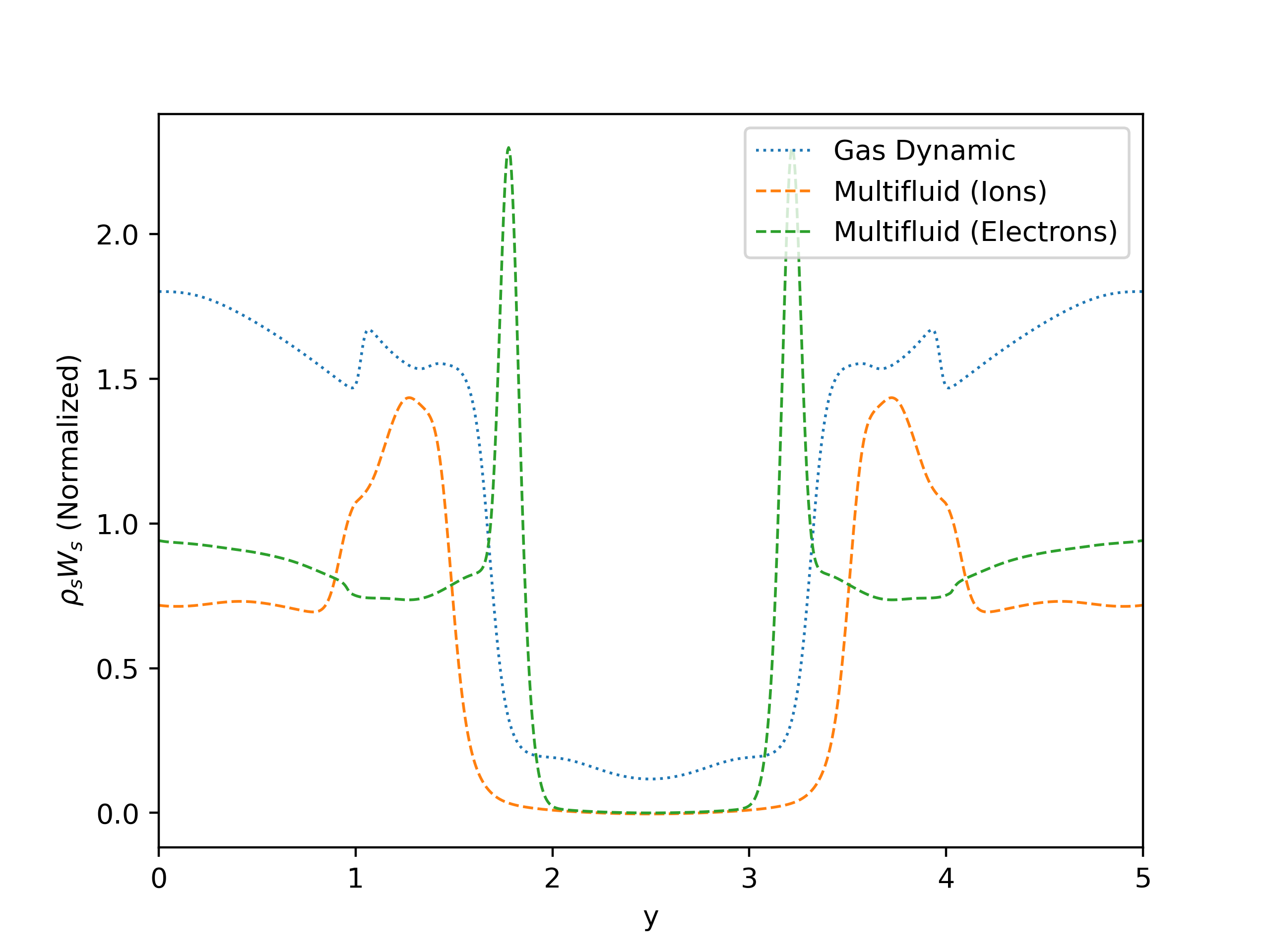}}
\caption{Cross-sectional profiles of the relativistic mass densities ${\rho_s W_s}$ for each fluid species at time ${t = 15}$ for the strongly magnetized accretion problem onto a static (non-rotating) neutron star with ${v_{\infty} = 0.3}$, ${B_{\infty}^{z} = 1.0}$, and ${j = 0}$, taken in the near-downstream region (${x = 3.5}$) on the left, and the far-downstream region (${x = 4.5}$) on the right, of the neutron star, obtained using the general relativistic multifluid solver with initial ion Larmor radius ${r_L = 0.1}$. The gas dynamic (${r_L \to \infty}$) asymptotic reference solution is also shown, and the relativistic mass density ${\rho_e W_e}$ of the electron fluid species has been multiplied by ${\frac{m_i}{m_e} = 1836.2}$ for normalization purposes. The profiles are taken in the equatorial plane ${z = 2.5}$ of the neutron star. In both regions, we see that the electron and ion densities near the neutron star are driven to zero (in contrast to the gas dynamic solution) by the presence of the very large magnetic fields, and that the lighter electrons cluster preferentially near to the neutron star surface.}
\label{fig:neutronstar_mhd_multifluid_stream}
\end{figure*}

The general relativistic two-fluid simulation (consisting of an electron fluid and an ion fluid) is initialized in the same way as before, i.e. with ${\rho_{i, \infty} = 3.0}$, ${\rho_{e, \infty} = 3.0 \frac{m_e}{m_i}}$, ${\Gamma_e = \Gamma_i = \frac{5}{3}}$, etc., ion-to-electron mass ratio ${\frac{m_i}{m_e} = 1836.2}$, and with elementary charges ${q_e = -q_i}$ set so as to yield an initial ion Larmor radius of ${r_L = \frac{m v_{\text{th}}}{\left\lvert q \right\rvert \left\lVert \mathbf{B} \right\rVert} = 0.1}$\footnote{By the end of the simulation, the ion Larmor radius near the neutron star surface is approximately ${r_L \approx 5 \times 10^{-5}}$.}, as shown in Appendix \ref{sec:code_units_appendix}. Figure \ref{fig:neutronstar_mhd_multifluid} shows the ion and electron relativistic mass density contours (i.e. ${\rho_i W_i}$ and ${\rho_e W_e}$) obtained from the general relativistic multifluid solver at time ${t = 15}$, taken through the equatorial plane ${z = 2.5}$ of the neutron star. We see that the qualitative discrepancies between the shapes of the mass density profiles for the ion and electron fluid species are much more significant than they were for the more weakly magnetized black hole case, and can be observed in both the upstream and downstream regions of the neutron star, indicating that the strong magnetic fields have resulted in very substantial charge separations, and likely other non-ideal effects. We also note that the mass densities overall are much smaller, indicating that a larger fraction of the accreting fluid has been ejected by the magnetic field. In Figure \ref{fig:neutronstar_mhd_multifluid_stream}, we show the cross-sectional profiles for the relativistic mass densities of the ion and electron fluids (i.e. ${\rho_i W_i}$ and ${\rho_e W_e}$) in the near-downstream (${x = 3.5}$) and far-downstream (${x = 4.5}$) regions of the neutron star at time ${t = 15}$, with profiles again taken through the equatorial plane ${z = 2.5}$. Although no numerical GRMHD solution exists, we still include the gas dynamic (${r_L \to \infty}$) reference solution, and also multiply the electron mass density ${\rho_e W_e}$ by ${\frac{m_i}{m_e} = 1836.2}$, for the purpose of facilitating comparisons. We find in the near-downstream region that the very large magnetic fields have driven both the electron and ion densities to zero, or near-zero, in the region surrounding the neutron star (approximately ${2.4 < y < 2.6}$ for the electron fluid, approximately ${2.25 < y < 2.75}$ for the ion fluid), in stark contrast to the gas dynamic solution, in which the mass density remains strictly positive everywhere. We also find that the lighter electrons begin to cluster preferentially near to the neutron star surface (with small peaks at around ${y \approx 2.2}$ and ${y \approx 2.8}$), with the greater inertia of the ions causing them to be pushed further out (with corresponding small peaks at around ${y \approx 1.9}$ and ${y \approx 3.1}$). In the far-downstream region, these relatively small charge separations have now become very substantive, with very sharp electron peaks near the neutron star surface (at around ${y \approx 1.8}$ and ${y \approx 3.2}$), and much shallower ion peaks further out (at around ${y \approx 1.3}$ and ${y \approx 3.7}$). The regions over which the magnetic fields have driven the mass densities to near-zero are also much larger here, spanning approximately ${2.0 < y < 3.0}$ for the electron fluid, and ${1.75 < y < 3.25}$ for the ion fluid. The peak ion magnetization near the neutron star remains stable at around ${\sigma \approx 10^6}$ throughout the duration of the simulation.

\begin{figure*}
\centering
\subfigure{\includegraphics[trim={1cm, 0.5cm, 1cm, 0.5cm}, clip, width=0.45\textwidth]{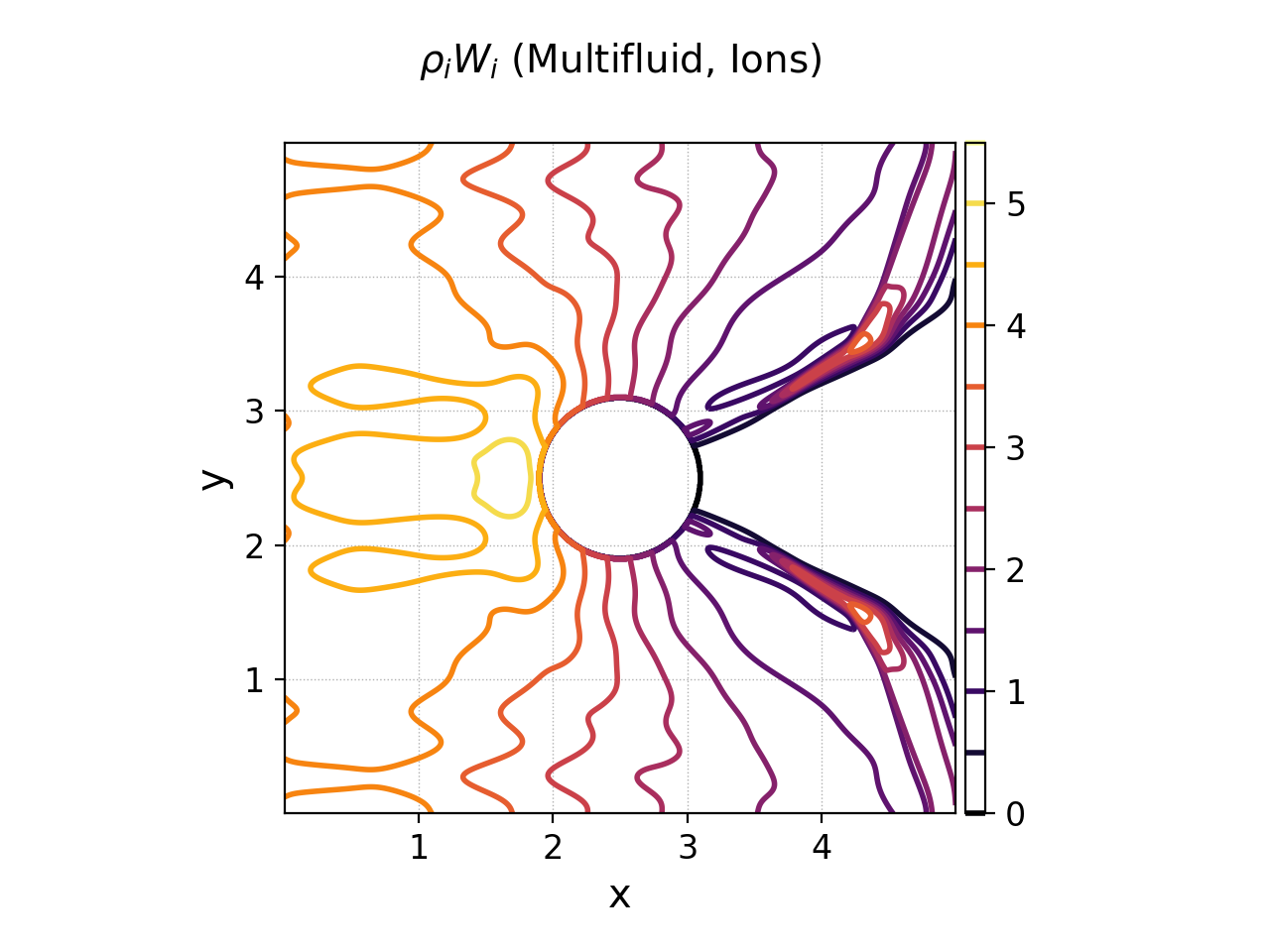}}
\subfigure{\includegraphics[trim={1cm, 0.5cm, 1cm, 0.5cm}, clip, width=0.45\textwidth]{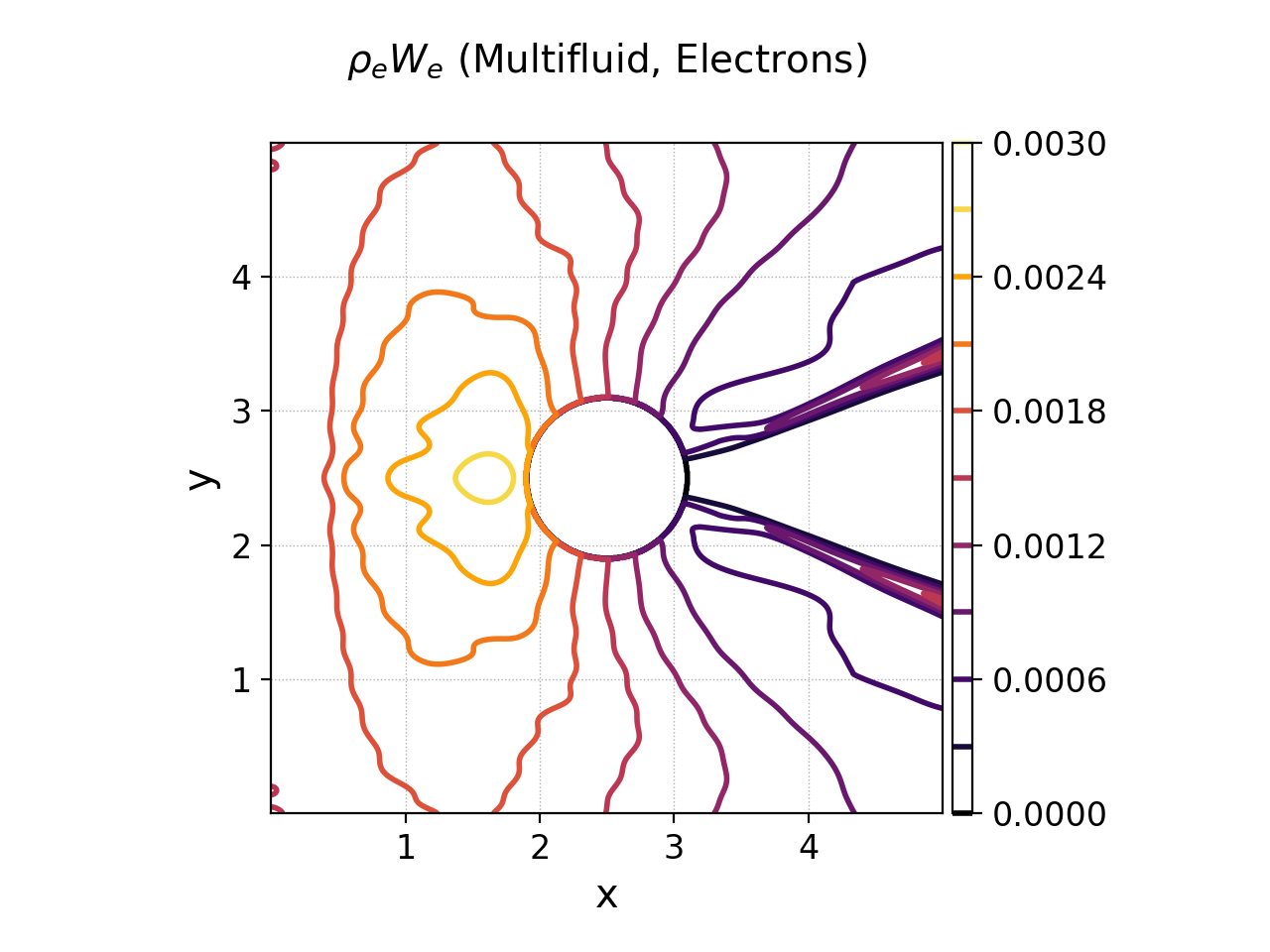}}
\caption{The relativistic mass density contours ${\rho_s W_s}$ at time ${t = 15}$ for the strongly magnetized accretion problem onto a static (non-rotating) neutron star with ${v_{\infty} = 0.3}$, ${B_{\infty}^{z} = 1.0}$, and ${j = 0}$, showing the mass density of the ion fluid species (${\rho_i W_i}$) on the left, and the electron fluid species (${\rho_e W_e}$) on the right, obtained using the general relativistic multifluid solver with initial ion Larmor radius ${r_L = 0.1}$. These accretion profiles have been taken through the equatorial plane ${z = 2.5}$ of the neutron star. We see substantive differences in the ion and electron fluid profiles, both upstream and downstream of the neutron star, indicating the presence of significant non-ideal charge separation effects arising from the large magnetic fields involved.}
\label{fig:neutronstar_mhd_multifluid}
\end{figure*}

\begin{figure*}
\centering
\subfigure{\includegraphics[trim={1cm, 0.5cm, 1cm, 0.5cm}, clip, width=0.45\textwidth]{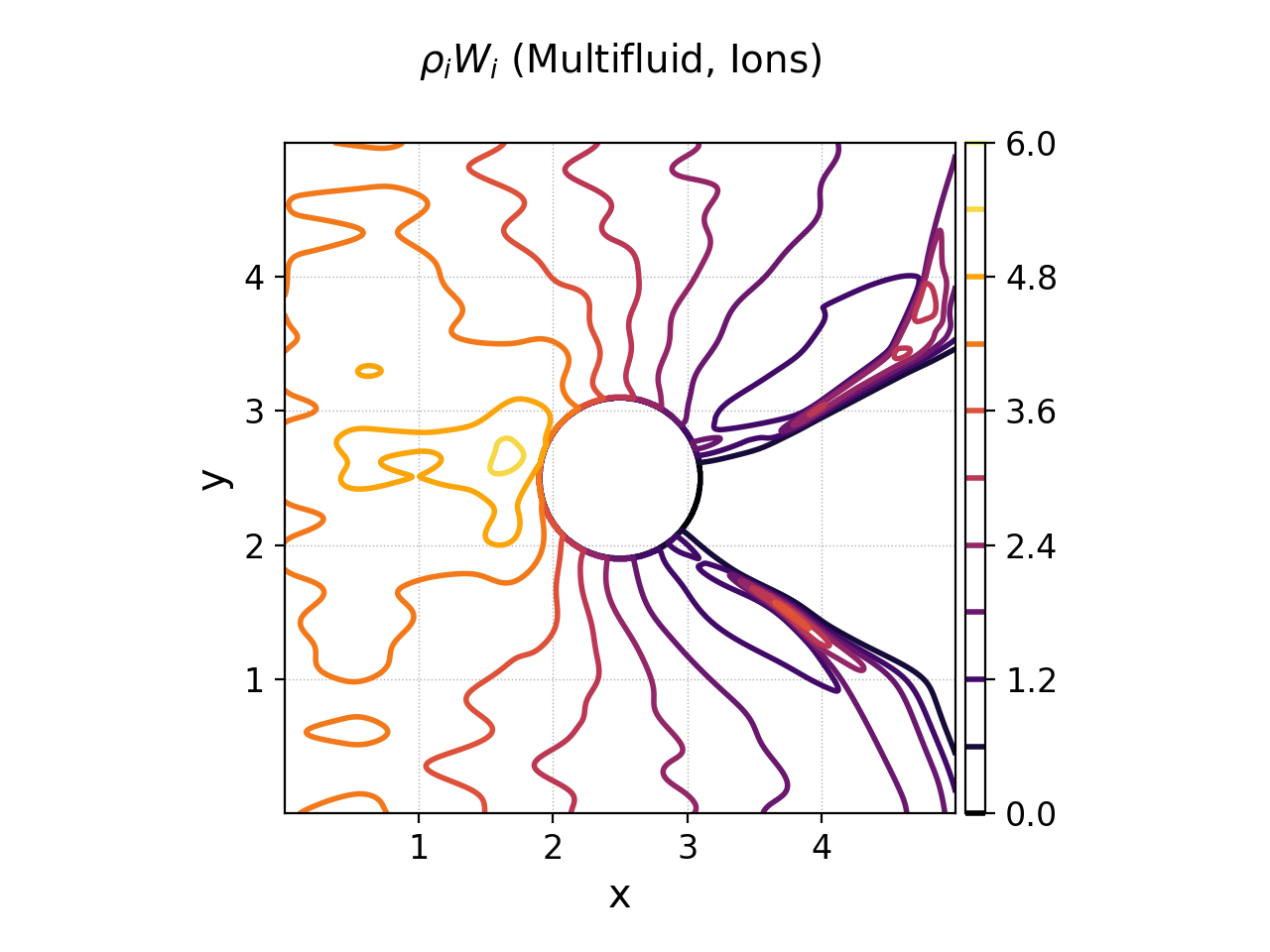}}
\subfigure{\includegraphics[trim={1cm, 0.5cm, 1cm, 0.5cm}, clip, width=0.45\textwidth]{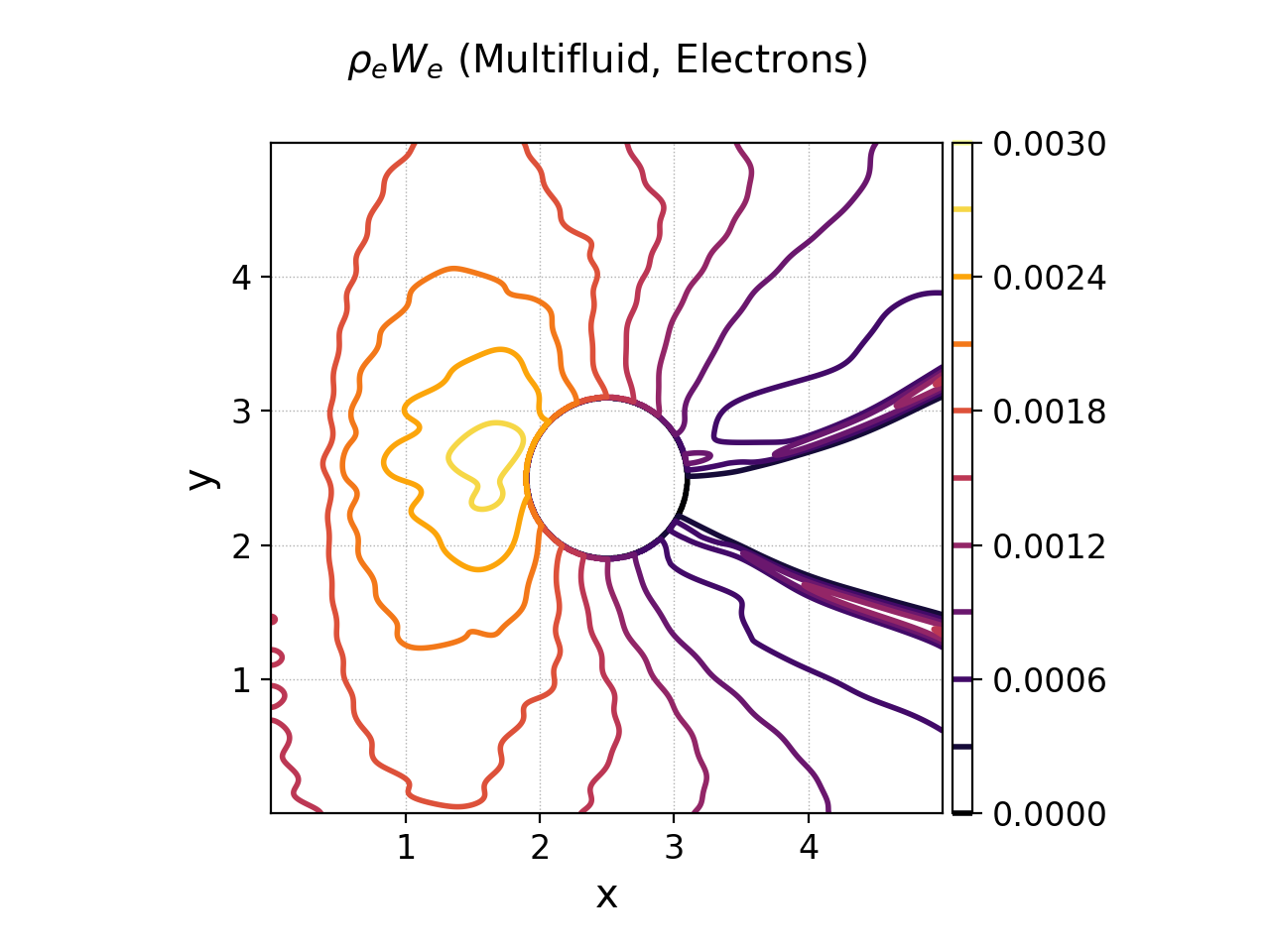}}
\caption{The relativistic mass density contours ${\rho_s W_s}$ at time ${t = 15}$ for the strongly magnetized accretion problem onto a rotating neutron star with ${v_{\infty} = 0.3}$, ${B_{\infty}^{z} = 1.0}$, and ${j = 0.4}$, showing the mass density contours for the ion fluid species (${\rho_i W_i}$) on the left, and the electron fluid species (${\rho_e W_e}$) on the right, obtained using the general relativistic multifluid solver with initial ion Larmor radius ${r_L = 0.1}$. These accretion profiles have been taken through the equatorial plane ${z = 2.5}$ of the neutron star. The discrepancies between the ion and electron fluid profiles, both upstream and downstream of the neutron star, have become even more pronounced than they were in the static (non-rotating) case, indicating larger non-ideal charge separation effects resulting from the stronger induced electric fields.}
\label{fig:neutronstar_spinning_mhd_multifluid}
\end{figure*}

\begin{figure*}
\centering
\subfigure{\includegraphics[width=0.45\textwidth]{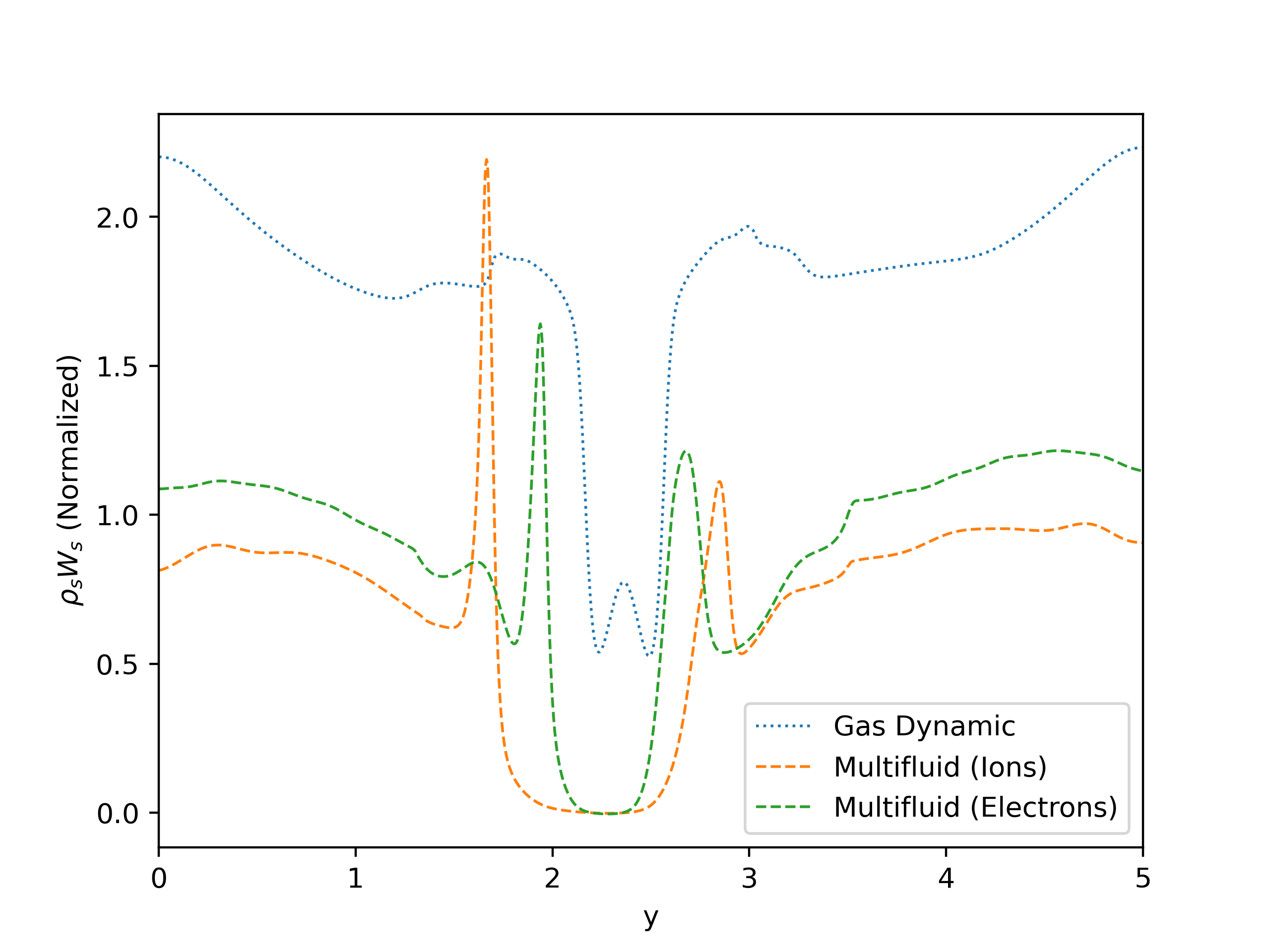}}
\subfigure{\includegraphics[width=0.45\textwidth]{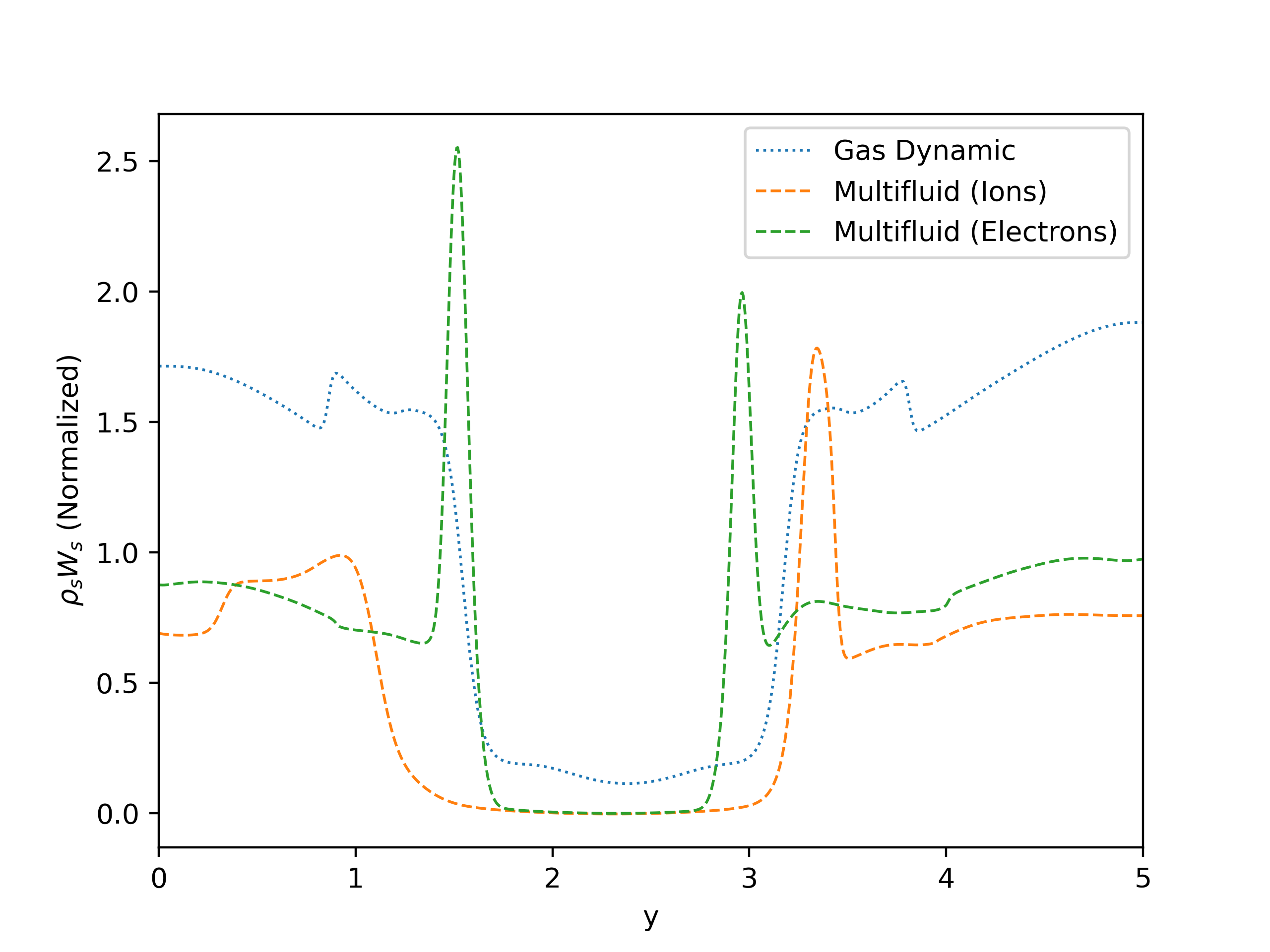}}
\caption{Cross-sectional profiles of the relativistic mass densities ${\rho_s W_s}$ for each fluid species at time ${t = 15}$ for the strongly magnetized accretion problem onto a rotating neutron star with ${v_{\infty} = 0.3}$, ${B_{\infty}^{z} = 1.0}$, and ${j = 0.4}$, taken in the near-downstream region (${x = 3.5}$) on the left, and the far-downstream region (${x = 4.5}$) on the right, of the neutron star, obtained using the general relativistic multifluid solver with initial ion Larmar radius ${r_L = 0.1}$. The gas dynamic (${r_L \to \infty}$) asymptotic reference solution is also shown, and the relativistic mass density ${\rho_e W_e}$ of the electron fluid species has been multiplied by ${\frac{m_i}{m_e} = 1836.2}$ for normalization purposes. The profiles are taken in the equatorial plane ${z = 2.5}$ of the neutron star. We again see the electron and ion densities near the neutron star being driven to zero, much like in the static (non-rotating) case, due to the presence of large magnetic fields, but now with a significant asymmetry in the electron and ion populations on the co-rotating and counter-rotating sides.}
\label{fig:neutronstar_spinning_mhd_multifluid_stream}
\end{figure*}

We finally consider wind accretion of a weakly magnetized ideal gas, obeying the same initial conditions as for the static (non-rotating) neutron star case, onto a strongly magnetized rotating neutron star of mass ${M = 0.3}$ and dimensionless spin ${j = \frac{J}{M^2} = 0.4}$, corresponding to the estimated spin of PSR J1748-2446ad (a pulsar with a measured frequency of 716 Hz, see \cite{hessels_radio_2006}). We select these parameters since such a spin is both astrophysically plausible and comfortably below the estimated breakup speed of ${j \approx 0.7}$ for a standard neutron star equation of state (\cite{wang_targeted_2024}). Our simulation setup is the same as before, with cubical domain ${\left( x, y, z \right) \in \left[ 0, 5 \right]^3}$ containing the rotating neutron star at ${\left( x, y, z \right) = \left( 2.5, 2.5, 2.5 \right)}$, with the same radial magnetic field with magnetization ${\sigma = \frac{\left\lVert \mathbf{B} \right\rVert^2}{4 \pi \rho} = 10^6}$ as for the static (non-rotating) case, using a discretization of ${1024^3}$ cells, an HLLC Riemann solver, and a CFL coefficient of 0.9. The general relativistic two-fluid solution is found in Figure \ref{fig:neutronstar_spinning_mhd_multifluid}, showing both the ion and electron relativistic mass density contours (i.e. ${\rho_i W_i}$ and ${\rho_e W_e}$) at time ${t = 15}$, again through the equatorial plane ${z = 2.5}$ of the neutron star. The large qualitative discrepancies between the ion and electron fluid profiles, both upstream and downstream of the neutron star, have become even more pronounced than they were in the static (non-rotating) case, indicating that the non-ideal charge separation effects have been exaggerated by the stronger parallel electric fields being induced by the neutron star's rotation. We can analyze this phenomenon in more detail by examining cross-sectional profiles of the relativistic mass densities of the ion and electron fluids (i.e. ${\rho_i W_i}$ and ${\rho_e W_e}$) in both the near-downstream (${x = 3.5}$) and far-downstream (${x = 4.5}$) regions of the neutron star at time ${t = 15}$, as shown in Figure \ref{fig:neutronstar_spinning_mhd_multifluid_stream}, with profiles taken through the equatorial plane ${z = 2.5}$ as usual. As previously, we also include the gas dynamic (${r_L \to \infty}$) reference solution, and multiply the electron mass density ${\rho_e W_e}$ by ${\frac{m_i}{m_e} = 1836.2}$ to facilitate comparisons. We find that the magnetic fields have again driven both the electron and ion fluid densities to zero or near-zero, with the near-vacuum region now shifted to approximately ${2.1 < y < 2.4}$ for the electron fluid and approximately ${1.9 < y < 2.5}$ for the ion fluid due to frame-dragging effects from the neutron star's rotation. We note that this shift in the fluid profiles due to frame-dragging is more extreme for the multifluid solution than for the gas dynamic one, due to magnetic field ``locking'' and consequent co-rotation of the magnetic field with the neutron star surface. As in the static (non-rotating) case, the lighter electrons cluster preferentially near to the neutron star surface, with a large peak on the counter-rotating side at around ${y \approx 1.9}$ and a small peak on the co-rotating side at around ${y \approx 2.6}$. The greater inertia of the ions pushes them further out from the neutron star, creating a very large peak on the counter-rotating side at around ${y \approx 1.6}$ and a much smaller peak on the co-rotating side at around ${y \approx 2.8}$. In the far-downstream region, these large charge separations become even more substantial, but with the aforementioned asymmetry in the electron fluid smoothed out, and the asymmetry in the ion fluid reversed (i.e. with a large peak in the electron fluid on the counter-rotating side at around ${y \approx 1.5}$ matched by a comparably large peak on the co-rotating side at around ${y \approx 2.9}$, and with a moderate peak in the ion fluid on the co-rotating side at around ${y \approx 3.3}$ unmatched by any corresponding peak on the counter-rotating side as the ion fluid profile has been considerably smoothed out in that region). The large magnetic fields have driven the mass densities to near-zero here over the approximate regions ${1.7 < y < 2.8}$ for the electrons and ${1.3 < y < 3.1}$ for the ions. The peak ion magnetization near the neutron star surface increases slightly, to around ${\sigma \approx 1.1 \times 10^6}$, over the course of the simulation, presumably due to increased twisting of the magnetic field lines from frame-dragging.

\begin{figure*}
\centering
\subfigure{\includegraphics[width=0.45\textwidth]{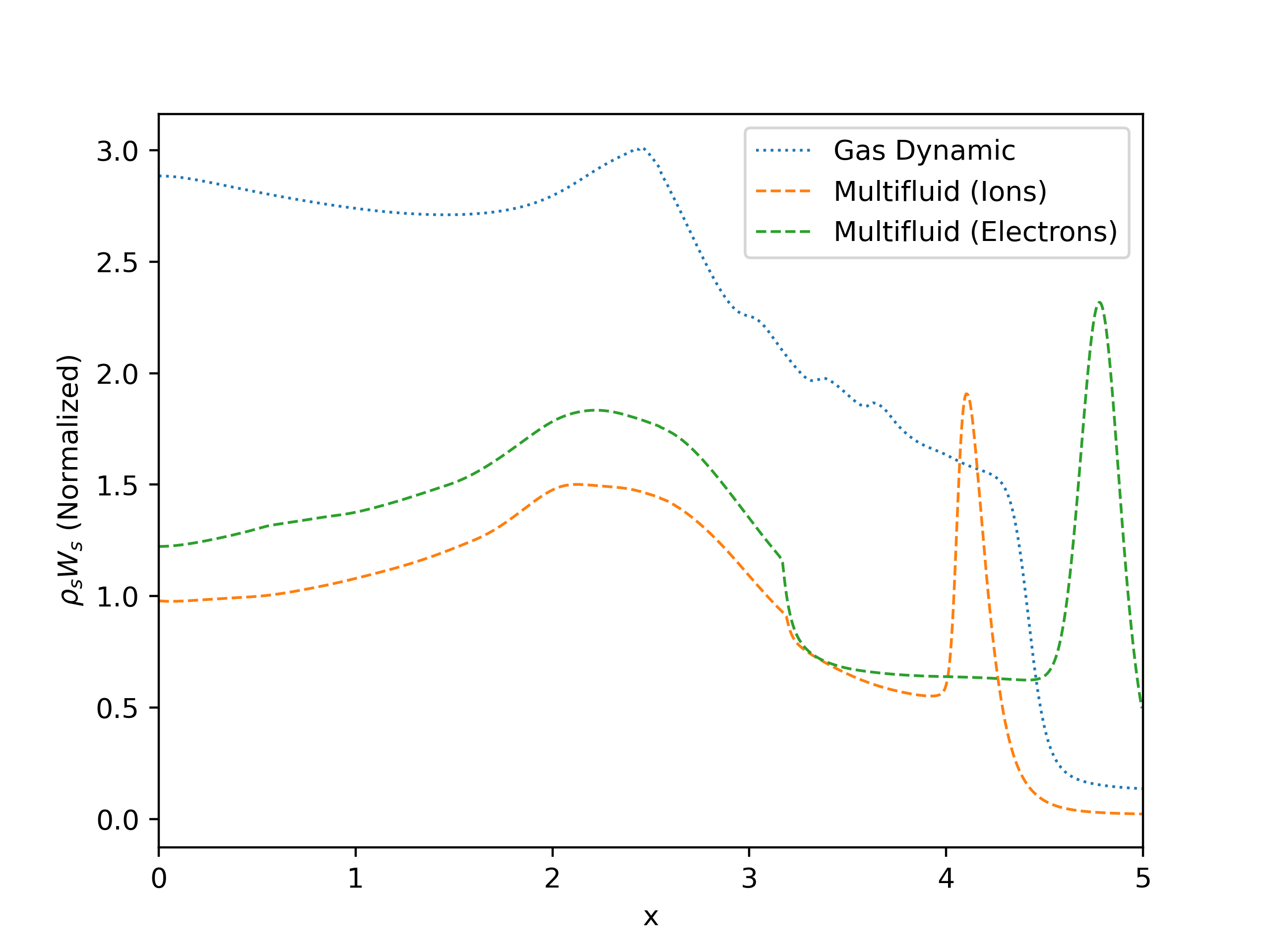}}
\subfigure{\includegraphics[width=0.45\textwidth]{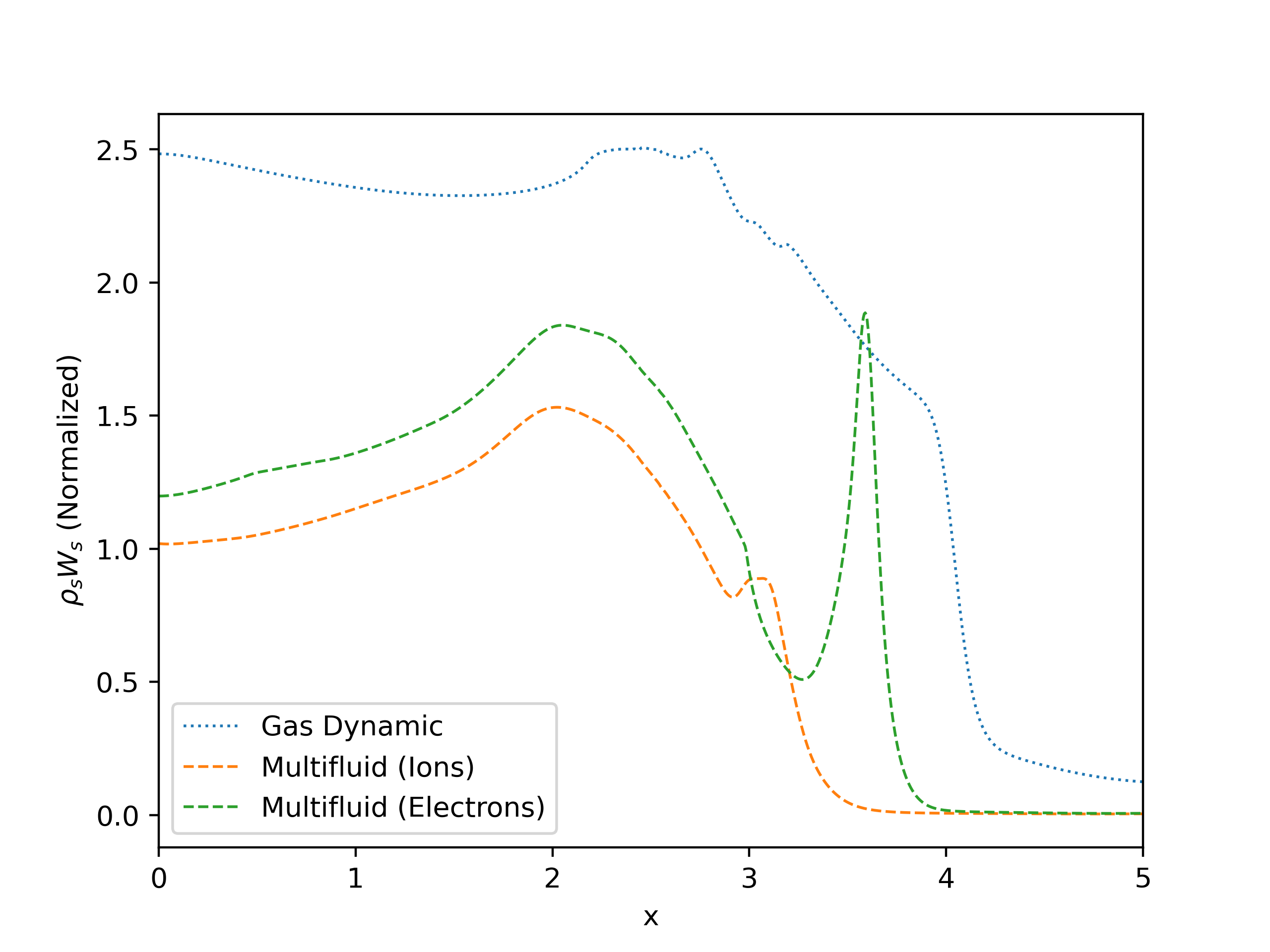}}
\caption{Cross-sectional profiles of the relativistic mass densities ${\rho_s W_s}$ for each fluid species at time ${t = 15}$ for the strongly magnetized accretion problem onto a rotating neutron star with ${v_{\infty} = 0.3}$, ${B_{\infty}^{z} = 1.0}$, and ${j = 0.4}$, taken on the co-rotating (${y = 3.4}$) and counter-rotating (${y = 1.6}$) sides of the neutron star, obtained using the general relativistic multifluid solver with initial ion Larmor radius ${r_L = 0.1}$. The gas dynamic (${r_L \to \infty}$) asymptotic reference solution is also shown, and the relativistic mass density ${\rho_e W_e}$ of the electron fluid species has been multiplied by ${\frac{m_i}{m_e} = 1836.2}$ for normalization purposes. The profiles are taken in the equatorial plane ${z = 2.5}$ of the neutron star. We see sizable charge separation in the electron and ion density peaks on the co-rotating side, and a significant dampening effect on the ion density on the counter-rotating side, due to greater ion inertia.}
\label{fig:neutronstar_spinning_mhd_multifluid_stream2}
\end{figure*}

\begin{figure*}
\centering
\subfigure{\includegraphics[trim={1cm, 0.5cm, 1cm, 0.5cm}, clip, width=0.45\textwidth]{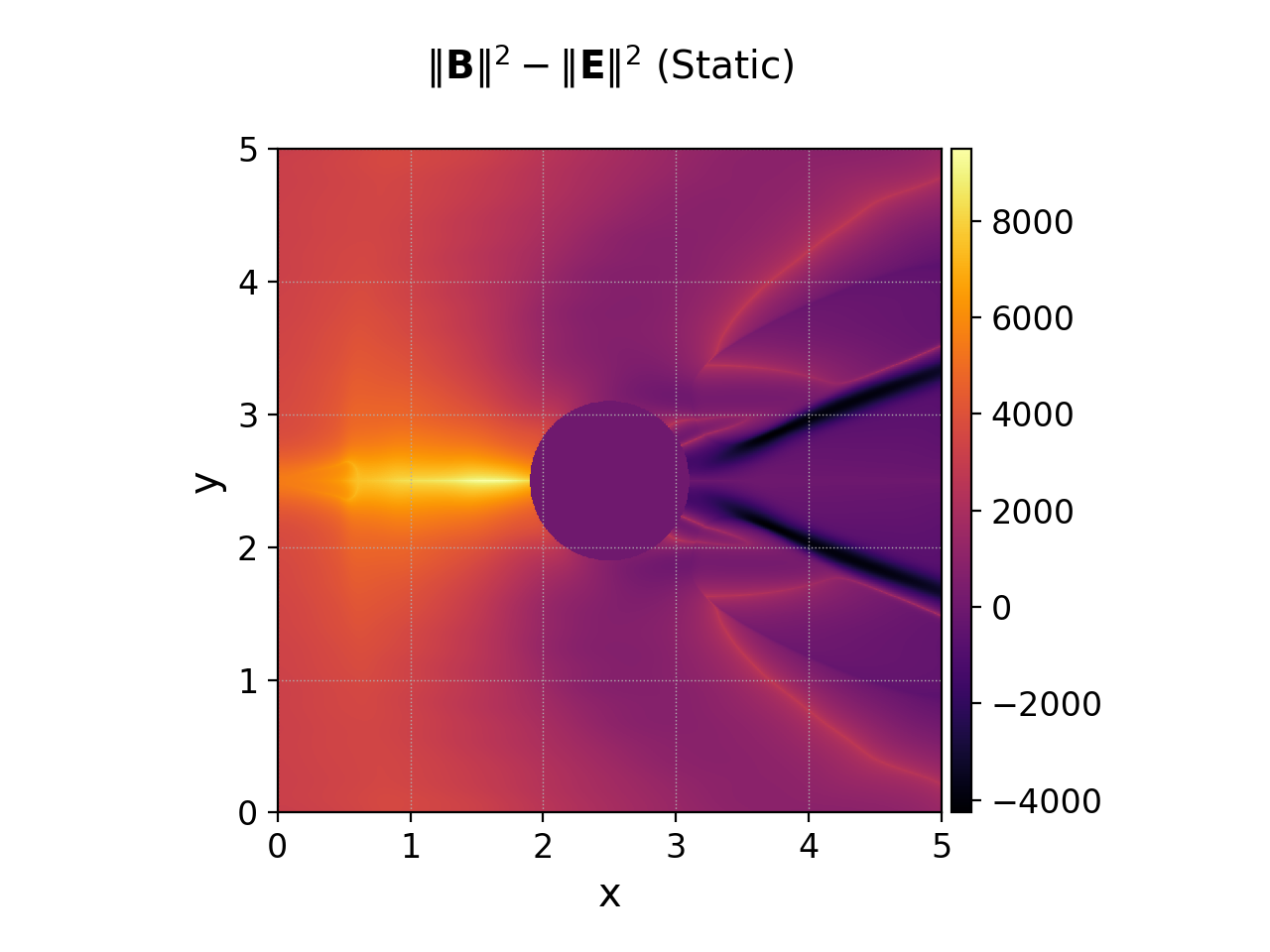}}
\subfigure{\includegraphics[trim={1cm, 0.5cm, 1cm, 0.5cm}, clip, width=0.45\textwidth]{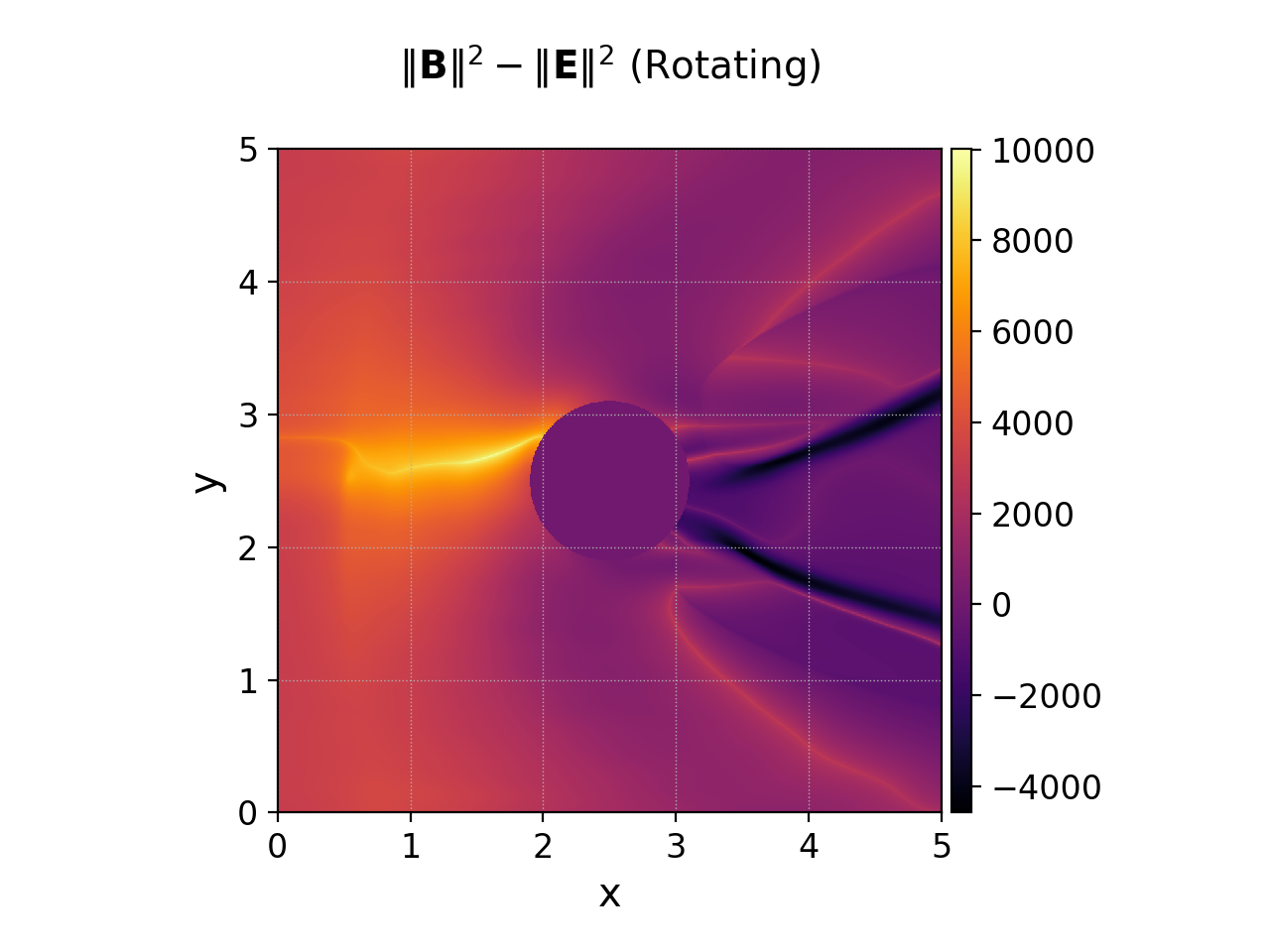}}
\caption{The Lorentz invariant scalar ${\left\lVert \mathbf{B} \right\rVert^2 - \left\lVert \mathbf{E} \right\rVert^2}$ at time ${t = 15}$ for the strongly magnetized accretion problem with ${v_{\infty} = 0.3}$ and ${B_{\infty}^{z} = 1.0}$ onto a static neutron star (non-rotating, ${j = 0}$) on the left, and onto a rotating neutron star (${j = 0.4}$) on the right, obtained using the general relativistic multifluid solver with initial ion Larmor radius ${r_L = 0.1}$. These field profiles have been taken through the equatorial plane ${z = 2.5}$ of the neutron star. In the static (non-rotating) case, we see very dark regions in the downstream wake of the neutron star where ${\left\lVert \mathbf{B} \right\rVert^2 - \left\lVert \mathbf{E} \right\rVert^2}$ is being driven strongly negative, indicating the presence of very large parallel electric fields. In the rotating case, we see that ${\left\lVert \mathbf{B} \right\rVert^2 - \left\lVert \mathbf{E} \right\rVert^2}$ becomes slightly more negative within these regions, indicating that the electric field induction has become marginally more efficient as a result of the neutron star's rotation.}
\label{fig:neutronstar_spinning_mhd_multifluid_field}
\end{figure*}

\begin{figure*}
\centering
\subfigure{\includegraphics[trim={1cm, 0.5cm, 1cm, 0.5cm}, clip, width=0.45\textwidth]{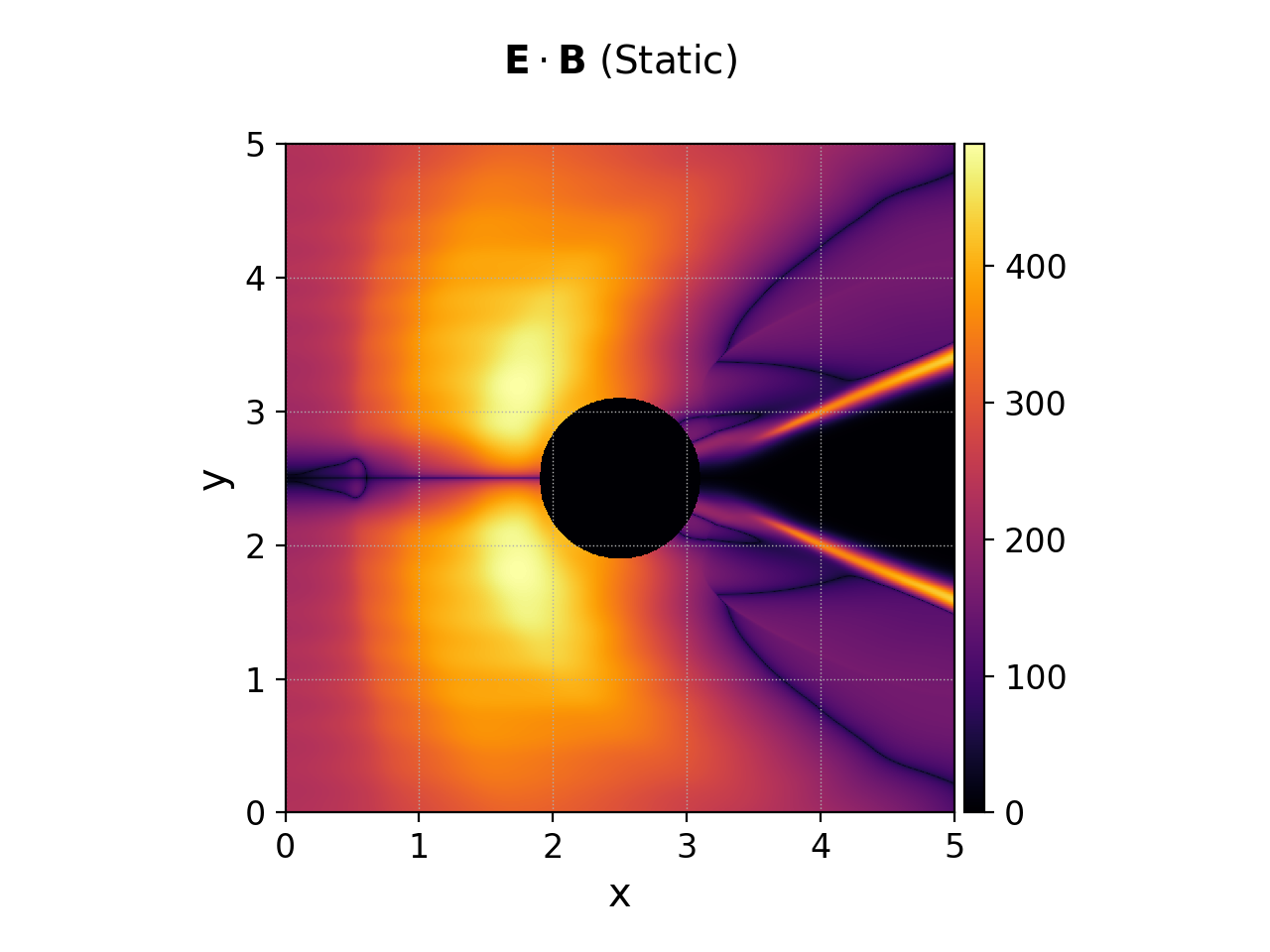}}
\subfigure{\includegraphics[trim={1cm, 0.5cm, 1cm, 0.5cm}, clip, width=0.45\textwidth]{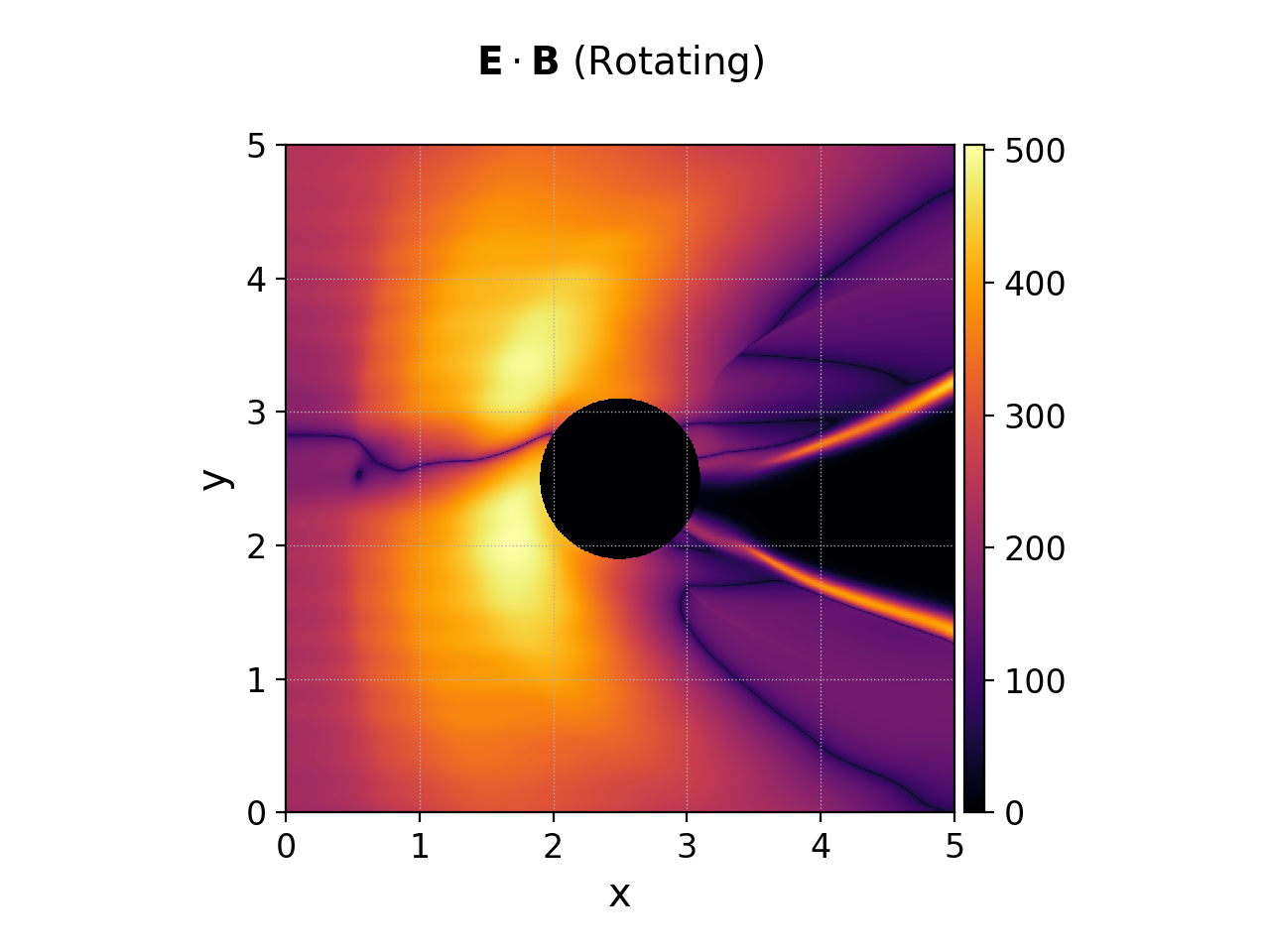}}
\caption{The Lorentz invariant scalar ${\mathbf{E} \cdot \mathbf{B}}$ at time ${t = 15}$ for the strongly magnetized accretion problem with ${v_{\infty} = 0.3}$ and ${B_{\infty}^{z} = 1.0}$ onto a static neutron star (non-rotating, ${j = 0}$) on the left, and onto a rotating neutron star (${j = 0.4}$) on the right, obtained using the general relativistic multifluid solver with initial ion Larmor radius ${r_L = 0.1}$. These field profiles have been taken through the equatorial plane ${z = 2.5}$ of the neutron star. The large values of ${\mathbf{E} \cdot \mathbf{B}}$ in the downstream wake of the neutron star, in both the static and rotating cases, confirm that the regions where ${\left\lVert \mathbf{B} \right\rVert^2 - \left\lVert \mathbf{E} \right\rVert^2}$ is driven negative are indeed due to large electric fields acting \textit{parallel} to the magnetic field.}
\label{fig:neutronstar_spinning_mhd_multifluid_field2}
\end{figure*}

These rotation-induced asymmetries in the accretion profiles are analyzed in more detail by taking corresponding cross-sections in the same ${z = 2.5}$ equatorial plane, now on the co-rotating (${y = 3.4}$) and counter-rotating (${y = 1.6}$) sides of the neutron star, as shown in Figure \ref{fig:neutronstar_spinning_mhd_multifluid_stream2}. We find, in contrast to the spinning black hole case, that the fluid profiles for the ions and electrons on the co-rotating side of the neutron star are in fact qualitatively different, since the co-rotation of the magnetic field with the neutron star surface causes the lighter electrons to be preferentially accelerated, resulting in a significant charge separation in the downstream region between the peak electron density at around ${x \approx 4.8}$ and the peak ion density at around ${x \approx 4.1}$. On the counter-rotating side, the ion profile has been almost completely smoothed out due to greater ion inertia, while the electron density peak persists but is pushed back to around ${x \approx 3.5}$. Finally, just as we did in the black hole case, we can confirm that these large charge separations are being sustained by the induction of correspondingly large parallel electric fields in the downstream region of the neutron star, by plotting the Lorentz invariant scalar quantity ${\left\lVert \mathbf{B} \right\rVert^2 - \left\lVert \mathbf{E} \right\rVert^2}$ through the equatorial plane ${z = 2.5}$ for both neutron stars, as shown in Figure \ref{fig:neutronstar_spinning_mhd_multifluid_field}. In the static (non-rotating) case, we see the formation of an extended current sheet across the entire downstream region (${x > 3.4}$) of the neutron star, in addition to very dark regions in the wake of the neutron star where ${\left\lVert \mathbf{B} \right\rVert^2 - \left\lVert \mathbf{E} \right\rVert^2}$ is being driven strongly negative by the presence of very large parallel electric fields. In the rotating case, the shape of the current sheet becomes slightly distorted due to frame-dragging effects (although the effect is nowhere near as significant as it is in the black hole case, due to the increased stability afforded by the significantly stronger magnetic fields), and we find that the rotation of the neutron star has also caused ${\left\lVert \mathbf{B} \right\rVert^2 - \left\lVert \mathbf{E} \right\rVert^2}$ to be driven even more strongly negative in the downstream region, due to the neutron star's rotation increasing the efficiency of induction of the parallel electric fields (although, again, the effect is much less significant than in the black hole case). The fact that these large electric fields are acting \textit{parallel} to the magnetic field can be confirmed by plotting the Lorentz invariant scalar quantity ${\mathbf{E} \cdot \mathbf{B}}$ through the equatorial plane ${z = 2.5}$, as shown in Figure \ref{fig:neutronstar_spinning_mhd_multifluid_field2}. We see, indeed, that the regions in the downstream wake of the neutron star where ${\left\lVert \mathbf{B} \right\rVert^2 - \left\lVert \mathbf{E} \right\rVert^2}$ is driven negative are associated with very high values of ${\mathbf{E} \cdot \mathbf{B}}$. In both cases, we therefore confirm very strong violation of the ${\left\lVert \mathbf{E} \right\rVert^2 \ll \left\lVert \mathbf{B} \right\rVert^2}$ assumption of ideal GRMHD, further demonstrating the necessity for something analogous to the general relativistic multifluid solver when correctly modeling the effects of the strong magnetic fields accompanying neutron stars in curved spacetime.

\section{Conclusions}

In this work, we have introduced a general relativistic multifluid formalism that may be considered a reduced form of the multifluid equations developed previously by \cite{denicol_resistive_2019}, \cite{most_modelling_2022}, and others. Our multifluid formalism can be derived as a limiting case of general relativistic kinetics, and conversely the general relativistic magnetohydrodynamics (GRMHD) equations can be derived as a limiting case of our formalism. We have developed a robust numerical scheme for solving the resulting multifluid equations by combining the tetrad-first approach of \cite{gorard_tetrad-first_2025} for handling Riemann problems in curved spacetime, the HLLC Riemann solver of \cite{mignone_hllc_2005} for solving the purely hydrodynamic Riemann problem in special relativity, the analytical reconstruction scheme of \cite{eulderink_general_1994} for reconstructing the primitive fluid variables from the conserved ones, a general relativistic extension of the hyperbolic cleaning scheme of \cite{munz_finite-volume_2000} for correcting divergence errors in the electric and magnetic fields, and the explicit integration scheme of \cite{gottlieb_strong_2001} for coupling the fluid species to each other (via the electromagnetic field), as well as to the underlying spacetime geometry, via their respective source terms. We have first validated this numerical scheme against the 1D flat spacetime Riemann problem of \cite{balsara_total_2001}, based on the non-relativistic MHD Riemann problem of \cite{brio_upwind_1988}, and demonstrated that we are able to recover the correct SRMHD solution in the limit ${r_L \to 0}$ as the ion Larmor radius within a relativistic two-fluid system goes to zero. We have also demonstrated, using a highly magnetized variant of the Riemann problem of \cite{noh_errors_1987}, that the multifluid solver is still able to produce stable solutions, even when Lorentz factors and magnetizations exceed the typical values that SRMHD and GRMHD primitive variable reconstruction methods are able to handle. We then validated the method against the magnetospheric variant of the \cite{wald_black_1974} problem of a spinning black hole in a uniform magnetic field in 2D axisymmetry, and showed that we are able to capture the correct current sheet formation and magnetic reconnection behavior within the equatorial plane of the black hole that was previously shown by \cite{komissarov_electrodynamics_2004} using force-free and resistive electrodynamics, but which is entirely absent within GRMHD simulations. Finally, we validated against weakly magnetized accretion problems involving static and spinning black holes, and strongly magnetized accretion problems involving non-rotating and rotating neutron stars, in full 3D. Throughout all of these validation tests in curved spacetimes, we have demonstrated the consistent ability of the general relativistic multifluid approach to capture the formation of significant charge separations around the compact object, the induction of large (parallel) electric fields, discrepancies between ion and electron inertias, and other non-ideal effects that go beyond the usual GRMHD approximation. We also demonstrate the ability of the multifluid formalism to model the highly magnetized environment surrounding a rotating neutron star without sacrificing either accuracy or numerical stability.

One important limitation of our numerical method stems from our use of an explicit ODE integrator (SSP-RK3) for handling both the geometric and electromagnetic source terms, which means that, due to the restrictive effect of the plasma frequency and cyclotron frequency on the stable time-step, significant subcycling is required in order to maintain numerical stability within simulations involving a small ion Larmor radius ${r_L}$. This fundamentally limits our ability to perform simulations involving realistic scale separations between the compact object radius ${r_g}$ and the Larmor radius ${r_L}$. A major topic of future algorithmic investigation will therefore be the development of either fully implicit source term coupling (as was done previously by \cite{wang_exact_2020} for the non-relativistic multifluid equations), or a hybrid implicit-explicit (IMEX) coupling scheme, wherein the magnetic field is handled implicitly but the electric field is still kept fully explicit. Such an extension is non-trivial, as unlike the non-relativistic multi-fluid source couplings which can be integrated implicitly with a straightforward linear solve of the momentum and electric field, in the relativistic limit the pressure and kinetic energy cannot be decoupled, and the update of the total fluid energy must be simultaneously solved alongside the update of the momentum and electric field. We thus obtain a non-linear term from the work done by the electromagnetic field on the plasma (and vice versa) which must be handled by e.g., a Newton solve, similar to recent work for the flat spacetime relativistic two-fluid solver of \cite{agnihotri_second_2025}. Nevertheless, implementation of an implicit non-linear source solve would dramatically improve both the efficiency and accuracy of general relativistic multifluid simulations, by eliminating the need for subcycling and allowing one instead to step over the fast plasma timescales using the same time-step as the homogeneous equations. The stable time-step would then be restricted only by the largest absolute eigenvalue of the flux Jacobian ${\frac{\partial \mathbf{F} \left( \mathbf{U} \right)}{\partial \mathbf{U}}}$, and not by the largest absolute eigenvalue of the source Jacobian ${\frac{\partial \mathbf{S} \left( \mathbf{U} \right)}{\partial \mathbf{U}}}$, which will be many orders of magnitude larger for any realistic scale separation. Another major direction of future development on the numerical side will be the proper integration of the general relativistic multifluid solver with a numerical solver for the Einstein field equations (analogous to what was previously done for the GRMHD equations within codes such as \texttt{WhiskyMHD} by \cite{giacomazzo_whiskymhd:_2007}, and \texttt{IllinoisGRMHD} by \cite{etienne_illinoisgrmhd:_2015}), thereby enabling fully dynamic spacetime simulations of magnetic field configurations resulting from multifluid gravitational collapse of neutron stars to black holes, the multifluid modeling of electromagnetic precursors within neutron star binary collisions, and many other scenarios in which gravitational effects are significant. 

Finally, we emphasize that the approach detailed here is genuinely \textit{multi-fluid}, with the algorithm being ultimately agnostic to the number of fluid species evolved. It will thus be an interesting extension of the formalism presented here to add models for pair production such as those detailed in \cite{parfrey_first-principles_2019}, which utilize the local parallel electric field to estimate the pair-loading from accelerated electrons. As with other advantages of the multifluid approach outlined in this study, there are both physics-based and numerics-based benefits of adding pair production. We know pairs are produced in regions of strong parallel electric fields that we obtain self-consistently from the evolution of the multifluid system, thus improving the physical fidelity of the model. We also avoid one of the other numerical reasons GRMHD codes often employ density floors in their algorithms, since we can self-consistently model the vacuum gaps where the pairs would be produced and capture the plasma loading from these QED processes. We anticipate the physics fidelity of the model, especially with regards to the simulation of non-ideal effects, to be improved yet further by the evolution of the full (potentially anisotropic) pressure tensor ${\Pi_{s}^{\mu \nu}}$ in place of the reduced scalar pressure ${p_s}$ used at present.

We have already demonstrated that the general relativistic multifluid approach is capable of sustaining much stronger poloidal fields, and of generating much larger Poynting fluxes, around spinning black holes immersed within uniform magnetic fields than the GRMHD approach as a consequence of the induction of much larger electric fields within the ergosphere and the forced co-rotation of magnetic field lines threading the equatorial current sheet around the black hole. This presents the exciting possibility of performing a fully self-consistent simulation of black hole jet-launching using the multifluid formalism, in a manner that has previously only been possible using fully kinetic methods such as GRPIC (e.g. \cite{parfrey_first-principles_2019}), since GRMHD has been unable to model the Blandford-Znajek mechanism of jet formation self-consistently due to insufficiently strong electric fields generated within the ergosphere (e.g. \cite{komissarov_magnetic_2007}). Since our multifluid simulations can be performed at higher resolution than full continuum relativistic kinetics, and with potentially greater accuracy (due to the absence of counting noise) than GRPIC, such an investigation could potentially place the first reasonable bounds on quantities such as ${\sigma}$ (ion magnetization) within realistic black hole jets, which current GRMHD-based methods have not been able to obtain. As we have also demonstrated the ability of the multifluid model to capture magnetic reconnection effects within black hole spacetimes, another crucial topic for further investigation is the quantification of the reconnection \textit{rate}, which is known to be underpredicted significantly within GRMHD models (e.g. \cite{bransgrove_magnetic_2021}). It seems likely, in light of the findings presented here, that global general relativistic multifluid simulations will yield a much closer approximation of the true magnetic reconnection rate than GRMHD, which motivates the further possibility of using multifluid simulations to investigate the relative significance of the Blandford-Znajek and magnetic reconnection mechanisms for black hole jet formation (e.g. \cite{de_gouveia_dal_pino_role_2010}), since analysis of the latter effect has hitherto been accessible using GRPIC-based methods only. The demonstrated ability of the multifluid approach to model the highly magnetized environments around neutron stars robustly also presents an exciting new potential frontier in the investigation of underlying mechanisms for neutron star jet-launching, which remain more poorly understood than the mechanisms underlying black hole jets (e.g. \cite{das_three-dimensional_2024}), and discrepancies between the driving mechanisms for the two types of compact object jets have not yet been systematically explored. 

Finally, the self-consistent evolution of electrons in this formalism presents unique opportunities for more realistic synthetic observations from the self-consistent electron temperature evolution. All of the dynamical consequences of this improved physics fidelity, such as the  self-consistent parallel electric fields and a more accurate model of magnetic reconnection, further improve the accuracy of the electron temperature evolution, and we will not have to rely on subgrid prescriptions of electron heating for mocking up the radiation from electrons when utilizing this model. This model thus presents a unique opportunity to connect with observations of some of the most extreme objects in the universe, and to give new insights into these exotic laboratories at the intersection complex plasma dynamics and general relativity.

\section*{Acknowledgements}

The authors thank Sasha Philippov for important feedback and suggestions on the manuscript, and in particular for suggesting the 2D axisymmetric test presented in Section \ref{sec:2d_axisymmetric}. The authors also thank Nikola Bukowiecka and Kyle Parfrey for useful discussions and suggestions at various points during the development of these ideas. Computations were performed using the Stellar and Della clusters at Princeton University. J.G. was partially funded by the Princeton University Research Computing group. J.G., J.J., and A.H. were partially funded by the U.S. Department of Energy under Contract No. DE-AC02-09CH1146 via an LDRD grant. The development of \textsc{Gkeyll} was partially funded, besides the grants mentioned above, by the NSF-CSSI program, Award Number 2209471.

\section*{Data Availability}

The requisite source code to reproduce all simulation results presented within this paper can be found at \url{https://github.com/ammarhakim/gkeyll}, with corresponding input files at \url{https://github.com/ammarhakim/gkyl-paper-inp}.

\bibliographystyle{mnras}
\bibliography{MultiFluidPaper}

\appendix

\section{Multifluid Source Terms in Primitive Variable Form}
\label{sec:primitive_variable_appendix}

Although the geometric source terms coupling each fluid species to the underlying spacetime geometry, as well as the electromagnetic source terms coupling the momentum density of each fluid to the current density of the shared electric field, are currently integrated explicitly in terms of the conserved variables ${S_{i, s}}$ (three-momentum density) and ${\tau_s}$ (relativistic energy density), in future work we intend to investigate the possibility of constructing either fully implicit or hybrid implicit-explicit (IMEX) integrators to circumvent the significant time-step restrictions currently incurred by our use of explicit methods. For this purpose, it will useful instead to integrate the source terms in terms of the primitive variables ${v_{s}^{i}}$ (three-velocity) and ${p_s}$ (pressure), namely:

\begin{multline}
\frac{d v_{s}^{i}}{d t} = \frac{1}{\rho_s h_s W_{s}^{2}} \left[ \frac{q_s}{m_s} \left( \rho_s W_s E^i + \rho_s W_s \left[ \varepsilon_{j k}^{i} v_{s}^{j} B^k \right] \right) \right.\\
+ T_{\text{Fluid}, s}^{t t} \left( \frac{1}{2} \beta^j \beta^k \gamma^{i l} \partial_l \gamma_{j k} - \alpha \gamma^{i l} \partial_l \alpha \right) + T_{\text{Fluid, s}}^{t k} \beta^k \gamma^{i l} \partial_l \gamma_{k j}\\
- \left( \frac{T_{\mu \nu, \text{Fluid}, s} n^{\mu} \perp_{j}^{\nu}}{\alpha} \right) \gamma^{i l} \partial_l \beta^j - \frac{q_s v_{s}^{i}}{m_s} \left( \rho_s W_s v_{s}^{j} E_j \right)\\
- T_{\text{Fluid, s}}^{t t} \left( \beta^j \beta^k K_{j k} + \beta^j \partial_j \alpha \right) v_{s}^{i}\\
\left. - T_{\text{Fluid}, s}^{t j} v_{s}^{i} \left( - \partial_j \alpha + 2 \beta^k K_{j k} \right) - v_{s}^{i} T_{\text{Fluid}, s}^{j k} K_{j k} \right],
\end{multline}
and:

\begin{multline}
\frac{d p_s}{d t} = \frac{\Gamma_s - 1}{W_s} \left[ \frac{q_s}{m_s} \left( \rho_s W_s v_{s}^{j} E_j \right) \right.\\
+ T_{\text{Fluid}, s}^{t t} \left( \beta^i \beta^j K_{i j} + \beta^i \partial_i \alpha \right) + T_{\text{Fluid}, s}^{t i} \left( - \partial_i \alpha + 2 \beta^j K_{i j} \right)\\
T_{\text{Fluid}, s}^{i j} K_{i j} - \frac{q_s v_{s}^{i}}{m_s} \left( \rho_s W_s E_i + \rho_s W_s \left[ \varepsilon_{i j k} v_{s}^{j} B^k \right] \right)\\
- T_{\text{Fluid}, s}^{t t} v_{s}^{i} \left( \frac{1}{2} \beta^j \beta^k \partial_i \gamma_{j k} - \alpha \partial_i \alpha \right) - T_{\text{Fluid}, s}^{t k} v_{s}^{i} \beta^j \partial_i \gamma_{k j}\\
\left. + v_{s}^{i} \left( \frac{T_{\mu \nu, \text{Fluid}, s} n^{\mu} \perp_{j}^{\nu}}{\alpha} \right) \partial_i \beta^j \right],
\end{multline}
respectively, where in the latter case we have assumed an ideal gas equation of state in which:

\begin{equation}
p_s = \left( \Gamma_s - 1 \right) \rho_s \varepsilon_s,
\end{equation}
and therefore that:

\begin{equation}
\frac{d p_s}{d t} = \left( \Gamma_s - 1 \right) \rho_s \frac{d \varepsilon_s}{d t},
\end{equation}
assuming that the rest mass density ${\rho_s}$ remains constant throughout the source term update. Unlike the conserved variable form, there is no way of integrating the multifluid source terms in primitive variable form that remains correct for an arbitrary equation of state.

\section{Multifluid Code Units}
\label{sec:code_units_appendix}

In various places throughout this paper, we fix a value of the (ion) magnetization ${\sigma}$, and use this value to determine the strength of the magnetic field ${\mathbf{B}}$, using:

\begin{equation}
\sigma = \frac{\left\lVert \mathbf{B} \right\rVert^2}{4 \pi \rho}, \qquad \implies \qquad \left\lVert \mathbf{B} \right\rVert = \sqrt{4 \pi \sigma \rho},
\end{equation}
where ${\rho = n_0 m}$ is the (ion) rest mass density, defined in terms of the (ion) mass $m$ and the reference (ion) number density ${n_0}$. Likewise, once the magnetic field strength is known, we can fix a value of the (ion) Larmor radius ${r_L}$ (in curved spacetime simulations, this is fixed relative to the Schwarzschild radius ${r_g = 2 M}$), and use this value to determine the (ion) charge $q$, using:

\begin{equation}
r_L = \frac{m v_{th}}{\left\lvert q \right\rvert \left\lVert \mathbf{B} \right\rVert} = \frac{\sqrt{2 T m}}{\left\lvert q \right\rvert \sqrt{4 \pi \sigma \rho}}, \qquad \implies \qquad \left\lvert q \right\rvert = \frac{\sqrt{2 T m}}{r_L \sqrt{4 \pi \sigma \rho}},
\end{equation}
where the temperature $T$ for an ideal gas is given by:

\begin{equation}
T = m \left( \Gamma - 1 \right) \frac{p}{\rho},
\end{equation}
in units where we have set the Boltzmann constant to unity, ${k_B = 1}$. Note that we select the (ion) mass $m$ such that the ratio ${\frac{q}{m} = 1}$, analogous to what is done in \cite{levinson_particle--cell_2018}, \cite{galishnikova_collisionless_2023}, etc. in the context of GRPIC methods.

We note that there is an implicit factor of ${\left( t_n \omega_p \right)^2}$ in front of the current density term on the right-hand side of equation \ref{eq:current_density}, where ${t_n}$ is the characteristic timescale of the simulation and ${\omega_p}$ is the (ion) plasma frequency:

\begin{equation}
\omega_p = \sqrt{\frac{4 \pi n q^2}{m}}.
\end{equation}
In the flat spacetime case, we can simply set the characteristic timescale ${t_n = \frac{1}{\omega_p}}$ such that ${\left( t_n \omega_p \right)^2 = 1}$. In the case of curved spacetime simulations, however, the characteristic timescale ${t_n}$ is fixed to be the crossing time ${t_g}$ of the compact object (either the black hole or the neutron star), and so we must enforce:

\begin{equation}
\left( t_g \omega_p \right)^2 = \left( t_g \sqrt{\frac{4 \pi n q^2}{m}} \right)^2 = 1,
\end{equation}
in some other way. In this case, we choose to fix the ratio ${\frac{q}{m} = 1}$ for the ions, thus simplifying this relation to:

\begin{equation}
\left( t_g \omega_p \right)^2 = \left( t_g \sqrt{4 \pi n m} \right)^2 = \left( t_g \sqrt{4 \pi \rho} \right)^2 = 1,
\end{equation}
where ${\rho = n m}$ is the (ion) mass density. This is why we fix the reference ion mass density to be ${\rho = \frac{1}{4 \pi}}$ within these simulations, in order to ensure that ${\left( t_g \omega_p \right)^2}$ remains 1.

\section{Neutron Star Metric Functions}
\label{sec:neutron_star_appendix}

The approximate exterior neutron star metric of \cite{pappas_accurate_2017}, for a neutron star of mass $M$, angular momentum $J$, mass quadrupole ${M_2}$, spin octupole ${S_3}$, and mass hexadecapole ${M_4}$, is defined by an axisymmetric Weyl-Lewis-Papapetrou line element in the cylindrical coordinate system ${\left( t, \rho, \varphi, z \right)}$:

\begin{multline}
d s^2 = g_{\mu \nu} d x^{\mu} d x^{\nu}\\
= - f \left( dt - \omega d \varphi \right)^2 - f^{-1} \left[ e^{2 \gamma} \left( d \rho^2 + d z^2 \right) + \rho^2 d \varphi^2 \right],
\end{multline}
with scalar metric functions ${f \left( \rho, z \right)}$, ${\omega \left( \rho, z \right)}$, and ${\gamma \left( \rho, z \right)}$ given by:

\begin{multline}
f \left( \rho, z \right) = 1 - \frac{2M}{\sqrt{\rho^2 + z^2}} + \frac{2 M^2}{\rho^2 + z^2}\\
+ \frac{\left( M_2 - M^3 \right) \rho^2 - 2 \left( M^3 + M_2 \right) z^2}{\left( \rho^2 + z^2 \right)^{\frac{5}{2}}}\\
+ \frac{2 z^2 \left( - J^2 + M^4 + 2 M_2 M \right) - 2 M M_2 \rho^2}{\left( \rho^2 + z^2 \right)^3} + \frac{A \left( \rho, z \right)}{28 \left( \rho^2 + z^2 \right)^{\frac{9}{2}}}\\
+ \frac{B \left( \rho, z \right)}{14 \left( \rho^2 + z^2 \right)^5},
\end{multline}
\begin{multline}
\omega \left( \rho, z \right) = - \frac{2 J \rho^2}{\left( \rho^2 + z^2 \right)^{\frac{3}{2}}} - \frac{2 J M \rho^2}{\left( \rho^2 + z^2 \right)^2} + \frac{F \left( \rho, z \right)}{\left( \rho^2 + z^2 \right)^{\frac{7}{2}}}\\
+ \frac{H \left( \rho, z \right)}{2 \left( \rho^2 + z^2 \right)^4} + \frac{G \left( \rho, z \right)}{4 \left( \rho^2 + z^2 \right)^{\frac{11}{2}}},
\end{multline}
and:

\begin{multline}
\gamma \left( \rho, z \right) = \frac{\rho^2 \left( J^2 \left( \rho^2 - 8 z^2 \right) + M \left( M^3 + 3 M_2 \right) \left( \rho^2 - 4 z^2 \right) \right)}{4 \left( \rho^2 + z^2 \right)^4}\\
- \frac{M^2 \rho^2}{2 \left( \rho^2 + z^2 \right)^2},
\end{multline}
respectively, where we have introduced the auxiliary scalar functions:

\begin{multline}
A \left( \rho, z \right) = 8 \rho^2 z^2 \left( 24 J^2 M + 17 M^2 M_2 + 21 M_4 \right)\\
+ \rho^4 \left( -10 J^2 M + 7 M^5 + 32 M_2 M^2 - 21 M_4 \right)\\
+ 8 z^4 \left( 20 J^2 M - 7 M^5 - 22 M_2 M^2 - 7 M_4 \right),
\end{multline}
\begin{multline}
B \left( \rho, z \right) = \rho^4 \left( 10 J^2 M^2 + 10 M_2 M^3 + 21 M_4 M + 7 M_{2}^{2} \right)\\
+ 4 z^4 \left( -40 J^2 M^2 - 14 J S_3 + 7 M^6 + 30 M_2 M^3 + 14 M_4 M + 7 M_{2}^{2} \right)\\
- 4 \rho^2 z^2 \left( 24 J^2 M^2 - 21 J S_3 + 7 M^6 + 48 M_2 M^3 + 42 M_4 M + 7 M_{2}^{2} \right)
\end{multline}
\begin{multline}
H \left( \rho, z \right) = 4 \rho^2 z^2 \left( J \left( M_2 - 2 M^3 \right) - 3 M S_3 \right)\\
+ \rho^4 \left( J M_2 + 3 M S_3 \right),
\end{multline}
\begin{multline}
G \left( \rho, z \right) = \rho^2 \left( J^3 \left( - \left( \rho^4 + 8 z^4 - 12 \rho^2 z^2 \right) \right) \right.\\
\left. + J M \left( \left( M^3 + 2 M_2 \right) \rho^4 - 8 \left( 3 M^3 + 2 M_2 \right) z^4 \right. \right.\\
\left. \left. + 4 \left( M^3 + 10 M_2 \right) \rho^2 z^2 \right) + M^2 S_3 \left( 3 \rho^4 - 40 z^4 + 12 \rho^2 z^2 \right) \right),
\end{multline}
and:

\begin{equation}
F \left( \rho, z \right) = \rho^4 \left( S_3 - J M^2 \right) - 4 \rho^2 z^2 \left( J M^2 + S_3 \right).
\end{equation}
\end{document}